\documentclass[aps,pra,twocolumn,superscriptaddress,10pt,article,nofootinbib,showpacs]{revtex4-1}
\usepackage{amsfonts}
\usepackage{amsmath}
\usepackage{amsthm}
\usepackage{braket}
\usepackage{graphicx}
\usepackage{hyperref}
\usepackage{color}
\usepackage{amssymb}
\usepackage{dsfont}
\usepackage{verbatim}
\usepackage{amsmath,amscd}

\newcommand{\peps}{\ensuremath{A}}
\newcommand{\hilbert}{\ensuremath{\mathbb{H}}}
\newcommand{\mathscr}[1]{\ensuremath{\mathcal{#1}}}

\newcommand{\fullgraph}{\ensuremath{\Lambda}}
\newcommand{\subgraph}{\ensuremath{\Gamma}}
\newcommand{\gauge}{\ensuremath{G}}
\newcommand{\gaugeoperator}{\ensuremath{\mathscr{G}}}
\newcommand{\spt}[1]{\ensuremath{\ket{\text{SPT}(#1)}}}
\newcommand{\bbra}[1]{\ensuremath{(#1|}}
\newcommand{\bket}[1]{\ensuremath{|#1)}}
\newcommand{\region}{\mathcal{R}}
\newcommand{\man}{\mathcal{M}}
\newcommand{\tr}[1]{\ensuremath{\text{Tr}[#1]}}
\newcommand{\st}{\ensuremath{a}}
\newcommand{\epath}{\ensuremath{p}}
\newcommand{\plaq}{\ensuremath{q}}
\newcommand{\isom}[1]{\ensuremath{\mathcal{W}^\peps_\region[#1]}}
\newcommand{\con}{\ensuremath{\phi}}
\newcommand{\hilb}{\ensuremath{\mathbb{H}}}
\newcommand{\conj}[1]{\ensuremath{\mathfrak{C}(#1)}}
\newcommand{\cc}{\ensuremath{c}}
\newcommand{\overlap}[2]{\ensuremath{M_{[#1],[#2]}}}
\newcommand{\puncture}{\ensuremath{{\Pi}}}
\newcommand{\bdry}{\ensuremath{\mathscr{V}}}

\newcommand{\mpod}{\ensuremath{\chi}}
\newcommand{\slant}[4]{\ensuremath{{#1}^{(#3)}(#4)}}
\newcommand{\slan}[3]{\ensuremath{{#1}^{(#3)}}}
\newcommand{\QW}[3]{\ensuremath{W_{#1}^{#2}(#3)}}
\newcommand{\QV}[3]{\ensuremath{Y_{#1}^{#2}(#3)}}

\newtheorem{claim}{Proposition}
\newtheorem{conjecture}{Conjecture}
\newtheorem{definition}{Definition}

\begin{document}

\title{Matrix product operators for symmetry-protected topological phases: 
Gauging and edge theories}

\author{Dominic J. \surname{Williamson}}
\affiliation{Vienna Center for Quantum Technology, University of Vienna, Boltzmanngasse
5, 1090 Vienna, Austria}
\author{Nick \surname{Bultinck}}
\affiliation{Department of Physics and Astronomy, Ghent University, Krijgslaan 281
S9, B-9000 Ghent, Belgium}
\author{Michael \surname{Mari\"en}}
\affiliation{Department of Physics and Astronomy, Ghent University, Krijgslaan 281
S9, B-9000 Ghent, Belgium}
\author{Mehmet B. \surname{\c{S}ahino\u{g}lu}}
\affiliation{Vienna Center for Quantum Technology, University of Vienna, Boltzmanngasse
5, 1090 Vienna, Austria}
\author{Jutho \surname{Haegeman}}
\affiliation{Department of Physics and Astronomy, Ghent University, Krijgslaan 281
S9, B-9000 Ghent, Belgium}
\author{Frank \surname{Verstraete}}
\affiliation{Vienna Center for Quantum Technology, University of Vienna, Boltzmanngasse
5, 1090 Vienna, Austria}
\affiliation{Department of Physics and Astronomy, Ghent University, Krijgslaan 281
S9, B-9000 Ghent, Belgium}

\begin{abstract}
Projected entangled pair states (PEPS) provide a natural ansatz for the ground states of gapped, local Hamiltonians in which global characteristics of a quantum state are encoded in properties of local tensors. We develop a framework to describe on-site symmetries, as occurring in systems exhibiting symmetry-protected topological (SPT) quantum order, in terms of virtual symmetries of the local tensors expressed as a set of matrix product operators (MPOs) labeled by distinct group elements.
These MPOs describe the possibly anomalous symmetry of the edge theory, whose local degrees of freedom are concretely identified in a PEPS. 
A classification of SPT phases is obtained by studying the obstructions to continuously deforming one set of MPOs into another, recovering the results derived for fixed-point models [X. Chen \textit{et al.}, Phys. Rev. B 87, 155114 (2013)]~\cite{GuWen}. Our formalism accommodates perturbations away from fixed point models, opening the possibility of studying phase transitions between different SPT phases.  
We also demonstrate that applying the recently developed quantum state gauging procedure to a SPT PEPS yields a PEPS with topological order determined by the initial symmetry MPOs.
The MPO framework thus unifies the different approaches to classifying SPT phases, via fixed-points models, boundary anomalies, or gauging the symmetry, into the single problem of classifying inequivalent sets of matrix product operator symmetries that are defined purely in terms of a PEPS.
\end{abstract}

\maketitle

\section{ Introduction}
\label{intro}
The phase diagrams of quantum many-body systems become much richer when global {symmetries} are imposed. It has become clear of late that in the presence of a global symmetry there exist distinct phases which cannot be distinguished via local order parameters. These phases are referred to as \emph{symmetry-protected topological} (SPT) phases~\cite{GuWen}. In contrast to topologically ordered systems~\cite{Wen90}, all SPT phases become trivial if the symmetry is allowed to be explicitly broken. While this implies that SPT ground states possess only short-range entanglement{,} they cannot be adiabatically connected to a product state without breaking the symmetry. Furthermore they exhibit interesting edge properties when defined on a finite system with nontrivial boundary. 

In recent years there has been a growing interplay between the theory of quantum many-body systems and quantum information. This has led to the development of tensor network ansatz for the ground states of local, gapped Hamiltonians~\cite{Fannes92, VerstraeteCirac06, VerstraeteMurgCirac08, GarciaVerstraeteWolfCirac08,bridgeman2016hand}. Tensor network methods have proven particularly useful in understanding the emergence of topological phenomena in quantum many-body ground states. In one dimension, Matrix Product States were used to completely classify SPT phases via the second cohomology group of their symmetry group~\cite{SchuchGarciaCirac11, 1Done, 1Dtwo}. In two dimensions, Projected-Entangled Pair States (PEPS) have been used to characterize systems with intrinsic topological order~\cite{Ginjectivity,transfermatrix,Buerschaper14,MPOpaper,chiral4}, symmetry-protected topological order~\cite{Chen} and chiral topological insulators~\cite{chiral1,chiral2,chiral3}.

{The first goal of this work is to present a general framework for the description of on-site symmetries within the PEPS formalism. The framework includes symmetry actions on states with topological order and thus provides a natural setting for the study of symmetry-enriched topological phases~\cite{turaev2000homotopy,kirillov2004g,drinfeld2010braided,etingof2009fusion,bombin2010topological,hung2013quantized,mesaros2013classification,barkeshli2014symmetry,tarantino2015symmetry,teo2015theory} with PEPS~\cite{michael}.
We then restrict to PEPS without topological order and provide a complete characterization of bosonic SPT order by formulating sufficient conditions to be satisfied by the individual PEPS tensors. Previously some powerful results for renormalization group (RG) fixed-point states with SPT order were presented by Chen \textit{et al.}~\cite{Chen,GuWen}}, the present work extends these results to systems with a finite correlation length. 
Furthermore, 
application of the quantum state gauging procedure of Ref.\cite{Gaugingpaper} within the framework presented here
illuminates the correspondence between SPT phases and certain topologically ordered phases {in the language of PEPS, providing a complementary description to the Hamiltonian gauging construction of Levin and Gu~\cite{LevinGu}}. This naturally brings together the classification of SPT phases via fixed-point models, gauging and anomalous boundary symmetries into a single unified approach that focuses only on MPOs which are properties of the ground states alone. 

To achieve these goals we have developed tools to deal with orientation dependent MPO tensors. These tools allow us to calculate the symmetry action on monodromy defected and symmetry twisted states and also modular transformations, pre- and post- gauging, in a local way that is governed by a single tensor.

{We first outline the general formalism for characterizing gapped phases in PEPS using matrix product operators (MPOs) in Section~\ref{formalism}. 
Section~\ref{globalsymmetry} presents a set of local conditions that lead to a large class of PEPS with global symmetries which fit within the general formalism.  Next, in Section~\ref{sptpeps}, we identify a class of short-range entangled PEPS and discuss how SPT order manifests itself in these models via their anomalous edge physics. Section~\ref{gaugingsptpeps} explains how gauging a SPT PEPS with a discrete symmetry group yields a long-range entangled PEPS with topological order. In Section~\ref{six} we study symmetry twists and monodromy defects of SPT PEPS. These concepts are then illustrated with a family of examples that fall within the framework of SPT PEPS in Section~\ref{exfpspt}. We show explicitly that gauging these states yields ground states of the twisted quantum double models~\cite{pasquier,tqd}, which are the Hamiltonian formulations of Dijkgraaf-Witten discrete gauge theories~\cite{DijkgraafWitten, Dewilde}.}

The appendices are organized into sections that review relevant background and others that provide technical details of results which are used throughout the paper. We first review the relevant properties of MPO-injective PEPS in Appendix~\ref{a}, provide an argument that a MPO-injective PEPS with a single block projection MPO is the unique ground state of its parent Hamiltonian in Appendix~\ref{b} and review the definition of the third cohomology of a single block MPO group representation in Appendix~\ref{c}.
In Appendix~\ref{newapp1} we present results concerning possible orientation dependencies of MPO group representations. In Appendix~\ref{newapp2} we discuss different crossing tensors, their composition and the effect of modular transformations. 
Appendix~\ref{d} contains a brief review of the quantum state gauging formalism and a proof that a gauged SPT PEPS is MPO-injective~\cite{MPOpaper}.
In Appendix~\ref{e} we present an extension of the quantum state gauging procedure of Ref.\cite{Gaugingpaper} to arbitrary flat $\mathsf{G}$-connections and use it to prove that the gauging procedure is gap preserving for arbitrary topologies and to furthermore construct the full topological ground space of a gauged SPT model. 
In Appendix~\ref{g} we develop a description of symmetry twisted states, topological ground states and monodromy defected states in terms of MPOs and calculate their transformation under the residual symmetry group.
Finally in Appendix~\ref{gaugingham} we demonstrate that the quantum state gauging procedure for finite groups is equivalent to the standard minimal coupling scheme for gauging Hamiltonians. 

\section{ Characterizing topological phases with matrix product operators}\label{formalism}
In this section we present a general framework for the classification of gapped phases with PEPS in terms of universal and discrete labels that arise directly from tensor network states. These discrete labels emerge from the set of MPO symmetries of the PEPS tensors and should remain invariant under continuous deformation of the MPOs.

\begin{figure}[ht]
\center
\begin{align*}
a) \vcenter{\hbox{
 \includegraphics[height=0.25\linewidth]{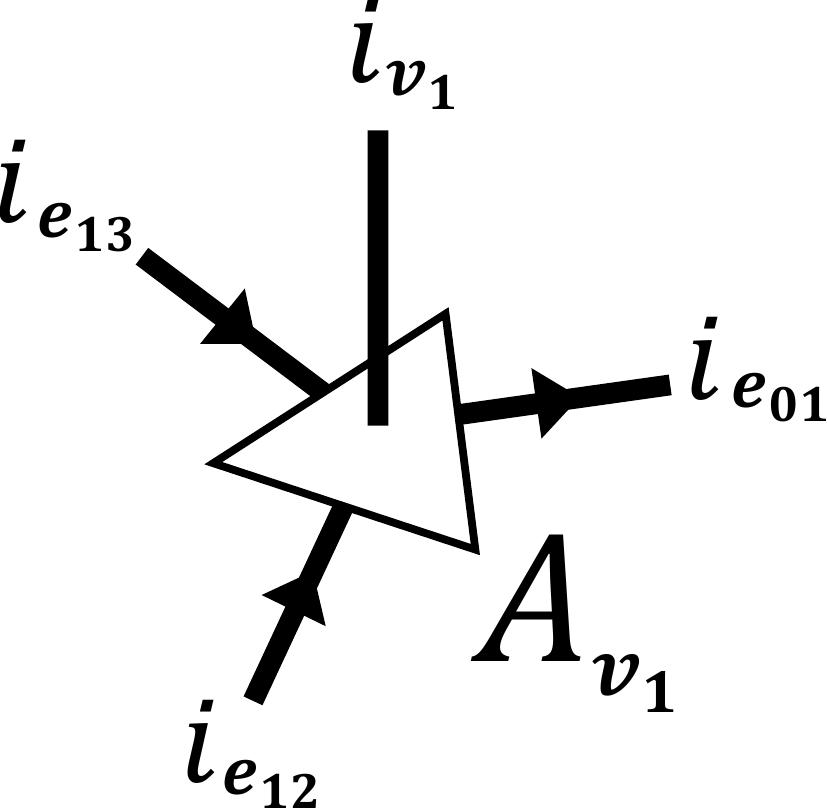}}} \hspace{1cm} b)
\vcenter{\hbox{
\includegraphics[height=0.25\linewidth]{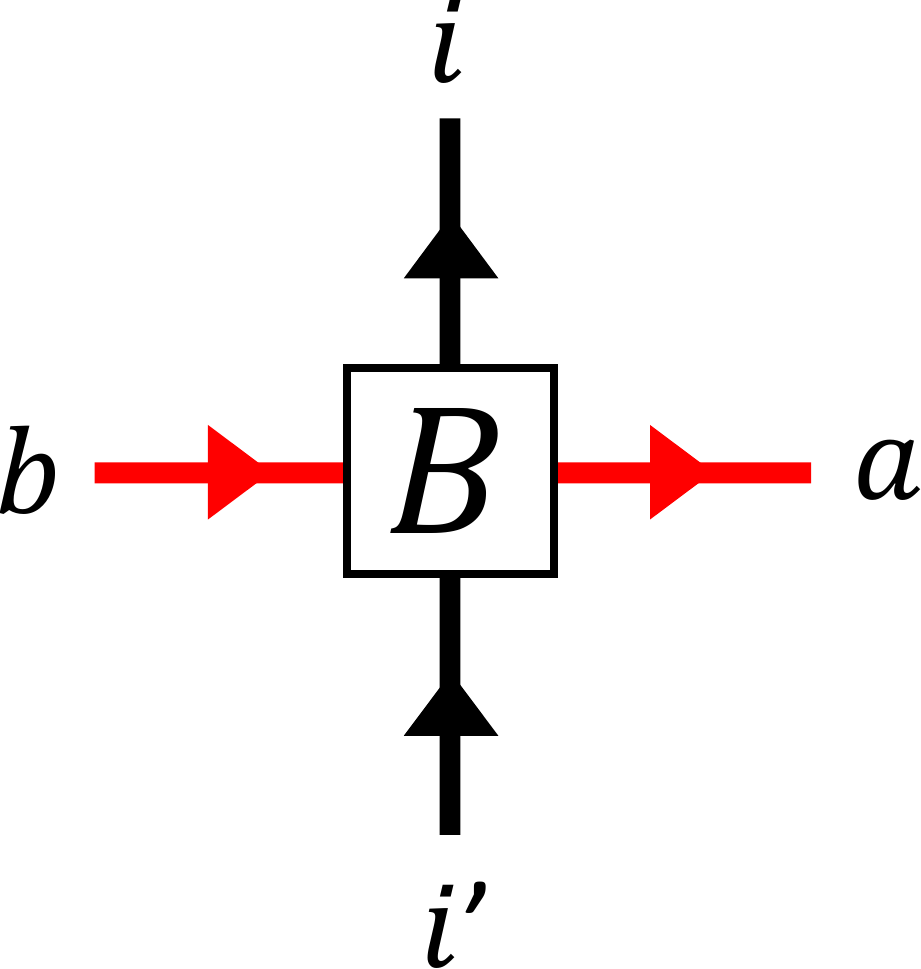}}} 
\end{align*}
\caption{a) A PEPS tensor on a trivalent vertex. b) A right handed MPO tensor.}\label{e1}
\end{figure}

A 2D PEPS can be defined on any directed graph $\fullgraph$  (most commonly a regular lattice) embedded in an oriented 2D manifold $\man$ given a tensor
$$\peps_v:=\sum_{i_v=1}^d\sum_{\{i_e\}=1}^D (\peps_v)_{{\{i_e\}}}^{i_v}\ket{i_v}\bigotimes_{e\in E_v}\bbra{i_e}$$
for every vertex $v\in \fullgraph$, where $E_v$ is the set of edges with $v$ as an endpoint, see Fig.\ref{e1}. Here $i_v$ is the physical index running over a basis for the Hilbert space of a single site $\mathbb{C}^d$ and each $i_e$ is a virtual index of dimension $D$ along an edge $e$ adjacent to $v$ in the graph $\fullgraph$. 

\begin{figure}[ht]
\center
\includegraphics[width=0.33\linewidth]{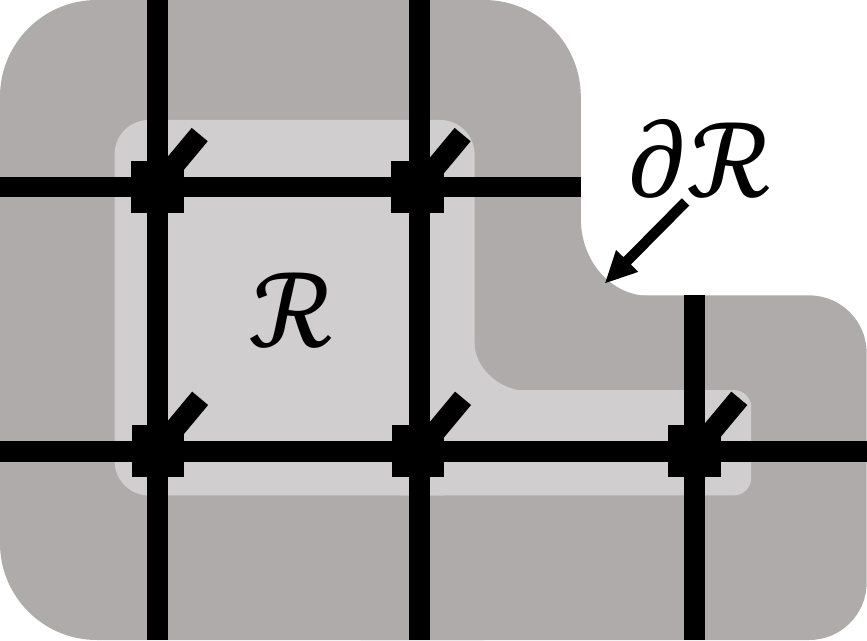} 
\caption{The PEPS map $\peps_\region$ from virtual indices on edges in $\partial \region$ to physical indices on vertices in $\region$.}\label{e2a}
\end{figure}

For any simply connected region $\region\subset\man$ whose boundary $\partial \region$ forms a contractible closed path in the dual graph $\fullgraph^*$ we define the PEPS map 
$$\peps_\region:(\mathbb{C}^D)^{\otimes |\partial \region|_e}\rightarrow (\mathbb{C}^d)^{\otimes |\region|_v},$$
from $|\partial \region|_e$ virtual indices on the edges that cross $\partial\region$ to $|\region|_v$ physical indices on the vertices in $\region$, by taking the set of tensors $\{ A_v \, |\, v\in\region\}$ and contracting each pair of indices that are assigned to an edge within $\region$, to yield
$$\peps_\region:=\sum_{\{i_v\}_{v\in \region}}\sum_{\{i_e\}_{e\in \overline{\region}}}\bigotimes_{v\in\region}(A_v)^{i_v}_{{\{i_e\}}_{e\in E_v}}\bigotimes_{v\in\region}\ket{i_v}\bigotimes_{e\in\partial\region}\bbra{i_e}$$
where $\overline{\region}:=\region\cup\partial\region$, see Fig.\ref{e2a}.

Universal properties of the phase of matter containing the PEPS wave function are manifest in the local symmetries of $\peps_\region$. The specific symmetries we consider are of the form $U^{\otimes|\region|_v}\peps_\region =\peps_\region O^{\partial \region}$, where $U$ is an on-site unitary corresponding to a physical symmetry that is respected in our classification of phases. Since physical symmetries necessarily form a group under multiplication, we henceforth use the notation $U(g),\ g\in\mathsf{G}$ (we do not consider non on-site symmetries such as lattice symmetries~\cite{jiang2015symmetric}).
$O^{\partial \region}$ is a MPO acting on the virtual space associated to the edges crossing ${\partial \region}$. In general,
$$O^{\partial \region}=\sum_{\{i_n\},\{i_n'\}=1}^D \tr{B_{\sigma_{i}}^{i_1,i_1'}\cdots B_{\sigma_N}^{i_{N},i_{N}'}} \ket{i_1\dots i_{N}}\bra{i_1'\dots i_{N}'}$$
where the edges crossing ${\partial \region}$ are ordered 1 to $N:=|{\partial \region}|_e$, by fixing an arbitrary base point and following the orientation of ${\partial \region}$ (specifically the orientation induced by $\man$). Each $(B_{\sigma_n}^{i,i'})_{a,b}$ is a $\chi\times\chi$ matrix, see Fig.\ref{e1}, which can depend on the handedness $\sigma_n=\pm$ of the crossing of $\partial\region$ and edge $n$ ($+$ for right, $-$ for left).

\begin{figure}[ht]
\center
\begin{align*}
\vcenter{\hbox{
\includegraphics[width=0.33\linewidth]{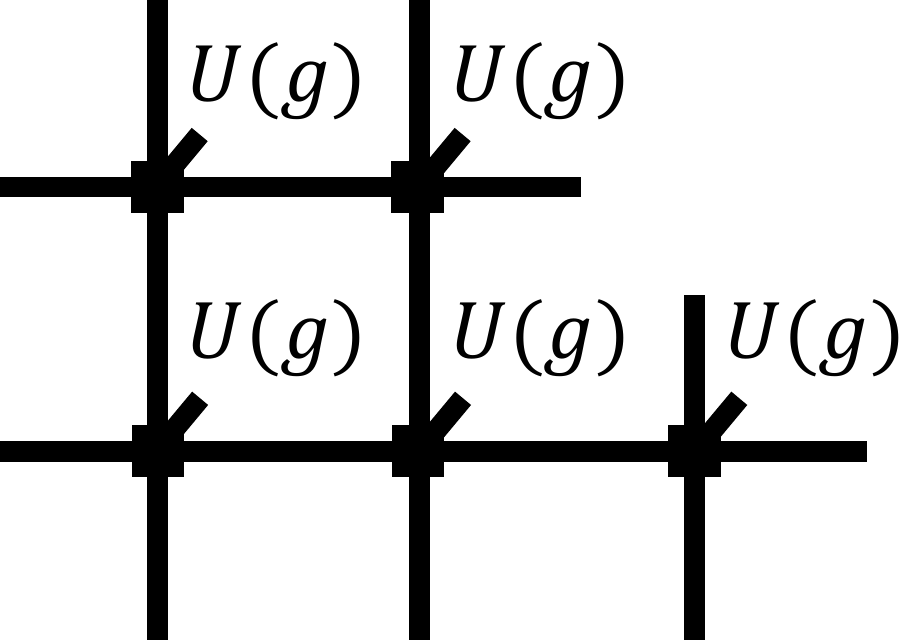}}}  \quad = \quad 
\vcenter{\hbox{
\includegraphics[width=0.33\linewidth]{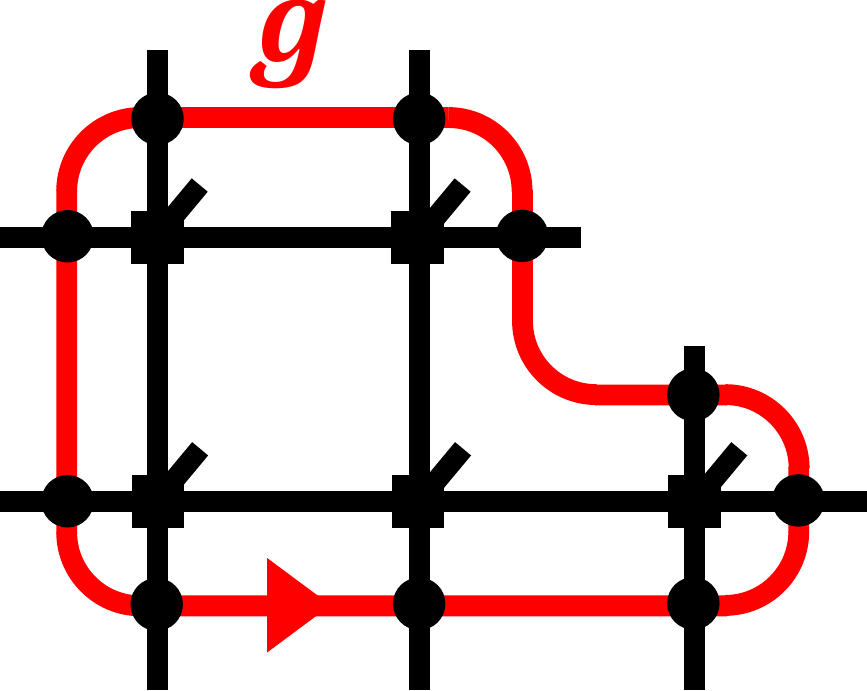}}} 
\end{align*}
\caption{The symmetry of the PEPS map $\peps_\region$ on a region $\region$ containing five sites.}\label{e2}
\end{figure}

Any truly topological symmetries should persist under arbitrary deformations of the region $\region$, hence the relevant task is to find a complete set $\mathcal{S}_g$ of linearly independent single block~\cite{MPSrepresentations} MPOs $O_{\alpha}^{\partial \region}(g) $ for every symmetry transformation $U(g)$ such that for every region $\region$ (satisfying the conditions outlined above) we have
\begin{equation}\label{singlelayer}
U(g)^{\otimes|\region|_v}\peps_\region=\peps_\region O^{\partial \region}_{\alpha}(g)
\end{equation}
see Fig.\ref{e2}.
There is an important subtlety in finding inequivalent MPOs that satisfy Eq.\eqref{singlelayer} since two linearly independent solutions $O^{\partial \region}_{1}(g),\, O^{\partial \region}_{2}(g)$ may coincide on the support of $\peps_\region$. This occurs precisely when they differ by an operator supported on the kernel of $\peps_\region$.
To remove this redundancy one must first find the set of all single block MPO symmetries $\mathcal{S}_1$ for $U(1)=\openone$. Assuming these MPOs are complete in the following sense $\sum_{\alpha}O_{\alpha}^{\partial\region}(1)=\peps_\region^+\peps_\region$, where $\peps_\region^+$ is a distinguished generalized inverse of $\peps_\region$, any MPO $\hat O^{\partial \region}$ can be projected onto the support of $\peps_\region$ to yield another MPO $\peps_\region^+\peps_\region \hat O^{\partial \region}$ with a (multiplicative) constant increase in the bond dimension.
 Hence the set of inequivalent single blocked MPO symmetries $\mathcal{S}_g:=\{O^{\partial \region}_{\alpha}(g)\}_\alpha$ can be found by taking all linearly independent MPOs satisfying Eq.\eqref{singlelayer}, projecting them onto the support subspace $\peps_\region^+\peps_\region$ and collecting the linearly independent single block MPOs that result. 

Eq.\eqref{singlelayer} implies that $\mathcal{S}:=\bigcup_g \mathcal{S}_g$ has a $\mathsf{G}$-graded algebra structure. This algebra structure and the number of elements in $\mathcal{S}$ must be independent of $\mathcal{R}$. Note the MPO matrices $B_{\sigma_e,\alpha}^{ij}(g)$ also do not depend on $\mathcal{R}$ hence for every region the MPO $O^{\partial \region}_{\alpha}(g)$ is constructed from the same local tensors. The symmetry relations of Eq.\eqref{singlelayer}, the graded algebra structure of $\mathcal{S}$ and any discrete labels of the MPO representation of this graded algebra provide universal labels of a quantum phase, independent of the details of the local tensors $\peps_v$. 
\begin{conjecture}[\cite{michael}]
A discrete set of labels that fully specify a symmetry-enriched topological phases of matter can be derived from $\mathcal{S}$, the MPO representation of $\mathsf{G}$, in a purely local fashion and these labels remain invariant under continuous physical perturbations.
\end{conjecture}
This set of labels can be calculated by following a similar approach to Ref.\cite{nick}, and they should describe the emergent symmetry defects and their $\mathsf{G}$-graded fusion and $\mathsf{G}$-crossed braiding properties. Note this data subsumes the underlying anyon theory and the possibly fractional symmetry transformation of the defects.

Intrinsic topological order is defined without reference to any symmetry and thus corresponds to the $\mathsf{G}=\{1\}$ case, in which PEPS are classified by $\mathcal{S}_1$. Injective PEPS~\cite{GarciaVerstraeteWolfCirac08} always posses trivial topological order and have $\mathcal{S}_1=\{\openone^{\otimes|\partial\region|}\}$ whereas all  known topological ordered PEPS~\cite{Ginjectivity, Buerschaper14, transfermatrix, MPOpaper, chiral4} satisfy Eq.\eqref{singlelayer} with a nontrivial $\mathcal{S}_1$. 
This was formalized in the framework of MPO-injectivity in Ref.\cite{MPOpaper}, which was shown to capture all Levin-Wen string-net models (the Hamiltonian version of Turaev-Viro state sum invariants~\cite{turaev1992state}). In Ref.\cite{MPOpaper} the independence of the MPO tensors from the region $\region$ was guaranteed by the intuitive \emph{pulling through} property and the more technical \emph{generalized} and \emph{extended inverse} properties, all of which were purely local conditions. 

By taking a global symmetry $\mathsf{G}$ into account, a finer classification is achieved in terms of $\mathcal{S}$ where $|\mathcal{S}_g|>0$ $\forall g\in G$. This classification contains symmetry-protected phases for $|\mathcal{S}_1|=1$ and symmetry-enriched topological phases for $|\mathcal{S}_1|>1$.
In the next section we demonstrate how solutions of Eq.\eqref{singlelayer} can be obtained for nontrivial elements $g\in\mathsf{G}$ in a similar fashion to Ref.\cite{MPOpaper}.

\section{ Global symmetry in PEPS}\label{globalsymmetry}
In this section we present a set of local conditions that lead to a general class of solutions to Eq.\eqref{singlelayer}.

Consider a PEPS on a trivalent directed graph $\fullgraph$ embedded in an oriented manifold $\man$, built from four index tensors $A$ which we interpret as linear maps from the virtual to physical indices $A:(\mathbb{C}^D)^{\otimes 3}\rightarrow\mathbb{C}^d$. Firstly, we require that the tensors $A$ satisfy the axioms of MPO-injectivity~\cite{MPOpaper}, a framework describing general gapped phases without symmetry. Thus (potentially after some blocking of lattice sites, which we assume has already been carried out) the projection $P:=A^+A$ onto the subspace within which the tensor $A$ is injective can be written as a matrix product operator
\begin{align}
\vcenter{\hbox{
\includegraphics[width=0.14\linewidth]{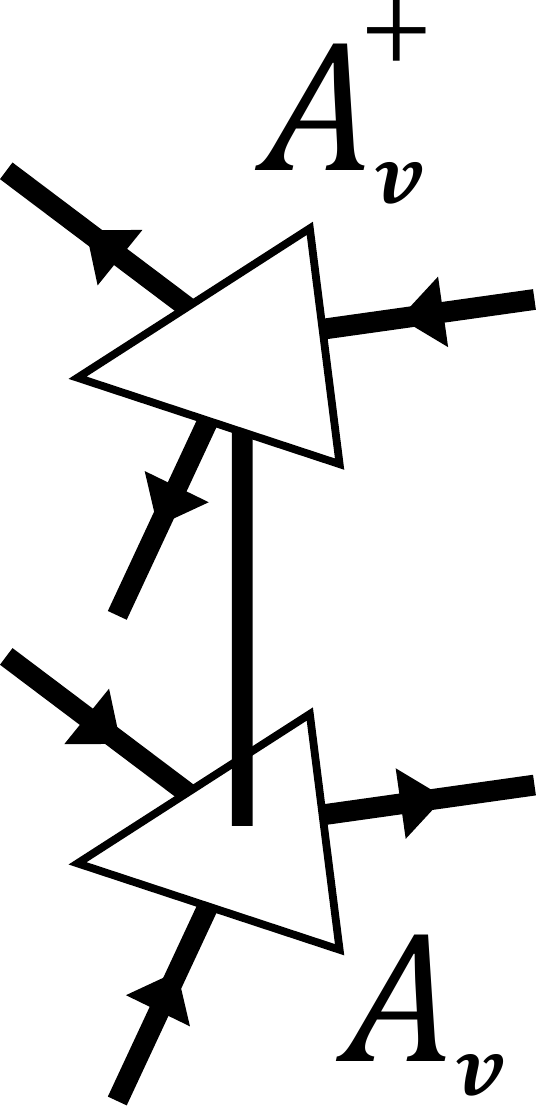}}} \label{n36}
 \ \ = 
\vcenter{\hbox{
\includegraphics[width=0.25\linewidth]{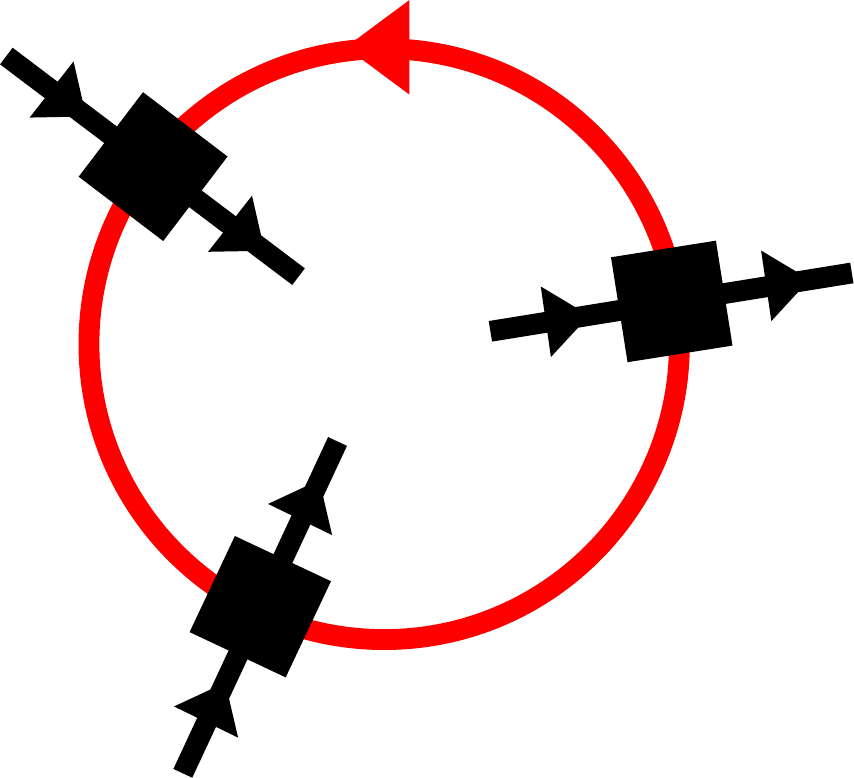}}}
\end{align}
here the MPO tensors are denoted as black squares and satisfy the axioms listed in~\cite{MPOpaper}, see Appendix~\ref{a} for a brief review. These axioms ensure that the same MPO is obtained for any larger region, independent of the order in which the generalized inverses are applied, and furthermore that this closed MPO is a projector independent of its length.

We now describe purely local sufficient conditions for a PEPS to be invariant under the on-site action $U(g)$ of a global symmetry group $\mathsf{G}$. Hereto, we introduce another set of closed MPOs $\{V^{\partial \region}(g)\, | \, g \in \mathsf{G}\}$ which inherit an orientation from $\partial \region$. These MPOs are composed of four index tensors that depend on a group element $g$. The tensors are depicted by filled circles in the following diagrams and are defined by conditions~\eqref{n23} and~\eqref{n25}
\begin{align}
\vcenter{\hbox{
\includegraphics[width=0.18\linewidth]{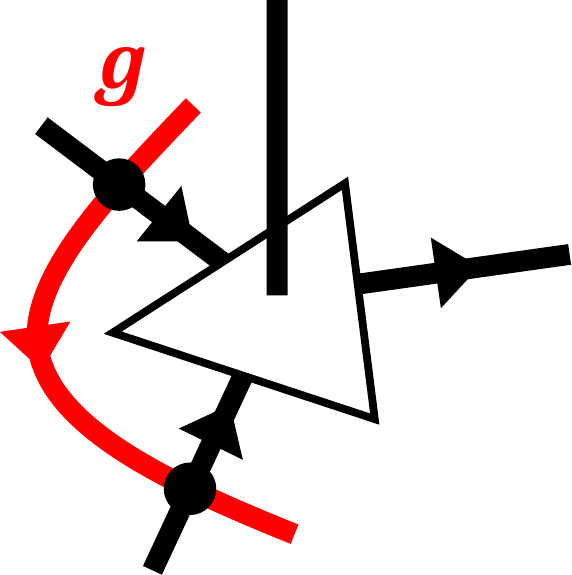}}} \label{n23}
\ = \vcenter{\hbox{\raisebox{0.25cm}{
\includegraphics[width=0.17\linewidth]{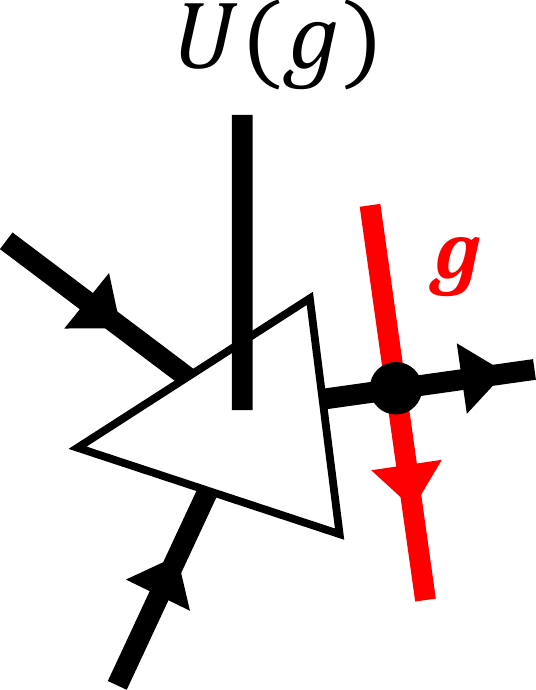}}}}
\end{align}
where $U(g)$ is a unitary representation of $\mathsf{G}$, and
\begin{align}
\vcenter{\hbox{
\includegraphics[width=0.18\linewidth]{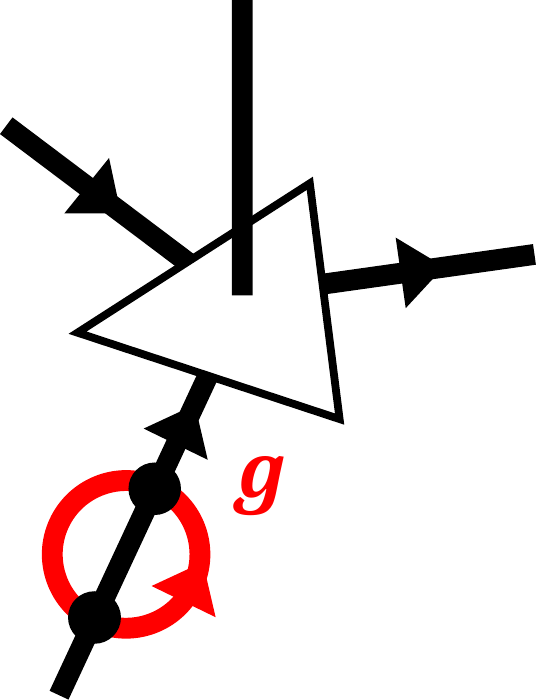}}} \label{n25}
\ = \vcenter{\hbox{
\includegraphics[width=0.18\linewidth]{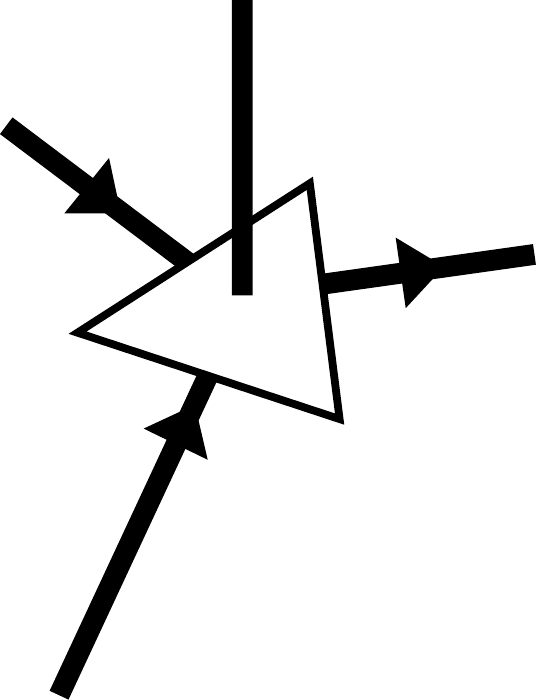}}}.
\end{align}
Note Eq.\eqref{n25} with the directions reversed is implied by the above conditions. The orientation of the MPO tensors is significant as pulling the MPO through a PEPS tensor in a right handed fashion, as in Eq.\eqref{n23}, induces an action $U(g)$ on the physical index while pulling through in a left handed fashion results in a physical action $U^\dagger(g)$, this follows directly from Eq.\eqref{n23} since $U$ is a unitary representation.

With these two properties, it is clear that the ground space of a MPO-injective PEPS constructed from the tensor $A$ on any closed system of arbitrary size is invariant under the global symmetry action $U(g)^{\otimes N}$. Hence such MPO-injective PEPS that are unique ground states must be eigenvectors of the global symmetry.
For the special case of injective PEPS~\cite{GarciaVerstraeteWolfCirac08} the MPO $P$ is simply the identity $P=\openone$ (i.e. a MPO with bond dimension $1$), the symmetry MPOs $V(g)$ can always be factorized into a tensor product of local gauge transformations~\cite{canonicalPEPS} and the ground state is unique.

From Eqs.\eqref{n23} and~\eqref{n25} it immediately follows that the PEPS tensors are intertwiners, i.e.~$U(g)A = AV(g)$, where $V(g)$ denotes a closed MPO acting on the three virtual indices. Without loss of generality, and in accordance with the general framework of Section~\ref{formalism}, we impose that the MPOs $V(g)$ act within the support space of $A$ such that $PV(g)=V(g)$, i.e.
\begin{align}
\vcenter{\hbox{
\includegraphics[width=0.29\linewidth]{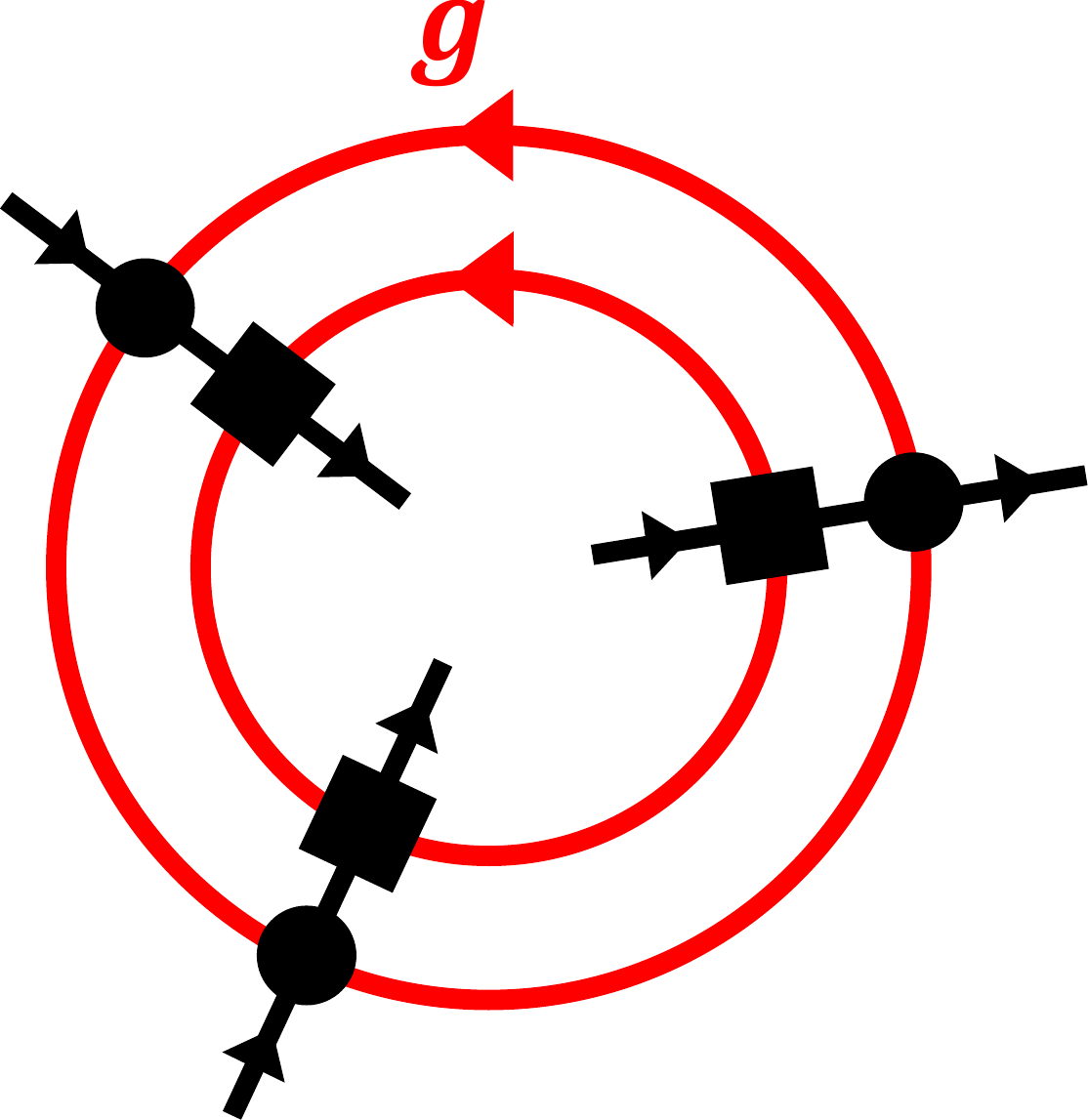}}} \label{n39}
\ = \vcenter{\hbox{
\includegraphics[width=0.25\linewidth]{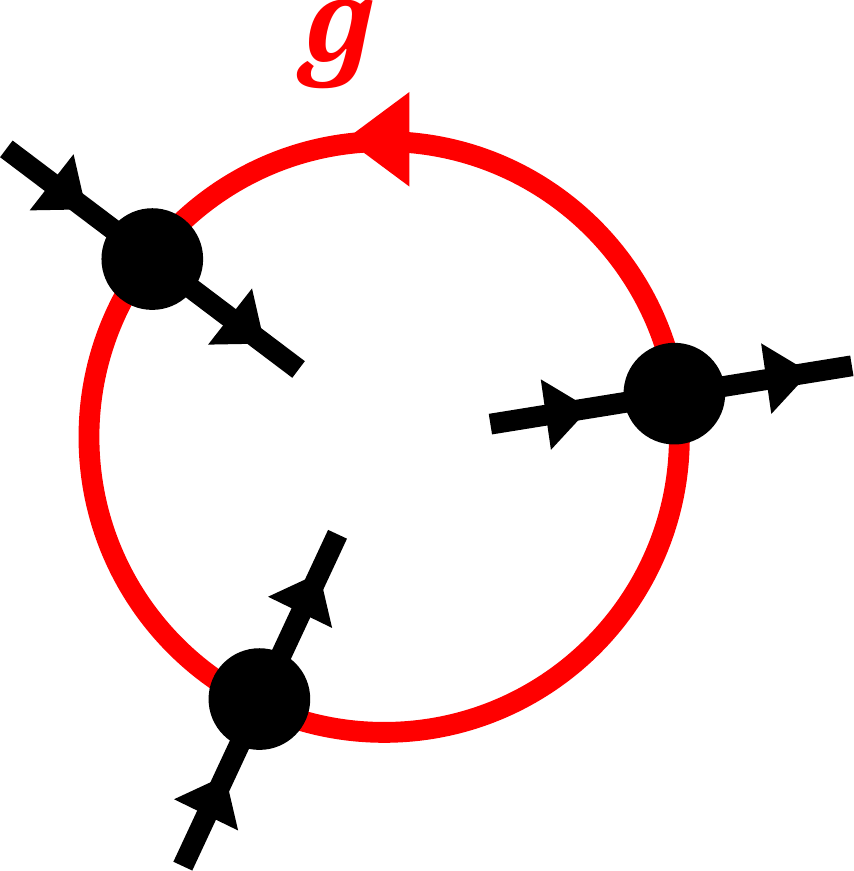}}}
\end{align}
and in particular $V(1)=P$, for $1$ the identity group element.  Hence the MPOs $V(g)$ form a representation of $\mathsf{G}$ since we have $AV(g_1g_2) = U(g_1g_2)A = U(g_1)U(g_2)A = AV(g_1)V(g_2)$, and thus $PV(g_1g_2) = PV(g_1)V(g_2)$ where $P:=A^+A$, see Eq.\eqref{n36}  (note $PV(g)P=\peps^+U(g)\peps P=PV(g)$). A similar argument shows that the symmetry MPO $V_\text{rev}(g)$ along the path with reversed orientation (inducing reversed arrows) equals $V(g^{-1})$ since $\peps V_{\text{rev}}(g)=U^\dagger(g)\peps=U(g^{-1})\peps=\peps V(g^{-1})$ which implies $ PV_{\text{rev}}(g)=PV(g^{-1})$. The above two arguments extend to arbitrary contractible regions $\region$ and boundary MPOs $P_{\partial \region},\,V^{\partial\region}(g)$.

If we do not project the boundary symmetries onto the support subspace of $A$ there are many equivalent choices for the symmetry action on the boundary. In particular, there might be choices for which the action is factorizable into a tensor product (see e.g.~Ref.\cite{Chen}), even if the support projector is not. However, the resulting boundary actions will generically not form a representation of the relevant symmetry group $\mathsf{G}$. The procedure we have outlined of projecting these actions onto the injectivity subspace provides an unambiguous recipe to identify the relevant set of boundary operators that form a MPO representation of the physical symmetry group $\mathsf{G}$. For the particular case of renormalization group fixed-point models, our recipe matches the results of Ref.\cite{Chen}, as illustrated in Section~\ref{exfpspt}.

With these properties it is clear that the class of symmetric PEPS satisfying Eqs.\eqref{n23} and \eqref{n25} constitute a special case of the general framework described in Section~\ref{formalism}. Let $V^{\partial\mathcal{R}}(g)$ denote the MPO corresponding to group element $g$ acting on the boundary of region $\mathcal{R}$ then we have
\begin{equation}
U(g)^{\otimes|\region|_v}\peps_\region=\peps_\region V^{\partial\mathcal{R}}(g)
\, .
\end{equation}
Note in the general case we may need to decompose $V^{\partial \mathcal{R}}(g)$ into a sum of single block MPOs to be consistent with Section~\ref{formalism}. 

This general class of solutions show that the formalism of Section~\ref{formalism} accommodates the description of both symmetries and topological order, and furthermore nontrivial actions of symmetries on states with topological order. Hence the formalism is well suited to describe symmetry-enriched topological phases within the PEPS framework. We plan to pursue this direction explicitly in future work\cite{michael}.

\section{ Symmetry-protected topological PEPS}\label{sptpeps}

Having discussed the general framework for gapped phases and global symmetries in PEPS, we now focus on the subclass corresponding to states with symmetry-protected topological order. In the first subsection we identify the characteristic properties of short-range entangled SPT PEPS. We proceed in the second subsection with an analysis of the edge properties of non-trivial SPT PEPS.

\subsection{ Identifying SPT PEPS}

First we must identify the relevant set of PEPS that accurately capture the short-range entanglement property characteristic of SPT phases.
As shown in Ref.\cite{MPOpaper} and argued in the previous sections, MPO-injective PEPS can describe topological phases with long-range entanglement. To single out the short-range entangled PEPS that are candidates to describe SPT states we require that the projection MPO $P$ has a single block when brought into its canonical form. Let $B_P^{ij}$ denote the MPO matrix with external indices $i$ and $j$, the single block property is equivalent to the transfer matrix $\mathbb{E}_P: = \sum_{ij}B_P^{ij}\otimes \bar{B}_P^{ij}$ having a unique eigenvalue of largest magnitude with a corresponding unique eigenvector of full rank. 
For RG fixed-point PEPS, which are injective on the support subspace of $P$, we argue that the single block property implies the topological entanglement entropy~\cite{KitaevPreskill,levin:topological-entropy} is zero.
\begin{claim}
For a RG fixed-point (zero correlation length) MPO-injective PEPS with a single blocked projector MPO $P$, the topological entanglement entropy of the PEPS is zero
\end{claim}
Note the rank of the reduced density matrix $\rho_\region$ on a finite homotopically trivial region $\region$ of a MPO-injective PEPS on a sphere equals the rank of the projection MPO surrounding that region, i.e.~$\mathrm{rank}({\rho_\region})=\mathrm{rank}(P_{\partial \region})$~\cite{MPOpaper}. Since the MPO $P$ is a projection, we have $\mathrm{rank}(P_{\partial \region}) = \mathrm{tr}(P_{\partial \region}) = \mathrm{tr}(P_{\partial \region}^2) = \mathrm{tr}(\mathbb{E}_P^L)$, where $L=|\partial\region|_e$ is the number of virtual bonds crossing the boundary of the region $\region$ under consideration. We then use the uniqueness of the largest eigenvalue $\lambda_{\text{max}}$ of $\mathbb{E}_P$ to conclude that, for large regions, the rank of the reduced density matrix scales as $\lambda_{\text{max}}^L$. This implies that the zero R\'{e}nyi entropy has no topological correction and for RG fixed-points this furthermore implies that the topological entanglement entropy is zero~\cite{topologicalrenyi}. We expect this property to hold throughout the gapped phase containing the fixed-point.

A further crucial property of a SPT phase without symmetry breaking is the existence of a unique ground state on any closed manifold. For a PEPS to be a unique ground state its transfer matrix must have a unique fixed-point. This excludes both symmetry-breaking and topological degeneracy~{\cite{boundarypaper,transfermatrix}}. By taking a PEPS sufficiently close to its isometric form~\cite{Ginjectivity,Buerschaper14,MPOpaper} we avoid the symmetry-breaking case (and assure the gap condition~\cite{SchuchGarciaCirac11}). Furthermore, in Appendix~\ref{b} we present an argument showing that MPO-injective PEPS with single block projection MPOs do not lead to topological degeneracy.

We have argued above that SPT PEPS should be MPO-injective on the support subspace of a single blocked projection MPO. In the language of Section~\ref{formalism} this implies $|\mathcal{S}_1|=1$ for SPT PEPS. 
We now show that in this case the symmetry MPOs are also single blocked.
\begin{claim} 
For any MPO-injective PEPS with a single blocked projection MPO, all symmetry MPOs of that PEPS can be chosen to be single blocked.
\end{claim}
Assume $V(g)$ contains multiple blocks when brought into canonical form $V(g)=\sum_i V_i(g)$, then we have $PV(g)=\sum_i V_{\pi(i)}(g)$ in canonical form (for some permutation $\pi$) since $V(g)=PV(g)$ for all lengths. This follows from the fact that a pair of MPOs which are equal for all lengths exhibit the same blocks when brought into canonical form~\cite{dpg}.
 Furthermore $\pi=1$ since $V_{{i}}(g)=P V_{\pi^{-1}({{i}})}(g)=P^2 V_{\pi^{-1}({{i}})}(g)=V_{\pi({{i}})}(g)$.

We have 
$$
P=V(g^{-1})V(g)=\sum_{{{i}}} V(g^{-1})V_{{{i}}}(g) 
$$
and since this equality holds for all lengths and $P$ has a single block, there can be only one block on the right hand side after bringing it into canonical form~\cite{dpg}. 
Hence one term in the sum gives rise to a $P$ block along with zero blocks in the canonical form and the others give rise only to zero blocks. Writing this out we have
$$
P=V(g^{-1})V_{{{i}}}(g) 
$$
multiplying by $V(g)$ from the left and making use of the invariance under $P$ implies
$$
V(g)=V_{{{i}}}(g) 
$$
which has a single block (after throwing away the trivial zero blocks).

The arguments in this subsection show that the subclass of symmetric, MPO-injective PEPS satisfying Eqs.\eqref{n23} and \eqref{n25} which accurately describe SPT phases are precisely those with a single blocked projection MPO, provided they are taken sufficiently close to an isometric form to discount the possibility of a phase transition.

Hence the framework of Section~\ref{formalism} yields a classification of SPT phases in terms of the discrete labels of the (necessarily single blocked) MPO group representation $V(g)$ of the physical symmetry group $\mathsf{G}$ which include the group structure and the third cohomology class $[\alpha] \in H^3(\mathsf{G},\mathsf{U(1)})$~\cite{Chen} (see Appendix~\ref{c} for a review).

\subsection{ Edge properties}

We now focus on how the MPO symmetries affect the edge physics of a SPT PEPS and discuss how this can be used to diagnose nontrivial SPT order.

A short-range entangled PEPS with MPO symmetries $V(g)$ that satisfy Eqs.\eqref{n23} and \eqref{n25} has non-trivial SPT order if the third cohomology class $[\alpha]$ of the MPO representation is non-trivial.  The existence of this non-trivial SPT order can be inferred by analyzing the edge physics when such a PEPS is defined on a finite lattice $\region$ with a physical edge (boundary) $\partial \region$. In this case the PEPS has open (uncontracted) virtual indices along the physical boundary and all virtual boundary conditions give rise to exact ground states of the canonical PEPS (bulk) parent Hamiltonian $H_{\text{PEPS}}$ (note boundary conditions orthogonal to the support of $P_{\partial \region}$ yield zero). Hence the ground space degeneracy scales exponentially with the length of the boundary, which is a generic property of any PEPS (bulk) parent Hamiltonian. 
The physically relevant question is whether the Hamiltonian can be perturbed by additional local terms $H_{\text{pert}}=\sum_v H_v$, which are invariant under $\mathsf{G}$, to gap out these edge modes and give rise to a unique symmetric ground state. 

\begin{figure}[ht]
\center
\begin{align*}
\vcenter{\hbox{\includegraphics[trim={0cm 9cm 26cm 0cm},clip,width=0.3\linewidth]{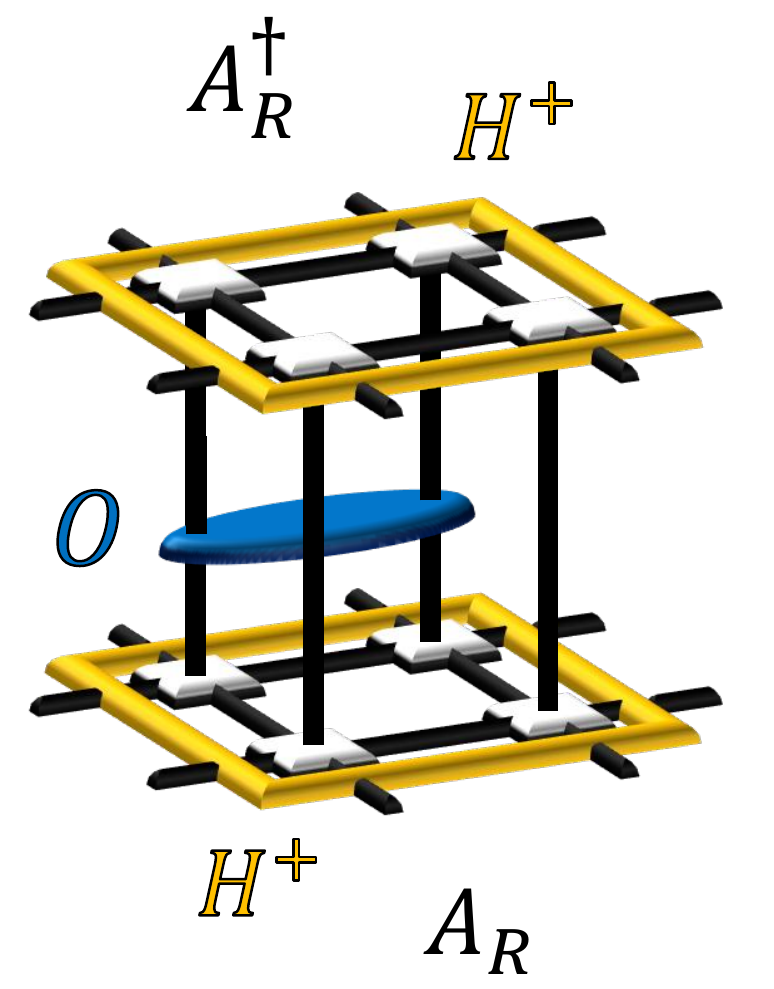}}}
\end{align*}
\caption{The bulk to boundary isometry, for a region $\region$ containing four sites, projected onto the injectivity subspace $P_{\partial \region}\isom{O}P_{\partial \region}=H^+\peps_\region^\dagger O \peps_\region H^+$.}\label{e3}
\end{figure}

In Ref.\cite{boundarypaper} an isometry $\mathcal{W}$ was derived that maps any operator $O$ acting on the physical indices of the PEPS to an effective operator acting on the virtual indices of the boundary $O\mapsto \isom{O}$. 
{Let $\peps_\region=WH$ be a polar decomposition of $\peps_\region$, where $W$ is an isometry from the virtual to physical level $(\mathbb{C}^D)^{\otimes |\partial \region|_e}\rightarrow (\mathbb{C}^d)^{\otimes |\region|_v}$. 
This induces the following isometry $\isom{O}:=W^\dagger O W$ that maps bulk operators to the boundary in an orthogonality preserving way.
Note there is some freedom in choosing $W$ precisely when $P_{\partial \region}$ is nontrivial, in this case we make the choice that best preserves locality.
Regardless of our choice of $W$ we always have $P_{\partial \region}\isom{O}P_{\partial \region}=H^+\peps_\region^\dagger O \peps_\region H^+$, where $H^+$ is defined to be the pseudoinverse of $H$, see Fig.\ref{e3}.}
Away from an RG fixed-point, however, it has not been proven that this isometry preserves locality. To this point we venture the following conjecture, {which was numerically illustrated for a particular non-topological PEPS in Ref.\cite{PEPSedges}},
\begin{conjecture}\label{conj1}
The boundary isometry of any PEPS with exponentially decaying correlations maps a local operator $O_v$ acting on the physical indices near the boundary to a (quasi-) local operator $\tilde O_e^v:=\isom{O_v}$ acting on the virtual degrees of freedom along the boundary. 
\end{conjecture}
From properties \eqref{n23} and \eqref{n25} it is clear that acting with $U(g)$ on every physical site is equivalent to acting with the MPO $V^{\partial \region}(g)$ on the virtual boundary indices of the PEPS, {hence a $\mathsf{G}$-symmetric local perturbation $H_v$ to the Hamiltonian at the physical level $H_{\text{PEPS}}$ is mapped to an effective (quasi-) local Hamiltonian term on the virtual boundary $\tilde{H}^v_e$ that is invariant under $V^{\partial \region}(g)$. The full symmetric edge Hamiltonian is given by
\begin{align}
\tilde{H}_{\text{edge}}&=P_{\partial \region}\, \isom{H_{\text{pert}} }\, P_{\partial \region}
\nonumber \\
&=V^{\partial \region}(1)\,  \left(\sum\limits_{e\in\partial\region}\sum\limits_{v\mapsto e} \tilde{H}^v_e \right)\, V^{\partial \region}(1)
\end{align}
where $v\mapsto e$ denotes that the bulk perturbation centered on site $v$ becomes a (quasi-) local boundary term centered on virtual bond $e$.
\\
Ground states of the perturbed physical Hamiltonian $H_{\text{bulk}}=H_{\text{PEPS}}+H_{\text{pert}}$ are given by contracting the virtual boundary indices of the ground state PEPS network with  ground states of the effective edge Hamiltonian, i.e. $\ket{\Psi_0^{\text{bulk}}}=\peps_\region\bket{\psi_0^{\text{edge}}}$.
If the edge Hamiltonian $\tilde{H}_\text{edge}$ is gapped and does not exhibit spontaneous symmetry breaking then its ground state $\bket{\psi_0^{\text{edge}}}$ is well approximated by an injective MPS that is invariant under $V^{\partial \region}(g)$.} However it was shown by Chen \textit{et al.} that this results in a contradiction, since an injective MPS cannot be invariant under the action of {a single blocked MPO group representation $V(g)$} with non-trivial third cohomology~\cite{Chen}.

{Consequently, the effective edge Hamiltonian $\tilde{H}_\text{edge}$ either exhibits spontaneous symmetry breaking, in which case the MPS is not injective, or must be gapless, in which case its ground state cannot be well approximated by a MPS. In the former case, the physical state $\peps_\region\bket{\psi_{0,i}^\text{edge}}$ obtained by contracting the virtual boundary indices of the PEPS network with $\bket{\psi_{0,i}^\text{edge}}$, one of the symmetry breaking ground states of $\tilde{H}_\text{edge}$, also exhibits symmetry breaking and hence does not qualify as a symmetric state. The latter case, on the other hand, implies that a local symmetric perturbation to the physical Hamiltonian is unable to gap out the gapless edge modes, which is one of the hallmarks of non-trivial SPT order. }

{
Here we have again relied on a form of Conjecture~\ref{conj1}, specifically that a PEPS with exponentially decaying correlations has a gapped transfer matrix, which implies that the gapless modes on the virtual boundary of the PEPS network are approximately identified, via the PEPS map $\peps_\region$, with physical degrees of freedom that are an order of the correlation length from the boundary.} 
Note this explicit identification of the gapless edge mode degrees of freedom is a major strength of the PEPS framework~\cite{PEPSedges}. Our conjecture is consistent with the intuition that as a SPT PEPS is tuned to criticality the gap of the transfer matrix shrinks and the edge modes extend further into the bulk, and is also supported by the results of Ref.\cite{SPTphasetransition} concerning phase transitions between symmetry-protected and trivial phases.

In this section we have identified a subclass of symmetric PEPS with short-range entanglement that are MPO-injective with respect to a single blocked projection MPO. This led to a classification of SPT phases within the framework of Section~\ref{formalism} in terms of the third cohomology class of the MPO symmetry representation. Finally we described the influence of the possibly anomalous MPO symmetry action on the boundary physics of the PEPS. In the next section we explore an alternative approach to classifying SPT phases with PEPS via gauging.

\section{ Gauging SPT PEPS}\label{gaugingsptpeps}

In this section we discuss how gauging a SPT PEPS yields a long range entangled PEPS whose topological order is determined by the symmetry MPOs. We then proceed to show that the gauging procedure preserves the energy gap of a symmetric Hamiltonian. 
Our approach explicitly identifies how the symmetry MPOs that determine the boundary theory of a SPT model are mapped to topological MPOs that describe the anyons of a topological theory~\cite{nick}.

\subsection{ Gauging SPT PEPS to topologically ordered PEPS}

We first outline the application of the gauging procedure from Ref.\cite{Gaugingpaper} to SPT PEPS and the effect this has upon the MPO symmetries.

Conditions \eqref{n23} and \eqref{n25} ensure that the SPT PEPS described in Section~\ref{sptpeps} are invariant under the global action $U(g)^{\otimes|\man|_v}$ of a symmetry group $\mathsf{G}$, hence the quantum state gauging procedure of Ref.\cite{Gaugingpaper} is applicable. 
{It was shown in Ref.\cite{Gaugingpaper} that the virtual boundary action of the physical symmetry in an injective PEPS becomes a purely virtual topological symmetry of the gauged tensors, with a trivial physical action. 
More precisely, it was shown that the gauging procedure transforms an injective PEPS, with virtual bonds in $\mathbb{C}^D$ and a virtual symmetry representation that factorizes as $V^{\partial\mathcal{R}}(g)=v(g)^{\otimes L}$ (with $v(g):\mathbb{C}^D\to\mathbb{C}^D$), into a $\mathsf{G}$-injective PEPS, with virtual bonds in $\mathbb{C}^D\otimes \mathbb{C}[\mathsf{G}]$, 
that is injective on the support subspace of the projector $\sum_{g\in \mathsf{G}} \left[ v(g) \otimes R(g) \right]^{\otimes L}$. Here, $L := |\partial \region|_e$ is the number of virtual bonds crossing the boundary of the region $\mathcal{R}$ under consideration and $R(g)\ket{h}:=\ket{hg^{-1}}$ denotes the right regular representation of $\mathsf{G}$ on the new component $\mathbb{C}[\mathsf{G}]$ of the virtual bonds.} Let us recast this in the framework of Section~\ref{formalism}. The ungauged symmetric injective PEPS map satisfies 
\begin{equation}
\peps_\region V^{\partial\mathcal{R}}(g)=U(g)^{\otimes|\region|_v}\peps_\region
\end{equation}
for any region $\region\subset\man$ and $g\in\mathsf{G}$. Now let $O^{\partial\mathcal{R}}(g) := \left[ v(g) \otimes R(g)\right ]^{\otimes L}$, then the gauged PEPS map $\peps^\mathsf{g}_\region$ for any region $\region$ satisfies 
\begin{equation}
\peps_\region^\mathsf{g} O^{\partial\mathcal{R}}(g) =\peps_\region^\mathsf{g}
\end{equation}
for all $g\in\mathsf{G}$, which implies that the gauged PEPS $\peps^\mathsf{g}$ is in the same phase as a quantum double model constructed form $\mathsf{G}$, provided it is sufficiently close to a fixed-point to ensure there is no symmetry breaking~\cite{qdouble,Ginjectivity}.

The result of Ref.\cite{Gaugingpaper} can be extended to the general case outlined in Section~\ref{sptpeps} and Appendix~\ref{b} where the PEPS map $\peps_\region$ in region $\mathcal{R}$ has a non-factorizable MPO representation of the symmetry on the virtual level, given by $V^{\partial\mathcal{R}}(g):(\mathbb{C}^D)^{\otimes L} \to (\mathbb{C}^D)^{\otimes L}$, and is only injective on the support subspace of the projection MPO $P_{\partial \mathcal{R}}=V^{\partial \mathcal{R}}(1)$ which is required to be single blocked. Hence we have
\begin{align}
\peps_\region P_{\partial \mathcal{R}}&= \peps_\region \\
\peps_\region V^{\partial\mathcal{R}}(g)&=U(g)^{\otimes|\region|_v}\peps_\region 
\end{align}
for all $g\in\mathsf{G}$; note we have explicitly separated the $g=1$ case for emphasis. 
In the language of Section~\ref{formalism} we have $\mathcal{S}_g=\{ V^{\partial \region}(g)\},\ \forall g\in\mathsf{G}$.

The gauged PEPS $A^\mathsf{g}$ obtained by applying the procedure of Ref.\cite{Gaugingpaper} to $A$ has virtual bonds in $\mathbb{C}^D\otimes \mathbb{C}[\mathsf{G}]$ and satisfies the axioms of MPO-injectivity~\cite{MPOpaper}, but is now injective on the support subspace of the projection MPO $P^\mathsf{g}_{\partial \mathcal{R}} :=    \frac{1}{|\mathsf{G}|}\sum_{g \in \mathsf{G}}O^{\partial\mathcal{R}}(g)$, where $ O^{\partial\mathcal{R}}(g):= V^{\partial\mathcal{R}}(g) \otimes R(g)^{\otimes L}$, see Appendix~\ref{d} for a detailed proof. Writing these conditions out, we have
\begin{equation}
\peps^\mathsf{g}_\region  O^{\partial\mathcal{R}}(g) = \peps^\mathsf{g}_\region
\end{equation}
for all $g\in\mathsf{G}$, which implies $\peps^\mathsf{g}_\region P^\mathsf{g}_{\partial \mathcal{R}}=\peps^\mathsf{g}_\region$.
Note every MPO $O^{\partial\mathcal{R}}(g)$ is one of the original MPO symmetries $V^{\partial\mathcal{R}}(g)$ tensored with a tensor product representation on the new component $\mathbb{C}[\mathsf{G}]$ of the virtual space that was introduced by gauging. 
The MPO representation of $P^\mathsf{g}_{\partial\mathcal{R}}$ thus has a canonical form with multiple blocks labeled by $g\in\mathsf{G}$ that correspond to the single block MPOs $O^{\partial\mathcal{R}}(g)$. 
Hence for the gauged PEPS $\mathcal{S}_1=\{ O^{\partial \region}(g) \,|\, g\in\mathsf{G}\}$.
Importantly, tensoring with a local action $R(g)$ on the additional virtual space $\mathbb{C}[\mathsf{G}]$ does not change the bond dimension nor the third cohomology class of the MPO representation. 

The topological order of the gauged SPT PEPS is a twisted Dijkgraaf-Witten model (provided it is sufficiently close to a fixed-point to ensure there is no symmetry breaking) which is shown explicitly in Section~\ref{gaugingfixptspt}. We emphasize that up to the trivial operators $R(g)^{\otimes L}$ the same MPOs determine both the gapless edge modes of the SPT phase and, as argued in~\cite{Buerschaper14,MPOpaper}, the topological order of the gauged model. This realizes the gauging map from SPT models with a finite symmetry group to models with intrinsic topological order, explored at the level of Hamiltonians by Levin and Gu~\cite{LevinGu}, explicitly on the level of states. In Appendix~\ref{gaugingham} we apply the gauging procedure of Ref.\cite{Gaugingpaper} to families of SPT Hamiltonians with an arbitrary finite symmetry group, which yields an unambiguous gauging map to families of topologically ordered Hamiltonians.

We note that the PEPS gauging procedure can equally well be applied to gauge any normal subgroup $\mathsf{N}\unlhd\mathsf{G}$ of the physical symmetry group $\mathsf{G}$. This gives rise to states with symmetry-enriched topological order, where the topological component corresponds to a gauge theory with gauge group $\mathsf{N}$ and the global symmetry is given by the quotient group $\mathsf{G}/\mathsf{N}$; we plan to investigate this direction further in future work~\cite{michael}. 

\subsection{ Gauging preserves the gap} \label{gptg}

We now show that the gauging procedure of Ref.\cite{Gaugingpaper} preserves the energy gap of a symmetric Hamiltonian, which implies by contrapositive that two SPT PEPS are in different phases when the corresponding gauged PEPS lie in distinct topological phases. 

Let $H_\mathsf{m}$ denote a local gapped symmetric `matter' Hamiltonian, which captures the particular case of parent Hamiltonians for SPT PEPS. The Hamiltonian is a sum of local terms $H_\mathsf{m} := \sum_v h_v$, where each $h_v$ acts on a finite region within a constant distance of vertex $v$. Without loss of generality we take the Hamiltonian to satisfy $[h_v,U(g)^{\otimes|\man|_v}]=0,\,\forall g\in\mathsf{G}$ and shift the lowest eigenvalue of $H_\mathsf{m}$ to 0. The gap to the first excited energy level is denoted by $\Delta_m>0$.
We now apply the gauging procedure of Ref.\cite{Gaugingpaper} to obtain the gauged matter Hamiltonian defined by $H^\mathscr{G}_\mathsf{m}:=\sum_v\mathscr{G}_{\subgraph_v}[ h_v]$, for $\mathscr{G}_{\subgraph_v}$ given in Eq.\eqref{gaugingoperator}. 
This Hamiltonian is also local since each $\mathscr{G}_{\subgraph_v}$ is locality preserving.

The gauging procedure introduces gauge fields on the links of the PEPS network and the full Hamiltonian of the gauged system contains local flux constraint terms $H_\mathcal{B}:=\sum_p (\openone-\mathcal{B}_p)$ acting on these gauge fields by adding an energy penalty when the flux through a plaquette $p$ is not the identity group element. Each local term $\mathcal{B}_p$ is a Hermitian projector acting on the edges around plaquette $p$ which has eigenvalue 1 on any gauge field configuration ($\mathsf{G}$-connection) that satisfies the flux constraint and 0 otherwise, see Eq.\eqref{bp}. Furthermore $\mathcal{B}_p$ is diagonal in the group basis on the edges, hence $[\mathcal{B}_p,\mathcal{B}_{p'}]=0$.

The full Hamiltonian may also contain a sum of local commuting projections onto the gauge invariant subspace $H_P:=\sum_v (\openone-P_v)$, see Eq.\eqref{constraint}, this corresponds to a model with an effective low energy gauge theory rather than a strict gauge theory.
Hence the full Hamiltonian on the gauge and matter system is given by the following sum 
$$ H_\text{full}=H^\mathscr{G}_\mathsf{m}+\Delta_\mathcal{B} H_\mathcal{B}+\Delta_P H_P$$
where $\Delta_\mathcal{B},\Delta_P\geq 0$. Note a strictly gauge invariant theory is recovered in the limit $\Delta_P\rightarrow \infty$.
It is easy to verify that the components of the full Hamiltonian commute, i.e. $[H^\mathscr{G}_\mathsf{m},H_\mathcal{B}]=[H^\mathscr{G}_\mathsf{m},H_P]=[H_\mathcal{B},H_P]=0$, and hence are simultaneously diagonalizable. Furthermore, $H_\mathcal{B}$ and $H_P$ each have lowest eigenvalue $0$ and gap $1$.

Assuming $\Delta_P$ is sufficiently large, the low energy subspace of $H_\text{full}$ lies within the ground space of $H_P$ and hence is spanned by states of the form $P[\, \ket{\lambda}_{\fullgraph_v}\otimes\ket{\phi}_{\fullgraph_e}]$, with $P = \prod_{v\in\Lambda}P_v$, for a basis $\ket{\lambda}$ of the matter (vertex) degrees of freedom (we will consider the eigenbasis of $H_\mathsf{m}$) and a basis $\ket{\phi}$ of the gauge (edge) degrees of freedom (we will consider the group element basis).

Similarly, assuming $\Delta_\mathcal{B}$ is sufficiently large, the low energy subspace of $H_\text{full}$ lies within the ground space of $H_\mathcal{B}$ which is spanned by states whose gauge fields form a flat $\mathsf{G}$-connection on the edge degrees of freedom. Since we additionally have $[\mathcal{B}_p,P]=0$ the common ground space of $H_\mathcal{B}$ and $H_P$ is spanned by states of the form $P[\, \ket{\lambda}_{\fullgraph_v}\otimes\ket{\phi_\text{flat}}_{\fullgraph_e}]$, for a basis $\ket{\phi_\text{flat}}$ of the flat $\mathsf{G}$-connections on the edge degrees of freedom (note these are product states).

$\mathsf{G}$-connections form equivalence classes under the local gauge operations $\st_v^g:=\bigotimes_{e\in E_v^+} R_e(g)\bigotimes_{e\in E_v^-} L_e(g)$ (see appendix~\ref{d} for a more detailed definition of $a^g_v$). On a 1-homotopy trivial manifold (no noncontractible loops) there is only 1 such equivalence class given by all connections of the form $\ket{\phi_\text{flat}}=\prod_i a_{v_i}^{g_i}\ket{1}_{\fullgraph_e}$, where $\ket{1}_{\fullgraph_e}:=\ket{1}^{\otimes |\fullgraph_e|}$.
\begin{claim}\label{prop3}
For a 1-homotopy trivial manifold, the states $G\ket{\lambda}$ (for a basis $\ket{\lambda}$) span the common ground space of both $H_\mathcal{B}$ and $H_P$, where $G$ is the quantum state gauging map defined in Eq.\eqref{gaugingmap}. 
\end{claim}
Since $P_v=\int\mathrm{d} g\, U_v(g) \otimes \st_v^g$ one can easily see $P_v \st_v^g = P_v U^\dagger_v(g)$ and hence for any state in the intersection of the ground spaces of $H_\mathcal{B}$ and $H_P$ we have
\begin{align}
P[\, \ket{\psi}_{\fullgraph_v}\otimes\ket{\phi_\text{flat}}_{\fullgraph_e}]&=P[\, \ket{\psi}_{\fullgraph_v}\otimes\prod_i a_{v_i}^{g_i}\ket{1}_{\fullgraph_e}]
\nonumber \\
&=
P[\, [\prod_i U_{v_i}(g_i)]^\dagger\ket{\psi}_{\fullgraph_v}\otimes\ket{1}_{\fullgraph_e}]
\nonumber \\
&= G [\prod_i U_{v_i}(g_i)]^\dagger\ket{\psi}_{\fullgraph_v}
\end{align}
where we have started from our above characterization of the common ground space.

We now proceed to show that any eigenstate of $H_\mathsf{m}$ is mapped to an eigenstate of $H^\mathscr{G}_\mathsf{m}$ by the quantum state gauging map $G$.  See appendix~\ref{d} for the details about the operator and state gauging maps $\mathcal{G}$ and $G$ as constructed in~\cite{Gaugingpaper}. 
\begin{claim}[\cite{Gaugingpaper}] \label{p4}
The identity $\mathscr{G}_\subgraph[O]G = GO$ holds for any symmetric operator $O$.
\end{claim}
Suppose $O$ acts on the sites $v\in\subgraph\subset\fullgraph$ where $\subgraph$ is a subgraph of the full lattice which contains all the edges between its vertices, then we have
\begin{align}
&\mathscr{G}_\subgraph[O]G = \int\prod_{v\in\subgraph} \mathrm{d} h_v\bigotimes_{v\in\subgraph} U_v(h_v) O \bigotimes_{v\in\subgraph} U_v^\dagger(h_v) 
\nonumber \\
&\bigotimes_{e\in\subgraph} \ket{h_{v_e^-}h_{v_e^+}^{-1}}\bra{h_{v_e^-}h_{v_e^+}^{-1}} \int\prod_{v\in\fullgraph} \mathrm{d} g_v\bigotimes_{v\in\fullgraph} U_v(g_v)
\bigotimes_{e\in\fullgraph} \ket{g_{v_e^-}g_{v_e^+}^{-1}}
\nonumber \\
&\phantom{\mathscr{G}[O]G}=\int\prod_{v\in\fullgraph} \mathrm{d} g_v \prod_{v\in\subgraph} \mathrm{d} h_v  \bigotimes_{v\in\fullgraph} U_v(g_v) \bigotimes_{v\in\subgraph} U_v(g_v^{-1} h_v)  \, 
\nonumber \\
& O\, \bigotimes_{v\in\subgraph} U^{\dagger}_v(g_v^{-1}h_v) 
\prod_{e\in\subgraph} \delta_{(g_{v_e^-}^{-1}h_{v_e^-}),\, (g_{v_e^+}^{-1}h_{v_e^+})}
\bigotimes_{e\in\fullgraph} \ket{g_{v_e^-}g_{v_e^+}^{-1}}
\nonumber \\
&\phantom{\mathscr{G}[O]G}=G\, O
\end{align}
where edge $e$ runs from vertex $v_e^+$ to $v_e^-$. The last equality follows since the $\delta$ condition forces $(g_v^{-1}h_v)$ to be equal for all $v\in\subgraph$ (assuming $\subgraph$ is connected) and the operator $O$ is symmetric under the group action $[ O, \bigotimes_{v\in\subgraph} U_v(g)] = 0$.

This implies that any eigenstate $\ket{\psi_\lambda}$ of $H_\mathsf{m}$ with eigenvalue $\lambda$ gives rise to an eigenstate $G\ket{\psi_\lambda}$ of $H^\mathscr{G}_\mathsf{m}$ with the same eigenvalue. Note we have assumed that $G\ket{\psi_\lambda}\neq0$, which is the case when the representation under which $\ket{\psi_\lambda}$ transforms contains the trivial representation. This always holds for a unique ground state (possibly after redefining the matrices of the group representation by multiplicative phases $U(g)\mapsto e^{i\theta(g)}U(g)$\,).

If $H_\mathsf{m}$ has a unique ground state $\ket{\lambda_0}$ the ground state of the full Hamiltonian is given by $G\ket{\lambda_0}$ (since $H^\mathscr{G}_\mathsf{m}\geq0$ for $H_\mathsf{m}\geq 0$) and its gap satisfies $\Delta_\text{full}\geq\min(\Delta_\mathsf{m},\Delta_\mathcal{B},\Delta_P)$.

Hence if two local SPT Hamiltonians are connected by a gapped, continuous and symmetric path of local Hamiltonians then the gauged models are also connected by a gapped and continuous path of local Hamiltonians. 

In Appendix~\ref{e} we extend this proof to SPT Hamiltonians on topologically nontrivial manifolds where the gauging procedure leads to a topological degeneracy of the ground space. Orthogonal topological ground states are obtained by gauging distinct symmetry twisted SPT states, which are the subject of the next section.

\section{ Symmetry twists and monodromy defects} \label{six}

In this section we argue that symmetry twists and monodromy defects have a natural description in the tensor network formalism in terms of symmetry MPOs that correspond to anyons in the gauged model. 
We harness this description to calculate the effect that modular transformations have upon symmetry twisted and topological ground states via their effect on a four index \emph{crossing tensor}. Similarly we calculate the projective transformation of a monodromy defect by composing two \emph{crossing tensors}.
Our approach explicitly identifies how the symmetry MPOs that describe defects of a SPT model become topological MPOs that describe the anyons of a topological model~\cite{nick}. 

\subsection{Symmetry twists in SPT PEPS}

We first describe the construction of a symmetry twisted SPT PEPS in terms of the original SPT PEPS, symmetry MPOs and a crossing tensor. We then calculate the transformation of this state under the residual symmetry group.

\begin{figure}[ht]
\center
\begin{align*}
a) \vcenter{\hbox{
 \includegraphics[width=0.4\linewidth]{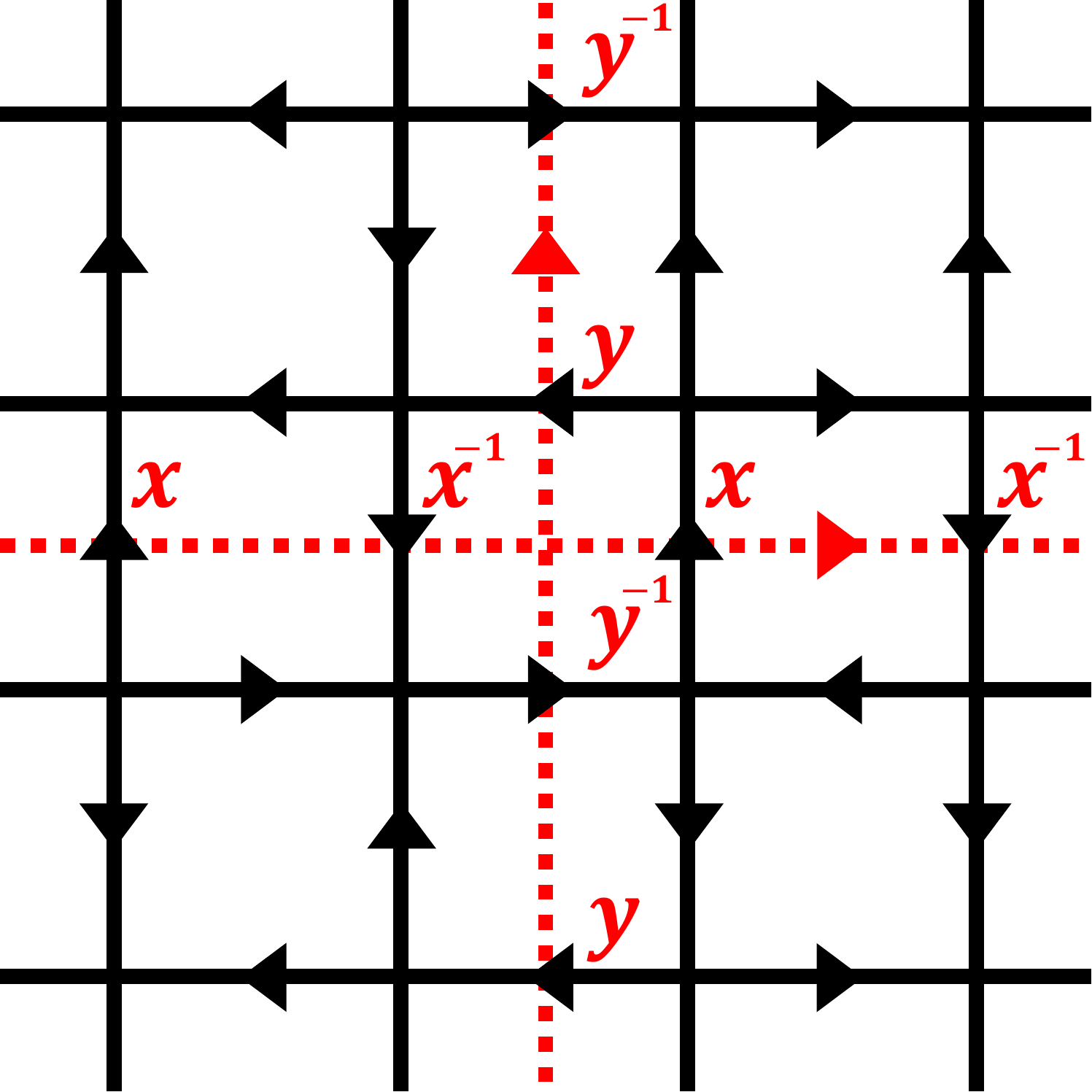}}} \quad b)
\vcenter{\hbox{
\includegraphics[width=0.4\linewidth]{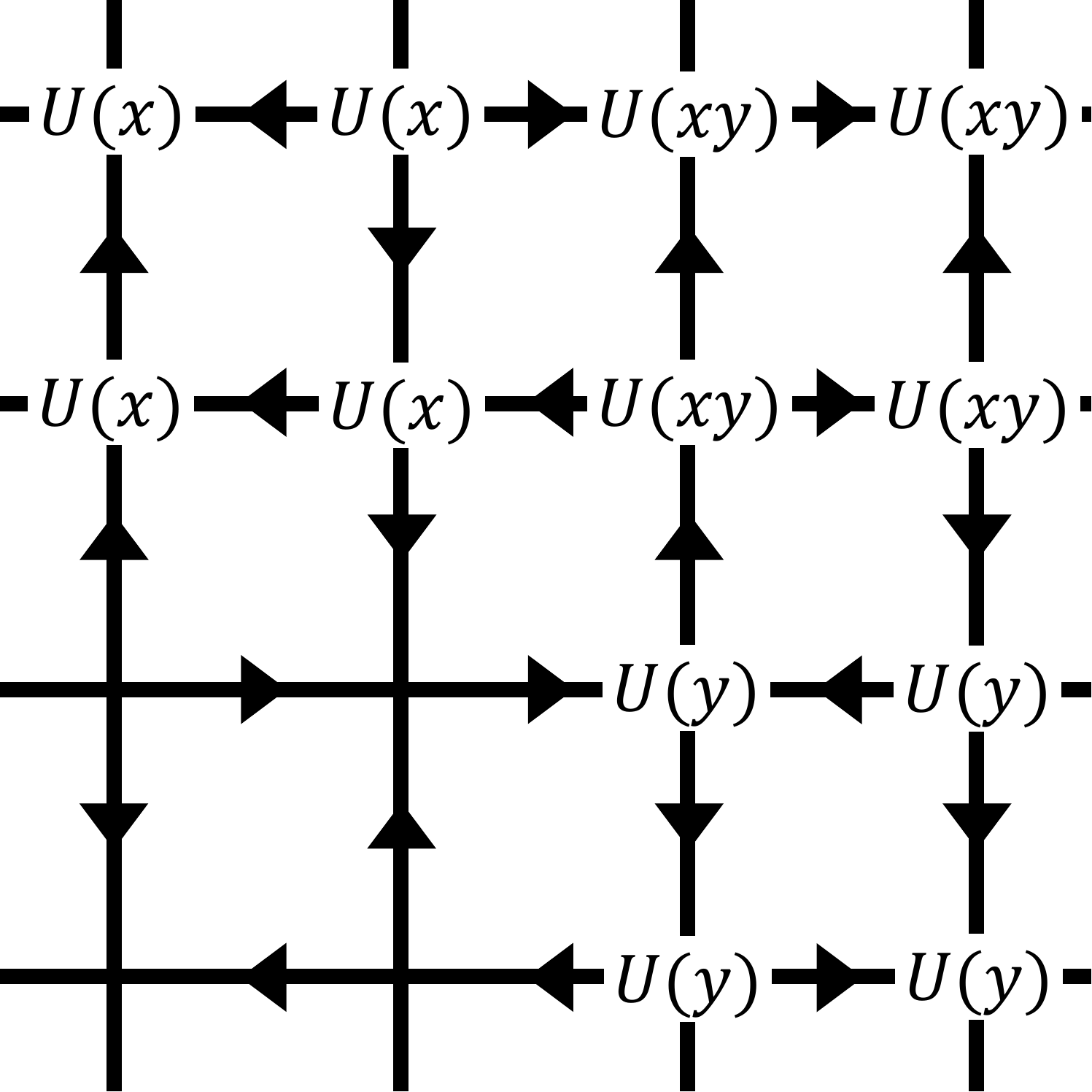}}} 
\end{align*}
\caption{a) A symmetry twist $(x,y)$ on an infinite plane. b) Physical action of the aforementioned symmetry twist.}\label{e4}
\end{figure}

For a flat gauge field configuration there is a well defined procedure for applying a corresponding symmetry twist to a local symmetric Hamiltonian, given by conjugating each local term by a certain product of on-site symmetries (see Appendix~\ref{e}). 
On a trivial topology such a symmetry twist can be applied directly to a symmetric state by acting with a certain product of on-site symmetries. 
For example a symmetry twist on an infinite plane, specified by a pair of commuting group elements $(x,y)\in\mathsf{G}\times \mathsf{G}$ and oriented horizontal and vertical paths $p_x,p_y$ in the dual lattice, acts on a state $\ket{\psi}$ in the following way 
\begin{align*}
\ket{\psi}^{(x,y)}:= \bigotimes_{v\in \mathcal{U}} U_v(x) \bigotimes_{v\in\mathcal{R}} U_v(y) \ket{\psi}
\end{align*}
where $\region$ is the half plane to the right of $p_y$, $\mathcal{U}$ the half plane above $p_x$, see Fig.\ref{e4}. 
Note $x$ and $y$ must commute for the relevant gauge field configuration to be flat. One can also understand why they must commute by first applying the $x$-twist which reduces the symmetry group to $\mathsf{C}(x)$ (the centralizer of $x$) and hence it only makes sense to implement a second twist for $y\in\mathsf{C}(x)$.
With this definition applying a symmetry twist to an eigenstate of a symmetric Hamiltonian (on a trivial topology) yields an eigenstate of the symmetry twisted Hamiltonian with the same eigenvalue. 

\begin{figure}[ht]
\center
\includegraphics[width=0.43\linewidth]{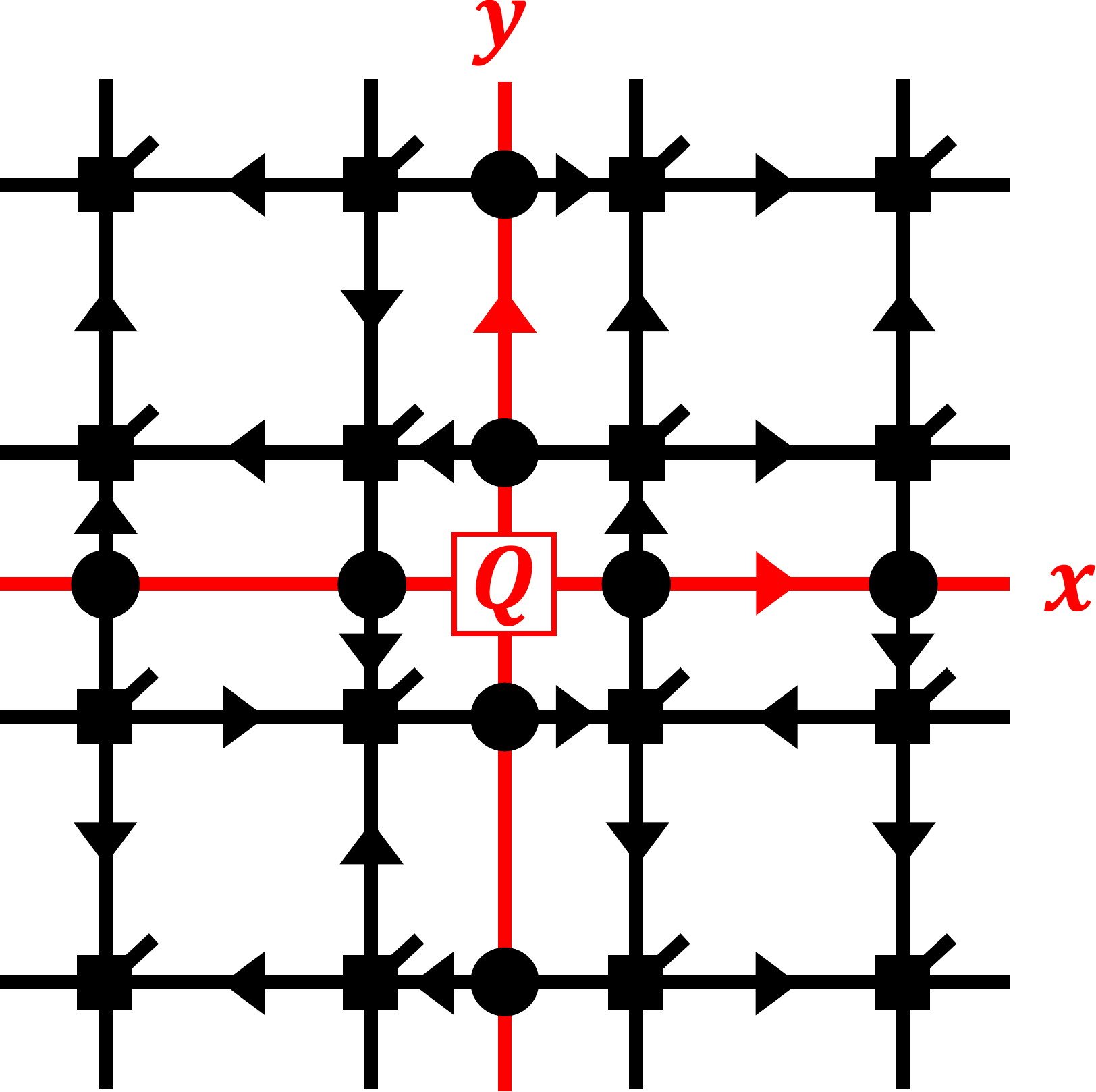}
\caption{$(x,y)$ symmetry twisted PEPS, for infinite or periodic boundary conditions.}\label{e5}
\end{figure}

The framework of SPT PEPS provides a natural prescription for the application of a symmetry twist directly to a PEPS on any topology, given by acting with symmetry MPOs on the virtual level of the PEPS. 
In the above example, assuming $\ket{\psi}$ is a SPT PEPS with local tensor $A$ and symmetry MPOs $V(g)$, Eq.\eqref{n23} implies that the symmetry twisted state $\ket{\psi}^{(x,y)}$ is given by acting on the virtual level of the PEPS $\ket{\psi}$ with the MPOs $V^{\epath_x}(x),V^{\epath_y}(y)$ (with inner indices contracted with the four index \emph{crossing tensor} $Q_{x,y}$~\eqref{cxy} where $\epath_x,\epath_y$ intersect) see Fig.\ref{e5}. 

The \emph{crossing tensor} $Q_{x,y}$ is defined in terms of the local reduction tensor of the MPO representation $X(x,y)$ (see Eqs.(\ref{n27},\ref{na21a})) 
\begin{align}\label{cxy}
Q_{x,y}:&=X(x,y) X^{+}(y,x)
\\
&= \QW{R}{x}{y}
\nonumber
\\
\vcenter{\hbox{
 \includegraphics[width=0.2\linewidth]{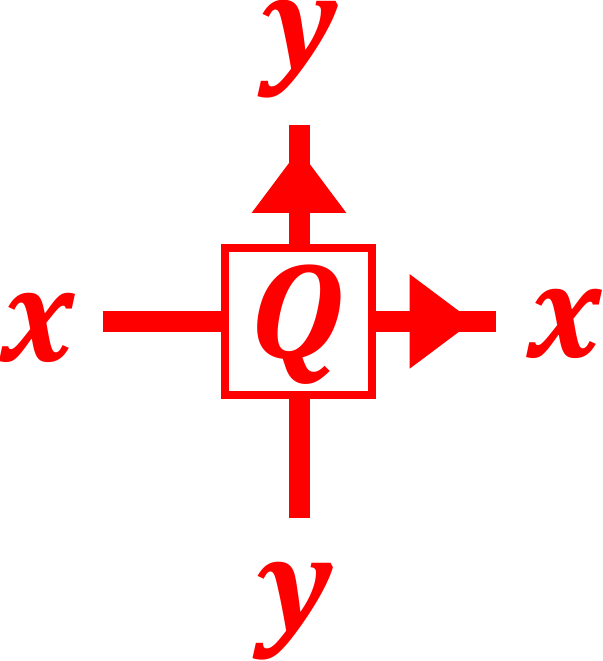}}}
\ :&=
\vcenter{\hbox{
 \includegraphics[width=0.25\linewidth]{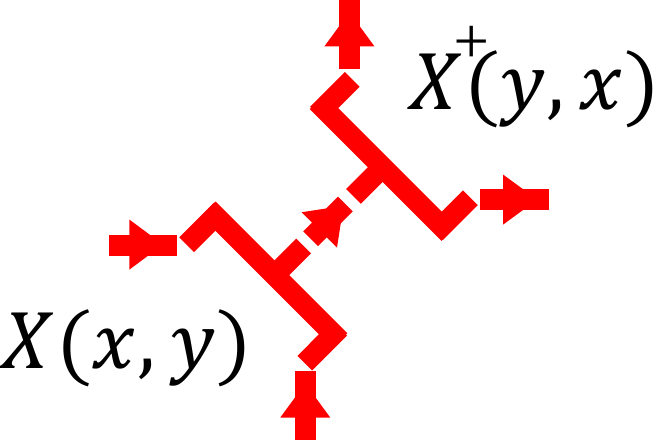}}} 
\nonumber
\end{align}
where $X^+(y,x)$ is the pseudoinverse of $X(y,x)$. 
Eq.\eqref{n23} and the \emph{zipper} condition \eqref{a3} for $X(x,y)$ imply that the $Q_{x,y}$ tensor contracted with MPOs $V^{\epath_x}(x),V^{\epath_y}(y)$ can be moved through the PEPS on the virtual level by applying appropriate on-site symmetries to the physical level.

This prescription extends straightforwardly to an arbitrary topology (see Appendix~\ref{e}) as we now demonstrate with the example of a symmetry twist on a torus for a pair of commuting group elements $(x,y)$ along distinct noncontractible cycles $p_x,p_y$. The symmetry twisted SPT PEPS $\ket{\psi}^{(x,y)}$ is again given by applying the MPOs $V^{\epath_x}(x),V^{\epath_y}(y)$ (with inner indices contracted with the crossing tensor $Q_{x,y}$) to the virtual level of the untwisted PEPS $\ket{\psi}$.
Importantly this prescription fulfills the condition that applying a symmetry twist to a PEPS groundstate of a symmetric frustration free Hamiltonian yields a groundstate of the symmetry twisted Hamiltonian due to Eq.\eqref{n23}. We note that similar tensor network techniques allow a construction of symmetry twists for time reversal symmetry~\cite{Gaugingtime}.

A symmetry twisted state with conjugated group elements $(x^g,y^g)$ is related, up to a phase, to the symmetry twisted state with group elements $(x,y)$ via a global  symmetry action as follows $\theta^{x,y}_g \ket{\psi}^{(x^g,y^g)}=U(g)^{\otimes |\man|_v}\ket{\psi}^{(x,y)}$. Similarly a symmetry twisted state for a local deformation of the paths $(p_x,p_y)\mapsto(\tilde p_x,\tilde p_y)$ is related to the symmetry twisted state for $(p_x,p_y)$ by a product of on-site symmetries corresponding to the deformation via Eq.\eqref{n23}. 
Hence the number of distinct classes of symmetry twisted states on a torus, under local operations, is given by the number of conjugacy classes of commuting pairs of group elements, which equals the number of irreducible representation of the quantum double ${D}(\mathsf{G})$~\cite{Ginjectivity}. 

It is apparent that a symmetry twisted state $\ket{\psi}^{(x,y)}$ forms a 1D representation under the physical action of the residual symmetry group $\mathsf{C}(x,y)$, where $\mathsf{C}(S)$ denotes the centralizer of a subset $S\subseteq G$. Assuming that the untwisted ground state $\ket{\psi}$ is symmetric under $\mathsf{G}$ (which can always be achieved after rephasing the physical representation) the symmetry twisted states may still form nontrivial 1D representations of their respective residual symmetry groups, this fact becomes important when counting the ground space dimension of the gauged model. Calculating these 1D representations explicitly within the PEPS framework yields the result $\theta^{x,y}_g=\slant{\alpha}{2}{x,y}{g}$ the second slant product of the 3-cocycle $\alpha$ that arose from the MPO group representation (see Appendix~\ref{newapp2}, Eq.\ref{na27}). 
Hence an $(x,y)$ symmetry twisted state is symmetric under $\mathsf{C}(x,y)$ iff $\slan{\alpha}{2}{x,y}\equiv 1$, in which case $y$ is called $\slan{\alpha}{1}{x}$-regular. If this property is satisfied by a given $y\in\mathsf{C}(x)$ it is also holds for all conjugates of $y$. Furthermore the number of $\slan{\alpha}{1}{x}$-regular conjugacy classes is known to be equal to the number of irreducible projective representations with 2-cocycle $\slan{\alpha}{1}{x}$~\cite{DijkgraafWitten}.

\subsection{Gauging the symmetry twisted SPT PEPS}

We now outline how the application of an appropriate gauging procedure to a symmetry twisted SPT PEPS yields a topological ground state.

There is a twisted version of the gauging procedure of Ref.\cite{Gaugingpaper} for each flat gauge field configuration which maps a symmetric Hamiltonian with the corresponding symmetry twist to a gauged Hamiltonian, the same one as obtained by applying the untwisted gauging procedure to the untwisted symmetric Hamiltonian (see Appendix~\ref{e} for more detail). 
For a fixed representative $(x,y)$ the twisted gauging operator $G_{x,y}$ is given by contracting the tensor product operators $R(x)^{\otimes |\epath_x|},R(y)^{\otimes |\epath_y|}$ with the virtual level of the original gauging operator $G$. 
The twisted versions of the state gauging map are orthogonal for distinct symmetry twists in general and furthermore the fixed representatives satisfy ${G_{x',y'}^\dagger G_{x,y}=\delta_{[x',y'],[x,y]}\int \mathrm{d}g \, U(g)^{\otimes |\man|_v} \delta_{g\in\mathsf{C}(x,y)}}$ (see Appendix~\ref{e}.3 for a proof of this). 
Hence each conjugacy class of symmetry twisted states that are symmetric under the residual symmetry group is mapped to an orthogonal ground state, while those that form a nontrivial 1D representation are mapped to 0. Consequently the dimension of the ground space for the gauged model is given by the number of irreducible representations of the twisted quantum double $D^\alpha(\mathsf{G})$ which can not be larger than the ground space dimension of a gauged trivial SPT model with the same symmetry group. 

Given a SPT PEPS ground state $\ket{\psi}$, the orthogonal ground states of the gauged model can be constructed by applying the gauging tensor network operator and acting with the SPT symmetry MPO and a product of on-site symmetry actions $[V(g) \otimes R(g)^{\otimes L}]$ along noncontractible cycles on the virtual level of the gauged tensor network $G\ket{\psi}$. 
 For a fixed representative $(x,y)$ of a symmetric class of symmetry twists the corresponding gauged ground state is given by contracting the MPOs $[V^{\epath_x}(x) \otimes R(x)^{\otimes |\epath_x|}],[V^{\epath_y} (y) \otimes R(y)^{\otimes |\epath_y|}]$ (with the crossing tensor $Q_{x,y}$ at the intersection point $\epath_x\cap\, \epath_y$~\cite{MPOpaper}) with the virtual level of the gauged PEPS $G\ket{\psi}$.

\subsection{Modular transformations}

We calculate the effect of modular transformation on symmetry twisted and topological ground states via their effect on a set of four index crossing tensors.

Symmetry twisted ground states have been used to identify non trivial SPT order via the matrix elements of modular transformations taken with respect to them~\cite{symmetrytwist,huang2015detecting}. 
We have calculated the SPT $\tilde S$ \& $\tilde T$ matrices, corresponding to a $\frac{\pi}{2}$ rotation and a Dehn twist respectively,  using our framework to find (see Eq.\eqref{na29})
\begin{align}
\bra{x',y'} \tilde S \ket{x,y} &= \slant{\alpha}{1}{y}{x^{-1},x}^{-1} \braket{x',y'|y,x^{-1}}
\\
\bra{x',y'} \tilde T \ket{x,y} &= \alpha(x,y,x) \braket{x',y'|x,xy}
\end{align}
where we have used the abbreviation $\ket{x,y}:=\ket{\psi}^{(x,y)}$, $\alpha^{(y)}$ is the slant product of $\alpha$ (see Appendix~\ref{newapp2}, Eq.\ref{slant}) and note $y\in\mathsf{C}(x)$. 
The gauging procedure elucidates the precise correspondence between these matrix elements and the $S$ \& $T$-matrix of the gauged theory~\cite{vishwanath,MPOpaper,moradi} which we have also calculated within the ground space (again see Eq.\eqref{na29})
\begin{align}
S &= \sum_{xy=yx} \slant{\alpha}{1}{y}{x^{-1},x}^{-1} \ket{[y,x^{-1}]}\bra{[x,y]} \\ T&= \sum_{ xy=yx} \alpha(x,y,x) \ket{[x,xy]}\bra{[x,y]} 
\end{align}
where $\ket{[x,y]}:=G_{x,y}\ket{\psi}^{(x,y)}$ denotes a ground state of the gauged model. 
Note in our framework we consider a fixed but arbitrary choice of representative for each conjugacy class, rather than group averaging over them. 

We have explicitly verified that $S$ \& $T$ generate a linear representation of the modular group in agreement with known results for lattice gauge theories (See Subsection~\ref{appmodular}).

\subsection{ Projective symmetry transformation of monodromy defects}

Here we describe an explicit construction of the projective representation that acts upon a monodromy defect. We calculate the 2-cocycle of this projective representation by considering the composition of pairs of crossing tensors.

\begin{figure}[ht]
\center
\begin{align*}
a) \vcenter{\hbox{
 \includegraphics[width=0.4\linewidth]{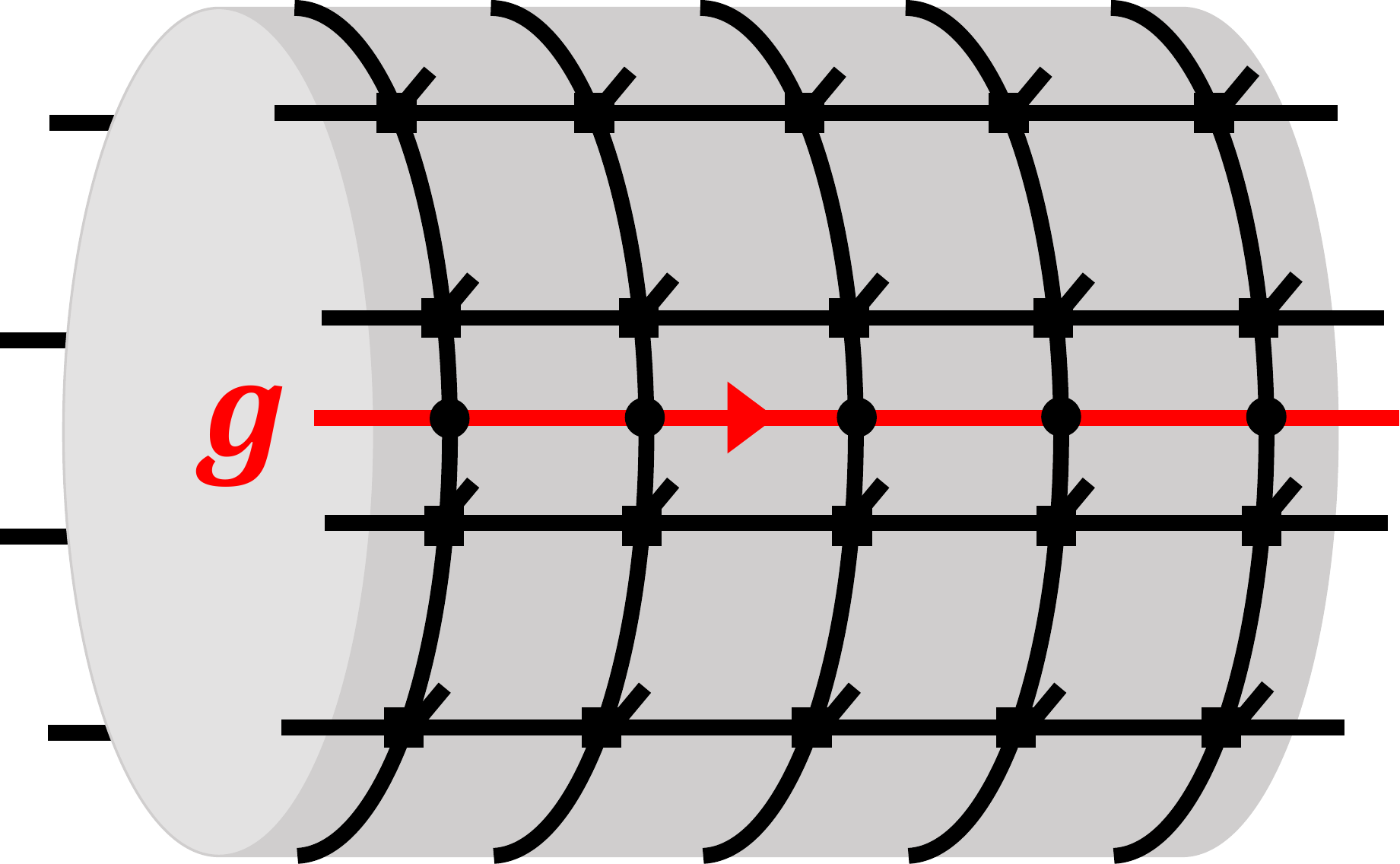}}} \quad b)
\vcenter{\hbox{
\includegraphics[width=0.4\linewidth]{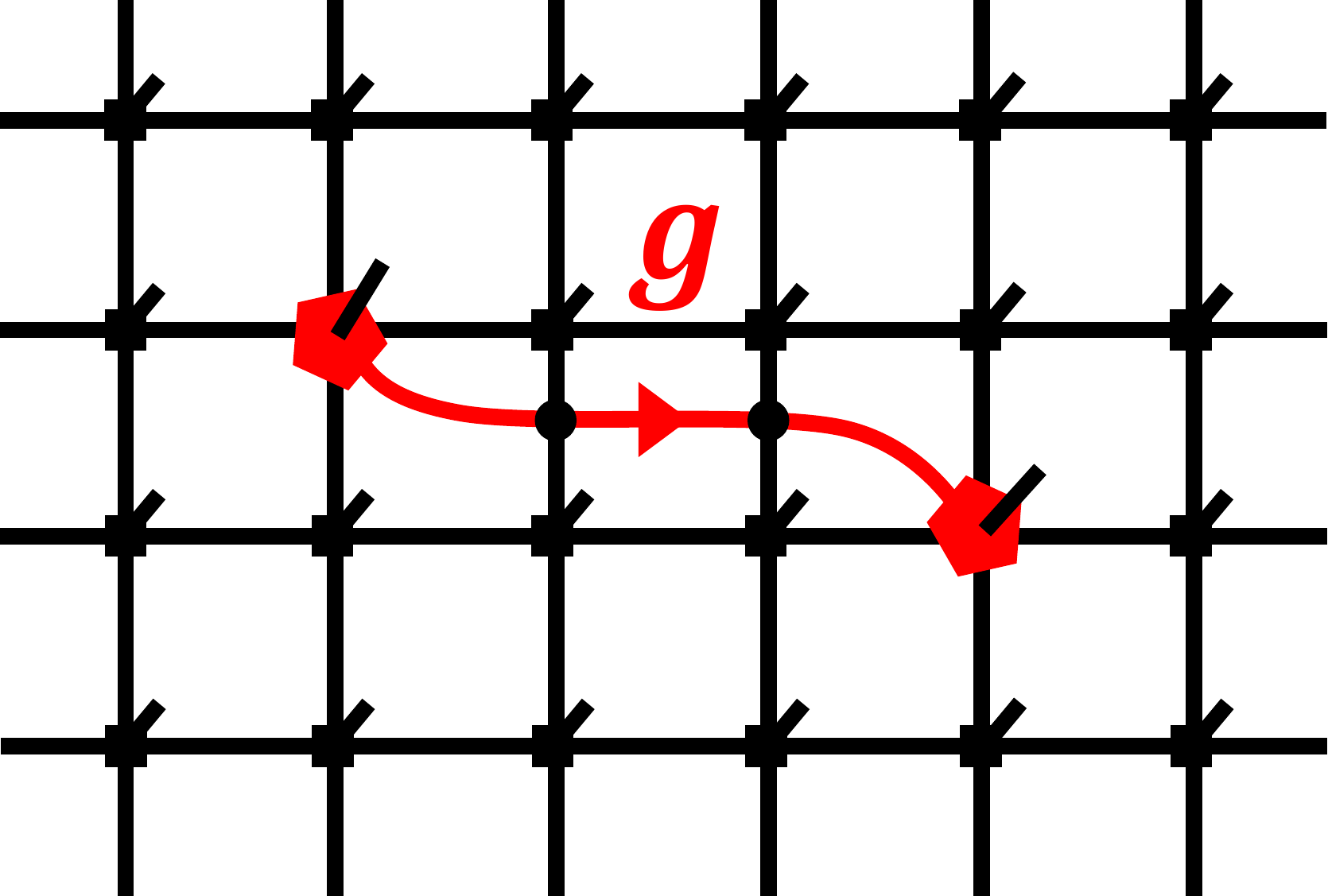}}} 
\end{align*}
\caption{a) A symmetry twist along a cylinder PEPS. b) A pair of monodromy defects  in a PEPS.}\label{e7}
\end{figure}

Monodromy defects can be understood as symmetry twists along paths with open end points and have proven useful for the identification of SPT phases~\cite{Wen1,Zaletel}. 
The prescription for applying symmetry twists to SPT PEPS extends naturally to a construction of a pair of monodromy defects at the ends of a path $\epath_g$, for $g\in\mathsf{G}$. This is given by applying a symmetry MPO $V^{\epath_g}(g)$ to the virtual level of the PEPS with an open inner index at either end of the path, which may be contracted with defect tensors replacing the PEPS tensors at each of the defects, see Fig.\ref{e7}. A defect tensor must lie in the support subspace of the projector $\bdry^g(1)$ acting on its virtual indices, see Eq.\eqref{e8}. This may leave some freedom in choosing the tensor which correspond to internal degrees of freedom of the defect, see Ref.\cite{nick} for further details. 
Applying the twisted gauging procedure for the corresponding gauge field configuration (which is flat except  near the defect points) explicitly maps the symmetry twisted PEPS to a PEPS that describes a pair of anyon excitations in the gauged theory, see Appendix~\ref{g} and Refs.\cite{Ginjectivity,nick}.

We now study a pair of monodromy defects on a twice punctured sphere topology, with a defect in each puncture, see Fig.\ref{e7}. This captures the case of a symmetry twist $g$ applied to a path $p_g$ along a cylinder, from one boundary to the other, and also the case of a pair of monodromy defects on a sphere, where each puncture is formed by removing a PEPS tensor and replacing it with a tensor that describes the defect. 

Treating a symmetry twisted SPT PEPS on a cylinder (of fixed radius) as a one dimensional system, it is clear that the bulk is invariant under the residual symmetry group $\mathsf{C}(g)$ since the symmetry twisted SPT PEPS on a torus formed by closing the cylinder (such that $p_g$ becomes a noncontractible cycle) is symmetric. In this case the PEPS can be interpreted as a MPS and standard results in this setting imply that the global symmetry $U(h)^{\otimes|\man|_v}$ is intertwined by the PEPS to a tensor product of projective symmetry representations on the left and right virtual boundaries $\bdry^g_L(h)\otimes \bdry^g_R(h)$. 

The projective boundary action $\bdry_R^g(h)$ of the symmetry can be explicitly constructed within the SPT PEPS framework. We find that it is given by a symmetry MPO acting on the PEPS virtual bonds entering the puncture, with its inner indices at the intersection of $p_g$ and the boundary of the puncture contracted with the tensor $\QV{R}{g}{h}$ (see Eq.\eqref{na21c}) that acts on the inner index of the symmetry twist MPO $V^{\epath_g}(g)$ entering the puncture.
\begin{equation}\label{e8}
\bdry_R^g(h)=\vcenter{\hbox{
 \includegraphics[width=0.21\linewidth]{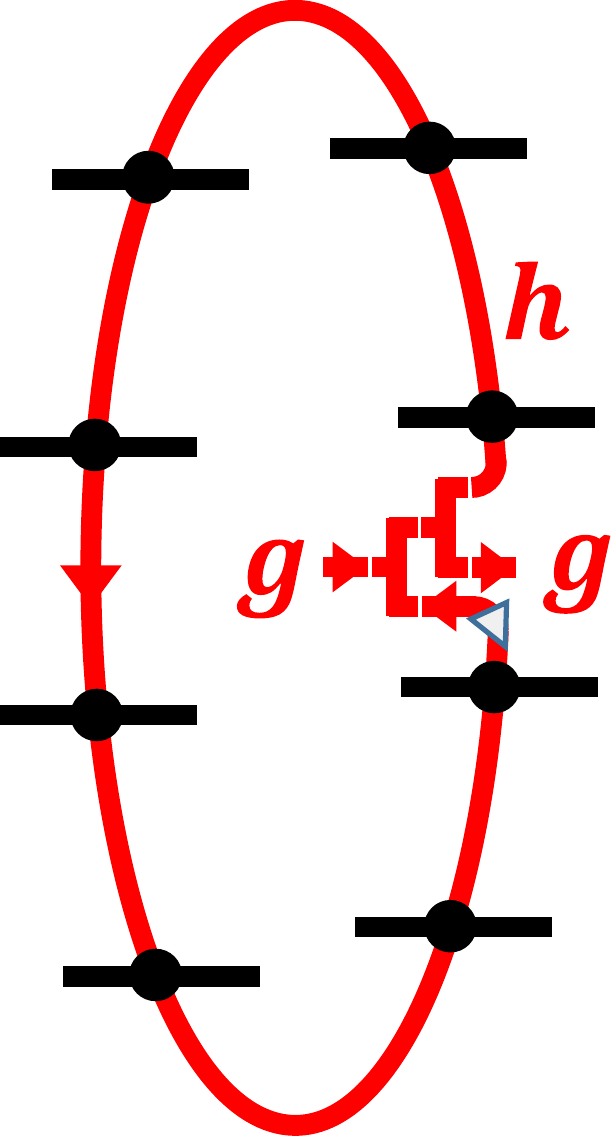}}}
\end{equation}

The multiplication of physical symmetries induces a composition rule for the $\QV{R}{g}{\cdot}$ tensors, see Appendix~\ref{g} for details. Explicit calculation of these products yields the 2-cocycle factor set $\omega^g$ of the projective boundary representation $\bdry_R^g(k)\bdry_R^g(h)=\omega^g(k,h)\bdry_R^g(kh)$ in terms of the 3-cocycle $\alpha$ of the MPO symmetry representation  $\omega^g(k,h)\sim \frac{\alpha(g,k,h)\alpha(k,h,g)}{\alpha(k,g,h)}$. This is consistent with the results of Ref.\cite{Zaletel}. Note that altering $\alpha$ by a 3-coboundary induces a 2-coboundary change to the 2-cocycle $\omega^g$, which hence forms a  robust label of the SPT phase. 
The projective symmetry action is closely related to the braiding of anyons in the gauged theory.

\section{ Example: fixed-point SPT states}\label{exfpspt}

Inspired by the illuminating examples in Refs.\cite{GuWen} and~\cite{Chen} we now present a family of SPT PEPS with symmetry group $\mathsf{G}$ and 3-cocycle $\alpha$ satisfying Eqs.\eqref{n23} and \eqref{n25}, and explicitly demonstrate that gauging these states~\cite{Gaugingpaper} yields MPO-injective PEPS that are the ground states of twisted quantum double Hamiltonians~\cite{tqd,DijkgraafWitten}.

\subsection{ Fixed-point SPT PEPS}

We describe our construction of fixed-point SPT PEPS and calculate the MPOs induced by the symmetry action on a site. We explicitly give the fusion tensors for these MPOs and verify that they satisfy the \emph{zipper} condition before determining the 3-cocycle of the MPO representation.

Our short-range entangled PEPS are defined on any trivalent lattice embedded in an oriented 2-manifold (dual to a triangular graph). They realize states equivalent to a standard SPT fixed-point construction on the triangular graph~\cite{GuWen,else2014classifying}. To this end we specify an ordering on the vertices of the triangular graph which induces an orientation of each edge, pointing from larger to smaller vertex. With this information we assign the following PEPS tensor $A_\triangle:\mathbb{C}(\mathsf{G})^{\otimes 6}\rightarrow\mathbb{C}(\mathsf{G})^{\otimes 3}$ to each vertex of the trivalent lattice
\begin{align}\label{n3}
\peps_\triangle :=
\int \prod_{v\in\triangle} \mathrm{d} g_v \ &\tilde{\alpha}_\triangle \bigotimes_{v\in \triangle} \ket{g_v}_{\triangle,v}
\nonumber \\
&\bigotimes_{e \in \triangle}( g_{v_e^-}|_{\triangle,e,v_e^-}( g_{v_e^+}|_{\triangle,e,v_e^+}
\end{align}
where edge $e$ is oriented from $v_e^+$ to $v_e^-$ (hence $v_e^- < v_e^+$) in the triangular graph. 
The phase $\tilde{\alpha}_\triangle$ is defined on a vertex of the trivalent PEPS lattice dual to plaquette $\triangle$ of the triangular lattice, whose vertices appear in the order $v,\ v',\ v''$ following the orientation of the 2-manifold (note the choice of starting vertex is irrelevant), by a 3-cocycle $\alpha$ as follows $\tilde{\alpha}_{\triangle}:=\alpha^{\sigma_\pi}(g_1g_2^{-1},g_2g_3^{-1},g_3)$. Where $(g_1,g_2,g_3):=\pi (g_v,g_{v'},g_{v''})$ with $\pi$ the permutation that sorts the group elements into ascending vertex order and $\sigma_\pi=\pm 1$ is the parity of the permutation (equivalently the orientation of $\triangle$ relative to the 2-manifold). In the following example the tensor $\peps_\triangle$, possessing six virtual and three physical indices, has non zero entries given by
\begin{equation}
\vcenter{\hbox{
\includegraphics[width=0.4\linewidth]{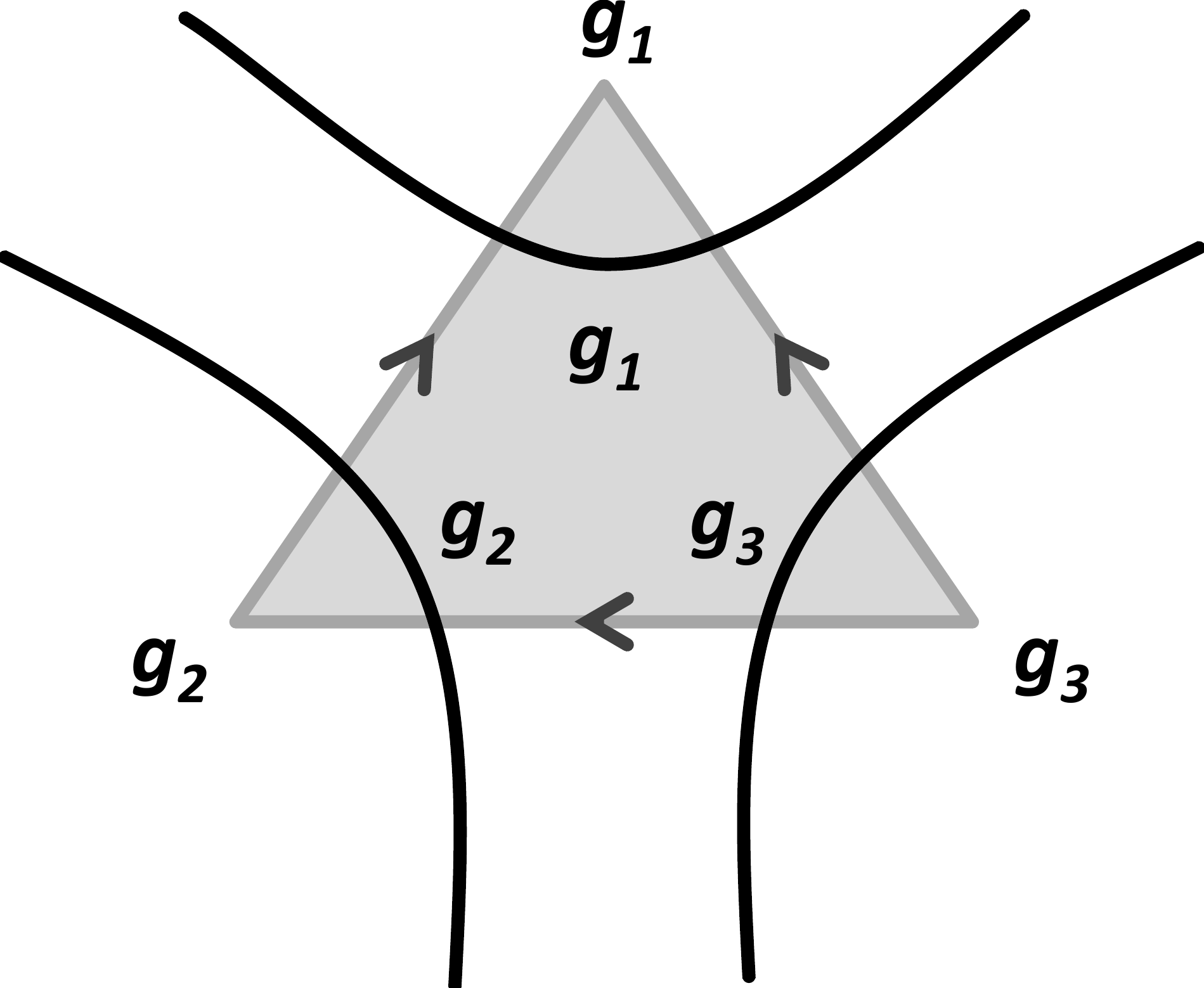}}} \label{d1} = \alpha(g_1g_2^{-1},g_2g_3^{-1},g_3)\, .
\end{equation}
Note the tensor diagrams in this section use the convention that physical vertex indices are written within the body of the tensor. Moreover we only depict the virtual and physical index combinations that give rise to non-zero values of the tensor. 

The global symmetry of the PEPS on a closed manifold is ensured by the following transformation property of each local tensor
\begin{align}\label{tensorsym}
R(h)^{\otimes 3} \peps_\triangle = \peps_\triangle \bigotimes_{e\in\triangle} [ Z^{\sigma_{\triangle,e}}_{e} (h) R(h)^{\otimes 2} ] \, ,
\end{align}
where  $ Z_{e} (h) := \int \mathrm{d} g_{v_e^-} \mathrm{d} g_{v_e^+} \alpha (g_{v_e^-}g_{v_e^+}^{-1},g_{v_e^+},h) \vert g_{v_e^-},g_{v_e^+}) \allowbreak (g_{v_e^-},g_{v_e^+}\vert$, and $\sigma_{\triangle,e}=\pm 1$ is $+1$ if $e$ is directed along the clockwise orientation of $\partial\triangle$, and $-1$ otherwise. 
With this definition one can check that Eq.\eqref{tensorsym} is equivalent to the cocycle condition \eqref{cocycle}. Note the boundary actions on the shared edge of two neighboring tensors $\peps_\triangle, \peps_{\triangle'}$, induced by group multiplication on the physical sites $\triangle,\triangle'$, cancel out since $\sigma_{\triangle,e}=-\sigma_{\triangle',e}$ from which it follows that the full PEPS (on a closed manifold) is invariant under the group action applied to all physical indices. In our example the symmetry property is\footnote{
Note the following subtlety, our tensor diagrams depict the coefficients of the map $A_\triangle$ and hence the group action $R(h)$ on the physical kets is equivalent to $R(h^{-1})$ on the coefficients, i.e. $R(h) \int \mathrm{d} g f(g) \ket{g}=\int \mathrm{d} g f(gh) \ket{g}$.}
\begin{equation}
\vcenter{\hbox{
\includegraphics[width=0.414\linewidth]{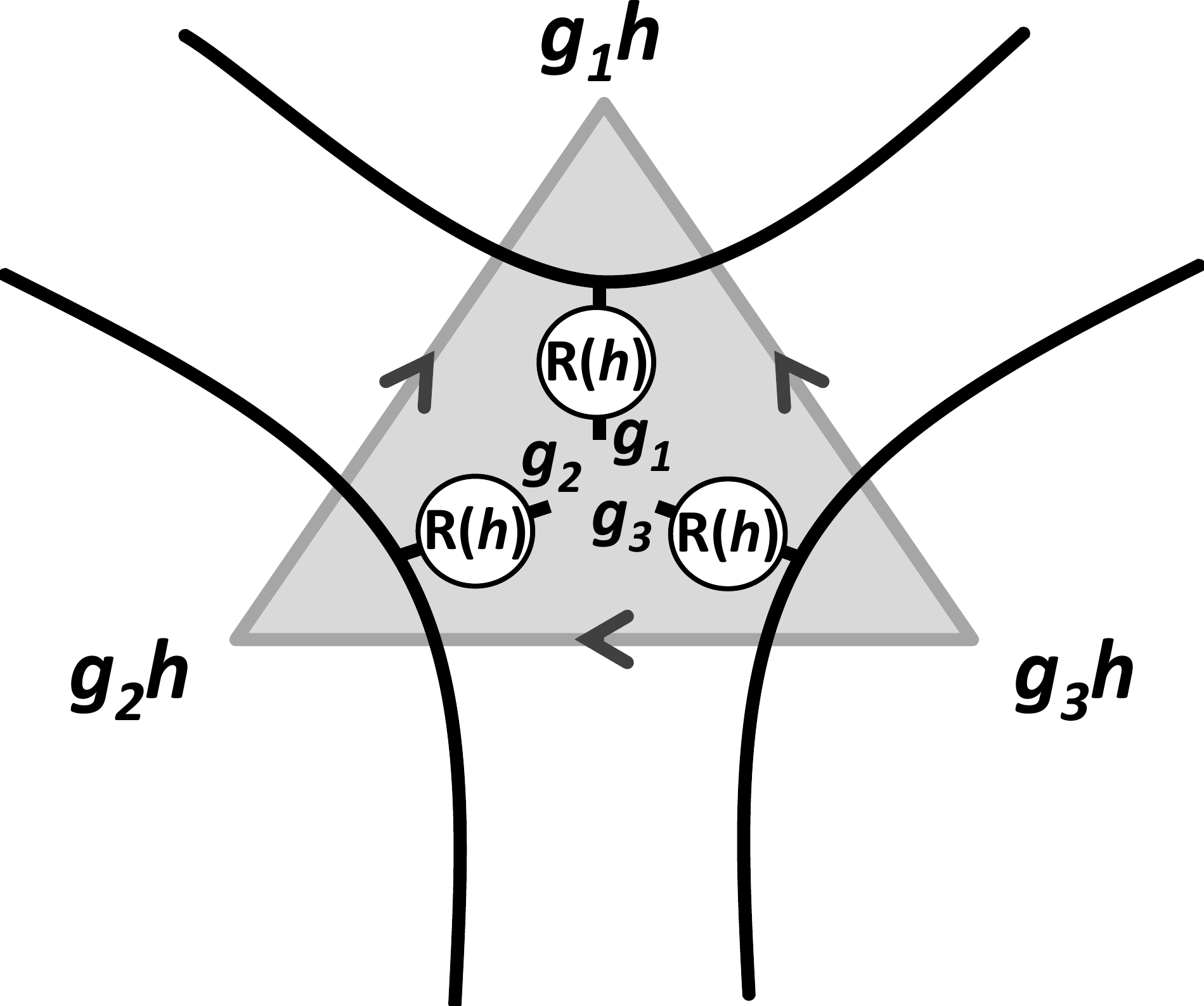}}} =\hspace{-0.2cm}\vcenter{\hbox{
\includegraphics[width=0.505\linewidth]{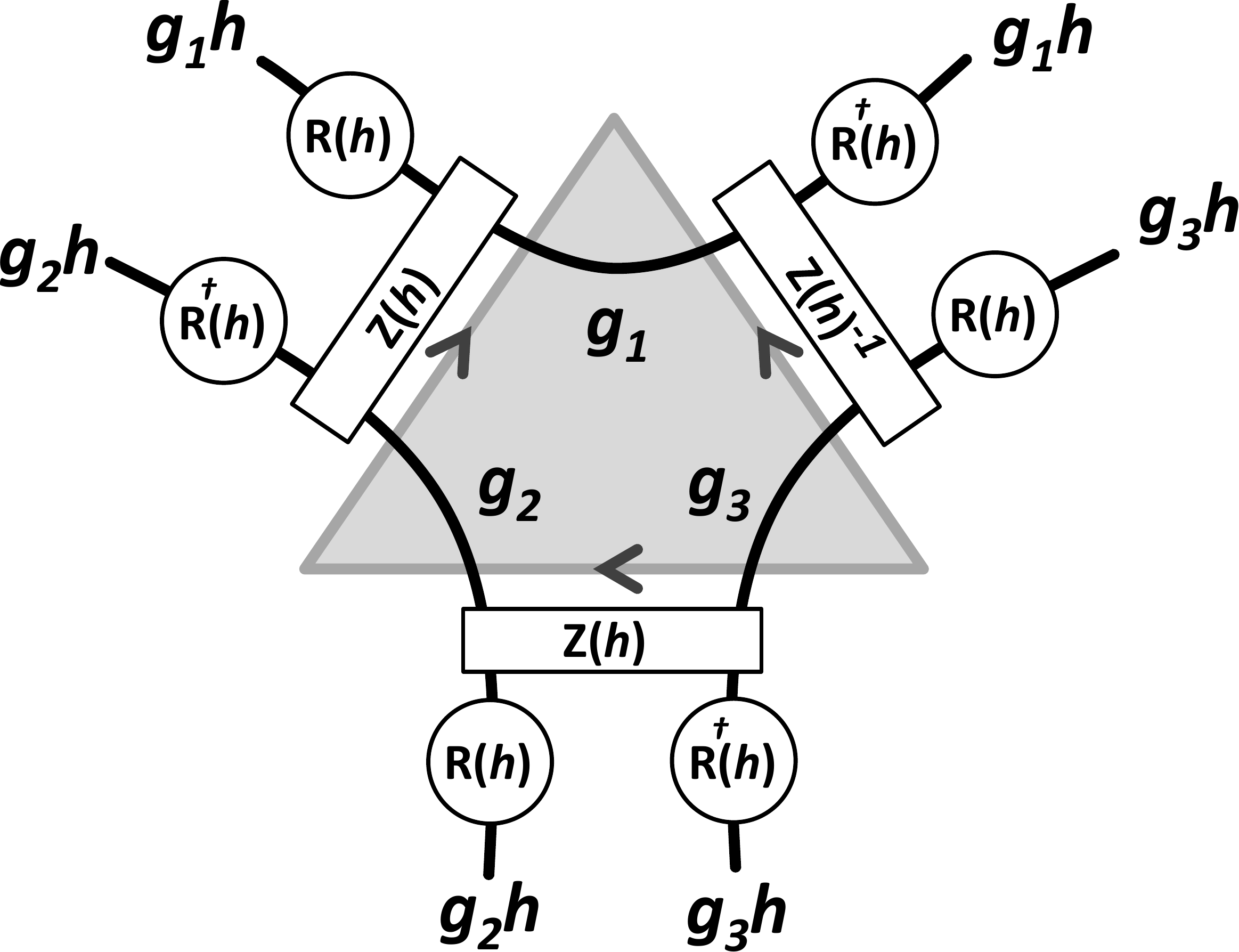}}} \label{d3} 
\end{equation}
Where the left side of the equality depicts the physical symmetry acting on a single tensor, and the right side depicts the virtual representation of the symmetry.

Note that a tensor product of the virtual symmetry matrices $[ Z^{\sigma_{\triangle,e}}_{e} (h) R(h)^{\otimes 2} ]$ in general do not constitute a representation of $\mathsf{G}$. A representation of $\mathsf{G}$ on the virtual level, $V(g)$, is obtained by projecting these matrices onto the subspace on which the PEPS tensor is injective. By doing so we construct MPOs that cannot be factorized as a tensor product. For the current fixed-point example we project $[ Z^{\sigma_{\triangle,e}}_{e} (h) R(h)^{\otimes 2} ]$ onto the subspace of virtual boundary indices corresponding to non-zero values of $\peps_\triangle$, Eq.\eqref{n3}. This yields a MPO $V(h)$ constructed from the following tensors
\begin{equation}\label{mpotensor}
\vcenter{\hbox{
\includegraphics[width=0.3\linewidth]{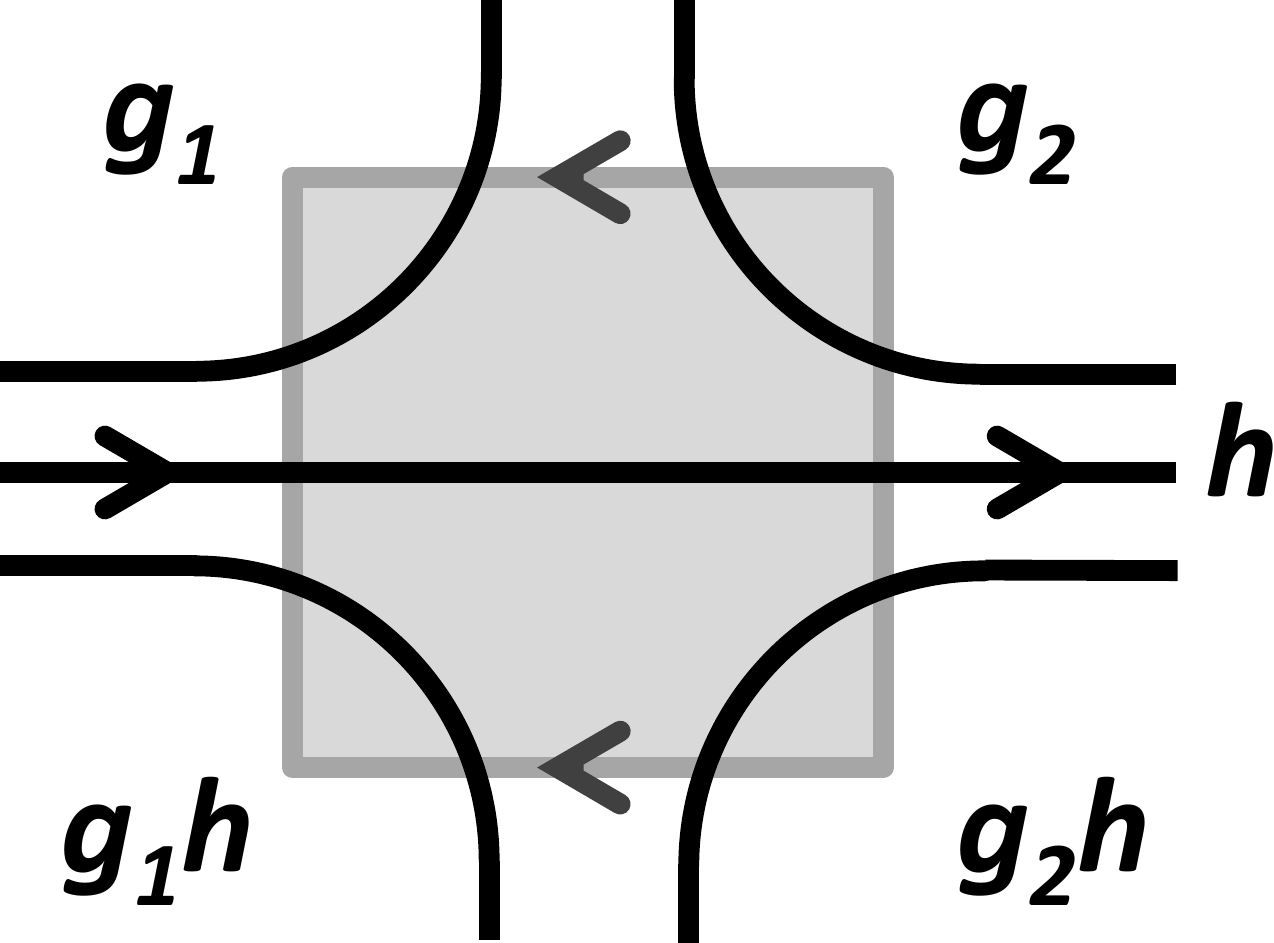}}} = \alpha (g_1g_2^{-1},g_2,h)
\end{equation}
note that for fixed $h$ these MPOs possess a single block. We introduce the isometry $X(h_1,h_2)$ 
\begin{equation}
\vcenter{\hbox{
\includegraphics[width=0.25\linewidth]{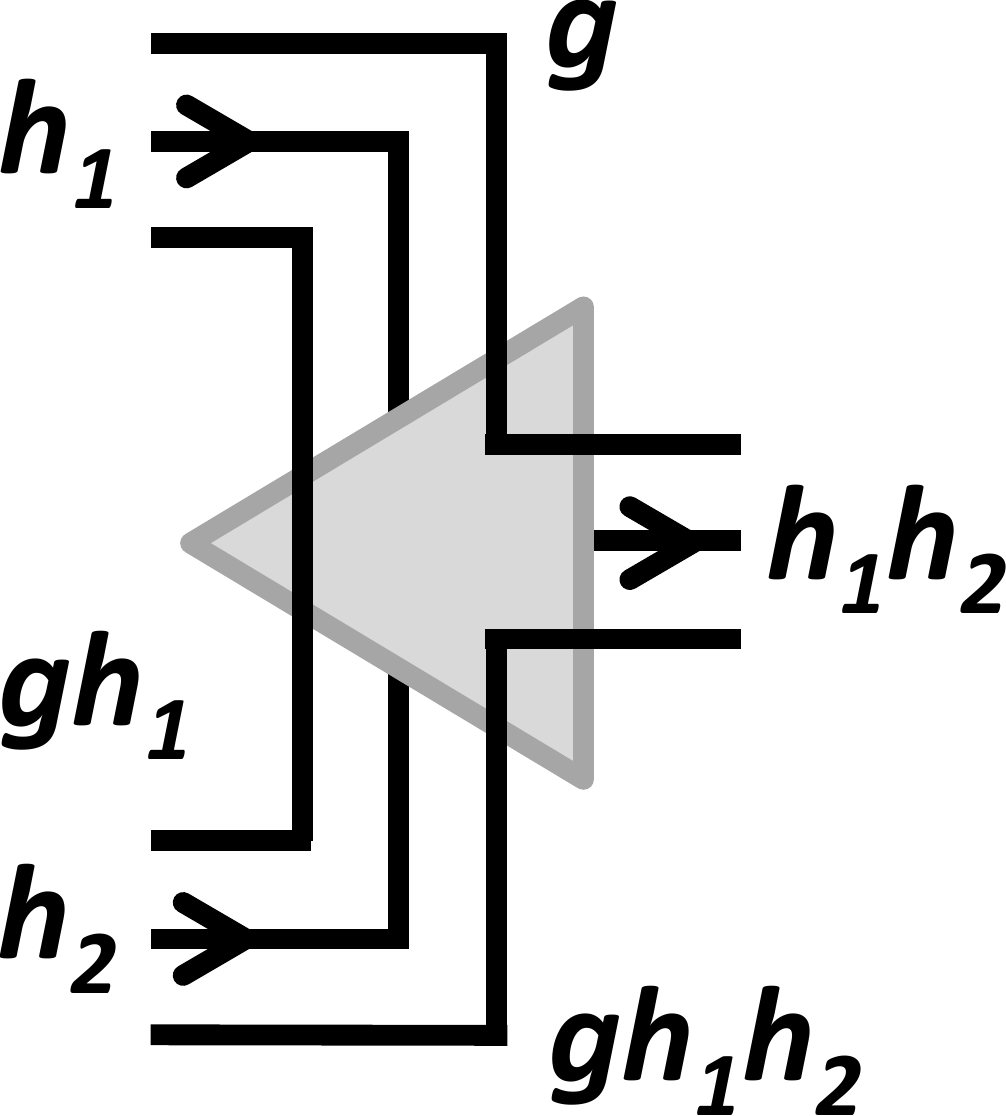}}} \label{d5} = \alpha(g,h_1,h_2)\, ,
\end{equation}
to describe the multiplication of two MPO tensors. With this isometry we have the following relation
\begin{equation}
\vcenter{\hbox{
\includegraphics[width=0.515\linewidth]{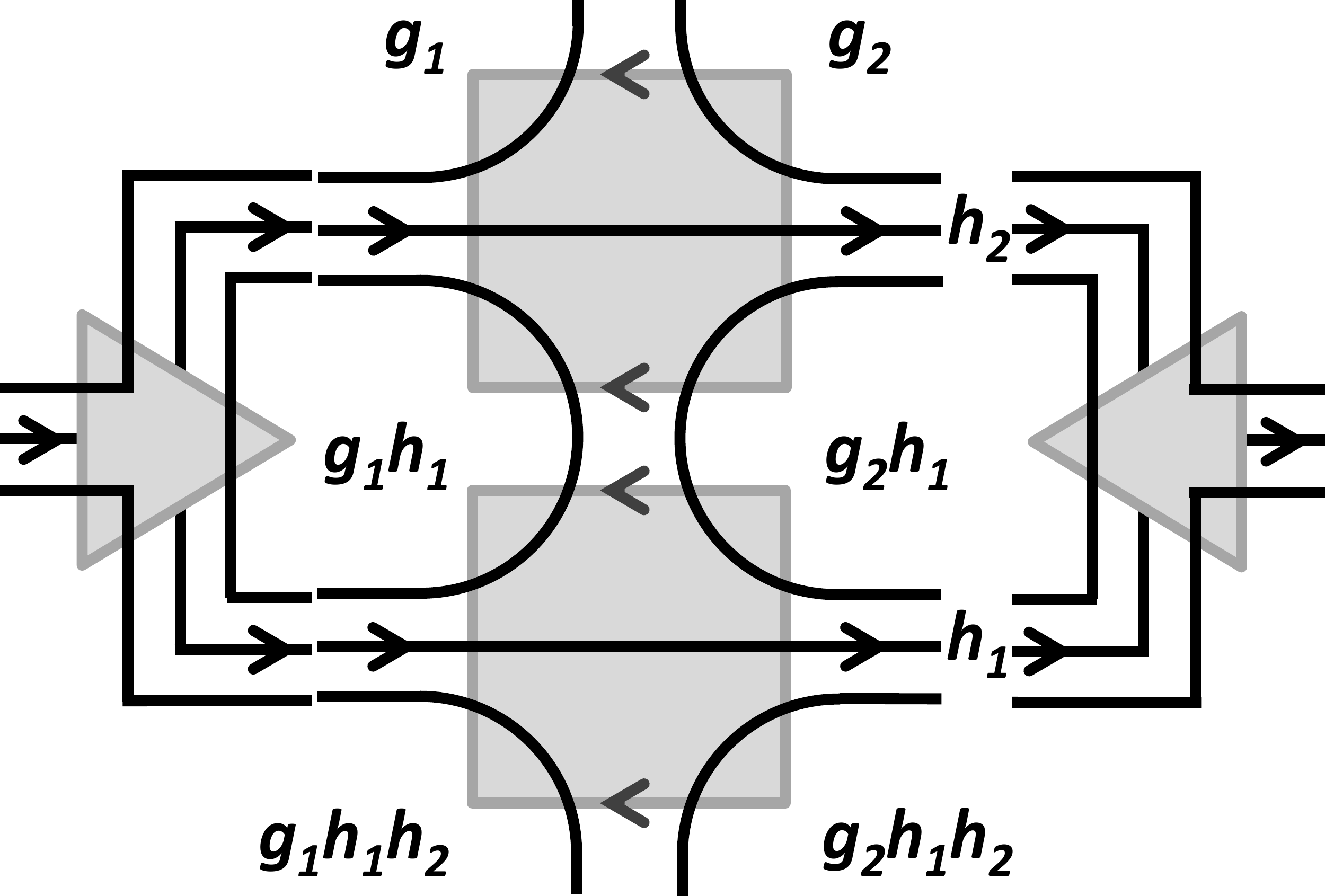}}} =\vcenter{\hbox{
\includegraphics[width=0.33\linewidth]{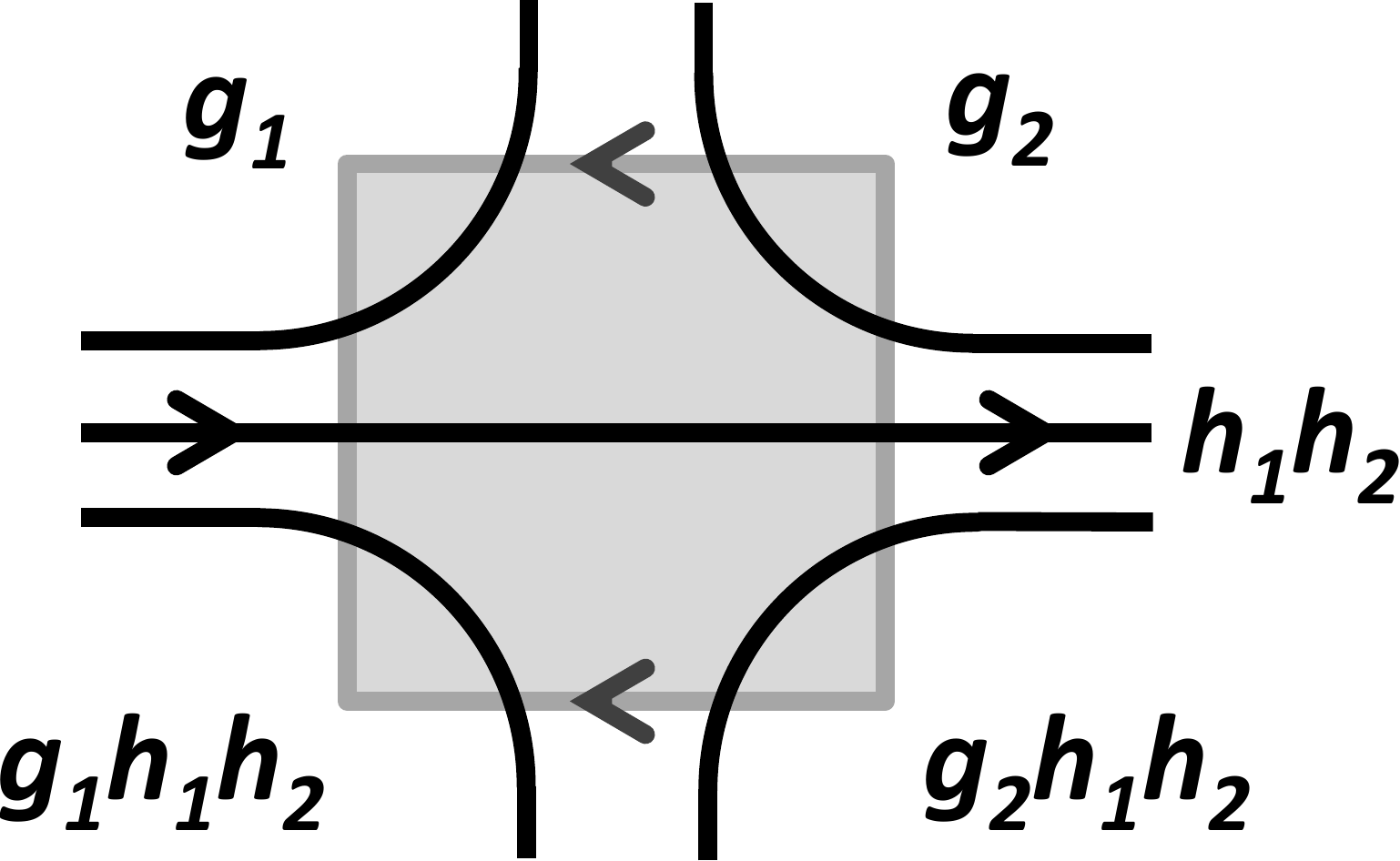}}} \label{d9}
\end{equation}
where the left most tensor of Eq.\eqref{d9} is $X^\dagger(h_1,h_2)$ and we have made use of the 3-cocycle condition~\eqref{cocycle}. This implies that the MPOs  $V(h)$ with fixed inner indices indeed form a representation of $\mathsf{G}$.
Note the stronger \emph{zipper} condition
\begin{equation}
\vcenter{\hbox{
\includegraphics[width=0.52\linewidth]{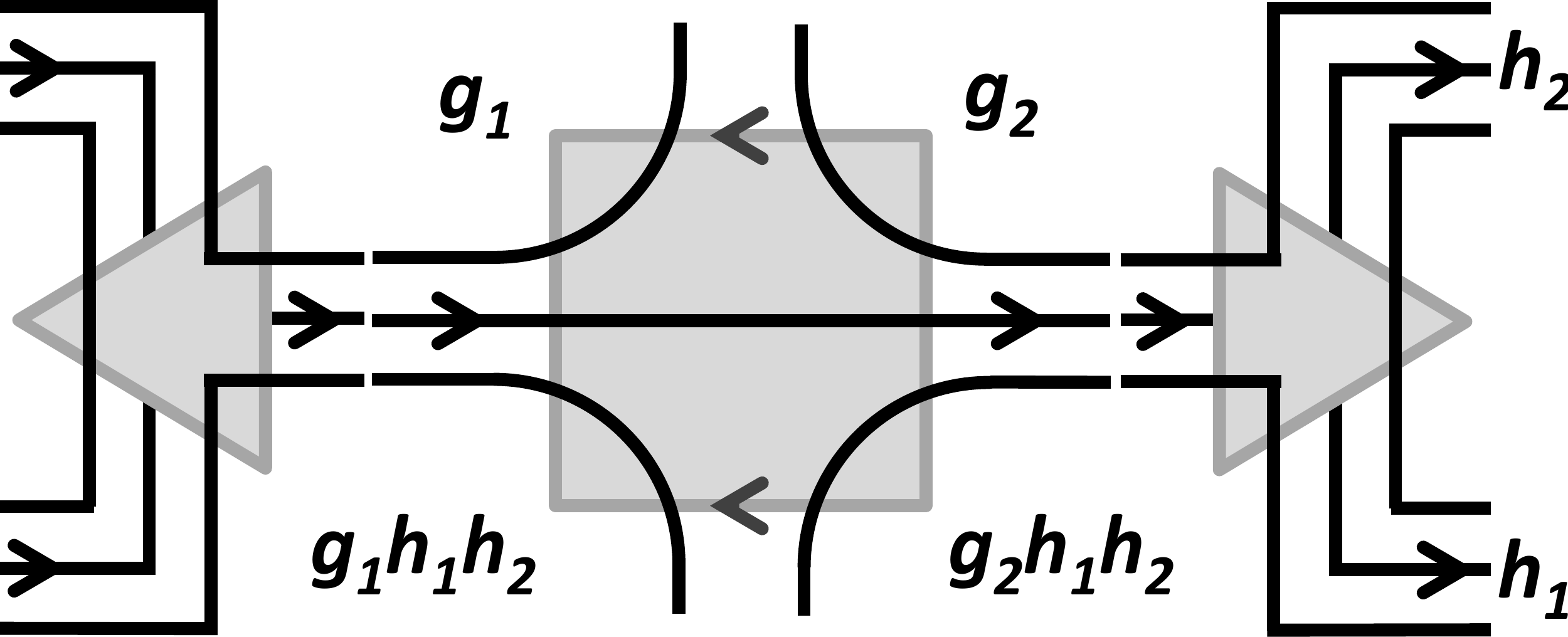}}} =\vcenter{\hbox{
\includegraphics[width=0.28\linewidth]{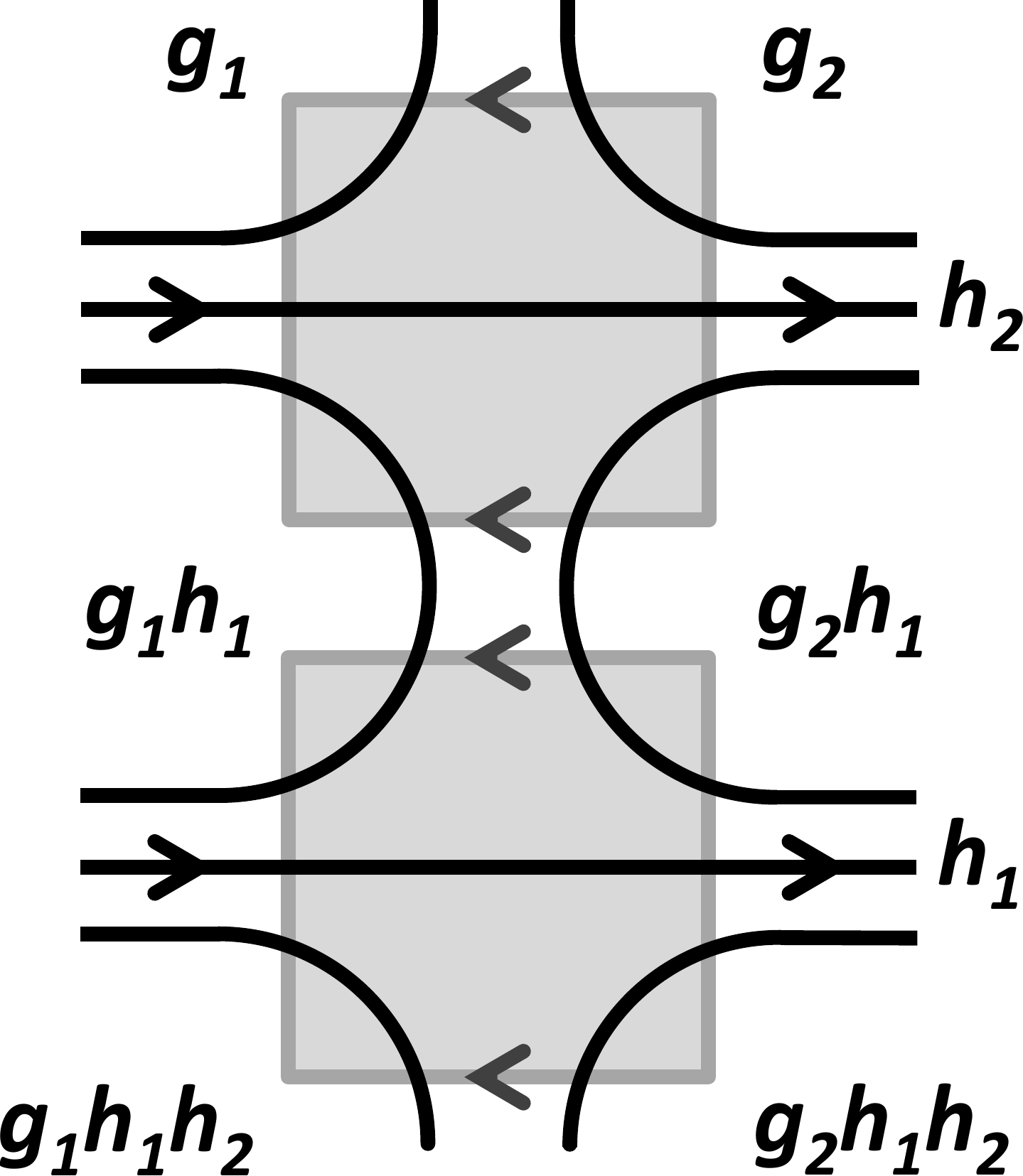}}} \label{e9}
\end{equation}
also holds for this MPO representation.

From Eq.\eqref{tensorsym} it is clear that the PEPS tensors $\peps_\triangle$, Eq.\eqref{n3}, together with the MPOs $V(h)$, defined by Eq.\eqref{mpotensor}, have SPT order described by the framework of Section~\ref{sptpeps}. We now calculate the third cohomology class of the MPOs to determine which SPT phase the model belongs to. For this we see that $X$ obeys the following associativity condition
\begin{multline}\label{n19}
X(h_1h_2,h_3) [X(h_1,h_2)\otimes \mathds{1}_{h_3}] = \\ \alpha^{-1}(h_1,h_2,h_3) \, X(h_1,h_2h_3) [\mathds{1}_{h_1}\otimes X(h_2,h_3)]\, ,
\end{multline}
which is again the 3-cocycle condition Eq.\eqref{cocycle}. From Eq.\eqref{n19} we thus conclude that the short-range entangled states described by the tensors of Eq.\eqref{n3} lie in a symmetry-protected topological phase labeled by the cohomology class $[\alpha^{-1}]\in H^3(G,U(1))$, see Appendix~\ref{a}.

One may be surprised to notice that one layer of strictly local unitaries (equivalent to the local unitary circuit $D_\alpha$ \eqref{dalpha}) acting on the vertices of the PEPS built from the tensors in Eq.\eqref{n3} can remove the 3-cocycles, thus mapping it to a trivial product state. Superficially this seems to contradict the fact that SPT states cannot be connected to the trivial product state by low-depth local unitary circuits that preserve the symmetry. However, this is not the case as this definition requires every individual gate of the circuit to preserve the symmetry~\cite{chen2010local}, which is not true for the circuit just described.

\subsection{ Gauging the fixed-point SPT PEPS}\label{gaugingfixptspt}

We now apply the quantum state gauging procedure of Ref.\cite{Gaugingpaper} to gauge the global symmetry of the fixed-point SPT PEPS defined in the previous subsection. For this we construct a gauging tensor network operator (matching that of Ref.\cite{Gaugingpaper} on the dual triangular graph) that couples gauge degrees of freedom to a given matter state. We proceed by applying a local unitary circuit to disentangle the gauge constraints and explicitly demonstrate that the resulting tensor describes the ground state of a twisted Dijkgraaf-Witten gauge theory.

The gauging map is defined by the following local tensors $G^\triangle:\mathbb{C}(\mathsf{G})^{\otimes 6}\otimes \mathbb{C}(\mathsf{G})^{\otimes 3} \rightarrow\mathbb{C}(\mathsf{G})^{\otimes 6}$
\begin{align}
G^\triangle:=\int \prod_{v\in\triangle} \mathrm{d} h_v \bigotimes_{v\in\triangle} &R_{\triangle,v}(h_v) \bigotimes_{e\in\triangle}[\, \ket{h_{v_e^-}h_{v_e^+}^{-1}}_{\triangle,e}
\nonumber \\
&\otimes( h_{v_e^+}|_{\triangle,e,v_e^+} ( h_{v_e^-}|_{\triangle,e,v_e^-}\, ] ,
\end{align}
note $G^\triangle$ introduces gauge degrees of freedom on the edges. For our example the gauging tensor is
\begin{equation}
\vcenter{\hbox{
\includegraphics[width=0.45\linewidth]{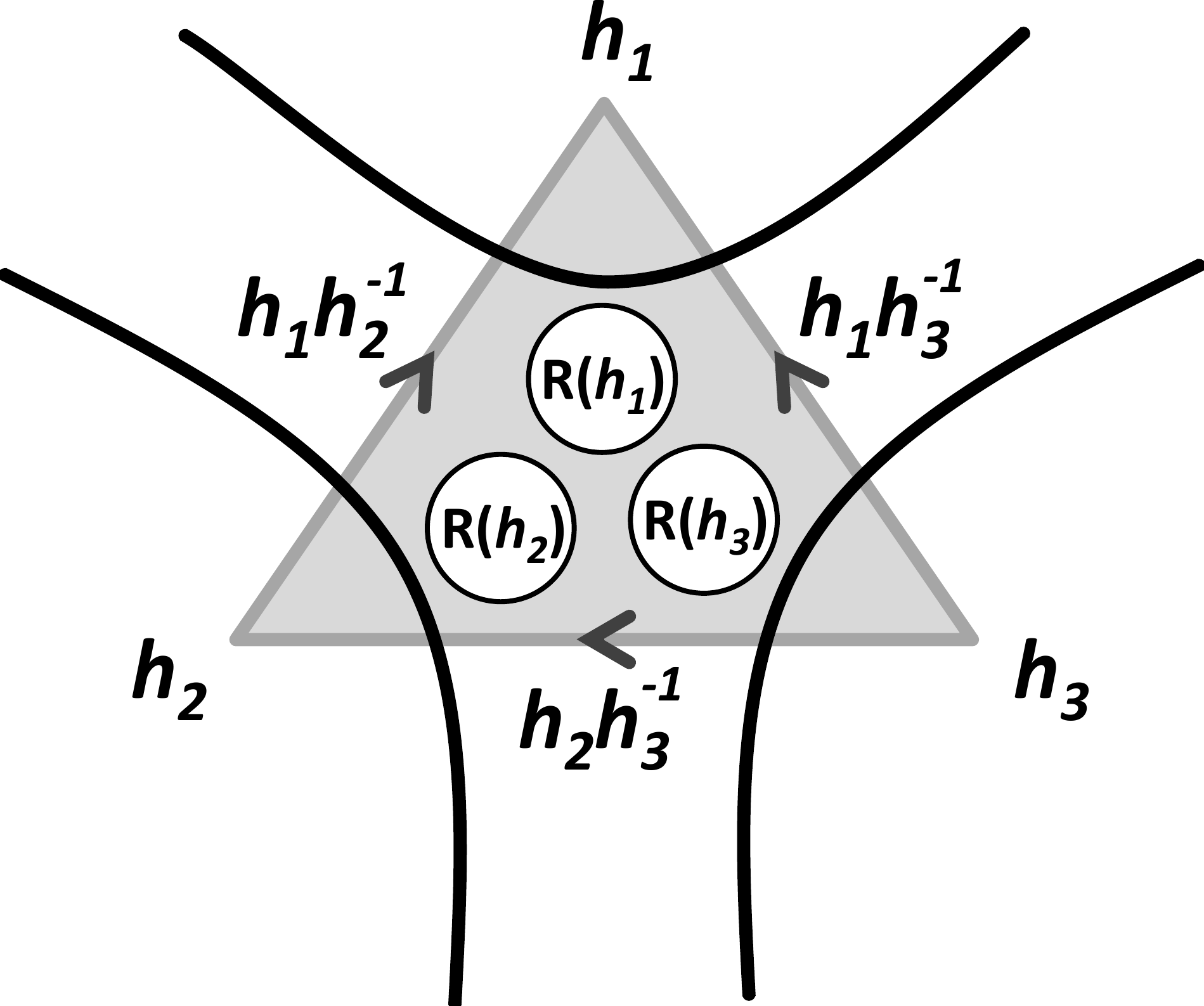}}} \label{d10}
\end{equation}
We can apply the gauging tensors locally to the SPT PEPS to form tensors for a gauge and matter PEPS
\begin{align}
&\bar{\peps}_\triangle:=\int \prod_{v\in\triangle} \mathrm{d} h_v \mathrm{d} g_v\, \tilde{\alpha}_\triangle \bigotimes_{v\in\triangle} \ket{g_vh_v^{-1}}_{\triangle,v} 
\nonumber \\
&\bigotimes_{e\in\triangle} \ket{h_{v_e^-}h_{v_e^+}^{-1}}_{\triangle,e}
 ( g_{v_e^+},h_{v_e^+}|_{\triangle,e,v_e^+}( g_{v_e^-}, h_{v_e^-}|_{\triangle,e,v_e^-}
\end{align}
in our example these are
\begin{equation}
\vcenter{\hbox{
\includegraphics[width=0.49\linewidth]{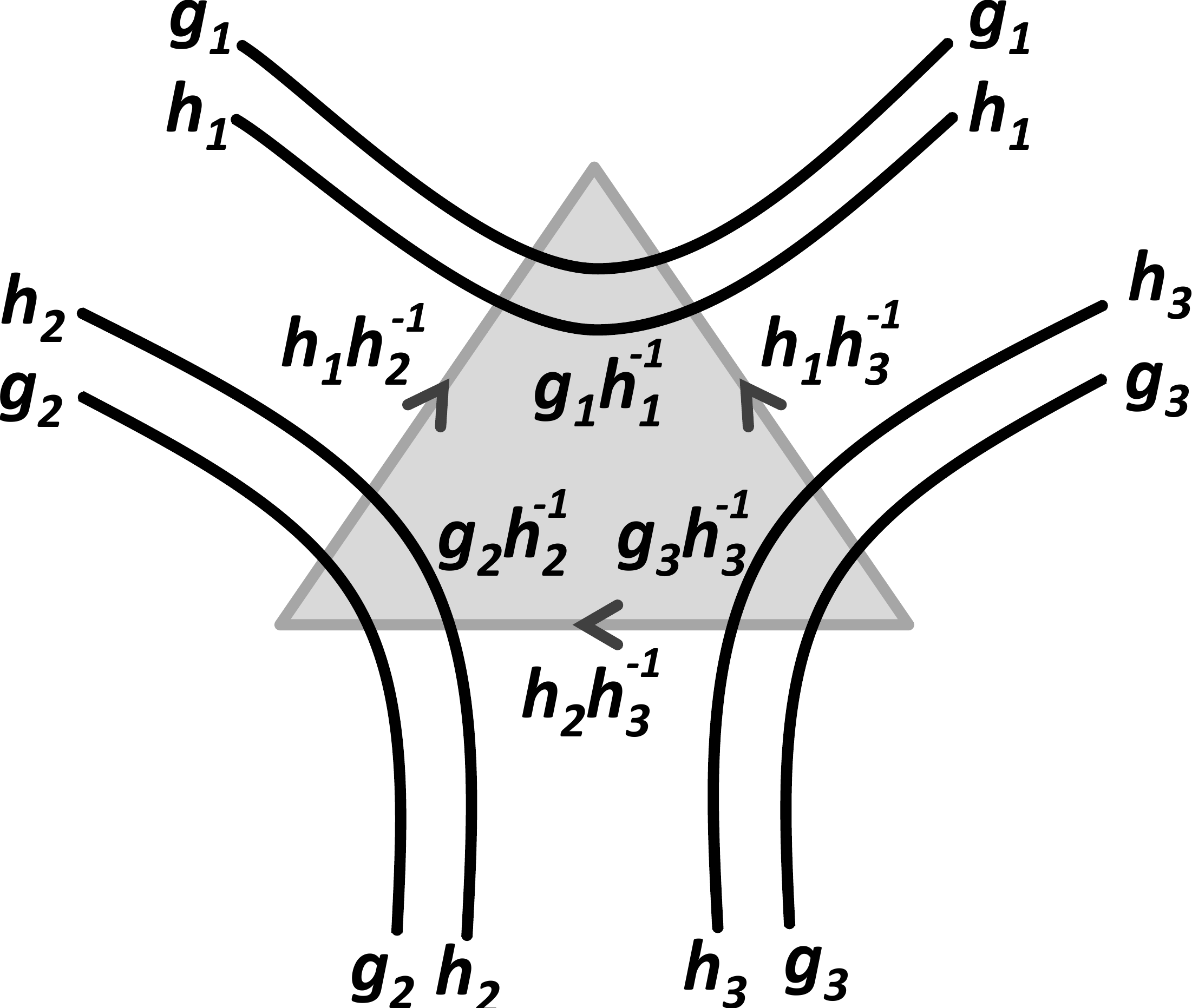}}} = \alpha(g_1g_2^{-1},g_2g_3^{-1},g_3)\, .
\label{d11}
\end{equation}
The gauged PEPS $\ket{\psi_\mathsf{g}}$, built from the tensors $\bar\peps_\triangle$, satisfies local gauge constraints $\tilde{P}_v\ket{\psi_\mathsf{g}} = \ket{\psi_\mathsf{g}}$ for every vertex $v$, where
\begin{align}
\tilde{P}_v:=\int \mathrm{d} g_v \bigotimes_{\triangle \ni v} [ R_{\triangle,v}(h) \bigotimes_{e\in E_v^+} R_{\triangle,e}(g_v) \bigotimes_{e\in E_v^-} L_{\triangle,e}(g_v) ] \nonumber
\end{align}
The gauge and matter tensor $\bar{\peps}_\triangle$ is MPO-injective with respect to a purely virtual symmetry inherited from the symmetry transformation of the SPT tensor $\peps_\triangle$ and it also intertwines a physical symmetry to a virtual symmetry due to the transformation of the gauging tensors
\begin{align}
\bar{\peps}_\triangle \bigotimes_{e\in\triangle} [ Z^{\sigma_{\triangle,e}}_{e} (h) R(h)^{\otimes 2} ] \otimes R(h)^{\otimes 2} =\bar{\peps}_\triangle
\\
\bigotimes_{v\in\triangle} R_{\triangle,v}(h) \bigotimes_{e\in\triangle} R_{\triangle,e}(h) L_{\triangle,e}(h) \bar{\peps}_\triangle 
\nonumber \\
= \bar{\peps}_\triangle  \bigotimes_{e\in\triangle} \openone^{\otimes 2} \otimes L(h)^{\otimes 2}
\end{align}
the latter symmetry reflects the invariance of the full PEPS under the gauge constraints $\tilde{P}_v$.

We next apply a local unitary circuit $\tilde{C}_\fullgraph$ to explicitly map the gauge and matter model to a twisted quantum double ground state on the gauge degrees of freedom alone. This circuit is given by the tensor product of the following local unitary on each site
\begin{align}
\tilde{C}_\triangle := \int \prod_{v\in\triangle} \mathrm{d} g_v \bigotimes_{v\in\triangle} \ket{g_v}\bra{g_v}_v \bigotimes_{e\in\triangle} L_e(g_{v_e^-}) R_e(g_{v_e^+})\, , \nonumber
\end{align}
which maps the gauge constraints to local rank one projectors on the matter degrees of freedom at each vertex $\tilde{C}_\fullgraph \tilde{P}_v \tilde{C}_\fullgraph = \int \mathrm{d} g_v \bigotimes_{\triangle \ni v} R_{\triangle,v}(h)$, fixing the state of the matter to be $\int \mathrm{d} g_v \bigotimes_{\triangle \ni v} \ket{g_v}_{\triangle,v}$. From this we infer that the circuit $\tilde{C}_\fullgraph$ disentangles the gauge from the matter degrees of freedom. To see this explicitly we apply the circuit locally to each PEPS tensor, along with a unitary change of basis on the virtual level (leaving the physical state invariant) to form the tensor $\bar{\bar{\peps}}_\triangle$ which is defined as follows
\begin{align}
& 
\bar{\bar{\peps}}_\triangle:= \tilde{C}_\triangle \bar{\peps}_\triangle \bigotimes_{e\in\triangle} U_{\triangle,e,v_e^+}\otimes U_{\triangle,e,v_e^-} 
\nonumber \\
& \phantom{\bar{\bar{\peps}}_\triangle}\ 
= \int \prod_{v\in\triangle} \mathrm{d} k_v \mathrm{d} g_v \,\tilde{\alpha}_\triangle \bigotimes_{v\in\triangle} \ket{k_v}_{\triangle,v}
\bigotimes_{e\in\triangle} [\, \ket{g_{v_e^-}g_{v_e^+}^{-1}}_{\triangle,e}
\nonumber \\
& \phantom{\bigotimes_{e\in\triangle} \ket{g_{v_e^-}g_{v_e^+}^{-1}} }
\otimes ( g_{v_e^+},k_{v_e^+}|_{\triangle,e,v_e^+}( g_{v_e^-}, k_{v_e^-}|_{\triangle,e,v_e^-} \, ]
\end{align}
where $U:=\int \mathrm{d} g \ket{g}\bra{g} \otimes S L^\dagger(g)$, with $S\ket{g}:=\ket{g^{-1}}$, satisfies $(g,h| U= (g,gh^{-1}| $. For our example this tensor is given by
\begin{equation}
\vcenter{\hbox{
\includegraphics[width=0.47\linewidth]{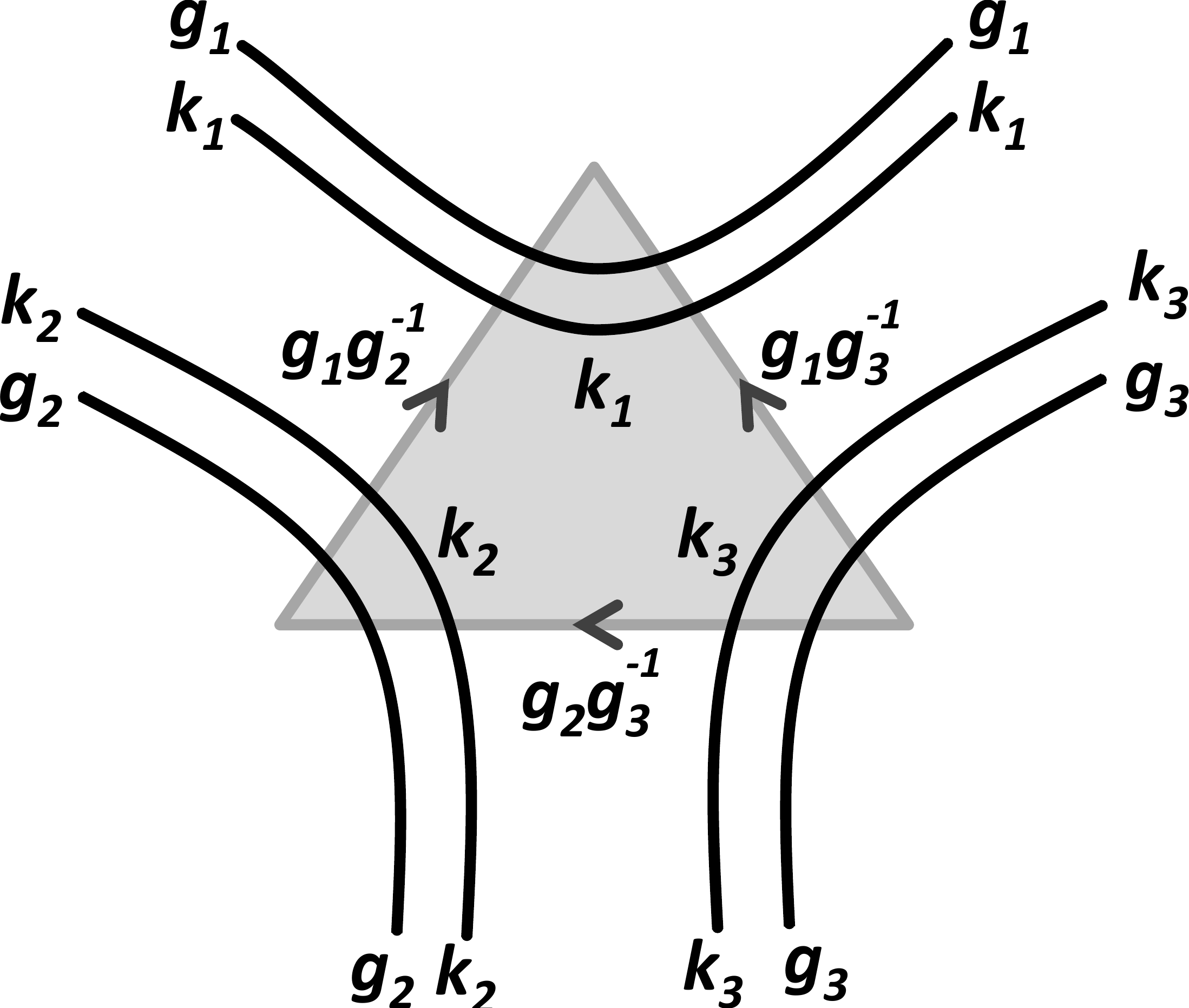}}} = \alpha(g_1g_2^{-1},g_2g_3^{-1},g_3)
\label{d12}
\end{equation}
This disentangled PEPS tensor $\bar{\bar{\peps}}_\triangle$ is now MPO-injective on the support subspace of the projection MPO given by a normalized sum of the symmetry MPOs from the SPT PEPS. Moreover the intertwining condition maps the physical vertex symmetry to a trivial action on the virtual space
\begin{align}
\bar{\bar{\peps}}_\triangle \bigotimes_{e\in\triangle}[ Z^{\sigma_{\triangle,e}}_{e} (h) R(h)^{\otimes 2} ] \otimes \openone ^{\otimes 2} = \bar{\bar{\peps}}_\triangle
 \\
\bigotimes_{v\in\triangle} R_{\triangle,v}(h) \bar{\bar{\peps}}_\triangle = \bar{\bar{\peps}}_\triangle \bigotimes_{e\in\triangle} \openone^{\otimes 2} \otimes R(h)^{\otimes 2}\, .
\end{align}
From this we see that $\bar{\bar{A}}_\triangle$ separates into a trivial local component on the matter degrees of freedom yielding the state $\bigotimes_v \int\mathrm{d} g_v \bigotimes_{\triangle \ni v} \ket{g_v}_{\triangle,v}$, and the following tensors on the gauge degrees of freedom
\begin{align}\label{n12}
 \int \prod_{v\in\triangle} \mathrm{d} g_v  \tilde{\alpha}_\triangle
 \bigotimes_{e\in\triangle} \ket{g_{v_e^-}g_{v_e^+}^{-1}}
 ( g_{v_e^+}|_{\triangle,e,v_e^+}( g_{v_e^-}|_{\triangle,e,v_e^-} .
\end{align}
These tensors define a PEPS on the gauge degrees of freedom that is a ground state of a 2D twisted quantum double with 3-cocycle $\alpha$. Note this PEPS matches the standard representation of the ground state on the subspace obtained by mapping $\bigotimes_{\triangle\ni v}\ket{g}_{\triangle,v}\mapsto \ket{g}_v$ and $\bigotimes_{\triangle\ni e}\ket{g}_{\triangle,e}\mapsto \ket{g}_e$. For our example this tensor is
\begin{equation}
\vcenter{\hbox{
\includegraphics[width=0.4\linewidth]{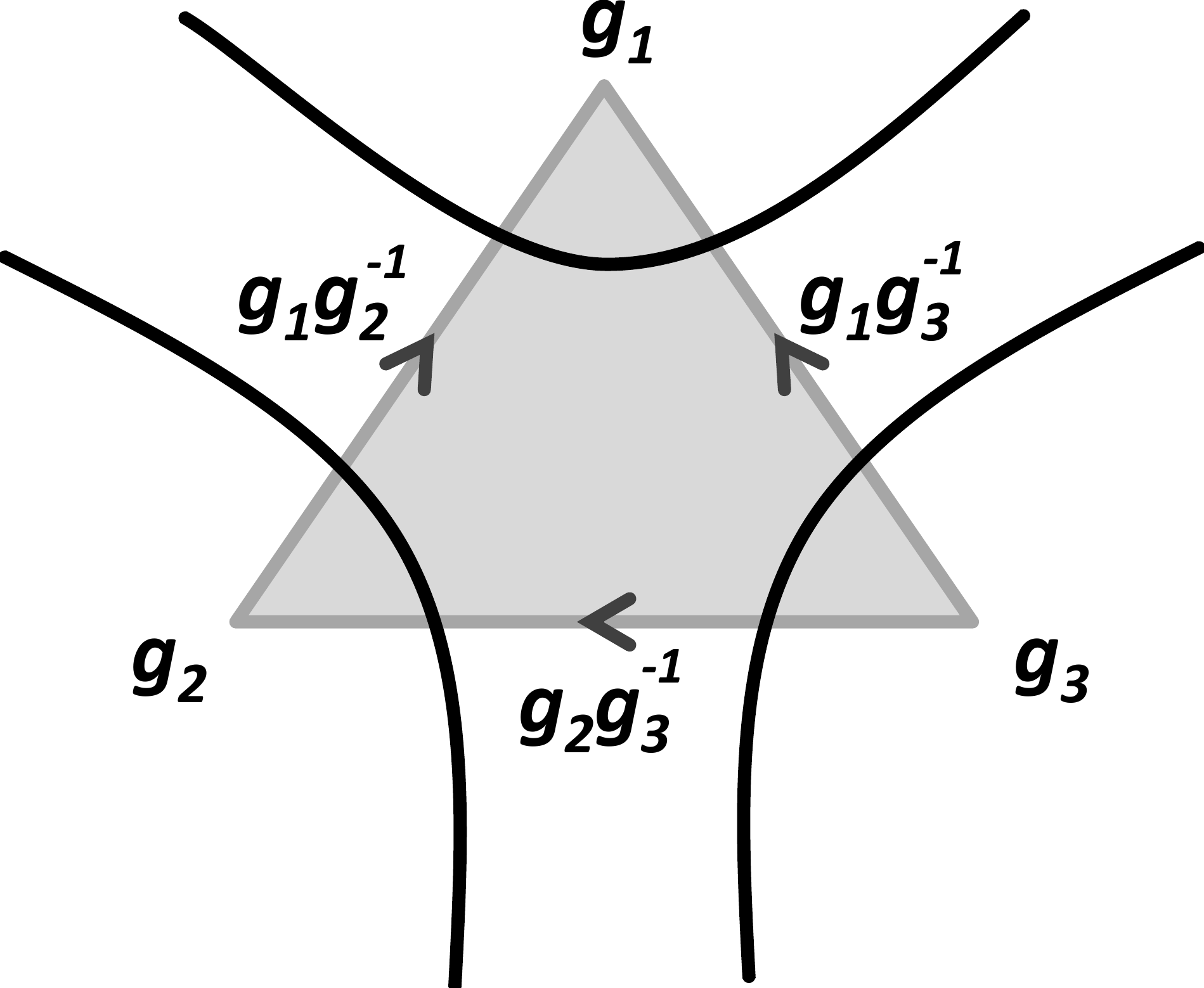}}} = \alpha(g_1g_2^{-1},g_2g_3^{-1},g_3)
\label{d13}
\end{equation}
note in the Abelian case the tensors in Eq.\eqref{d13} reduce to the string-net tensors~\cite{stringnet1,stringnet2} after a suitable mapping between 3-cocycles and $F$-symbols~\cite{ZNstringnet} (in the non-Abelian case one has to change to the basis of irreducible representations to make the identification). 

\subsection{Perturbations away from fixed-points}

The examples presented thus far in this section are all fixed-point states under a real space blocking renormalization group flow and have zero correlation length. This corresponds to the PEPS tensor that builds the state being of MPO-isometric type~\cite{MPOpaper}. 
More generally one could add an arbitrary perturbation that lies within the MPO-injectivity subspace (this can be constructed by applying the MPO projector to an arbitrary perturbation) to the MPO-isometric PEPS tensor to find a new MPO-injective PEPS that will generically have a finite correlation length. For a sufficiently small symmetric perturbation the resulting MPO-injective PEPS will lie in the same phase of matter as the fixed-point MPO-isometric PEPS~\cite{cirac2013robustness,Buerschaper14}. 

The simplest explicit perturbations away from fixed-point tensors are given by local filtering operations on the physical indices. For a given MPO-injective PEPS tensor $\peps$ local filtering by a projector $P$ generates a family of MPO-injective deformations $ \{P(\lambda)A \ |\ \lambda\in [0,1)\}$ where ${P(\lambda):=(1-\lambda)\openone + \lambda P}$. 
For topological PEPS $P$ can be an arbitrary projector on the physical index, while for SPT PEPS it must commute with the on-site symmetry action. This path of deformations can move from one phase of matter to another, for instance if we let $P=\ket{0}\bra{0}$ the deformation can induce an anyon condensation transition if $\peps$ describes a topologically ordered ground state~\cite{Gaugingpaper,shadows,marien2016condensation}. In the SPT case with on-site group action $R(g)$ one can consider ${P=\ket{\tilde e}\bra{\tilde e}}$, the projector onto the trivial representation, where ${\ket{\tilde e}=\frac{1}{|\mathsf{G}|} \sum\limits _{g\in \mathsf{G}} \ket{g} }$ to find a symmetric interpolation to the trivial phase. 
A framework to understand these transitions in terms of symmetry breaking of the virtual symmetry is described in Refs.~\cite{shadows,marien2016condensation}.

\section{ Conclusions}
{We have presented a unified picture for the characterization of all gapped phases, possibly with respect to certain physical symmetries, within the framework of PEPS in terms of virtual MPO symmetries.
 To achieve this we developed a characterization of global symmetry in the framework of MPO-injective PEPS~\cite{Buerschaper14,MPOpaper}. In contrast to the injective case~\cite{canonicalPEPS}, where the symmetry representation on the virtual indices factorizes into a tensor product, a MPO-injective PEPS tensors can have a virtual symmetry representation given by unfactorizable MPOs. We subsequently identified the short-range entangled PEPS to be those having a single block in the projection MPO onto the injectivity subspace. If the accompanying single block MPO virtual symmetry representation has a non-trivial third cohomology class it gives rise to unconventional edge properties and thus to symmetry-protected topological PEPS.} 
Our identification of the virtual entanglement structure of PEPS with SPT order opens new routes to study transitions between SPT phases by utilizing methods that have been developed to study anyon condensation transitions of topological phases~\cite{shadows,transfermatrix}.

{We demonstrated that applying the quantum state gauging procedure~\cite{Gaugingpaper} to a SPT PEPS transforms its MPO representation of $\mathsf{G}$ into a purely virtual symmetry of the gauged tensors. This implies that the resulting gauge-invariant PEPS also satisfies the axioms of MPO-injectivity, but with a projection MPO onto the injectivity subspace with a block structure labeled by the group elements $g\in\mathsf{G}$. 
This block structure of the projection MPO, together with the third cohomology class label, characterizes the phases of the twisted quantum double models which are known to have intrinsic topological order.} It was shown in Ref.\cite{MPOpaper} that the projection MPO determines all the topological properties of the gauged PEPS. This relation explains the mechanism behind the braiding statistics approach to SPT phases~\cite{LevinGu} at the level of the corresponding quantum states. It furthermore reveals  that both the gauging and boundary theory approaches to classifying SPT phases are recast in the PEPS framework as the classification of a common set of MPOs. We have illustrated these concepts for a family of RG fixed-point states, containing a representative for all two-dimensional bosonic SPT phases with a finite on-site symmetry group.

To prove these results we developed new tools to deal with orientation dependent MPO tensors and used them to calculate the symmetry action on monodromy defected and symmetry twisted states and also modular transformations, before and after gauging, in terms of a single tensor.

The general formalism presented in this paper describes both local physical symmetries and topological order of PEPS with virtual MPO symmetries. Furthermore, it captures the general action of a symmetry on a PEPS with topological order and hence yields a natural framework for the study of symmetry-enriched topological phases (SET). The quantum state gauging procedure can be adapted to gauge only a normal subgroup of the global symmetry group of a SPT PEPS, which allows one to explicitly construct families of SET PEPS. An open question is how the corresponding MPOs encode the discrete, universal labels of the SET phase and how to extract them. We further expect that a better understanding of excitations in MPO-injective PEPS~\cite{nick} will yield insights into the physical properties of SET phases such as symmetry fractionalization. We plan to study these matters in future work~\cite{michael}.

In this work we only explicitly consider finite on-site unitary symmetry actions. It is an interesting and relevant question to generalize this to time-reversal and continuous Lie group symmetries as well as lattice translation and point group symmetries. Progress has been made on incorporating these types of symmetries into PEPS in Ref.~\cite{jiang2015symmetric}. 
In particular since time-reversal can be realized as a local action on the PEPS tensors~\cite{Gaugingtime} a similar approach to that used here should apply, with some extra care necessary due to the possible action of time reversal on the symmetry MPOs. 

Another question which presents itself is how to generalize the constructions presented in this paper to fermionic systems. Partial progress has been made in the direction of applying the same principles to the formalism of fermionic PEPS~\cite{fpeps}. This has led to a (partial) classification of fermionic SPT phases~\cite{williamson2016fermionic,bultinck2016fermionic} based on supercohomology~\cite{supercohomology} and the existence of Majorana-type defects~\cite{cheng2015towards}. The quantum state gauging procedure works equally well for fermionic systems, but the gauge degrees of freedom are always bosonic. It would thus be interesting to see how fermionic SPT order can be probed in this way.

Our identification of SPT PEPS in 2D as being injective with respect to an injective MPO hints at a hierarchical definition of SPT PEPS in arbitrary dimension with an injective tensor network object associated to each codimension. This appears to recover the cohomological classification of bosonic SPT states in arbitrary dimensions by a generalization of the argument from~\cite{Chen}. We plan to explore this further in future work.
\\ \\
\emph{Acknowledgments -} We acknowledge helpful discussions with N. Schuch. This work was supported by EU grant SIQS and ERC grant QUERG, the Odysseus grant from the Research Foundation Flanders (FWO) and the Austrian FWF SFB grants FoQuS and ViCoM. M.M. and J.H. further acknowledge the support from the Research Foundation Flanders (FWO).

\bibliography{SPT}

\appendix

\section{ Axioms for MPO-injectivity}\label{a}
This section reviews {the axioms of MPO-injectivity as presented} in Ref.\cite{MPOpaper}.

We interpret the tensors $A$ of a MPO-injective PEPS as linear maps from the virtual to the physical space and apply a distinguished generalized inverse $A^+$, which gives rise to a projector that can be written as a MPO:
\begin{align}
\vcenter{\hbox{
\includegraphics[width=0.14\linewidth]{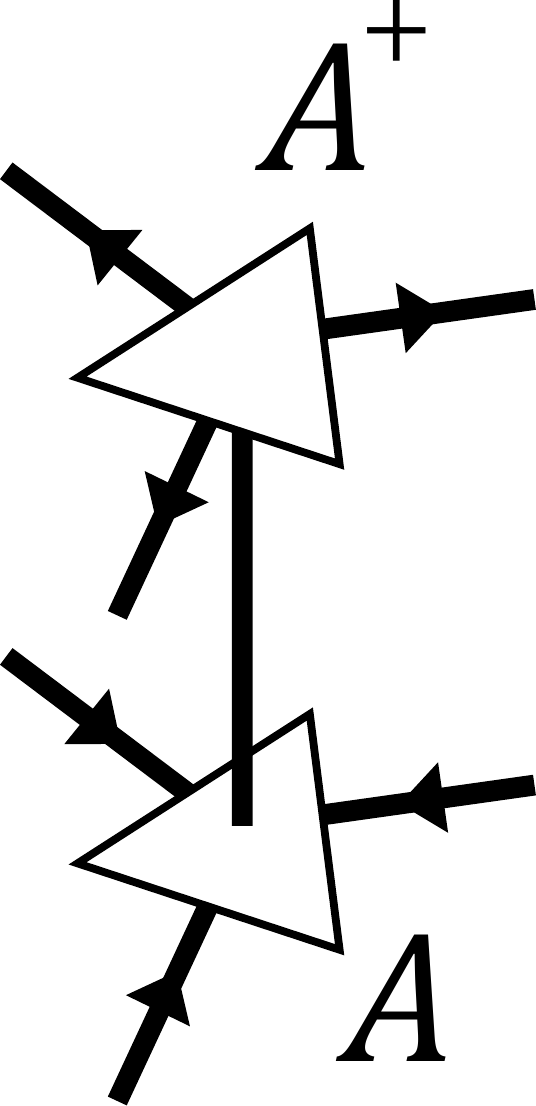}}} \label{n41}
\ = \vcenter{\hbox{
\includegraphics[width=0.25\linewidth]{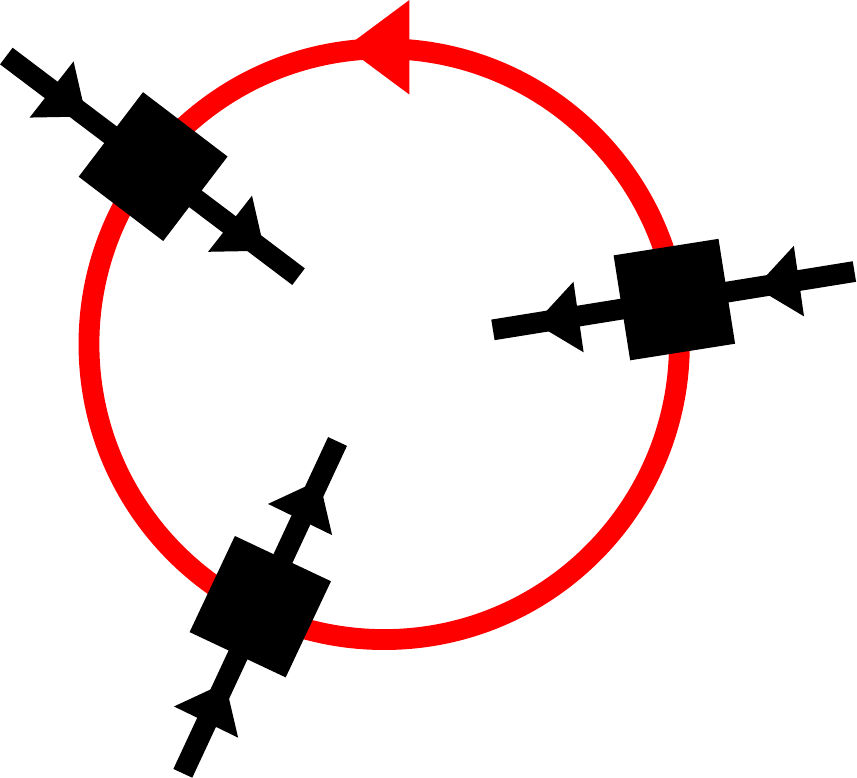}}}
\end{align}
We further require this MPO to satisfy the \emph{pulling through} property shown in Eq.\eqref{n42}.
\begin{align}
\vcenter{\hbox{
\includegraphics[width=0.18\linewidth]{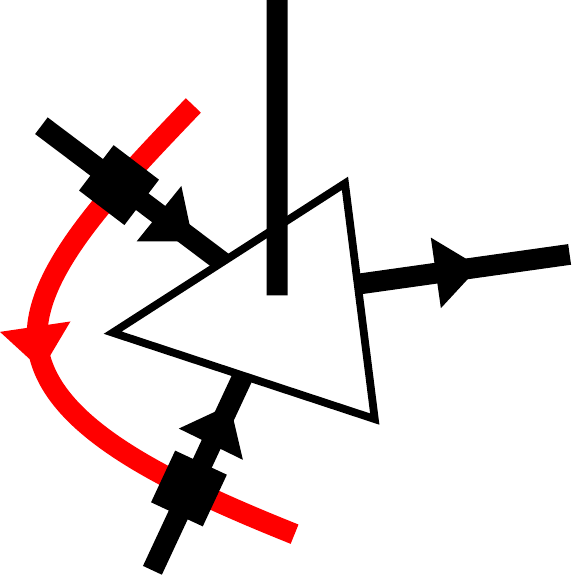}}} \label{n42}
\ = \vcenter{\hbox{
\includegraphics[width=0.18\linewidth]{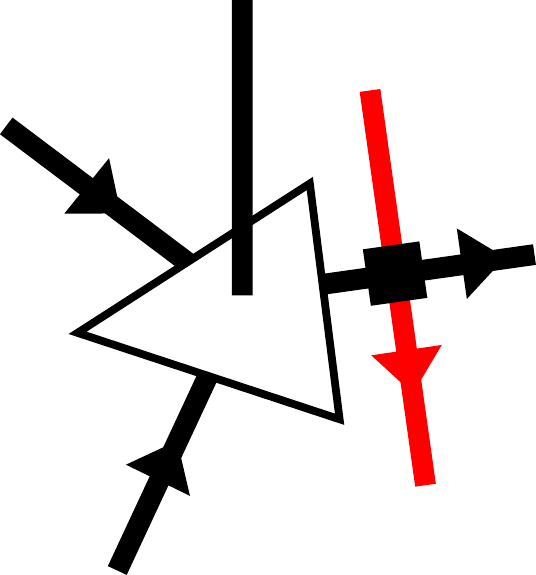}}}
\end{align}
The same property should also hold where the MPO gets pulled from three virtual indices to one or vice versa. This makes the presence of this MPO locally undetectable in the PEPS. Using the pulling through property, it is easy to check that the requirement for the MPO to be a projector is equivalent to the property shown in Eq.\eqref{n43}
\begin{align}
\vcenter{\hbox{
\includegraphics[width=0.13\linewidth]{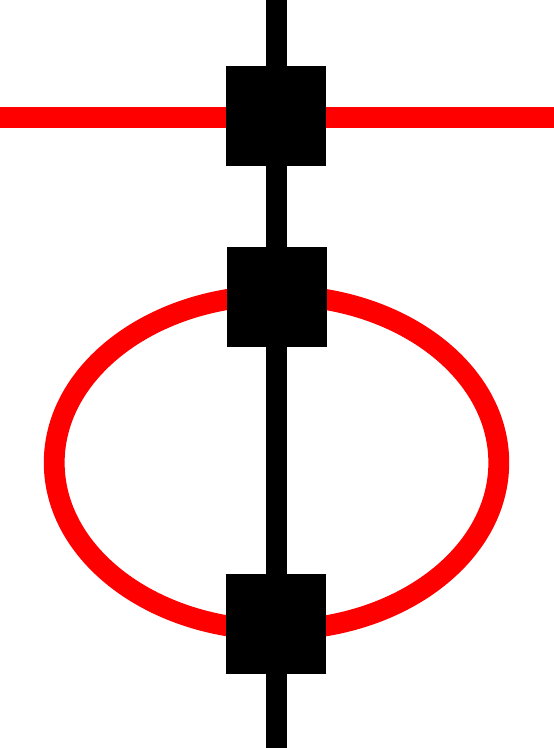}}} \label{n43}
\ = \vcenter{\hbox{
\includegraphics[width=0.13\linewidth]{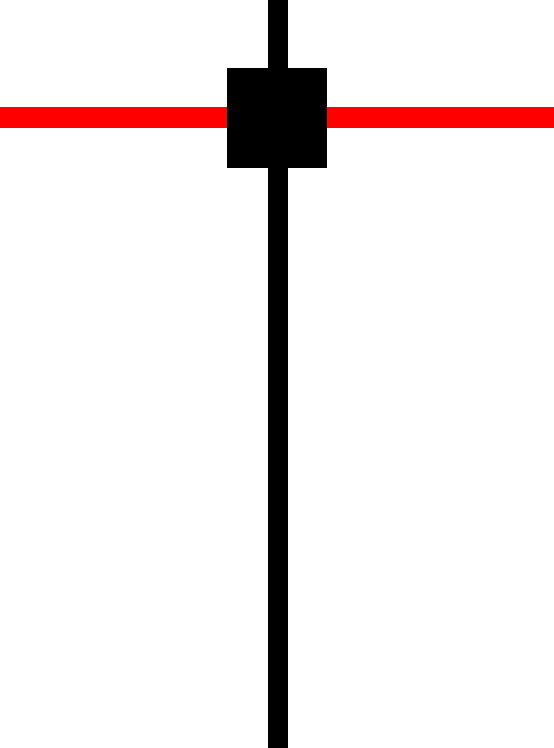}}}
\end{align}
We also need a technical requirement such that the properties of the PEPS grow in a controlled way with the number of sites. For example, we want two concatenated tensors to be injective on the support subspace of the projection MPO surrounding these two tensors. For this we need that there exists a tensor $E$, depicted in (\ref{n44}),
\begin{align}
E:= \label{n44}
\vcenter{\hbox{
\includegraphics[width=0.15\linewidth]{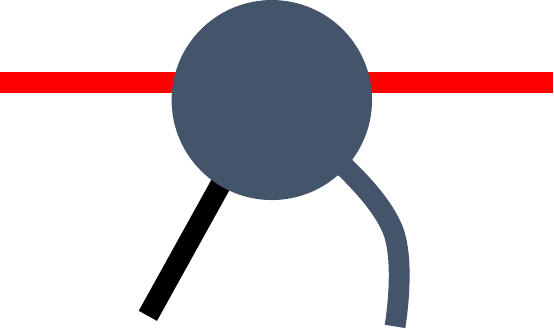}}}
\end{align}
such that we have the \emph{extended inverse} property (\ref{n45}).
\begin{align}
\vcenter{\hbox{
\includegraphics[width=0.25\linewidth]{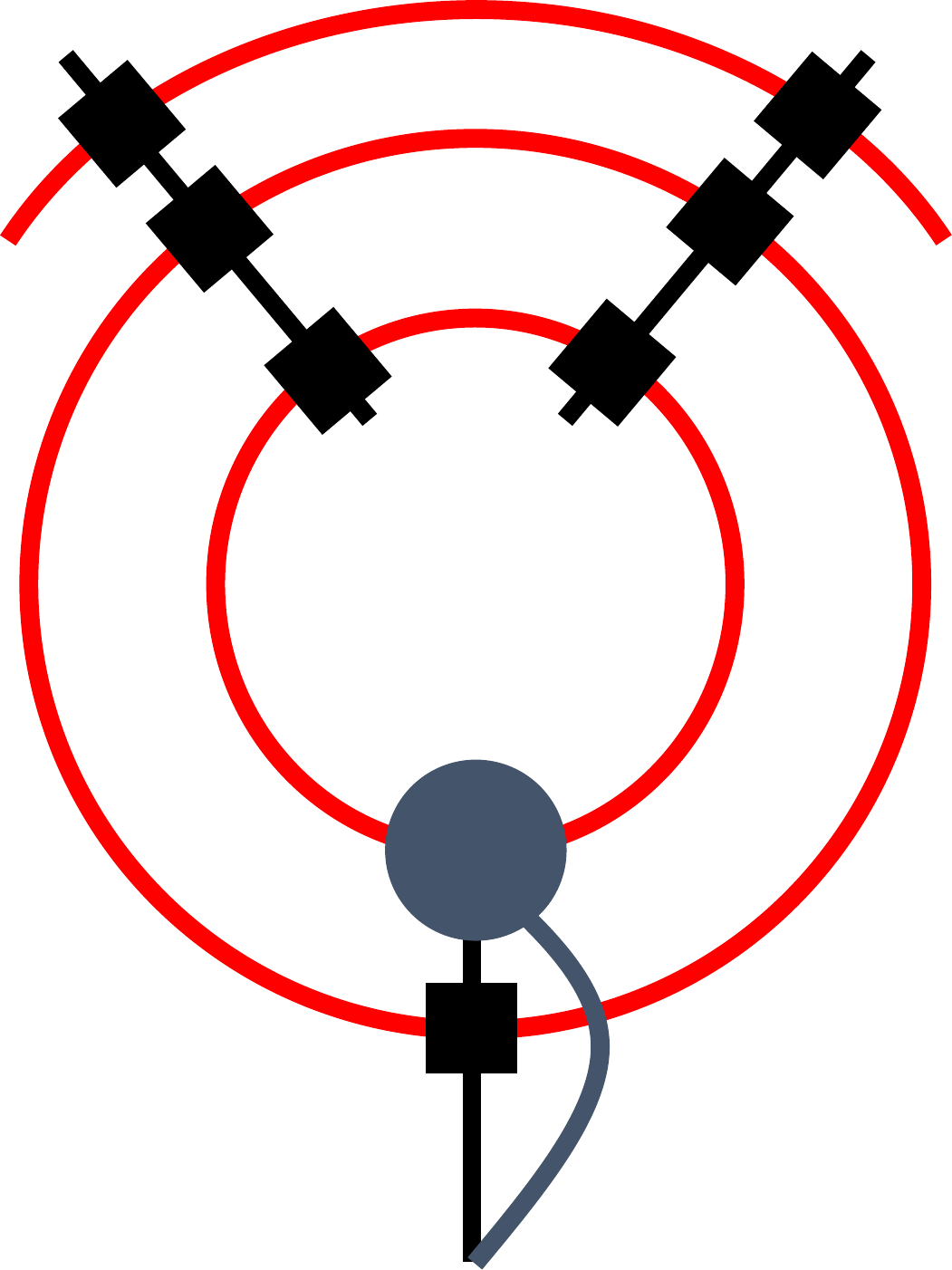}}} \label{n45}
 \ = \vcenter{\hbox{\raisebox{1.9 cm}{
\includegraphics[width=0.25\linewidth]{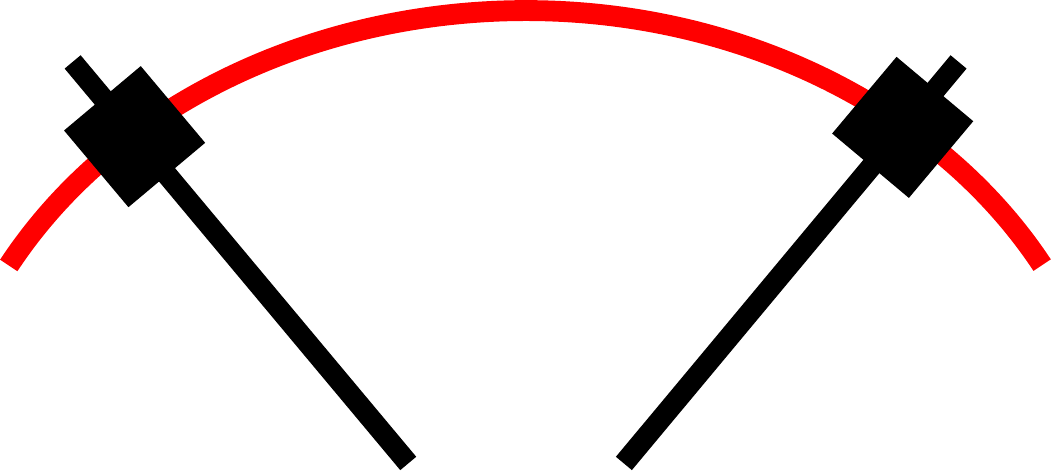}}}}
\end{align}
The extended inverse property allows one to prove many useful things such as the intersection property or an explicit expression for the ground state manifold on a torus~\cite{MPOpaper}. It turns out that under very reasonable assumptions about the projection MPO the extended inverse condition is automatically satisfied~\cite{nick}.

{\section{Uniqueness of SPT PEPS ground state}\label{b} 
In this appendix we demonstrate that the parent Hamiltonian of a MPO-injective PEPS with a single block projection MPO has a unique ground state on the torus (i.e. no topological degeneracy). A similar argument holds for higher genus surfaces.
\\
For a Hermitian projection MPO there is no need to keep track of a direction on the internal leg of the MPO, we also ignore the explicit directions on the edges of the PEPS as they are irrelevant to our arguments. We require the following condition (stronger than Eq.\eqref{n27}) 

We assume the projection MPO has been brought into a form satisfying the \emph{zipper} condition, i.e. there are no off diagonal blocks in the product of two MPO tensors after it has been brought into canonical form, equivalently 
\begin{align}\label{xinv}
\vcenter{\hbox{
\includegraphics[width=0.16\linewidth]{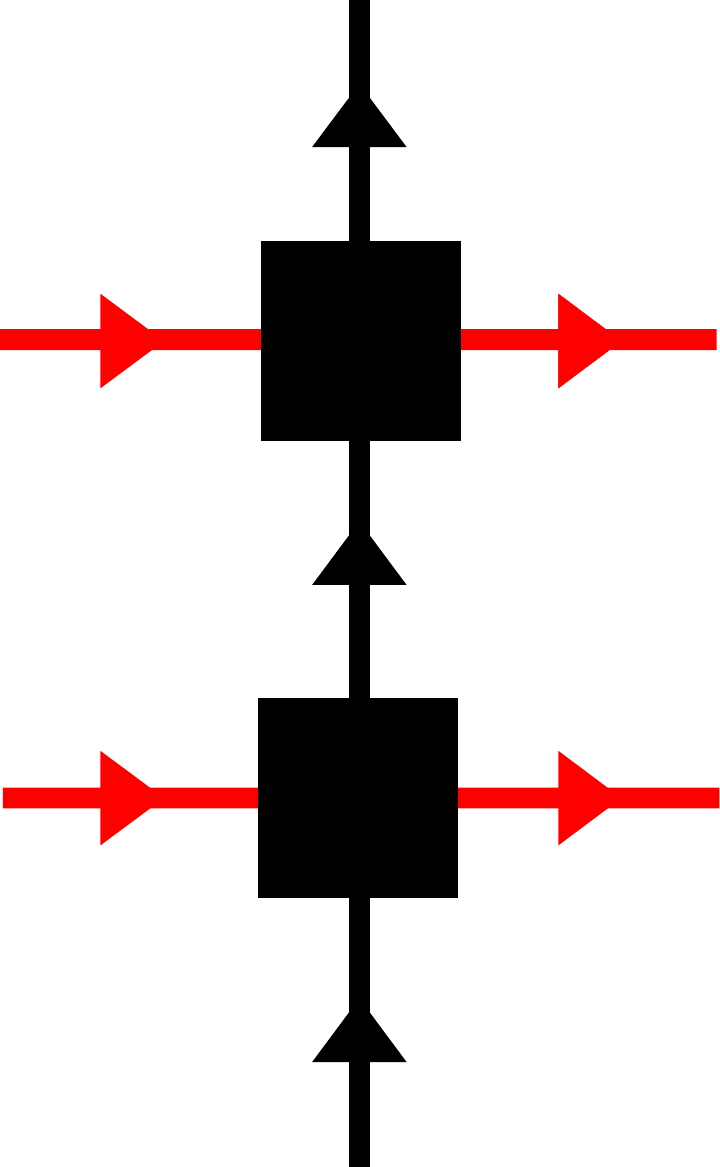}}}=\vcenter{\hbox{
\includegraphics[height=0.16\linewidth]{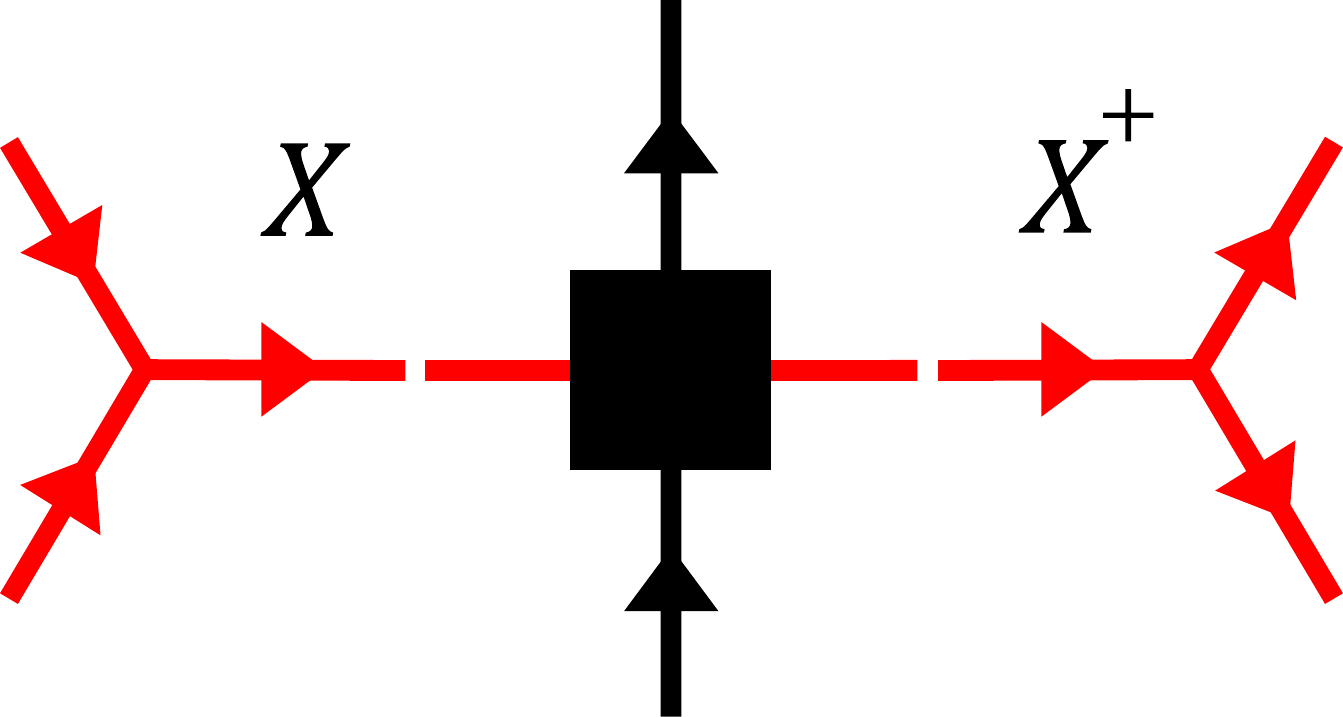}}}
\end{align}
where $X$ is the reduction tensor for multiplication of copies of the MPO which forms a single block representation of the trivial group. This is true of the MPOs arising from fixed-point models. 
For this representation we have the following version of Eq.\eqref{n29}
\begin{align}\label{amove}
\vcenter{\hbox{
\includegraphics[height=0.15\linewidth]{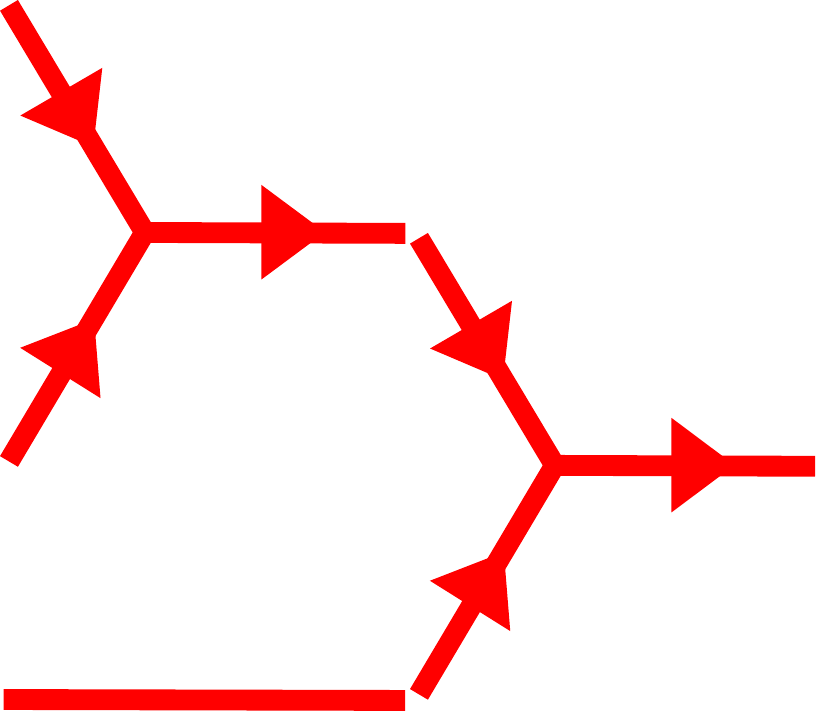}}}=
\alpha 
\vcenter{\hbox{
\includegraphics[height=0.15\linewidth]{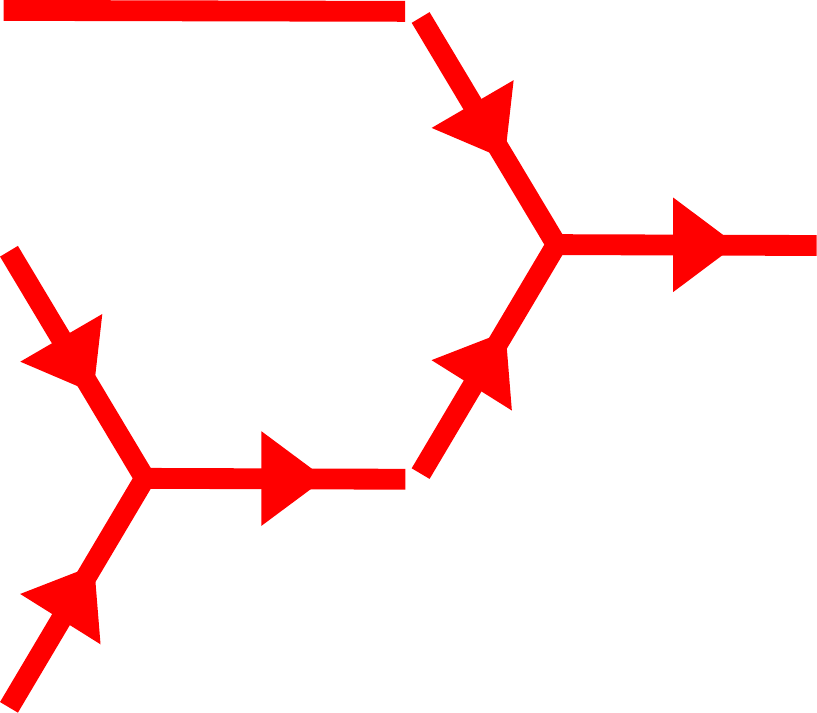}}}.
\end{align}
We now rewrite this equality in a more suggestive fashion
\begin{align}\label{amove}
\vcenter{\hbox{
\includegraphics[width=0.16\linewidth]{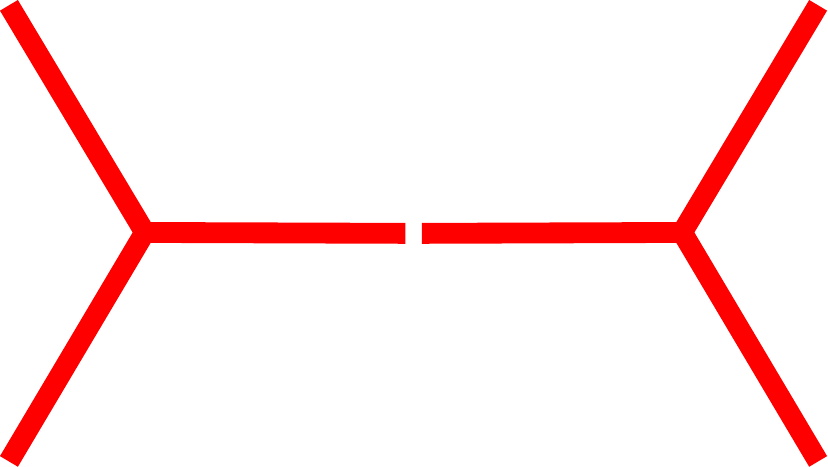}}}
\ = \, \alpha \vcenter{\hbox{
\includegraphics[width=0.16\linewidth,angle=90]{Figures/fig6}}} .
\end{align}
In the above, and throughout the remainder of this appendix, we ignore explicit direction dependence as it does not affect the arguments made.
\\
In the framework of MPO-injectivity different ground states of the PEPS parent Hamiltonian on the torus are spanned by tensor networks closed with different $Q$ tensor solutions (see Ref.\cite{MPOpaper}) connected to MPOs on the virtual level along the inequivalent noncontractible loops of the torus
\begin{align}
\vcenter{\hbox{
\includegraphics[width=0.2\linewidth]{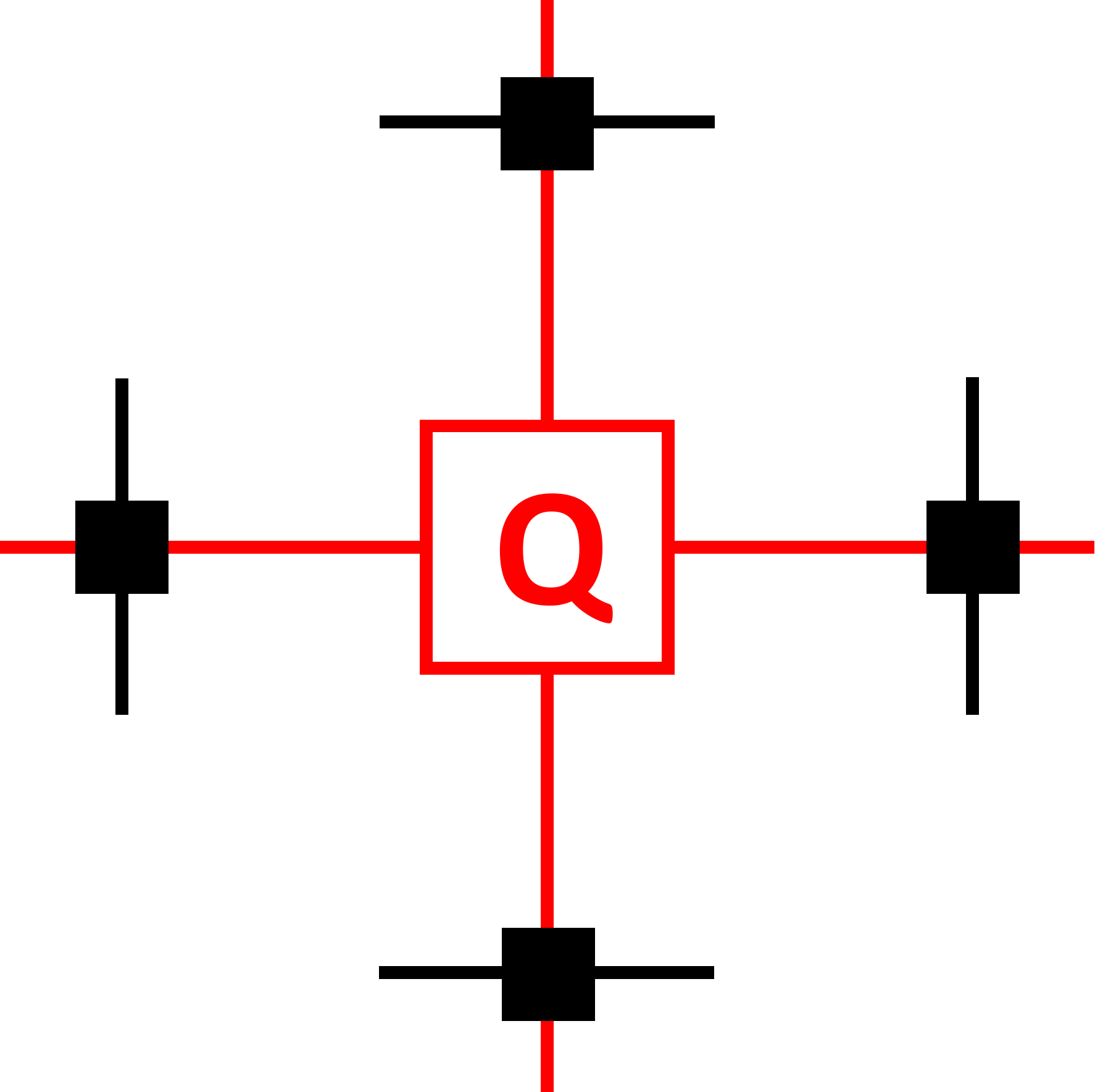}}}.
\end{align}
From the physical level one only has access to the $Q$ tensor projected onto the support subspace of a MPO loop along the closure of the system.
\begin{align}
\vcenter{\hbox{
\includegraphics[width=0.2\linewidth]{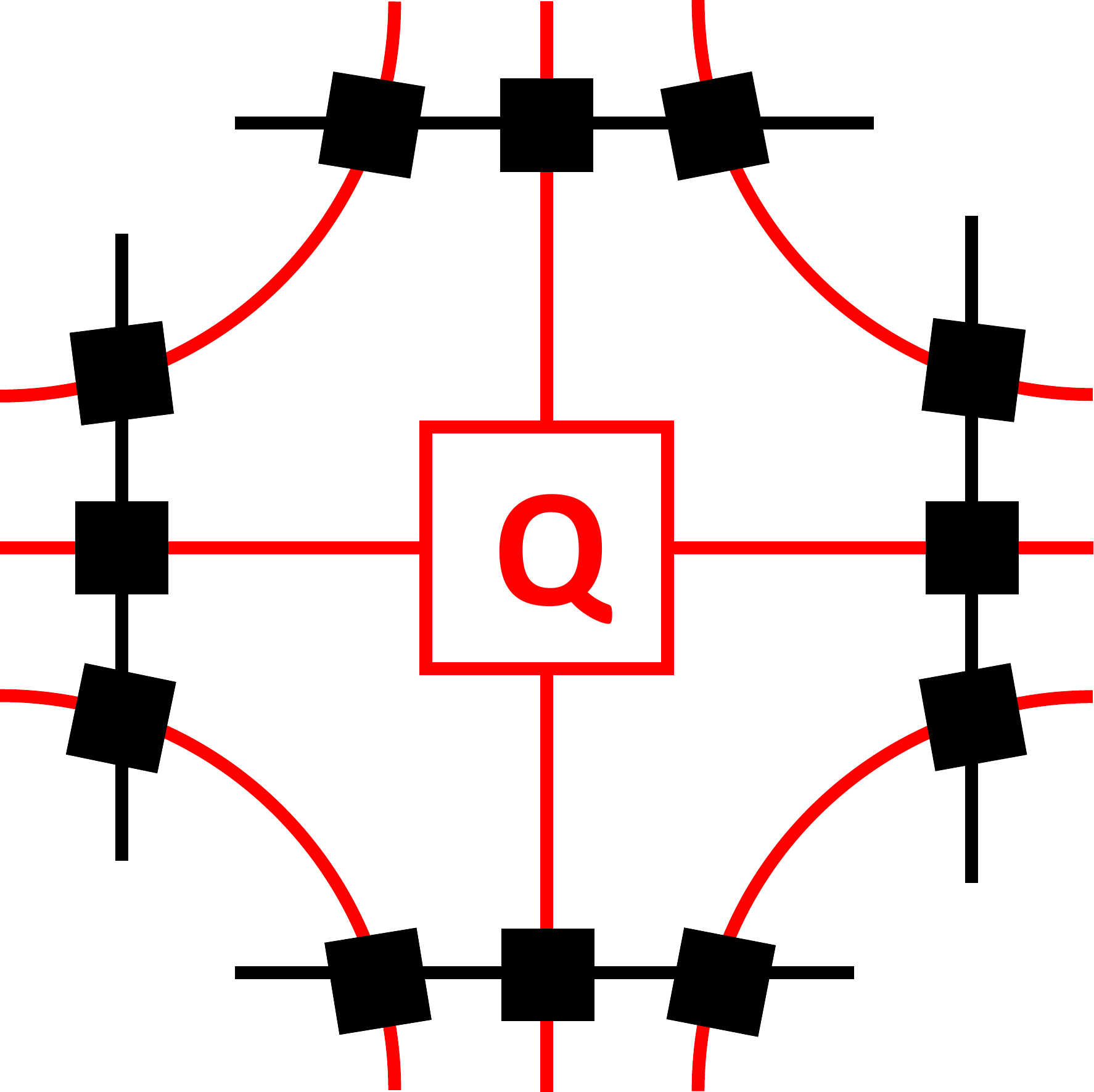}}}
\end{align}
Note this closure gives rise to the same ground state as the closed loop is a symmetry of the closed MPO-injective tensor network. Using condition \eqref{xinv} repeatedly (within the closed tensor network) leads to the following crossing tensor
\begin{align}
\vcenter{\hbox{
\includegraphics[width=0.2\linewidth]{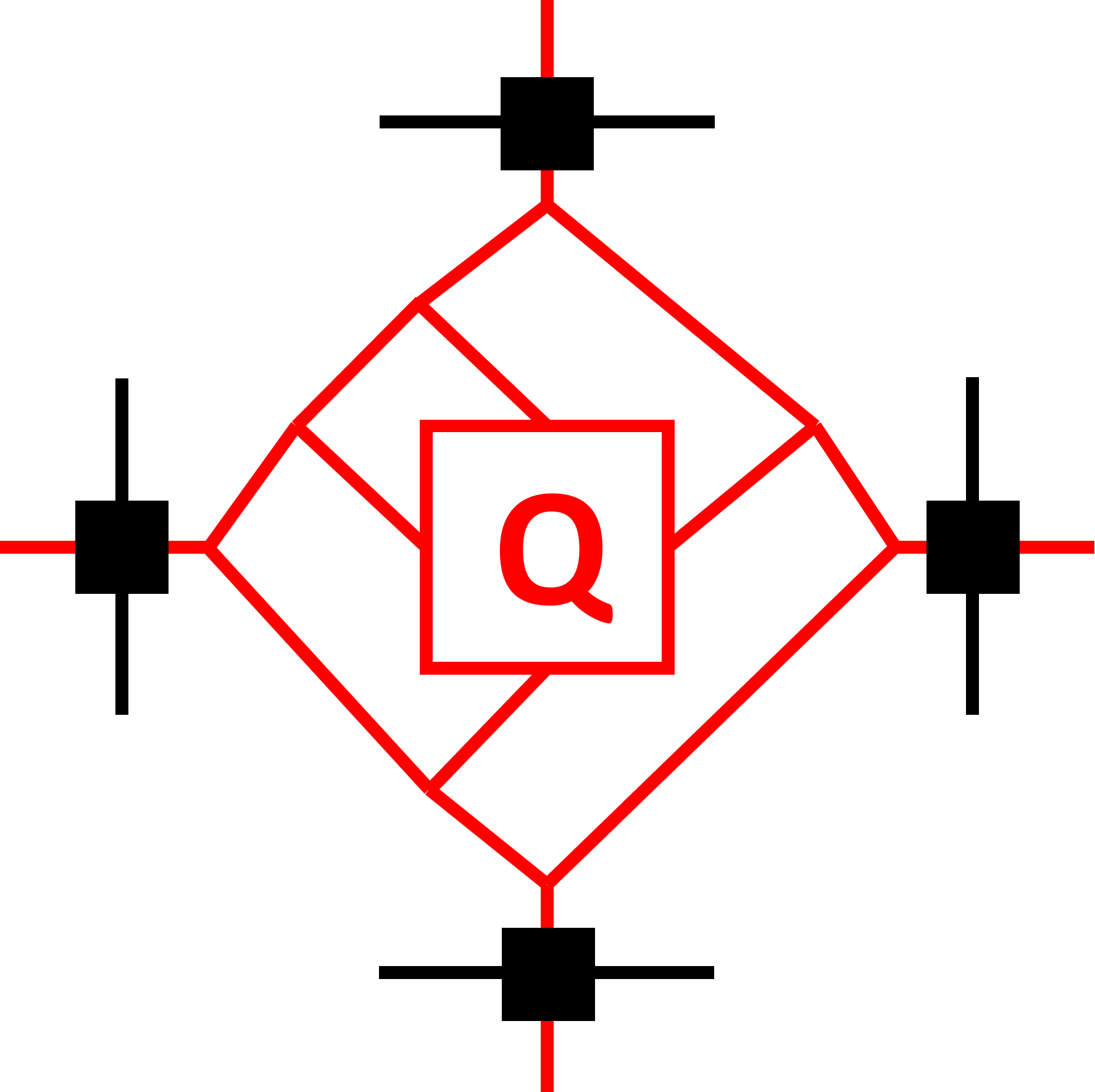}}}
\end{align}
which again gives rise to the same ground state. Following several more applications of Eqs.\eqref{xinv}~\&~\eqref{amove} we arrive at
\begin{align}
\vcenter{\hbox{
\includegraphics[width=0.2\linewidth]{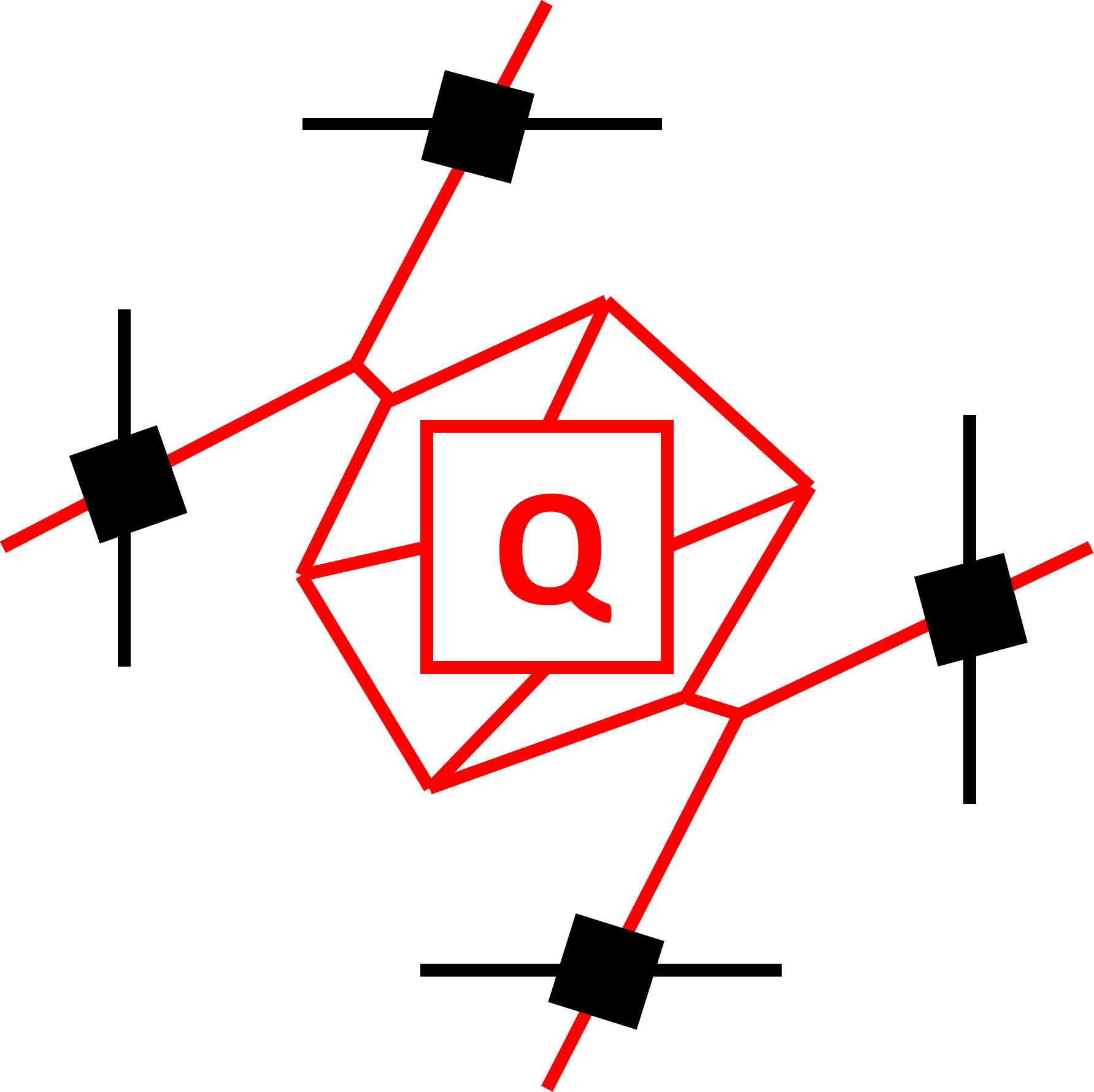}}}
\end{align}
Note the overall phase of the ground state is irrelevant.
Since the $Q$ tensor can be placed anywhere in the tensor network we have that the following matrix
\begin{align}
M_Q:=\vcenter{\hbox{
\includegraphics[width=0.2\linewidth]{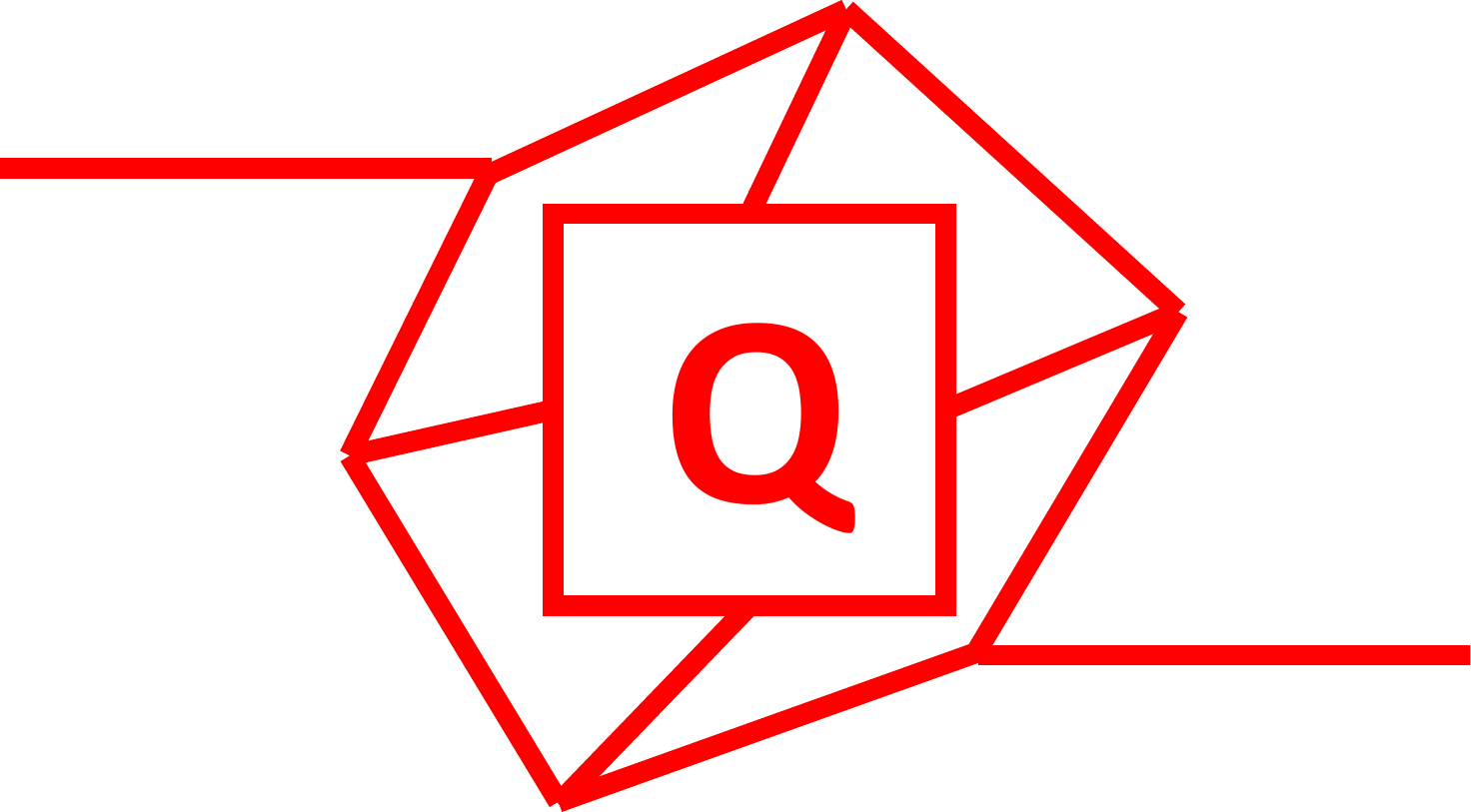}}}
\end{align}
commutes through the virtual level of the single block (injective) projection MPO and hence must be proportional to the identity $M_Q=1$.
Plugging this in we have the crossing tensor
\begin{align}
\vcenter{\hbox{
\includegraphics[width=0.2\linewidth]{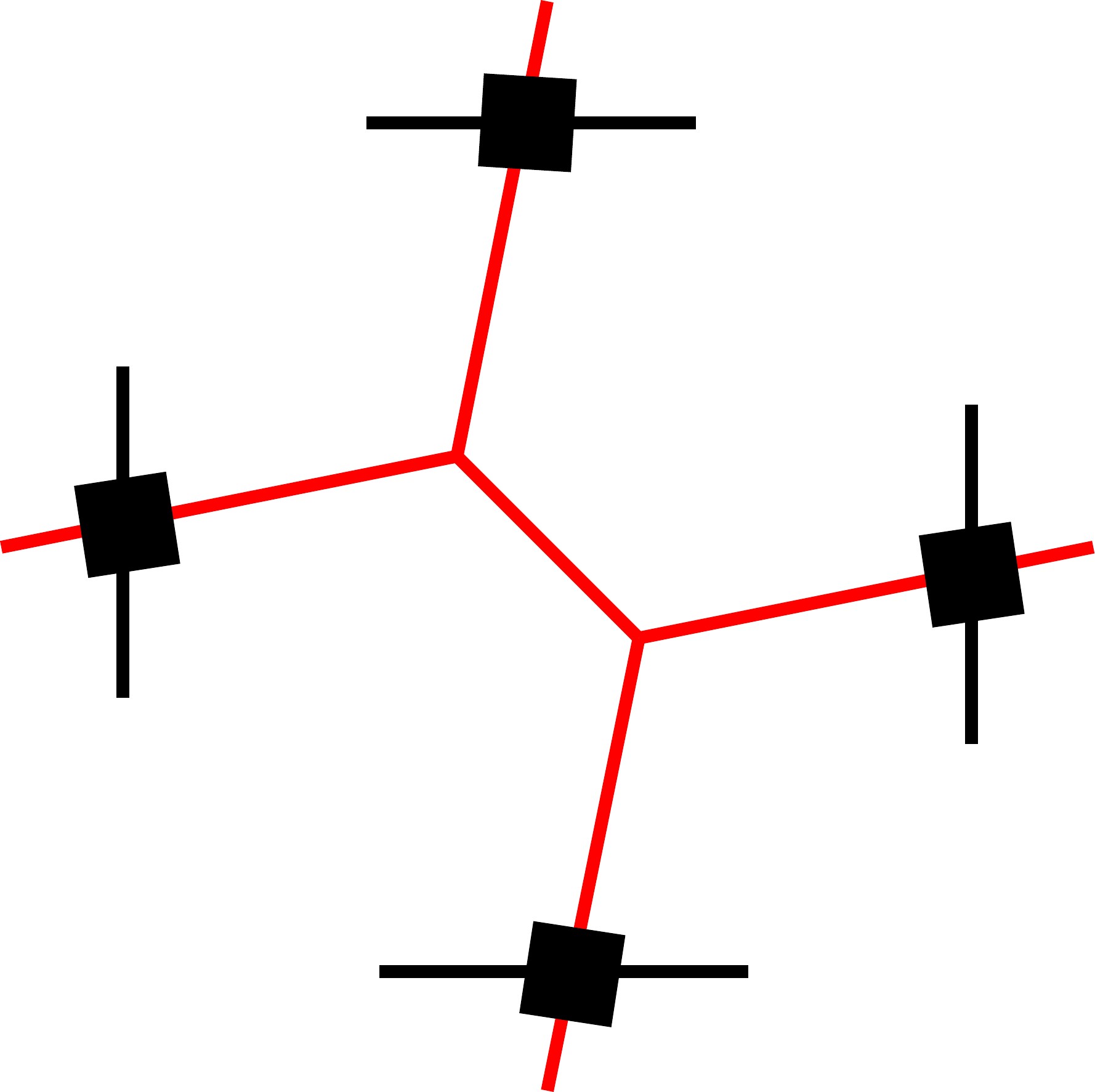}}}
\end{align}
which, by Eq.\eqref{xinv}, yields the same state as the following
\begin{align}
\vcenter{\hbox{
\includegraphics[width=0.2\linewidth]{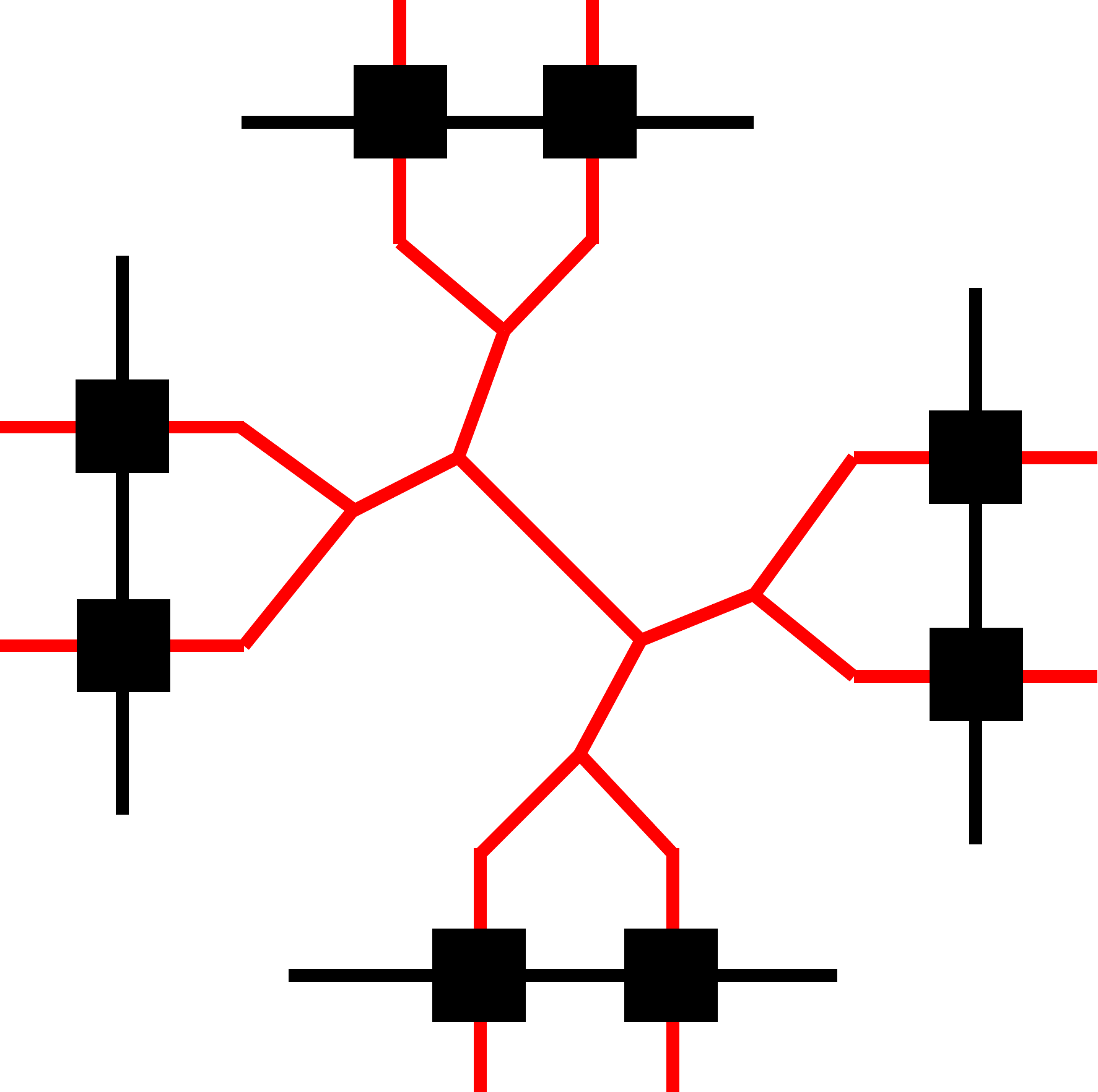}}}
\end{align}
and with several applications of Eqs.\eqref{xinv}~\&~\eqref{amove} one can verify that this is equivalent to 
\begin{align}
\vcenter{\hbox{
\includegraphics[width=0.2\linewidth]{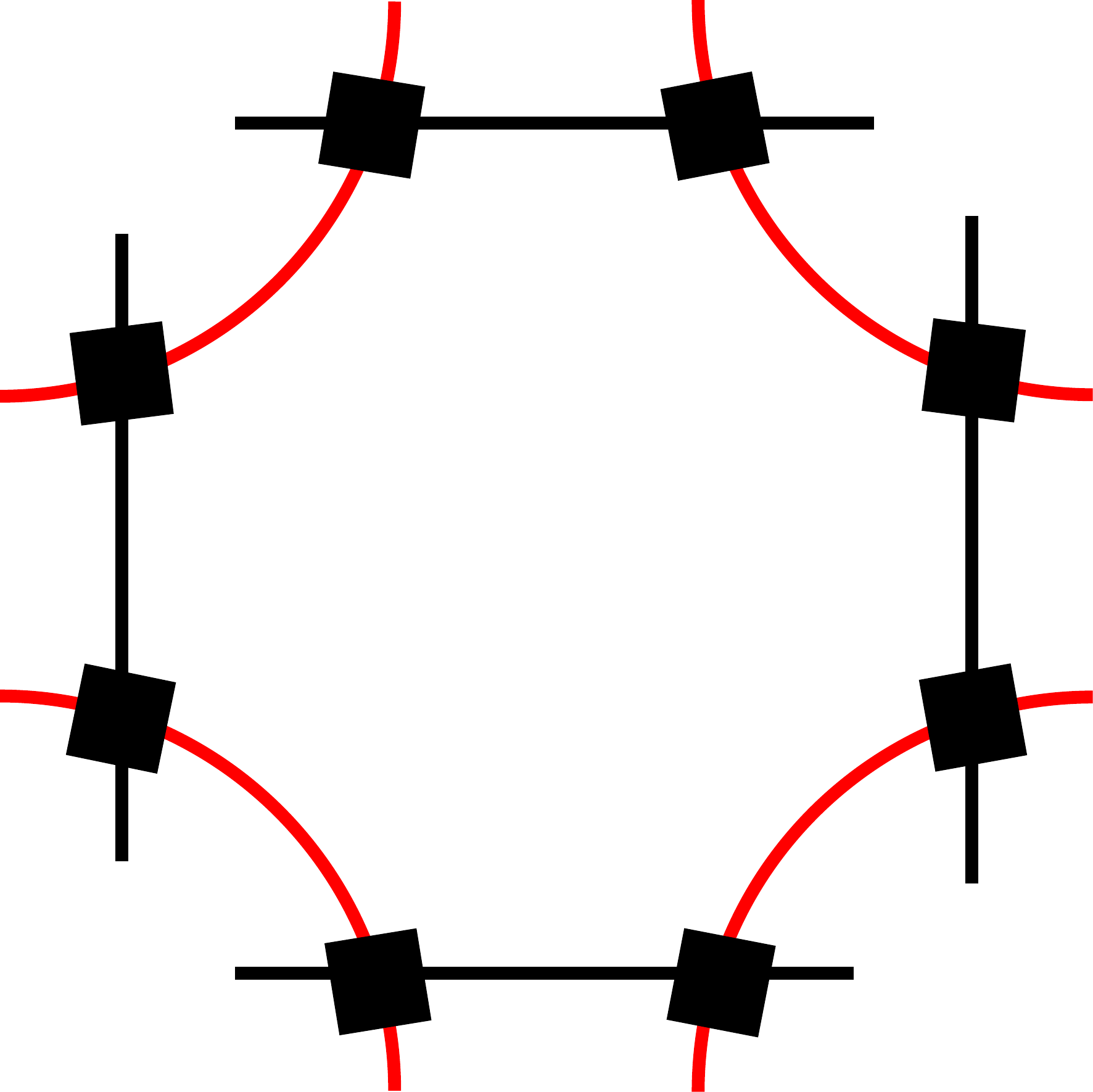}}}
\end{align}
which is easily seen to be a symmetry of a closed MPO-injective tensor network which hence yields the trivial ground state. 
To summarize we have seen that any $Q$ tensor solution gives rise to the unique ground state obtained by closing the tensor network without any MPOs on the virtual level.
}

\section{ Third cohomology class of a single block MPO group representation}\label{c}

In this appendix we recount the definition of the third cohomology class of an injective MPO representation of a finite group $\mathsf{G}$, as first introduced in Ref.\cite{Chen}. For details about group cohomology theory in the context of SPT order we refer the reader to Ref.\cite{GuWen}.

In a MPO representation of $\mathsf{G}$, multiplying a pair of MPOs labeled by the group elements $g_0$ and $g_1$ is equal to the MPO labeled by $g_0g_1$ for every length. Since the MPOs are injective we again know there exists a gauge transformation on the virtual indices of the MPO that brings both representations into the same canonical form~\cite{MPSrepresentations}. This implies that there exists an operator (the reduction tensor) $X(g_0,g_1):(\mathbb{C}^{\mpod})^{\otimes 2}\rightarrow \mathbb{C}^{ \mpod}$ such that Eq.\eqref{n27} holds.
\begin{align}
\vcenter{\hbox{
\includegraphics[width=0.35\linewidth]{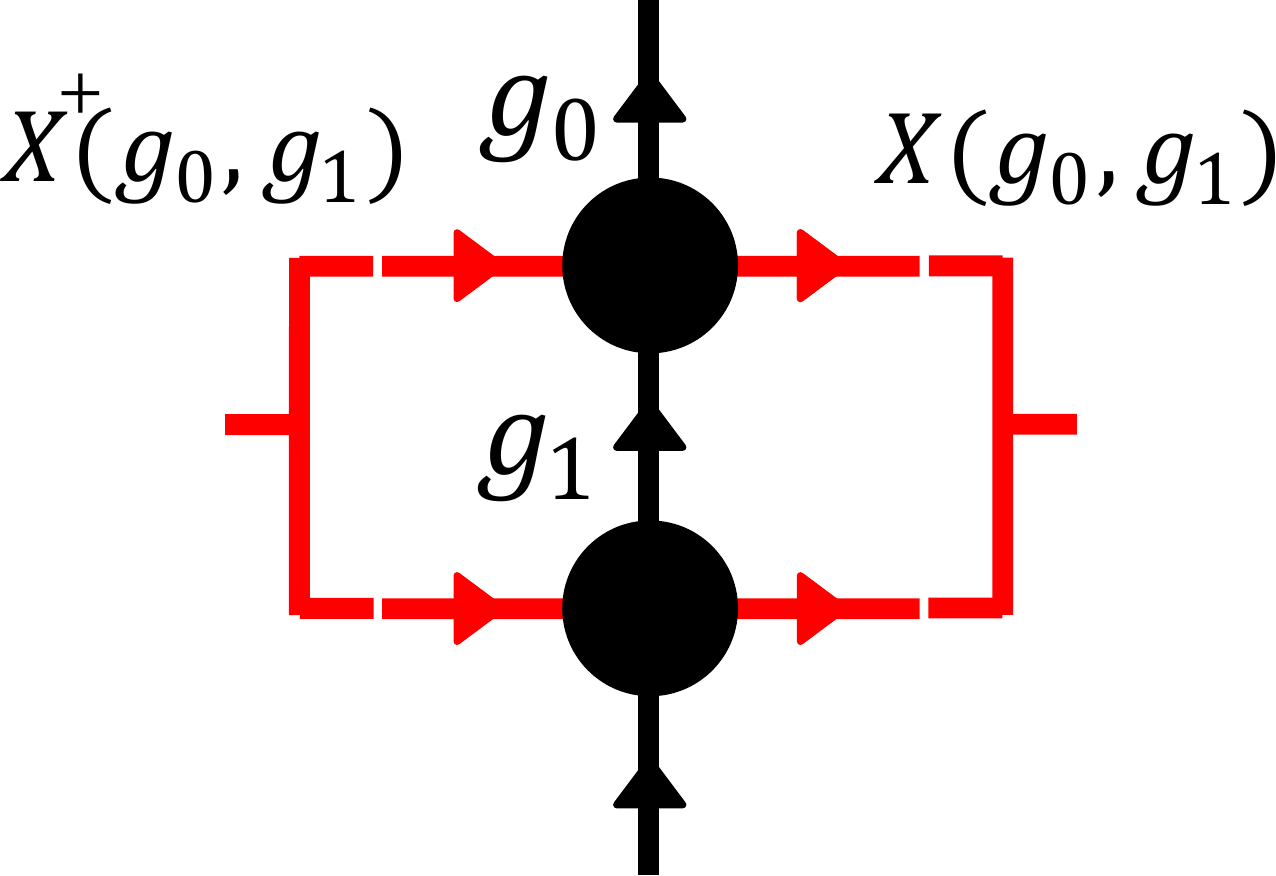}}} \label{n27}
= \vcenter{\hbox{
\includegraphics[height=0.16\linewidth]{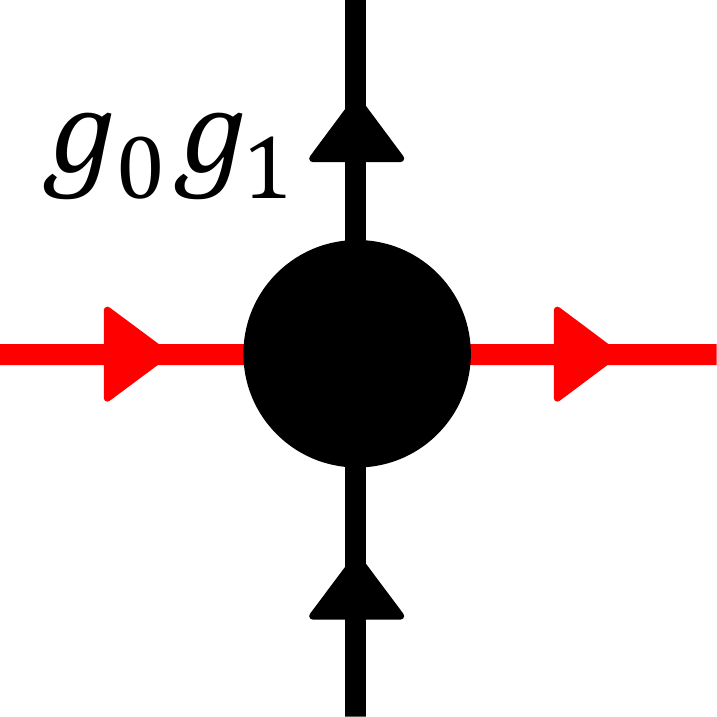}}}
\end{align}
note $X(g_0,g_1)$ is only defined up to multiplication by a complex phase $\beta(g_0,g_1)$.
If we now multiply three MPOs labeled by $g_0$, $g_1$ and $g_2$ there are two ways to reduce the multiplied MPOs to the MPO labeled by $g_0g_1g_2$. When only acting on the right virtual indices these two reductions are equivalent up to a nonzero complex number labeled by $g_0$, $g_1$ and $g_2$. This is shown in Eq.\eqref{n29}.
\begin{align}
\vcenter{\hbox{
\includegraphics[height=0.33\linewidth]{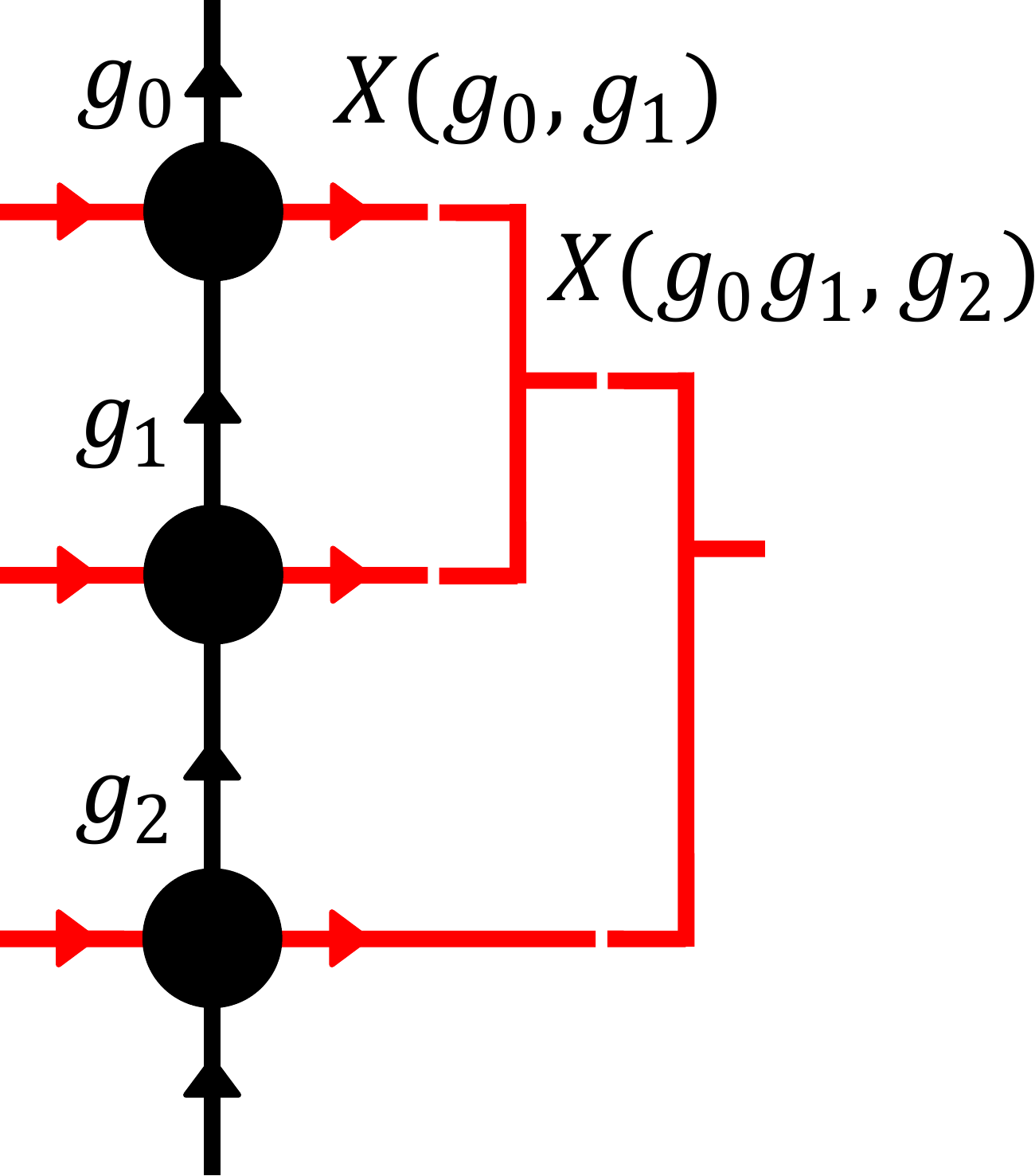}}} \label{n29}
= \alpha(g_0,g_1,g_2)\vcenter{\hbox{
\includegraphics[height=0.33\linewidth]{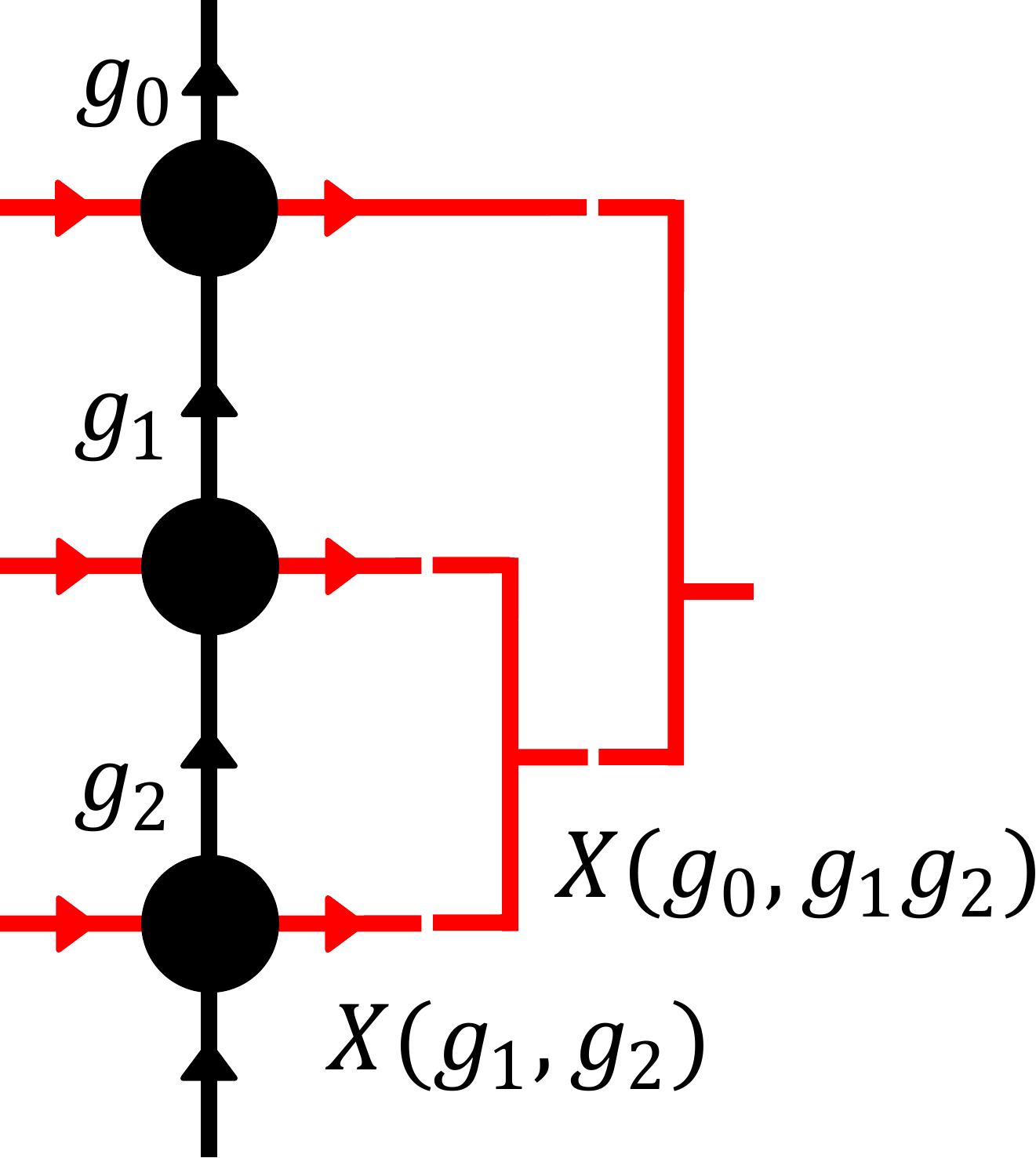}}}
\end{align}
By multiplying four MPOs one sees that $\alpha$ has to satisfy certain consistency conditions as the two different paths achieving the same reduction, shown in Eq.\eqref{n31}, should give rise to the same complex number. 
\begin{align}
\begin{split}
\vcenter{\hbox{
\includegraphics[height=0.35\linewidth]{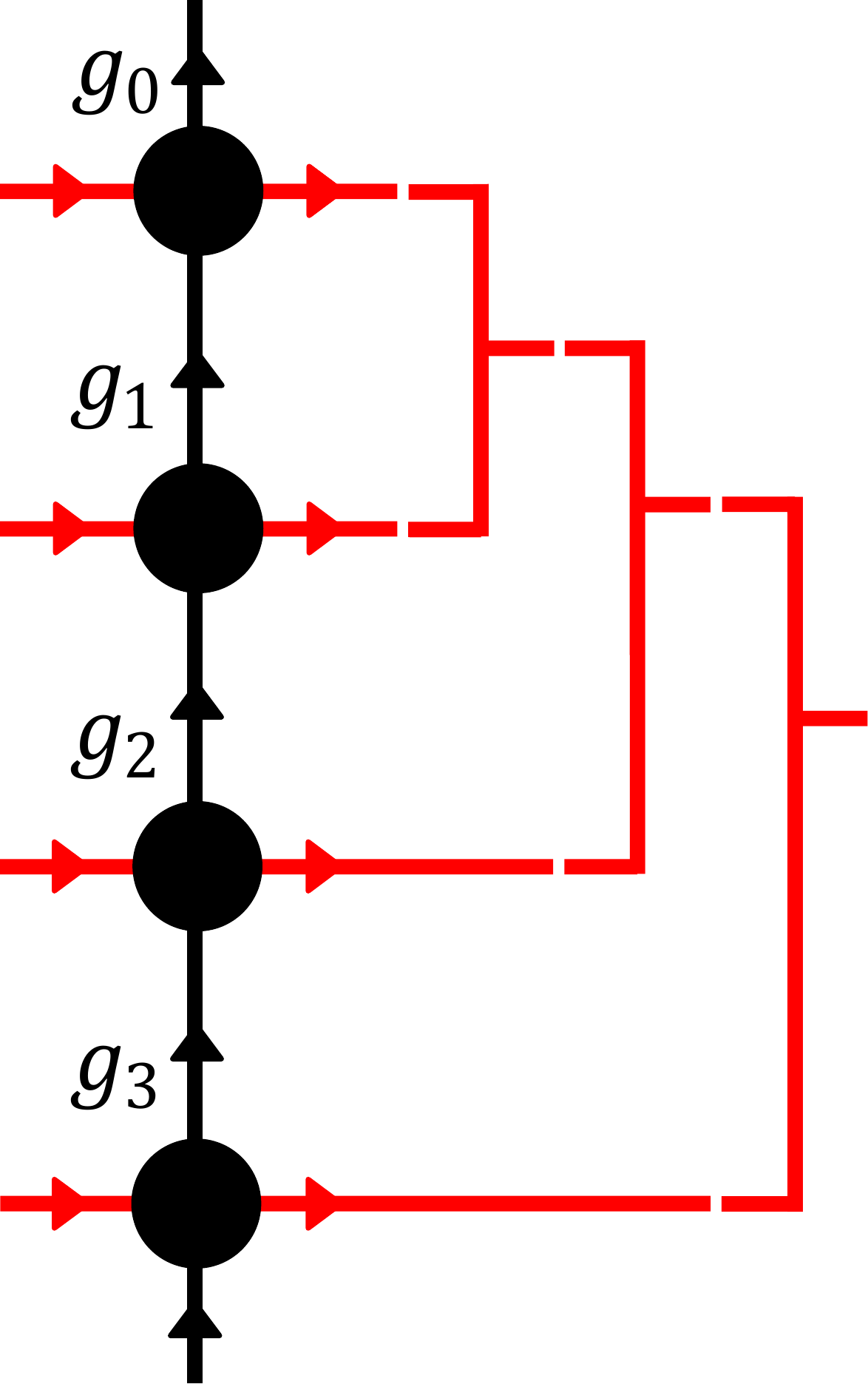}}} \label{n31}
\ \rightarrow \vcenter{\hbox{
\includegraphics[height=0.35\linewidth]{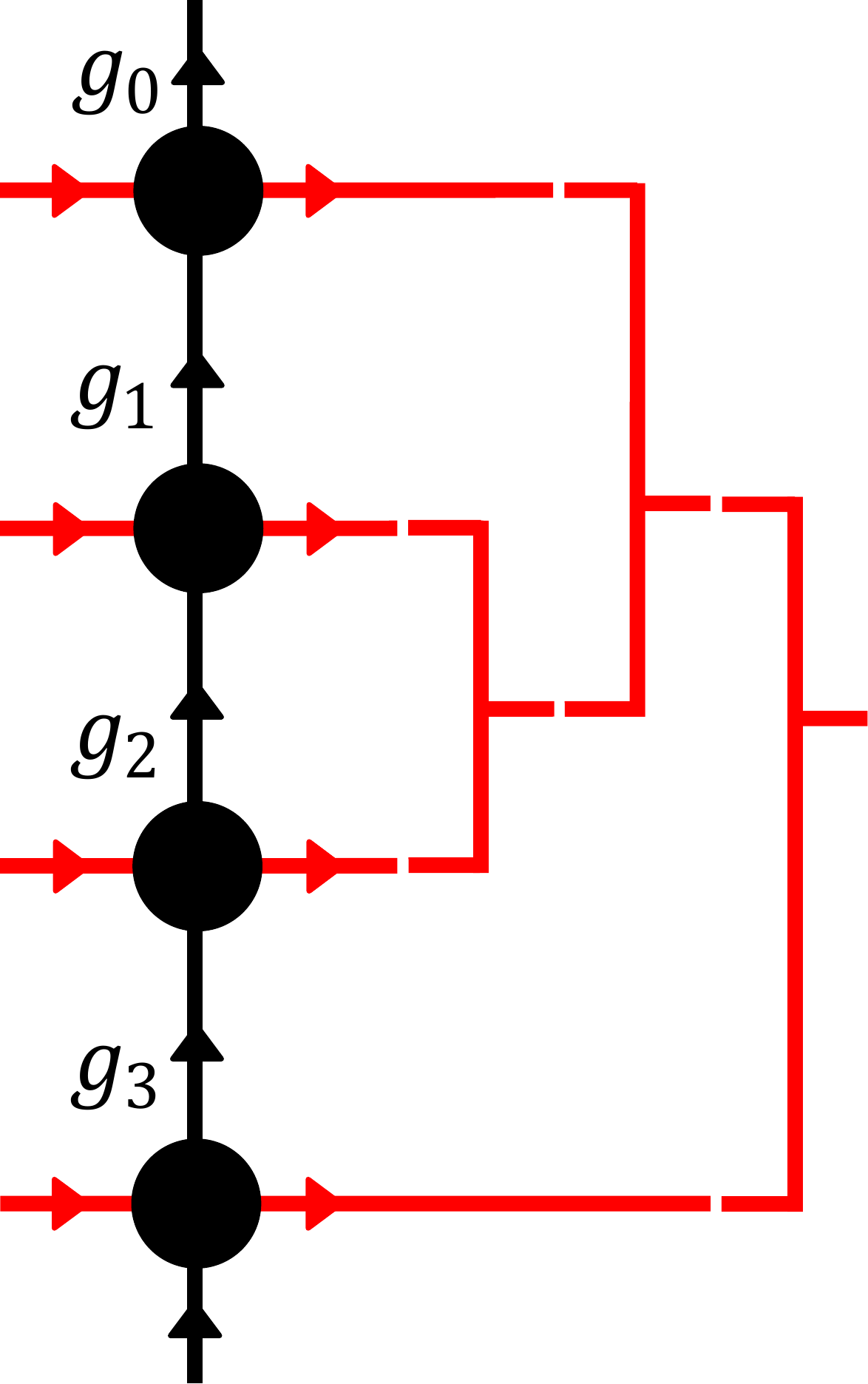}}}
\ \rightarrow \vcenter{\hbox{
\includegraphics[height=0.35\linewidth]{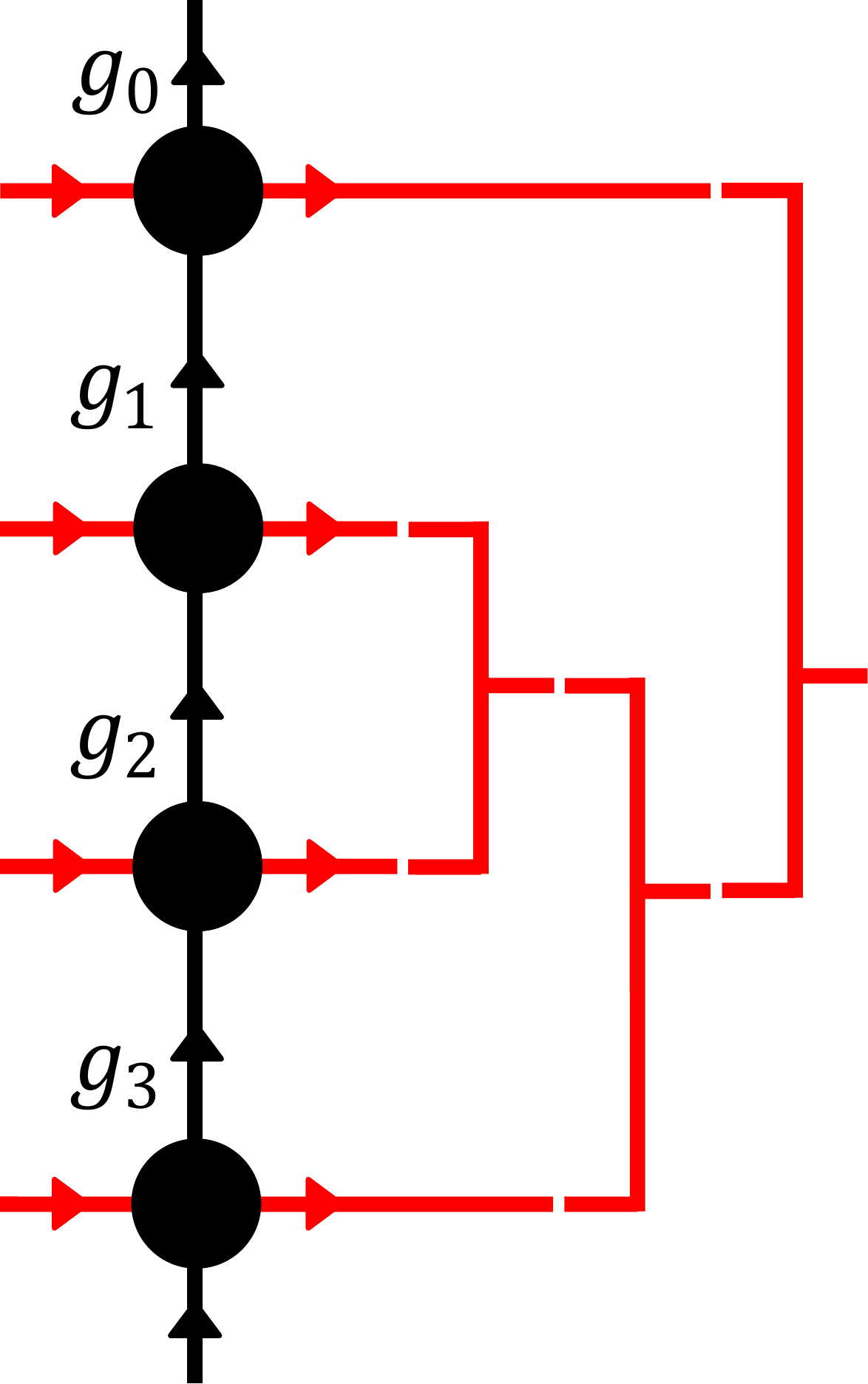}}}
 \\
\downarrow \hspace{5cm} \downarrow\hspace{1cm}
\\
 \vcenter{\hbox{
\includegraphics[height=0.35\linewidth]{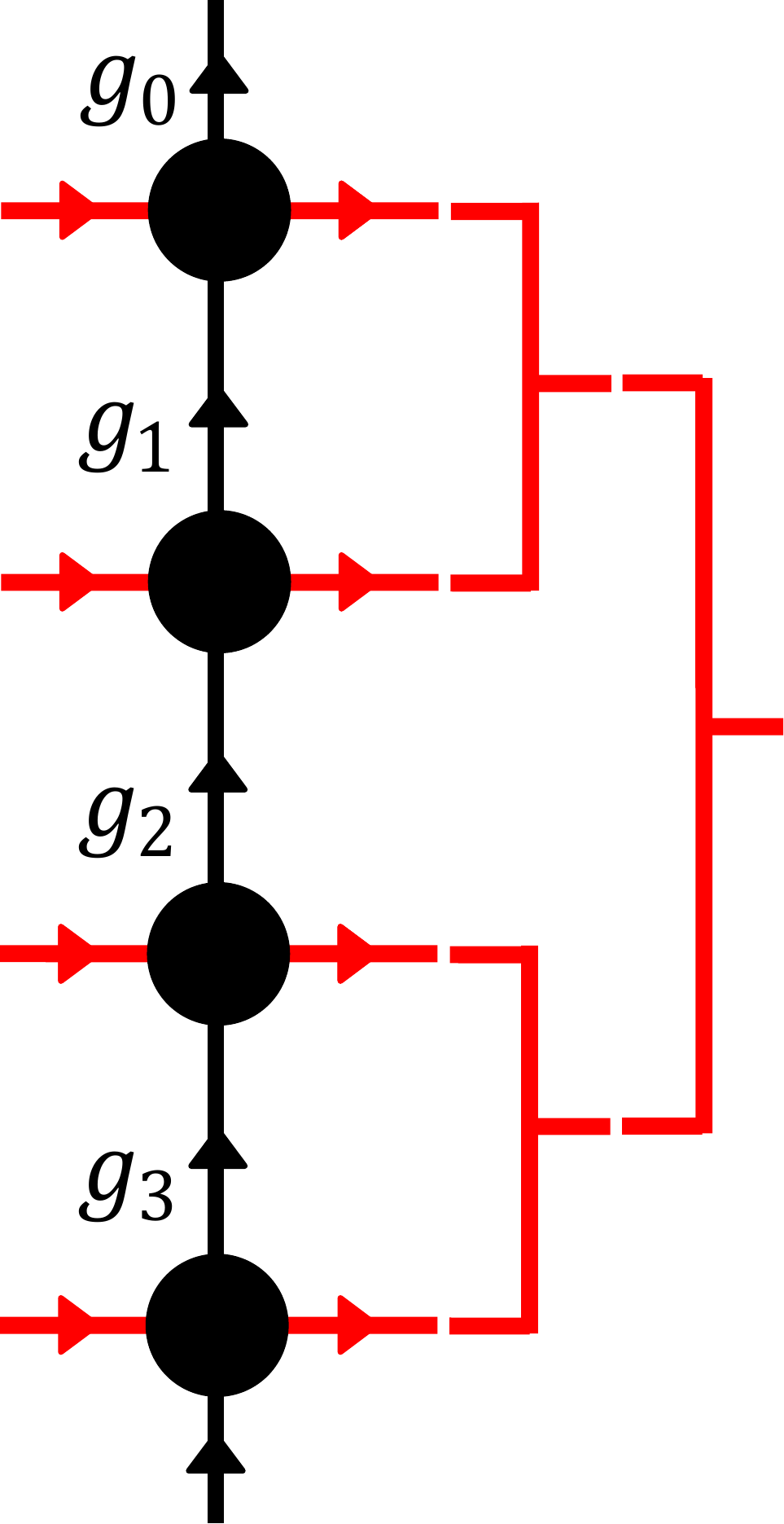}}}
\hspace{0.8cm} \rightarrow \hspace{0.8cm}
\vcenter{\hbox{
\includegraphics[height=0.35\linewidth]{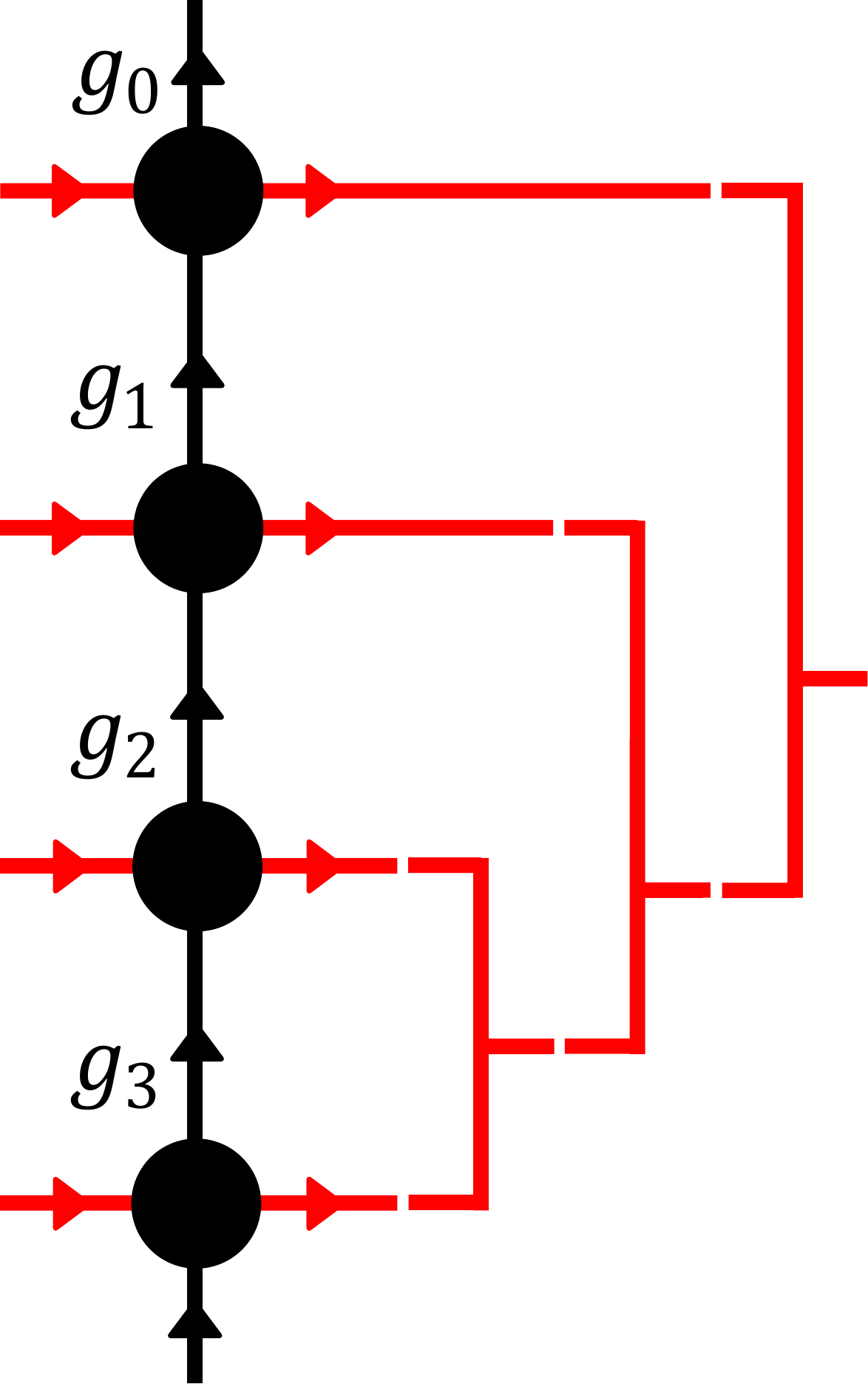}}}
\hspace{0.6cm}
\end{split}
\end{align}
Using Eq.\eqref{n29} one can easily verify that the consistency conditions are
\begin{equation}
\frac{\alpha(g_0,g_1,g_2)\alpha(g_0,g_1g_2,g_3)\alpha(g_1,g_2,g_3)}{\alpha(g_0g_1,g_2,g_3)\alpha(g_0,g_1,g_2g_3)} = 1 \label{cocycle}
\end{equation}
which are exactly the 3-cocycle conditions and hence $\alpha$ is a 3-cocycle. As mentioned above $X(g_0,g_1)$ is only defined up to a complex number $\beta(g_0,g_1)$. This freedom can change the 3-cocycle defined in Eq.\eqref{n29} by
\begin{equation}
\alpha'(g_0,g_1,g_2) = \alpha(g_0,g_1,g_2)\frac{\beta(g_1,g_2)\beta(g_0,g_1g_2)}{\beta(g_0,g_1)\beta(g_0g_1,g_2)}
\end{equation}
thus we see that $\alpha$ is only defined up to a 3-coboundary. For this reason the single block MPO group representation is endowed with the label $[\alpha]$ from the third cohomology group $H^3(\mathsf{G},\mathbb{C})$. Using the fact that $H^d(\mathsf{G},\mathbb{R}) = \mathbb{Z}_1$~\cite{GuWen} (and that $\mathbb{R}$ as an additive group is isomorphic to $\mathbb{R}^+$ as a multiplicative group), we thus obtain that the third cohomology class of the MPO representation $[\alpha]$ is an element of $H^3(\mathsf{G},\mathsf{U(1)})$. 

\section{Orientation dependencies of MPO group representations} \label{newapp1}

In this appendix we go beyond previous treatments of MPO group representations to consider subtleties that arise due to possible orientation dependencies of the tensors. 
We find a gauge transformation that reverses the orientation of MPO tensors, and use it to define the Frobenius-Schur indicator. We then find several \emph{pivotal} phases and relate them to the 3-cocycle of the MPO group representation.

\subsection{Orientation reversing gauge transformation}

To describe the most general bosonic SPT phases one must use lattices with oriented edges, the internal index of the MPO also carries an orientation which leads to the definition of a pair of possibly distinct MPO tensors which depend on the handedness of the crossing upon which they sit
\begin{align}
&B_+(g)=\vcenter{\hbox{
\includegraphics[height=0.16\linewidth]{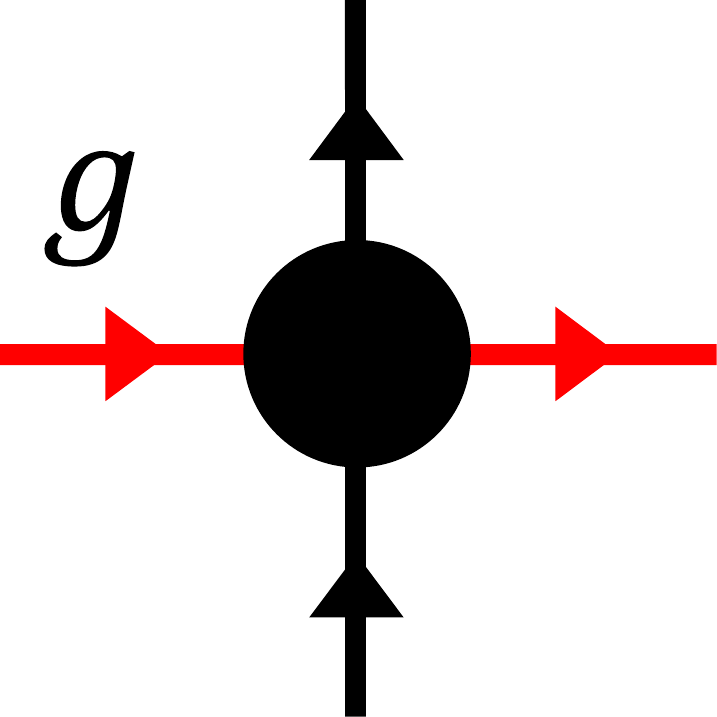}}}
\, ,
&B_-(g)=\vcenter{\hbox{
\includegraphics[height=0.16\linewidth]{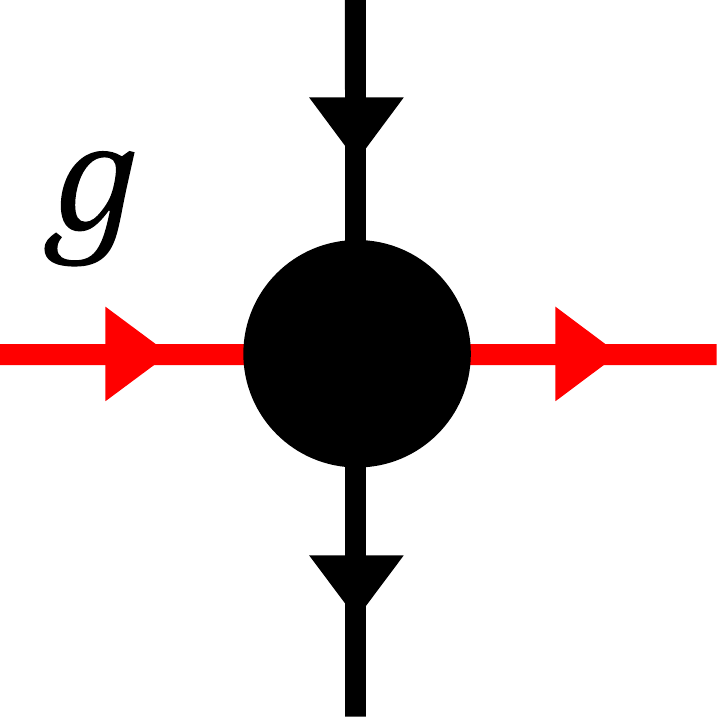}}} \label{na1}
\end{align}
As shown in Section~\ref{globalsymmetry} reversing the orientation of the internal MPO index corresponds to inverting the group element which the MPO represents, i.e. $V_{\text{rev}}(g)=V(g^{-1})$. Since this holds for any injective group MPO of arbitrary length standard results from the theory of MPS imply that the local tensors are related by an invertible gauge transformation which we denote $Z_g$
\begin{align}
\vcenter{\hbox{
\includegraphics[height=0.16\linewidth]{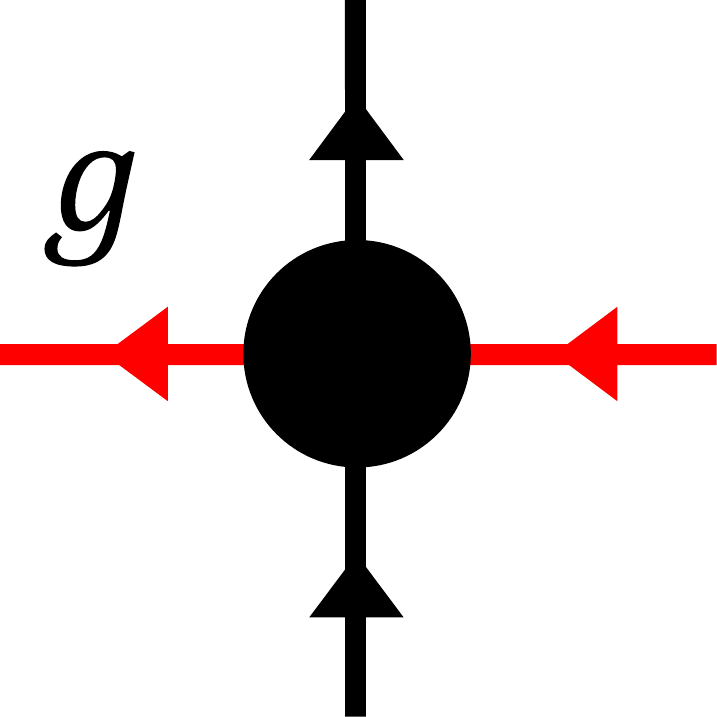}}}&=\vcenter{\hbox{
\includegraphics[height=0.16\linewidth]{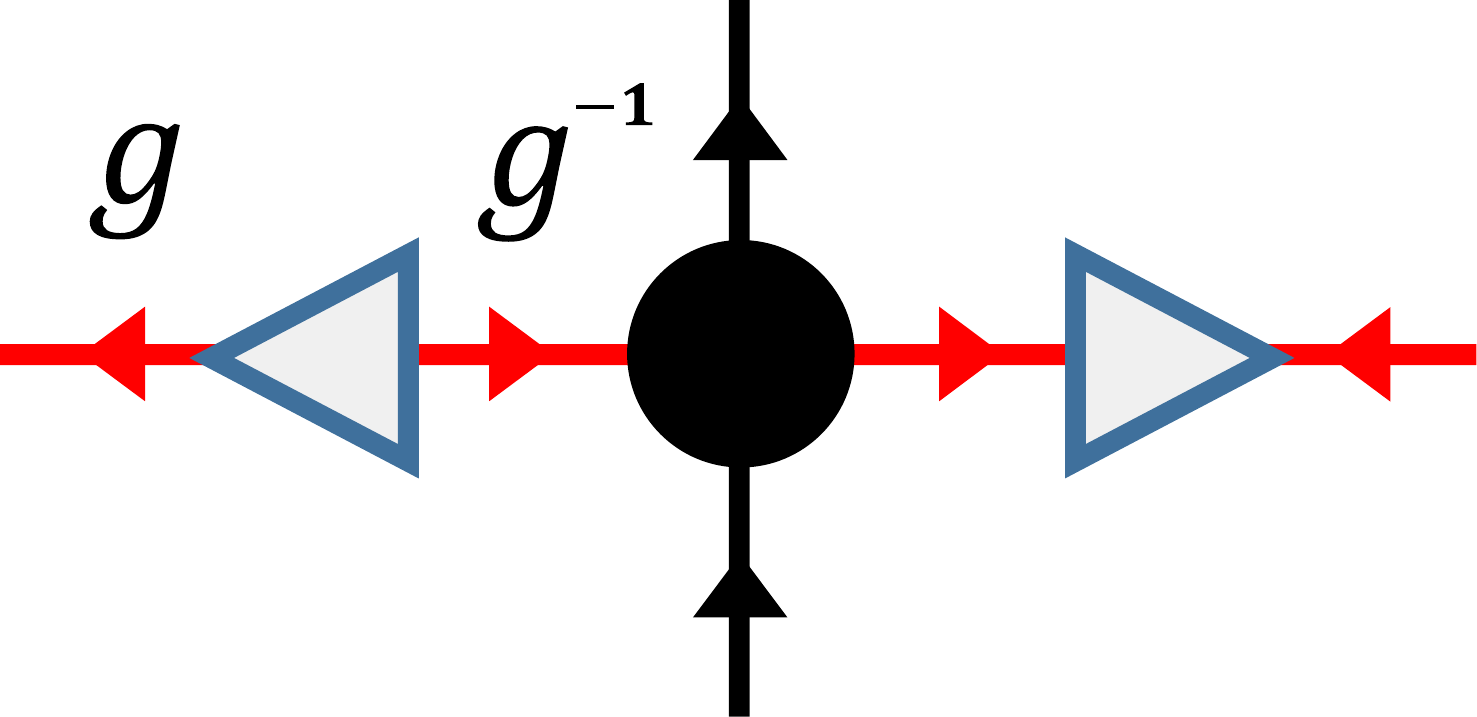}}} 
\\
\vcenter{\hbox{
\includegraphics[height=0.16\linewidth]{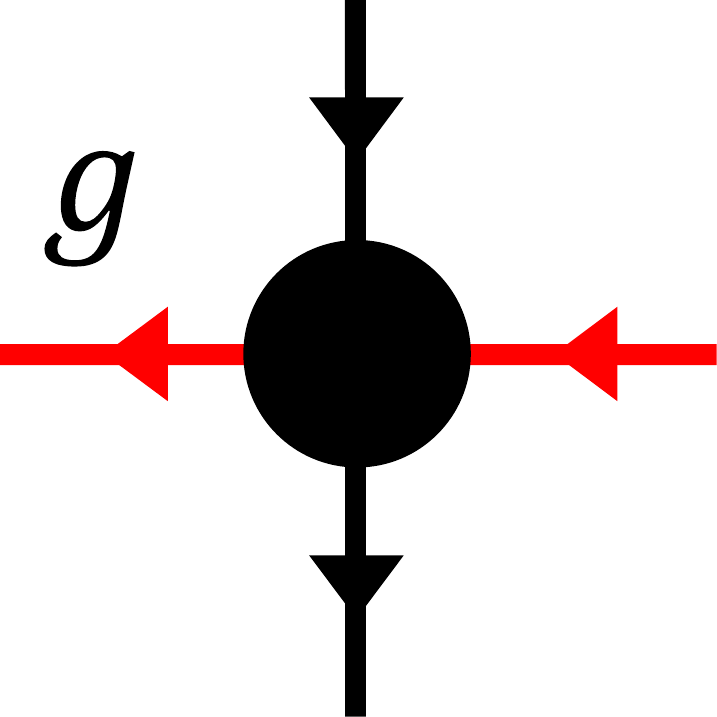}}}
&=\vcenter{\hbox{
\includegraphics[height=0.16\linewidth]{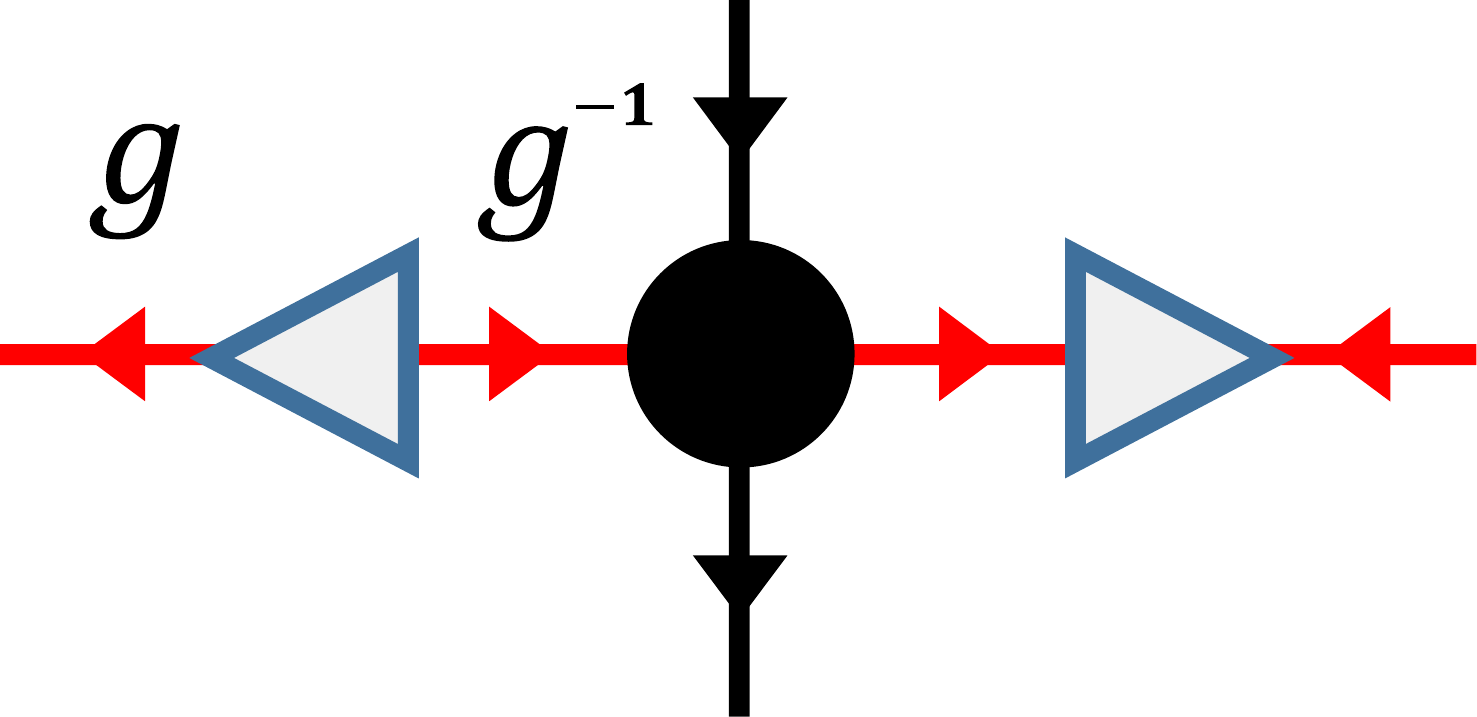}}} \label{na2}
\end{align}
where we use the following graphical notation for $Z_g$ and related matrices
\begin{align}
Z_g&=\raisebox{-0.15cm}{\hbox{\includegraphics[width=0.16\linewidth]{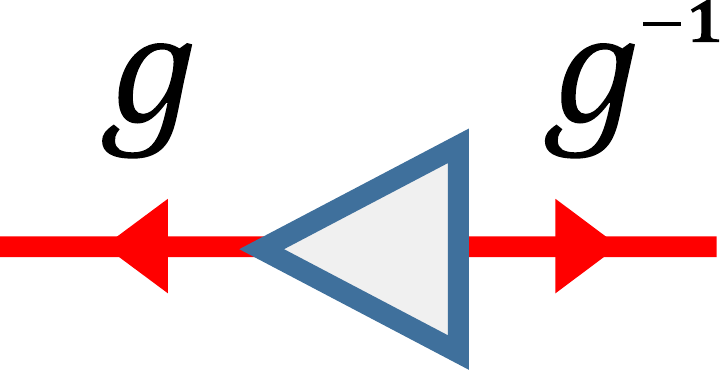}}} \, ,
&Z_g^T&=\raisebox{-0.15cm}{{\hbox{\includegraphics[width=0.16\linewidth]{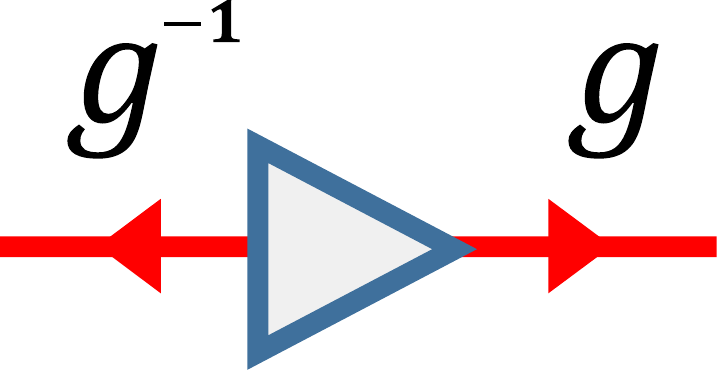}}}}
\\
Z_g^{-1}&=\raisebox{-0.15cm}{\hbox{\includegraphics[width=0.16\linewidth]{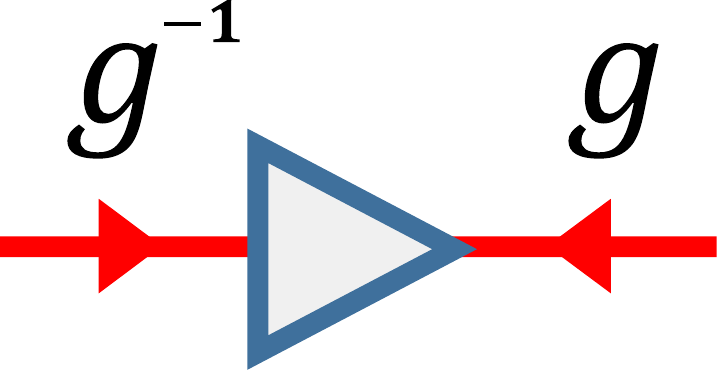}}} \, ,
&(Z_g^{-1})^T&=\raisebox{-0.15cm}{\hbox{\includegraphics[width=0.16\linewidth]{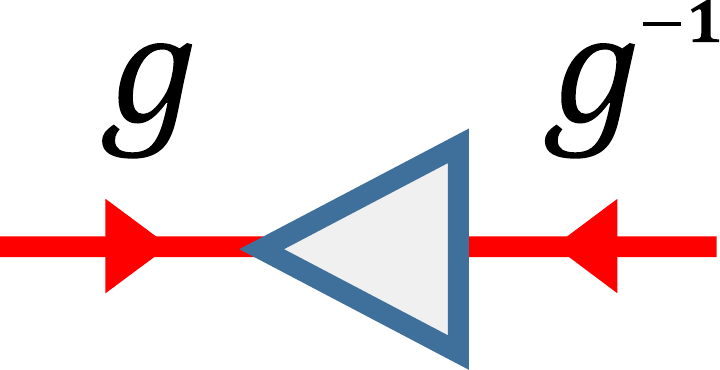}}}
\end{align}
which satisfy the relations 
\begin{align}
\raisebox{-0.1cm}{\hbox{\includegraphics[height=0.08\linewidth]{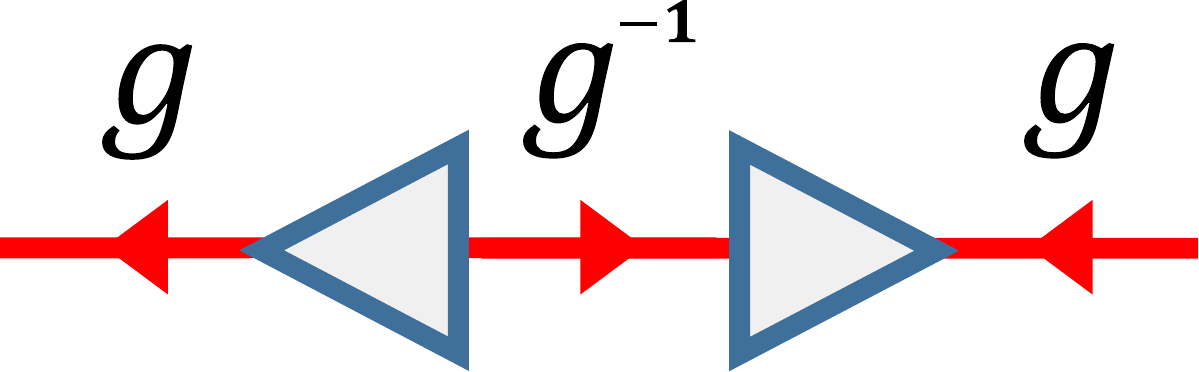}}}&=\raisebox{-0.0cm}{\hbox{\includegraphics[height=0.07\linewidth]{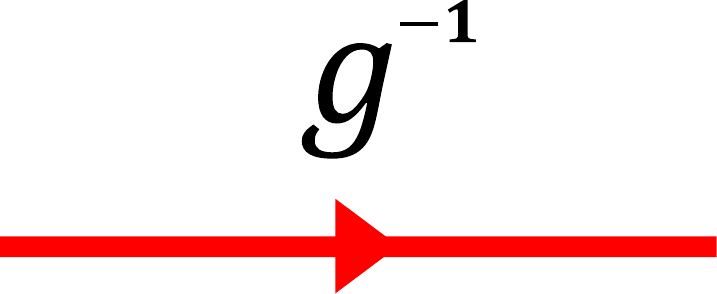}}}
\\
\raisebox{-0.1cm}{\hbox{\includegraphics[height=0.08\linewidth]{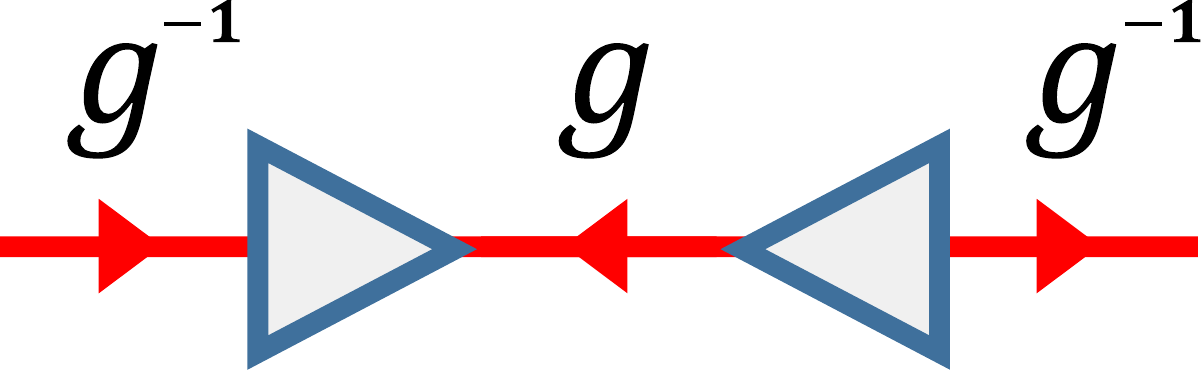}}}&=\raisebox{-0.0cm}{\hbox{\includegraphics[height=0.06\linewidth]{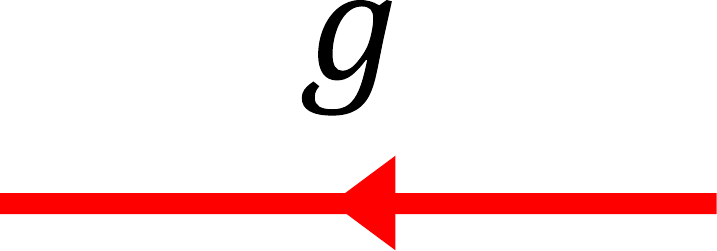}}}
\label{na3}
\end{align}
note while it seems \textit{a priori} that the gauge transformations in Eq.\eqref{na2} could be independent, the fact that the equation $V_{\text{rev}}(g)=V(g^{-1})$ holds for arbitrary orientations of the PEPS bonds implies that they can be chosen to be the same.

Applying the gauge transformation twice we arrive at the equality
\begin{align}
\vcenter{\hbox{\includegraphics[height=0.16\linewidth]{Figures/newfig16}}}=\vcenter{\hbox{\includegraphics[height=0.16\linewidth]{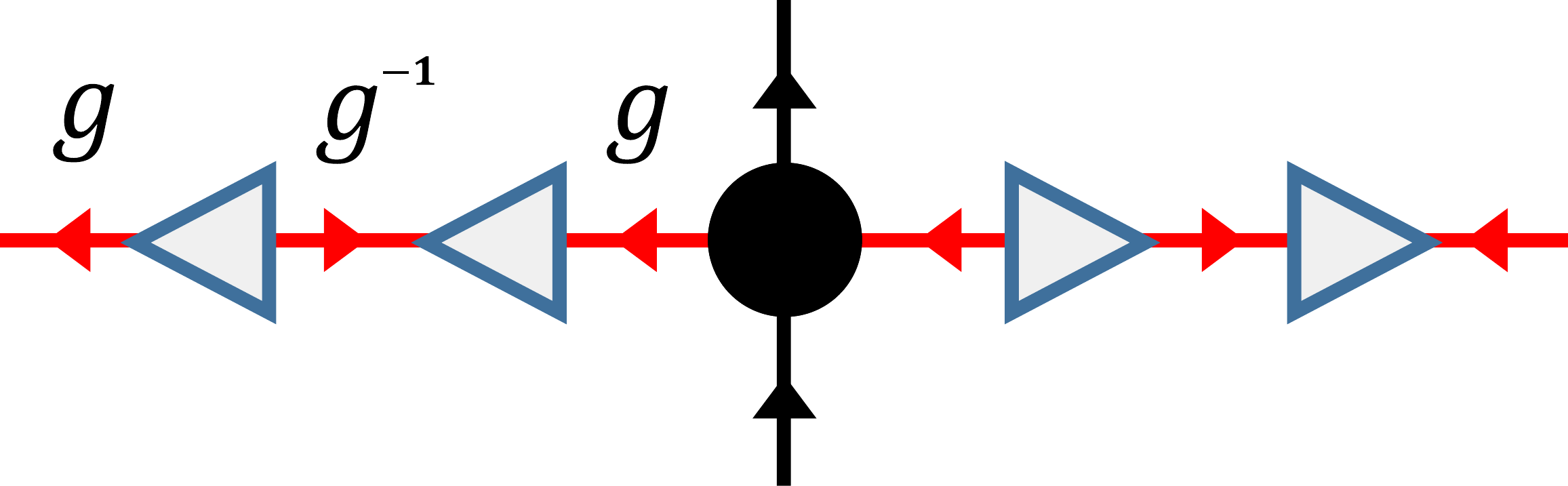}}} \label{na4}
\end{align}
which implies $Z_g(Z_{g^{-1}}^{-1})^T=\chi_g \openone$ for some $\chi_g\in U(1)$ since the MPO is injective. Hence $Z_g=\chi_g Z_{g^{-1}}^{T}$ i.e.
\begin{align}
\raisebox{-0.15cm}{\hbox{\includegraphics[height=0.08\linewidth]{Figures/newfig8}}}
=\chi_g \, \raisebox{-0.15cm}{\hbox{\includegraphics[height=0.08\linewidth]{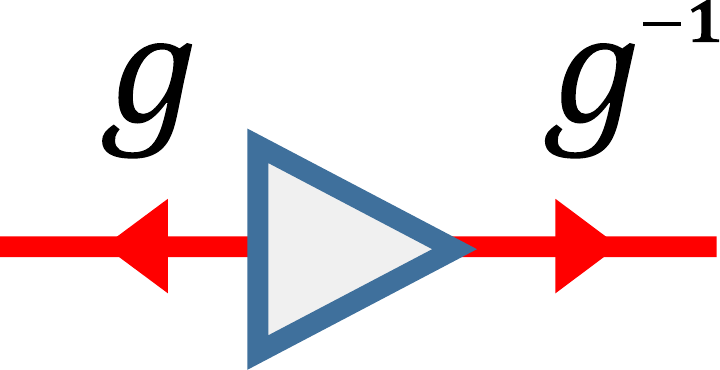}}} \label{na5}
\end{align}
where $\chi_g$ is analogous to the Frobenius--Schur indicator and can be seen to satisfy $\chi_g=\chi_{g^{-1}}^{-1}$. Note $\chi_g$ can be absorbed by redefinition of $Z_g$ whenever $g\neq g^{-1}$, but we will not do so at this point.

\subsection{Pivotal phases}

Since the multiplication of the injective MPOs forms a representation of $\mathsf{G}$ we have a local reduction as in Eq.\eqref{n27}. Again since this holds for arbitrary orientations of the PEPS bonds the reduction matrix $X(g_0,g_1)$ is the same for left and right handed MPOs. From here on we will work with a stronger restriction on the form of the MPOs such that the following \emph{zipper} condition holds 
\begin{align}
\raisebox{-.61cm}{\hbox{\includegraphics[height=0.2\linewidth]{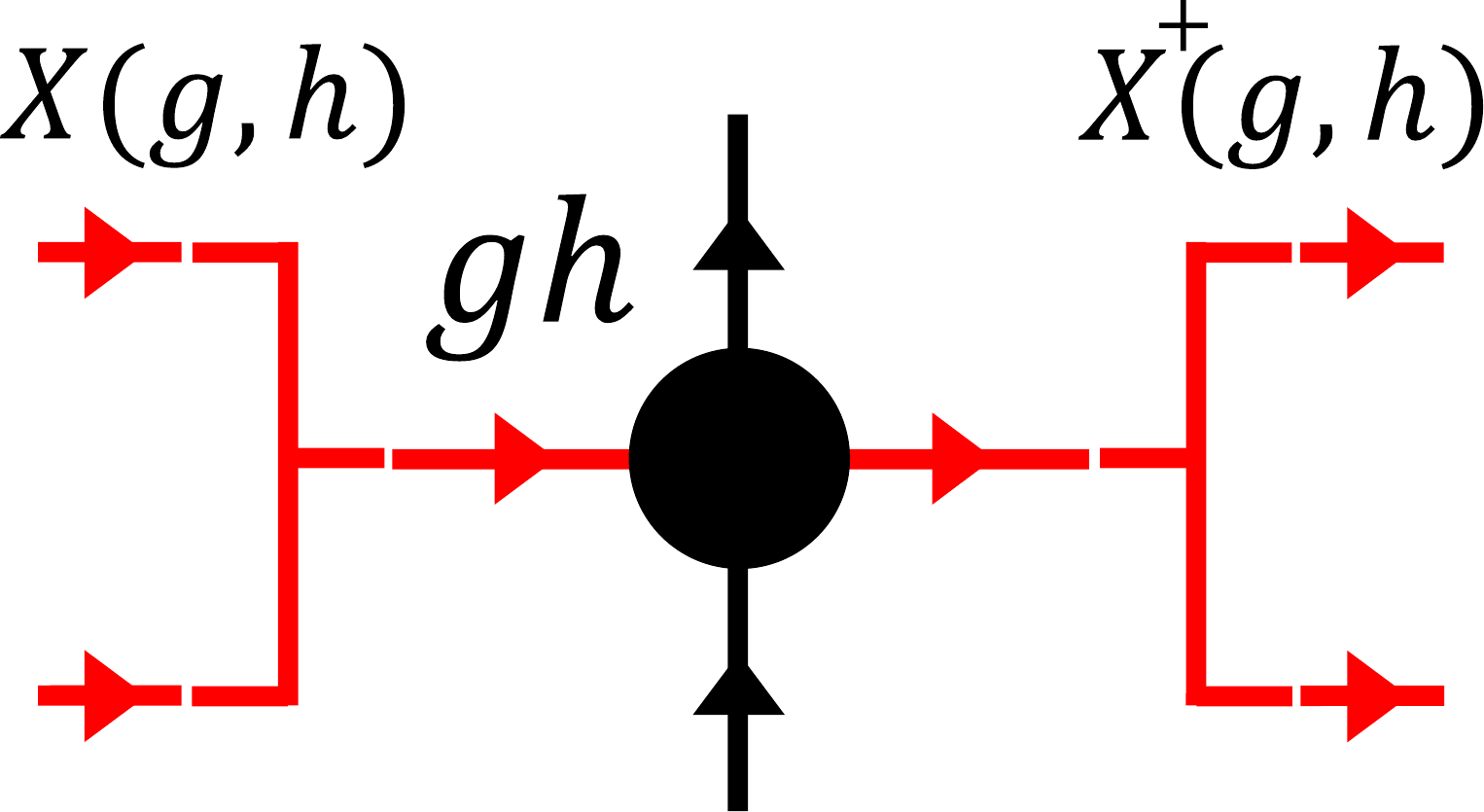}}}=\vcenter{\hbox{\includegraphics[width=0.16\linewidth]{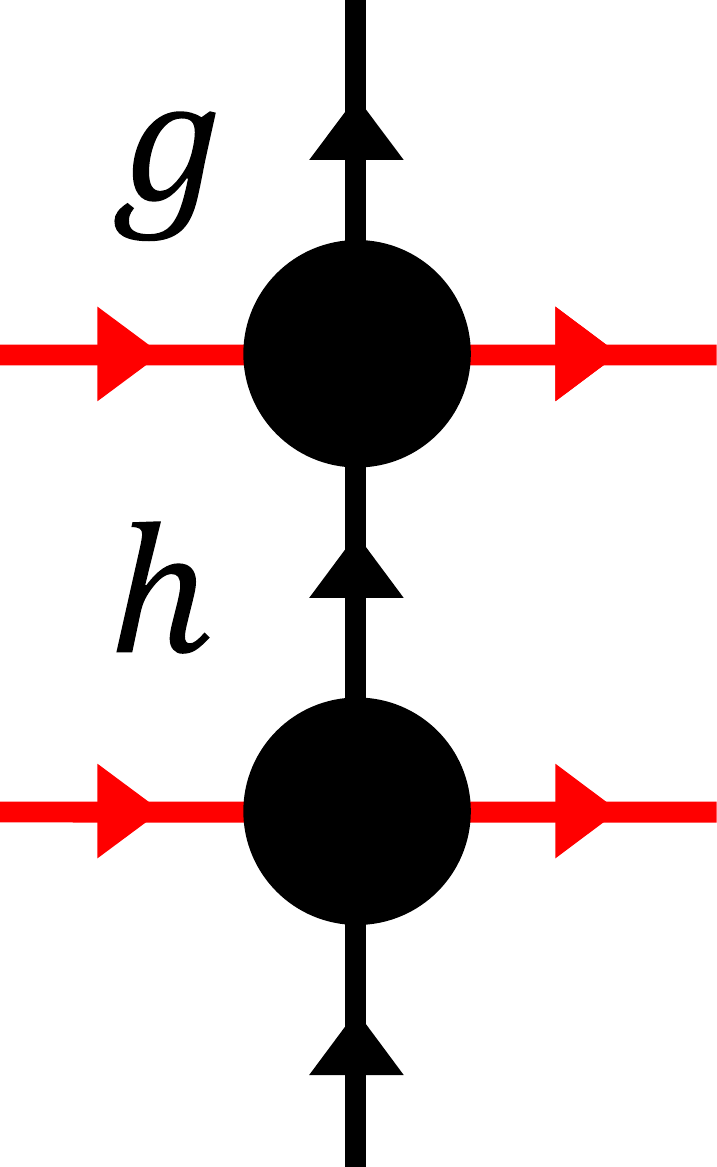}}} \label{a3}
\end{align}
this is equivalent to there being no off diagonal blocks in the product of two MPO tensors after it has been brought into canonical form, and is true for MPOs that arise from fixed-point models. 

Let us now derive a relation between ${\openone_{g}\otimes (Z_{h}^{-1})^T }\ X^{+}(g,h)$ and $X(gh,h^{-1})$ in terms of a \emph{one-line pivotal phase} which we then proceed to calculate in terms of the three cocycle $\alpha$ of the MPO group representation. Consider
\begin{align}
\vcenter{\hbox{\includegraphics[height=0.27\linewidth]{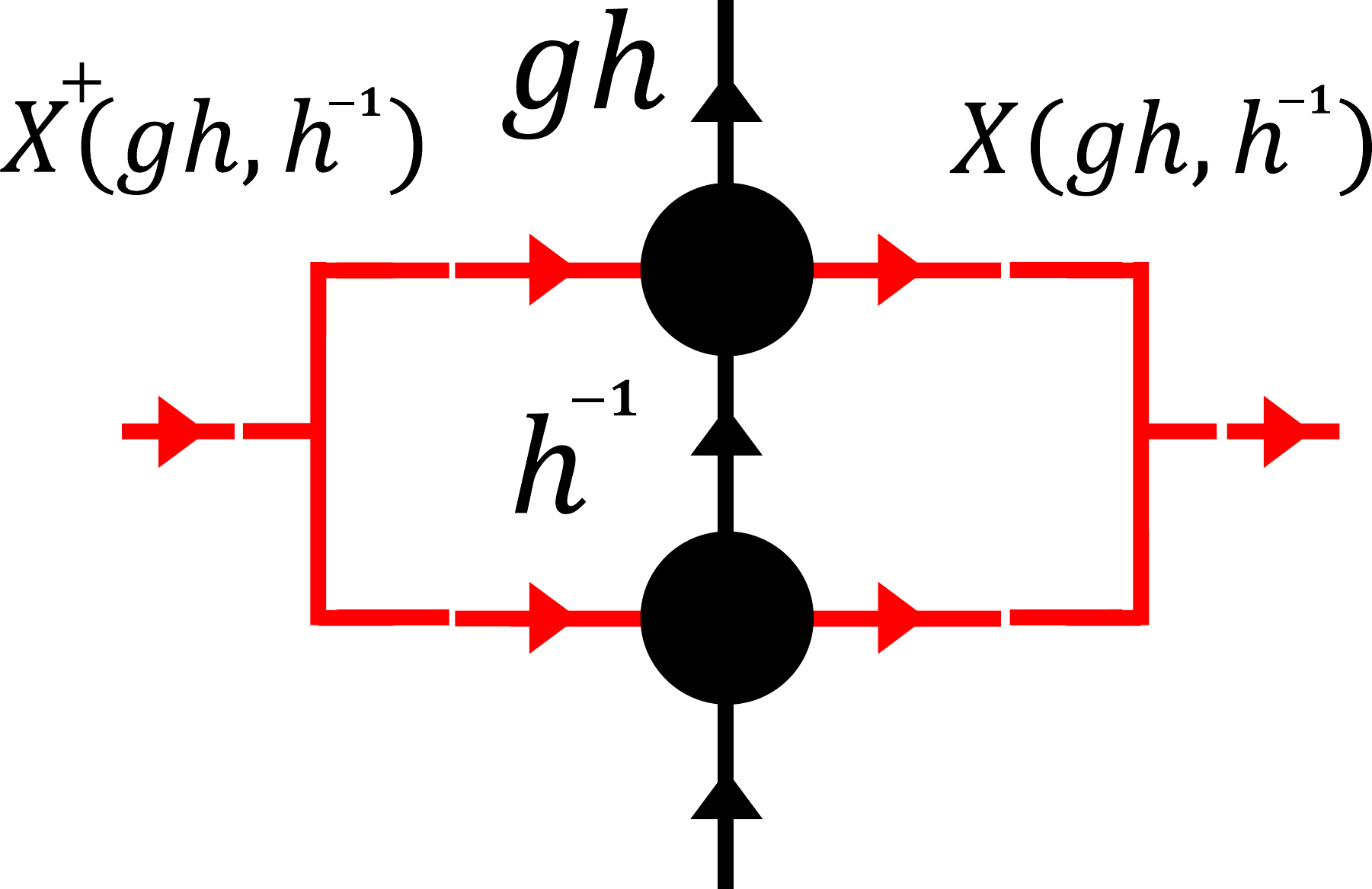}}}\ \
&=
\hspace{1cm}\vcenter{\hbox{\includegraphics[width=0.16\linewidth]{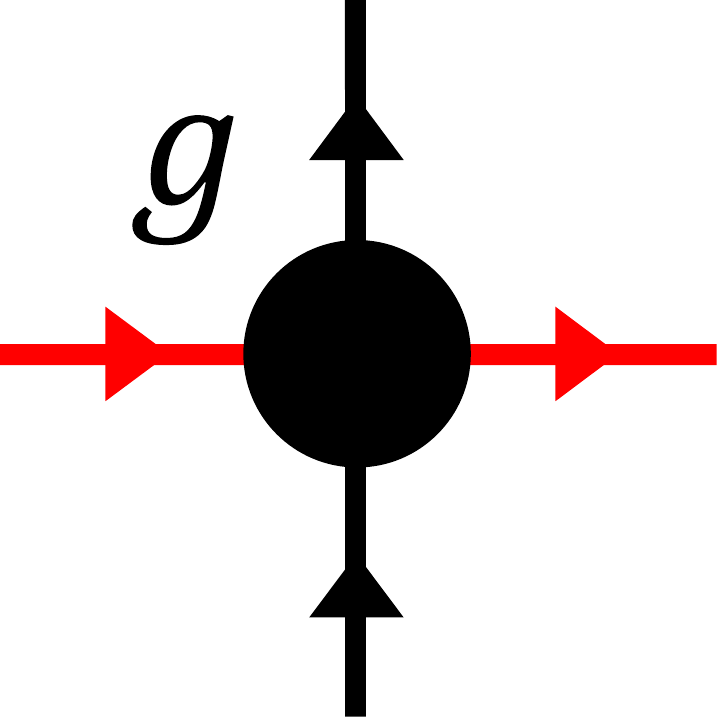}}}
\nonumber \\
&=\hspace{.53cm}\vcenter{\hbox{\includegraphics[height=0.37\linewidth]{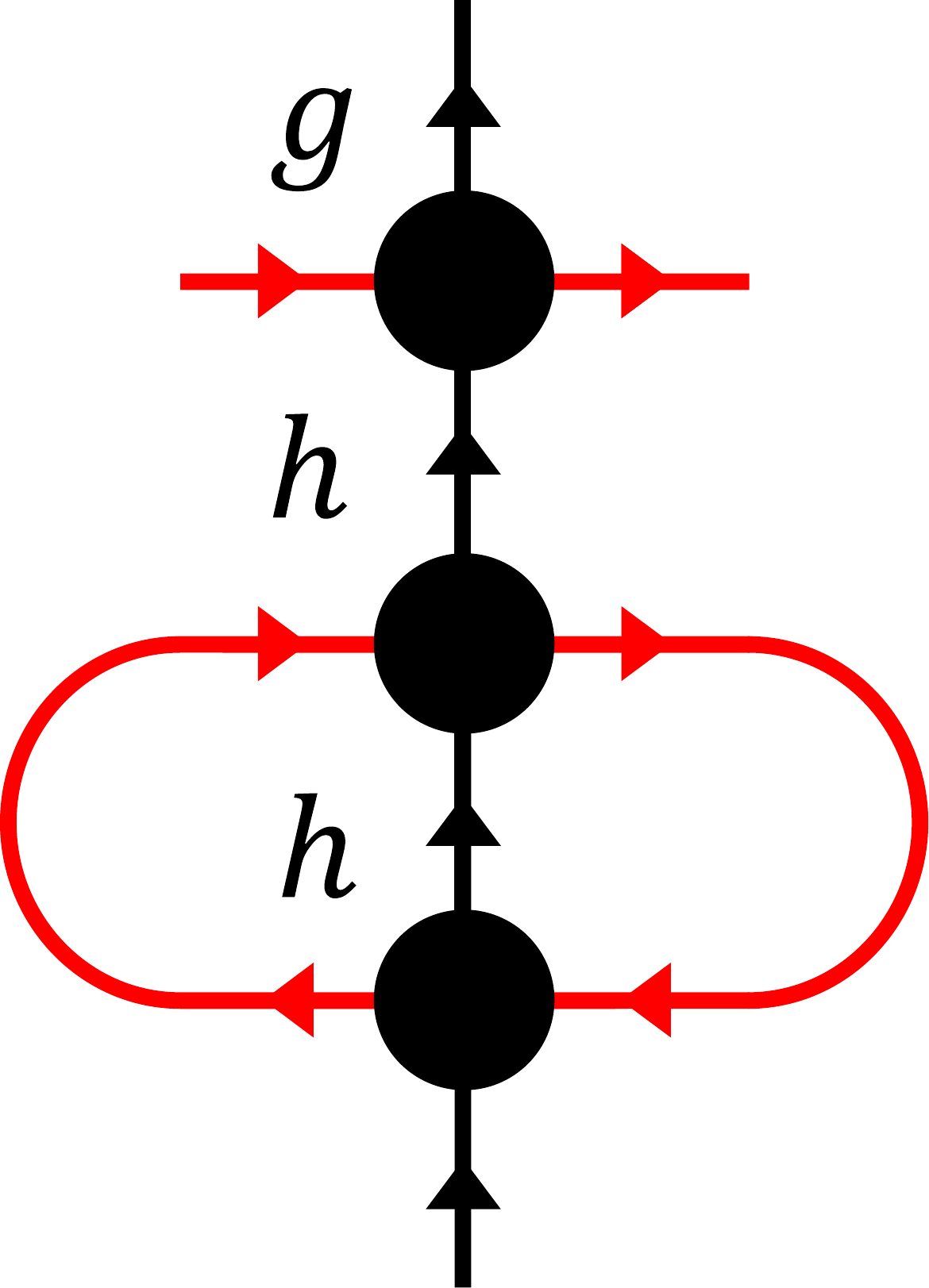}}}
\nonumber \\
&=\hspace{.35cm}\vcenter{\hbox{\includegraphics[height=0.36\linewidth]{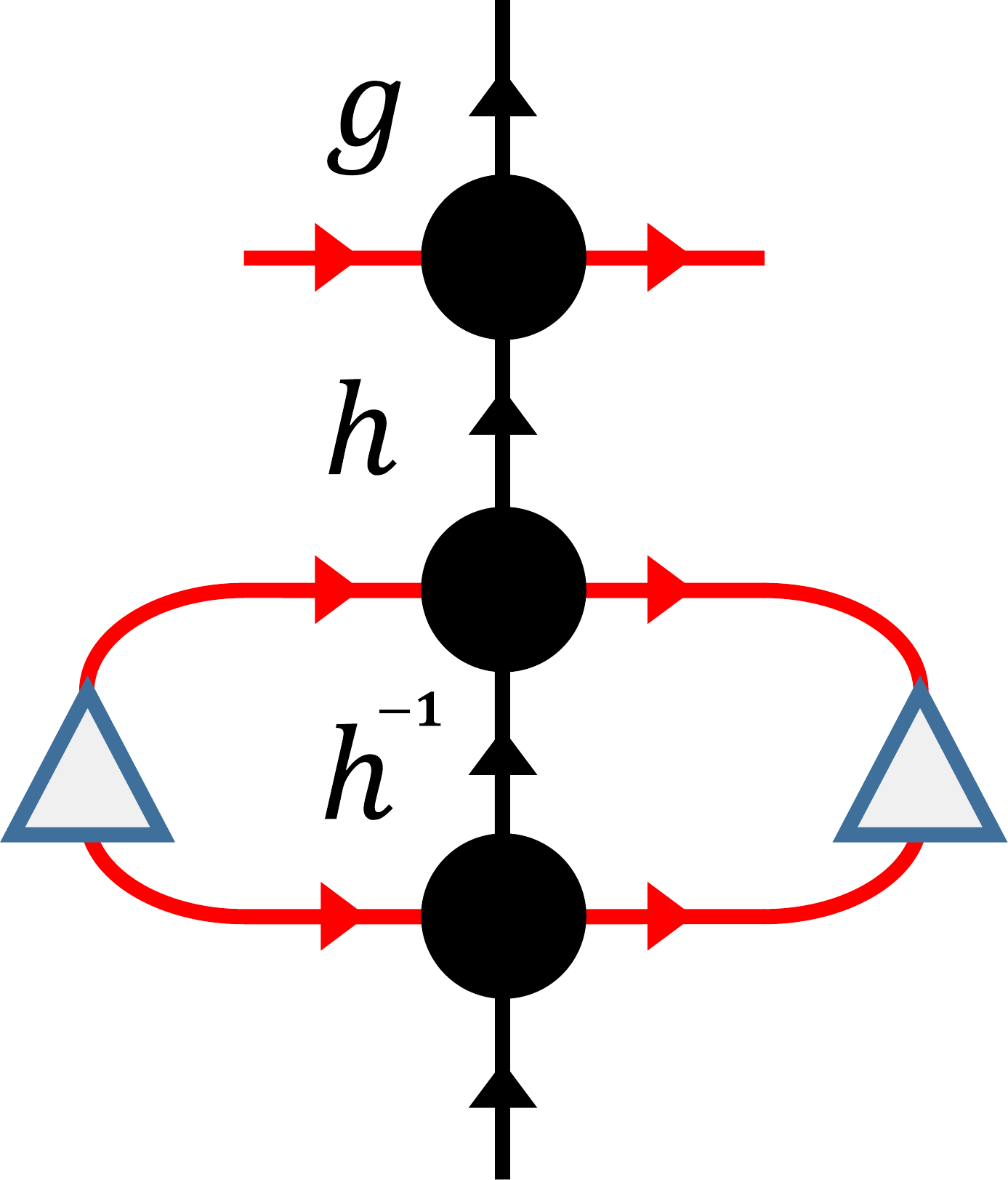}}}
\nonumber \\
&=\vcenter{\hbox{\includegraphics[height=0.35\linewidth]{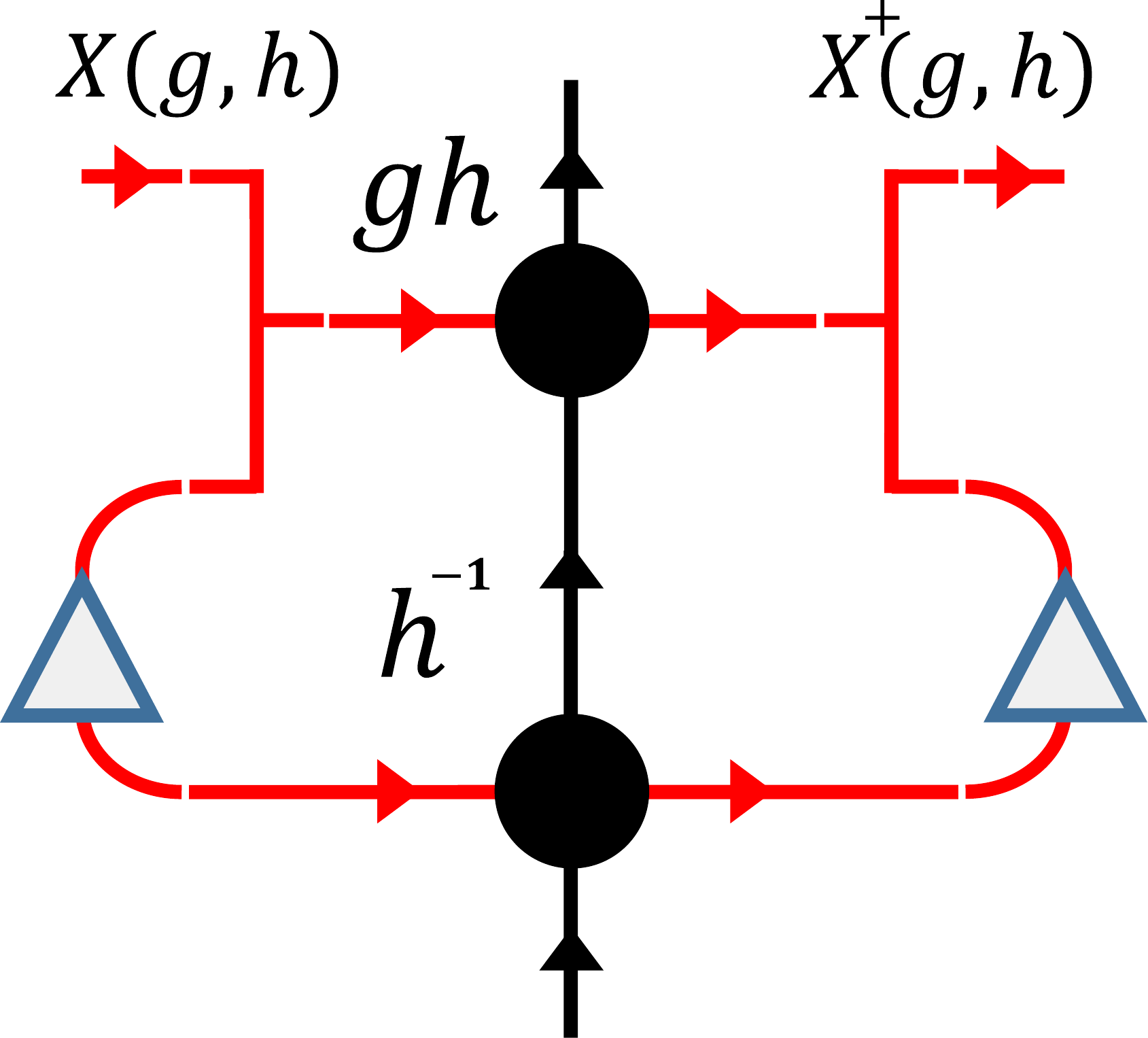}}}
 \label{na7}
\end{align}
 which yields the desired equality
\begin{align}
\raisebox{-.68cm}{\hbox{\includegraphics[height=0.25\linewidth]{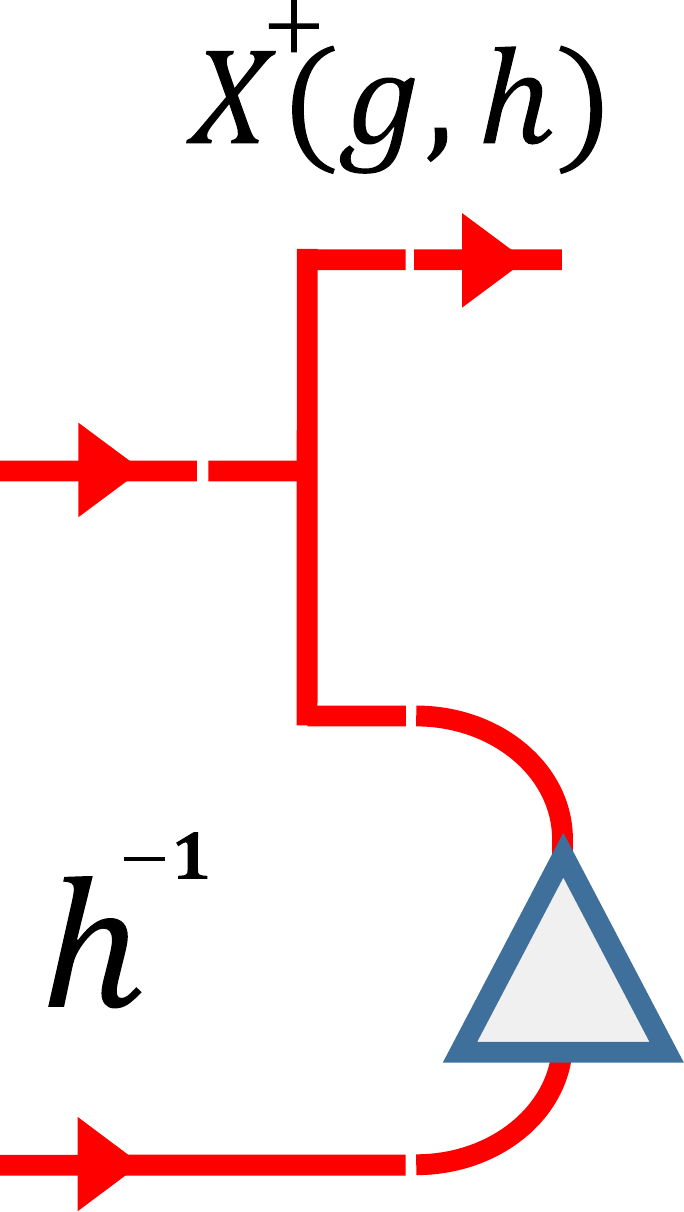}}}
= \gamma(gh,h^{-1}) \
\raisebox{-.6cm}{\hbox{\includegraphics[height=0.2\linewidth]{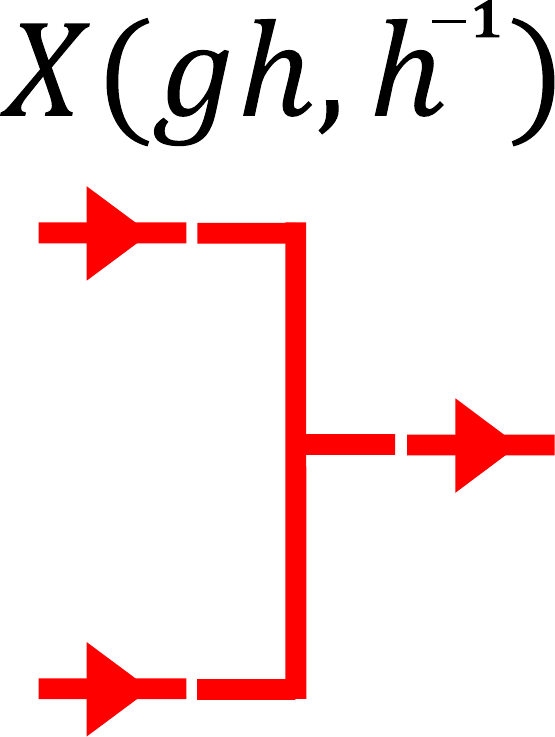}}}
 \label{na8}
\end{align}
where $\gamma(gh,h^{-1})$ is some yet to be determined one-line pivotal phase. We now separate $\gamma(gh,h^{-1})$ into a product of a phase specified by the cocycle $\alpha$ and another phase $b(g,h)$ which we show to be trivial. Multiplying Eq.\eqref{na8} by $X^{-1}(g_0g_1,g_1^{-1})$ yields
\begin{align}
\gamma(gh,h^{-1})  \, \raisebox{.01cm}{\hbox{\includegraphics[height=0.05\linewidth]{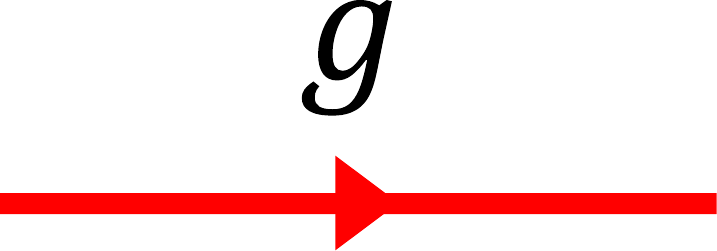}}}
\ = \raisebox{-0.6cm}{\hbox{\includegraphics[height=0.25\linewidth]{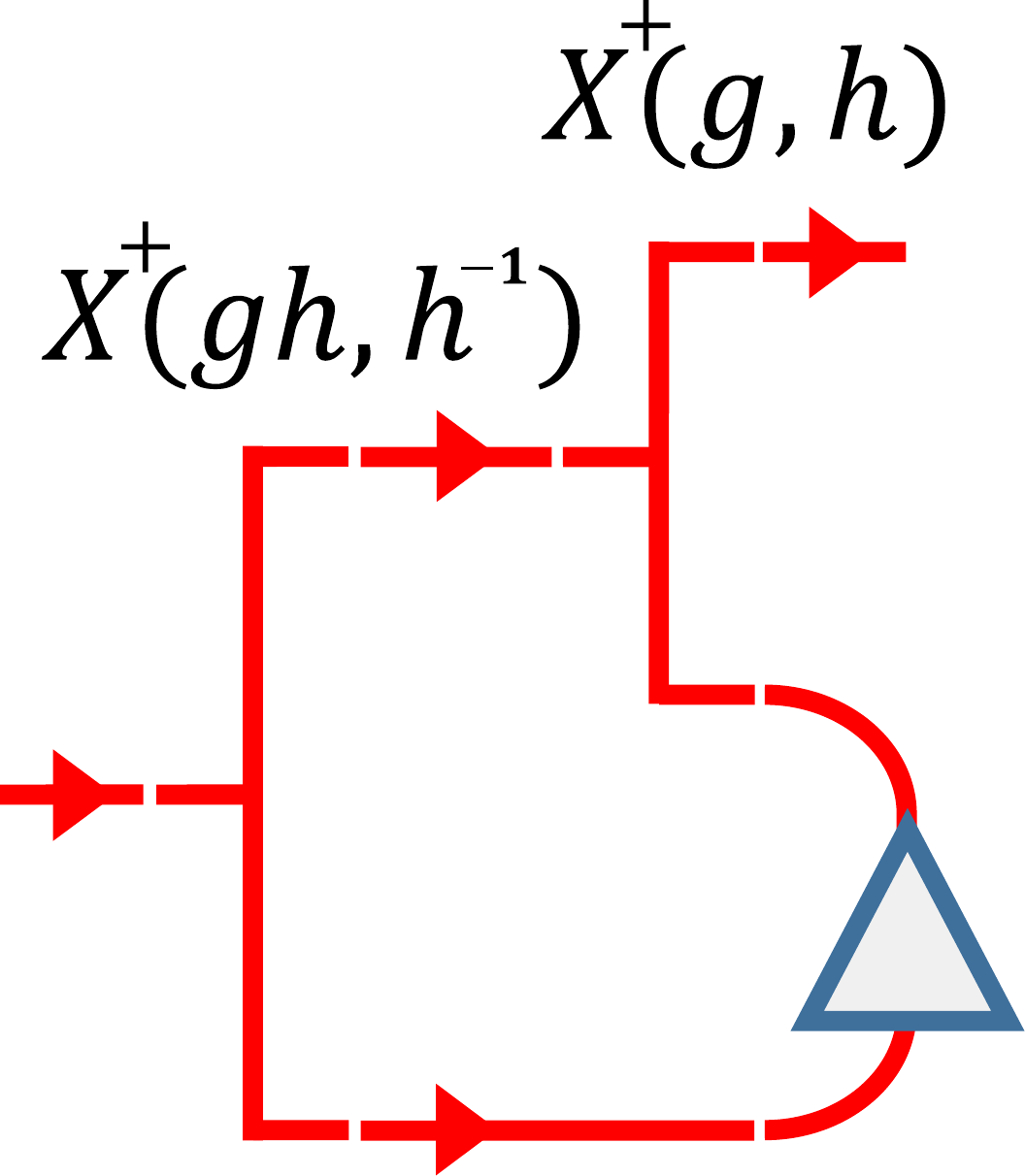}}}
\nonumber \\
=\alpha^{-1}(g,h,h^{-1}) \,  \raisebox{-1.02cm}{\hbox{\includegraphics[height=0.25\linewidth]{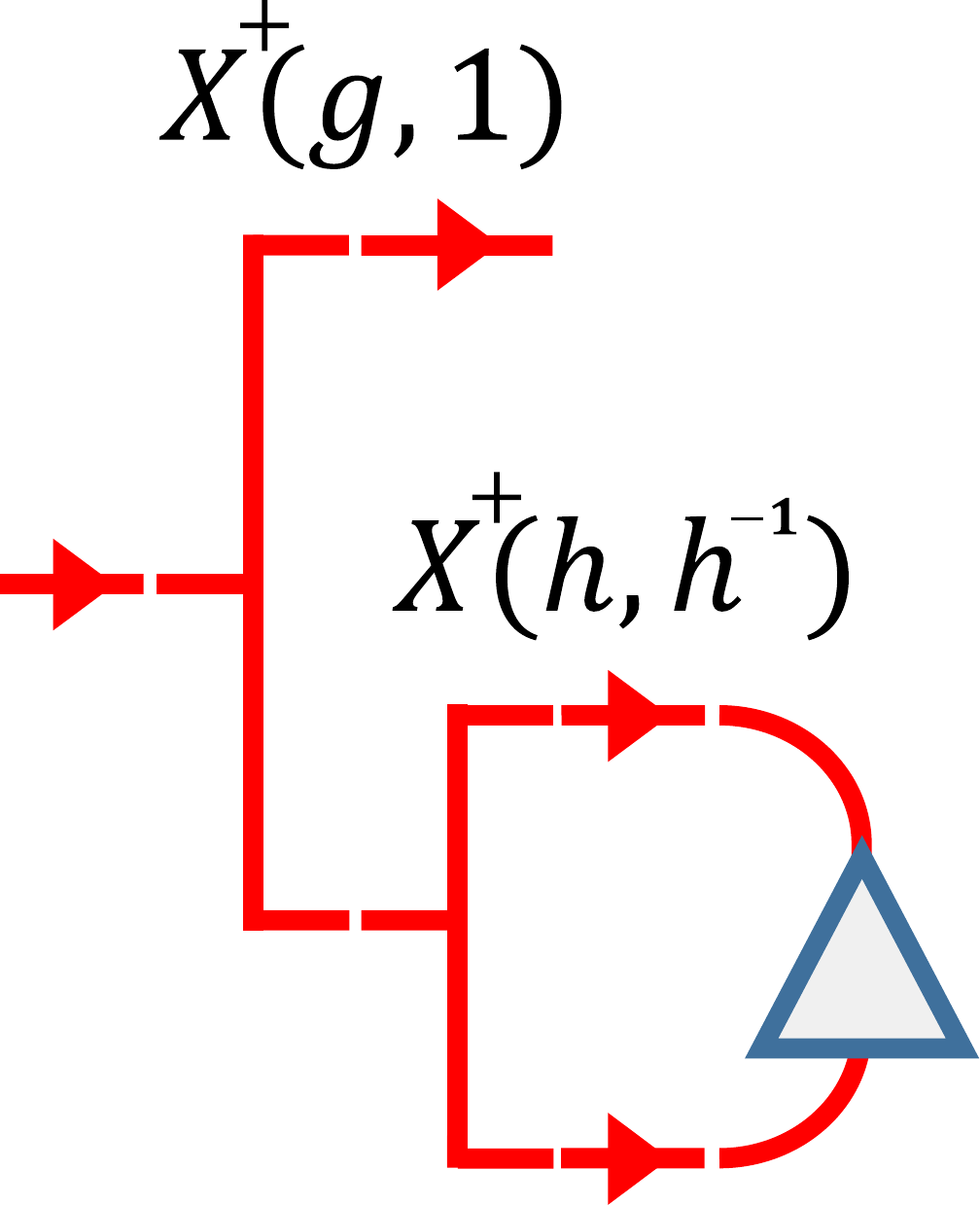}}}
\nonumber \\
=\alpha^{-1}(g,h,h^{-1}) \, b(g,h) \, \raisebox{.0cm}{\hbox{\includegraphics[height=0.05\linewidth]{Figures/newfig31}}}
 \label{na9}
\end{align}
Now considering 
\begin{align}
\vcenter{\hbox{\includegraphics[height=0.3\linewidth]{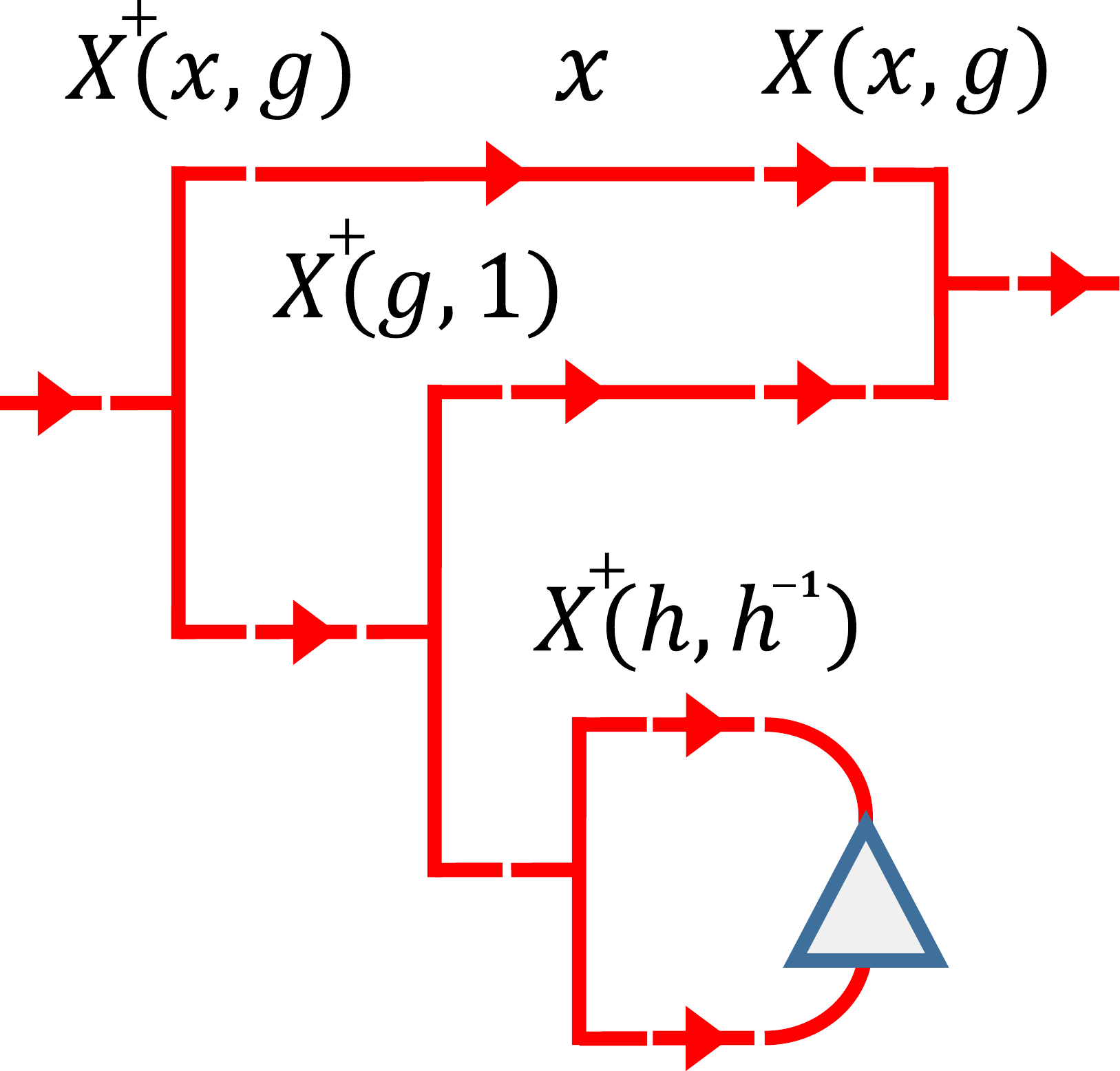}}}
= b(g,h) \, \raisebox{.01cm}{\hbox{\includegraphics[height=0.05\linewidth]{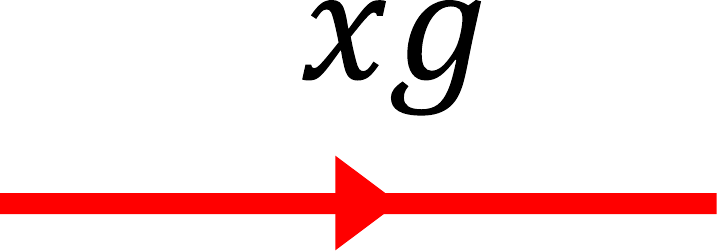}}}
 \label{na10}
\end{align}
after an application of Eq.\eqref{n29} to the left most reductions tensors we see that $b(g,h)=b(xg,h),\, \forall x$ and hence $b$ has no dependence on the first input and can be absorbed into the definition of $Z_h$. 
Similar reasoning yields another useful equality
\begin{align}
\vcenter{\hbox{\includegraphics[height=0.25\linewidth]{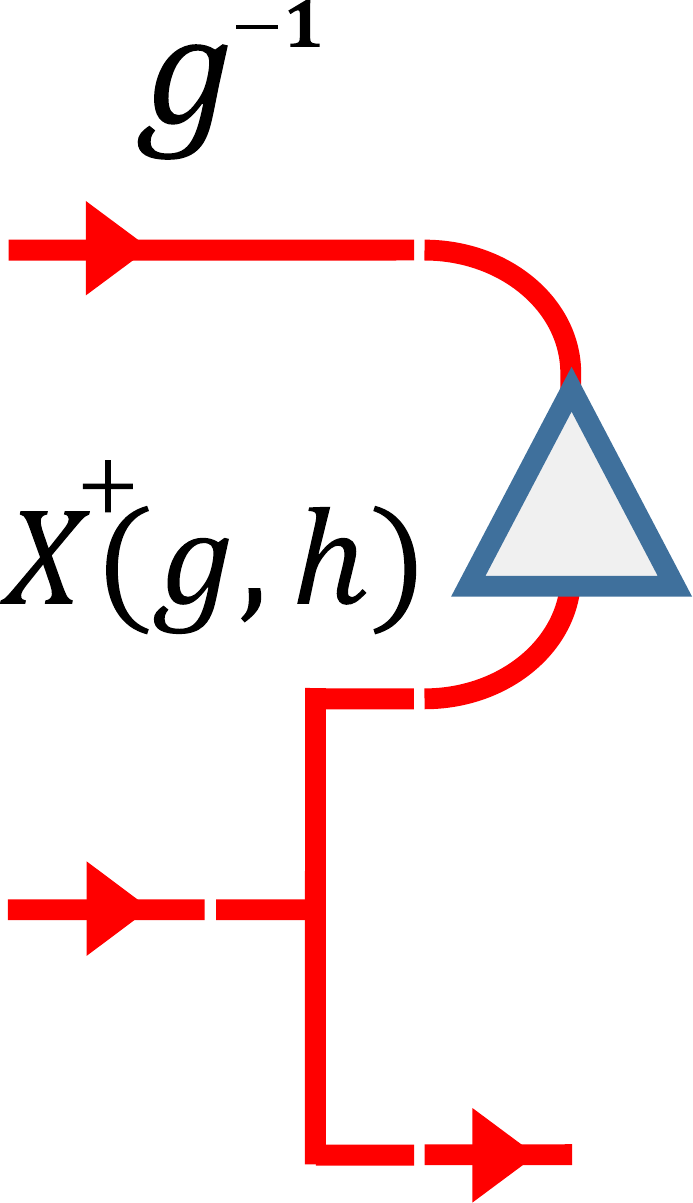}}}
= \alpha(g^{-1},g,h) \
\raisebox{-.6cm}{\hbox{\includegraphics[height=0.2\linewidth]{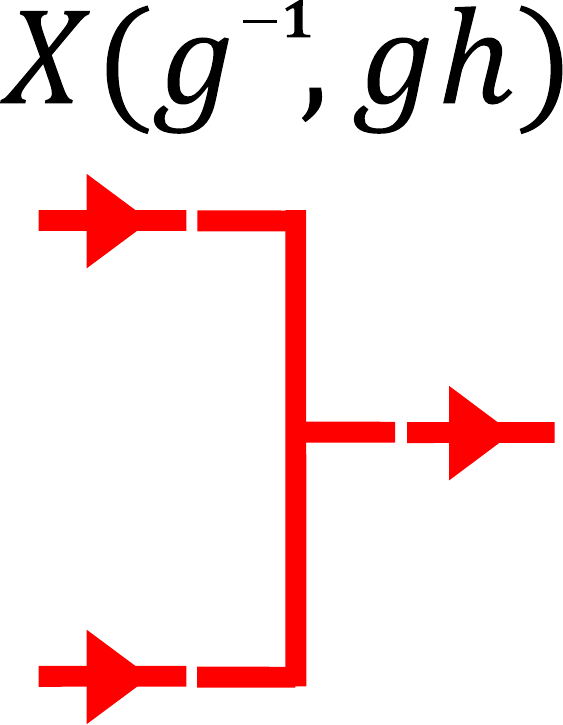}}}
 \label{na11}
\end{align}
In summary we have calculated the one-line pivotal phases 
\begin{align}
\gamma(gh,h^{-1})&=\alpha^{-1}(g,h,h^{-1}) \nonumber \\ 
\gamma'(gh,h^{-1})&=\alpha(g^{-1},g,h) \label{na12}
\end{align}

We now proceed to define a \emph{pivotal} phase relating the following different reductions of the same left handed MPO tensors
\begin{align}
\vcenter{\hbox{\includegraphics[width=0.16\linewidth]{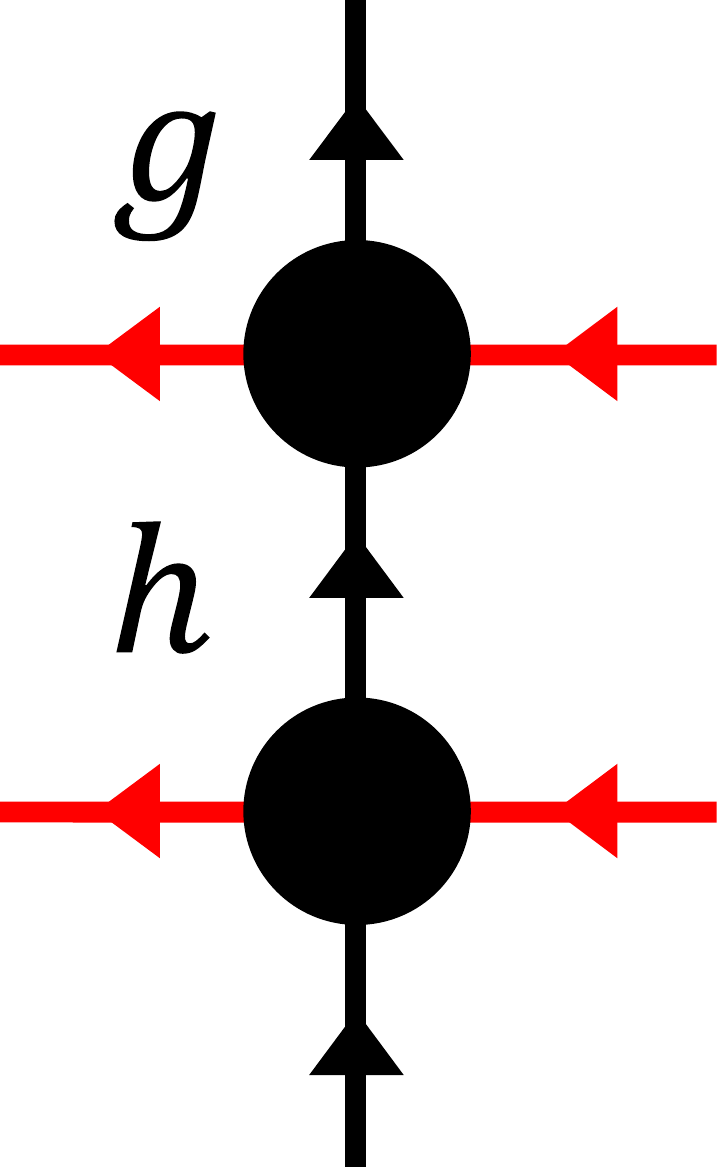}}}
&= \raisebox{-.63cm}{\hbox{\includegraphics[height=0.2\linewidth]{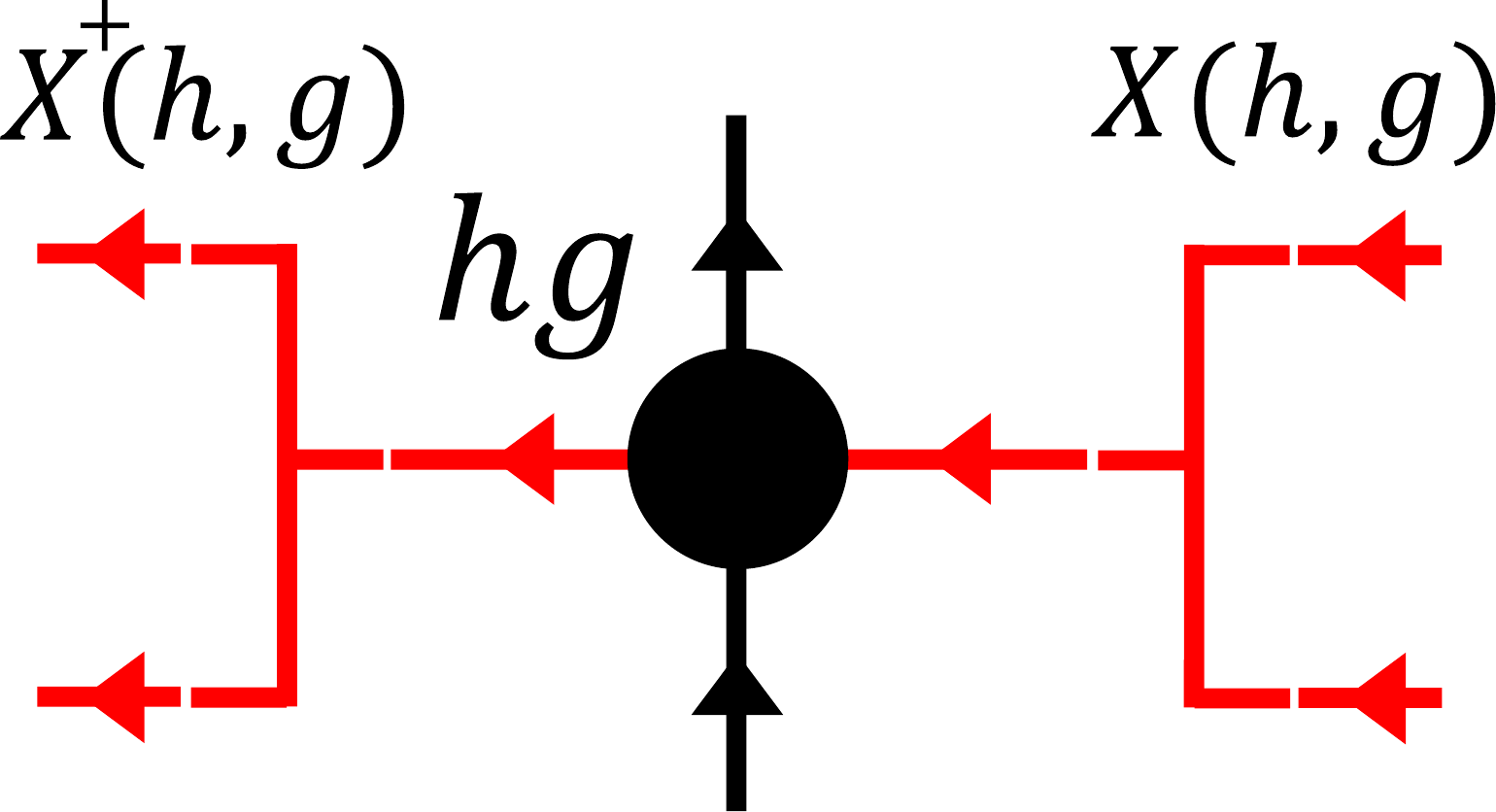}}}
 \label{na13}
\\
&= 
\raisebox{-.63cm}{\hbox{\includegraphics[height=0.2\linewidth]{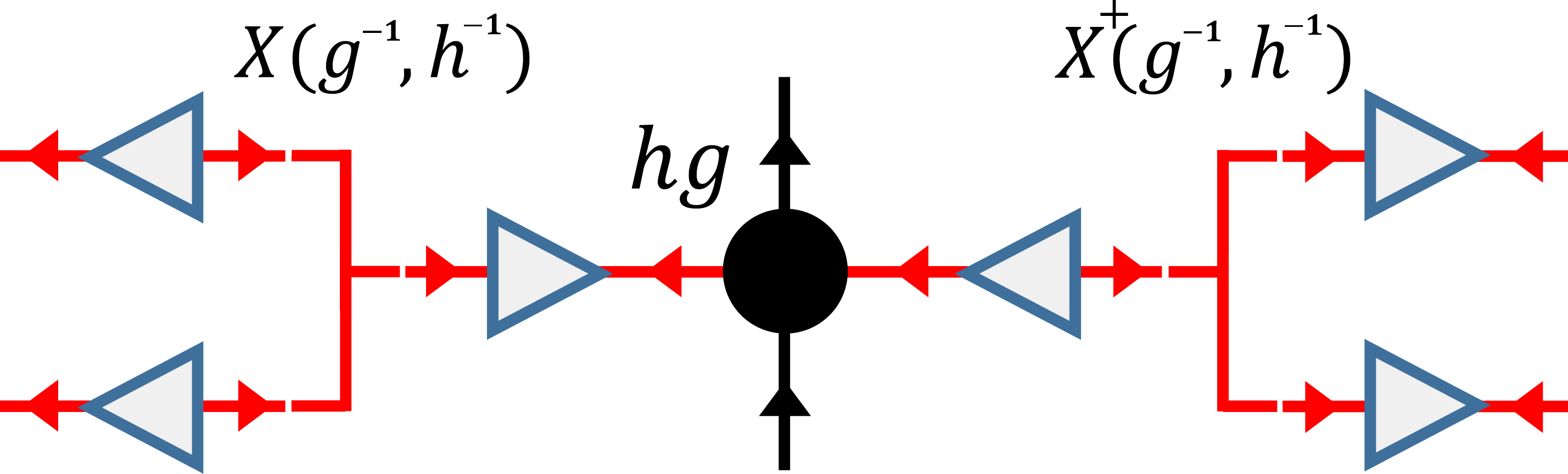}}}
\nonumber
\end{align}
Hence
\begin{align}
\raisebox{-.7cm}{\hbox{\includegraphics[height=0.21\linewidth]{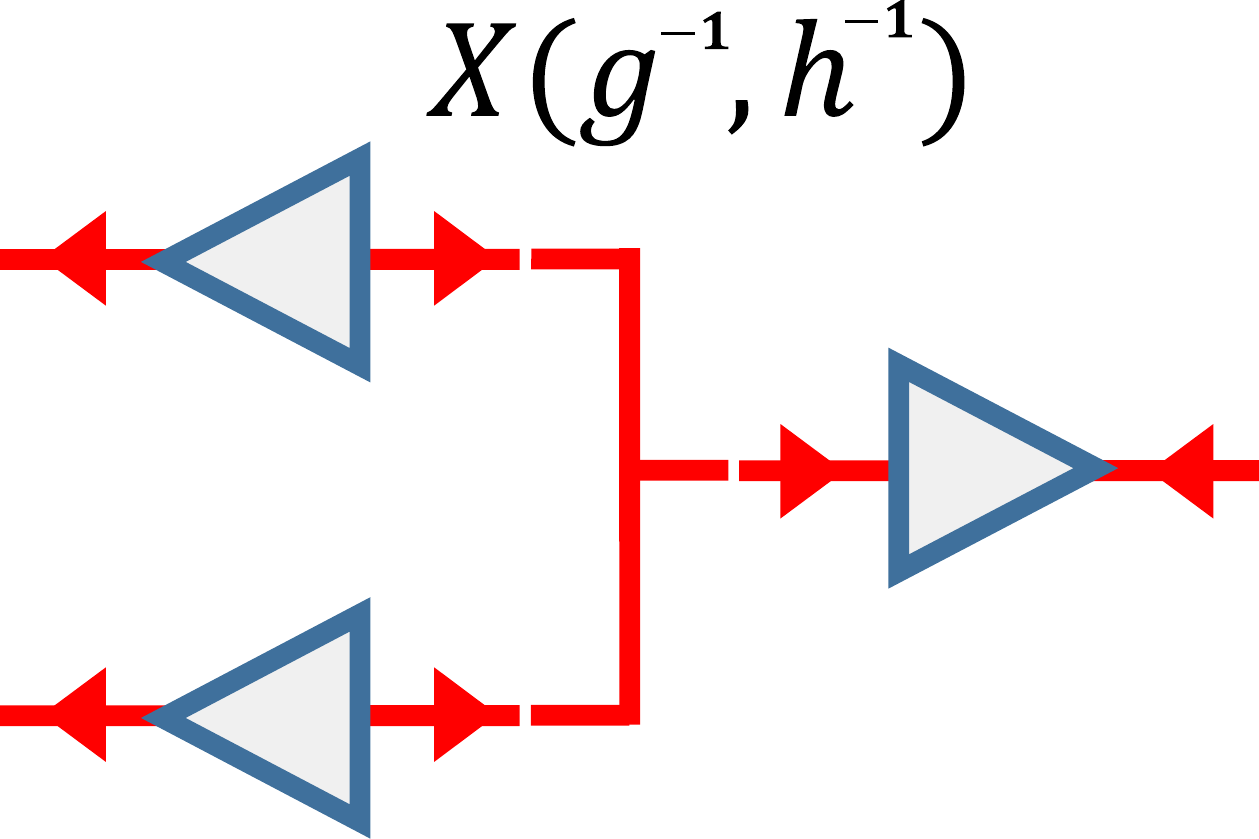}}}
= \beta(g,h) \ 
\raisebox{-.55cm}{\hbox{\includegraphics[height=0.2\linewidth]{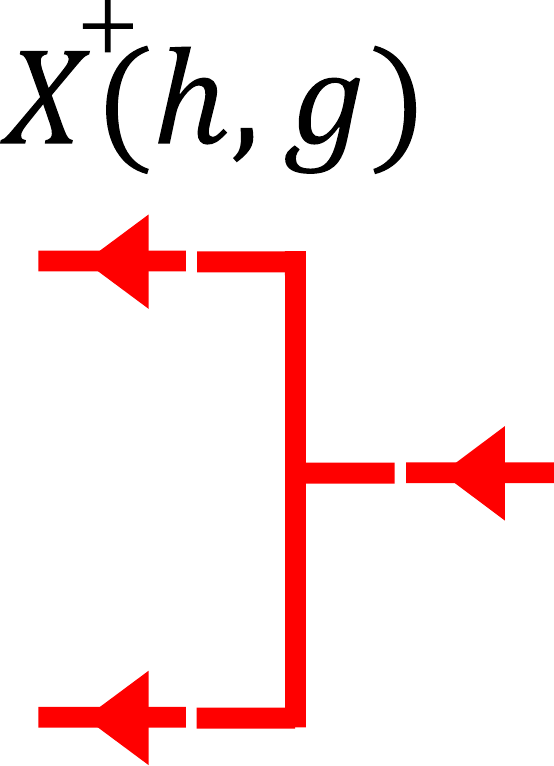}}}
 \label{na15}
\end{align}
for some pivotal phase $\beta(g,h) \in U(1)$. 
By making use of Eqs.(\ref{na8},\ref{na11},\ref{na12}) we calculate $\beta$ directly to find 
\begin{align}
\beta(g,h)=\varepsilon(g)\varepsilon(h) \tilde \beta(g,h) \label{na16}
\end{align}
where 
\begin{align*}
\varepsilon(g)&:=\chi_g\, \alpha(g,g^{-1},g) \\
\tilde \beta(g,h)&:=\frac{\alpha(h,g,g^{-1})}{\alpha(hg,g^{-1},h^{-1})}
\end{align*}
we proceed to show that $\varepsilon \cong 1$ and hence $\beta \cong \tilde \beta$.

Evaluating $\beta$ in two different ways as follows
\begin{align}
\beta(g,h) \ 
\raisebox{-.16cm}{\hbox{\includegraphics[height=0.1\linewidth]{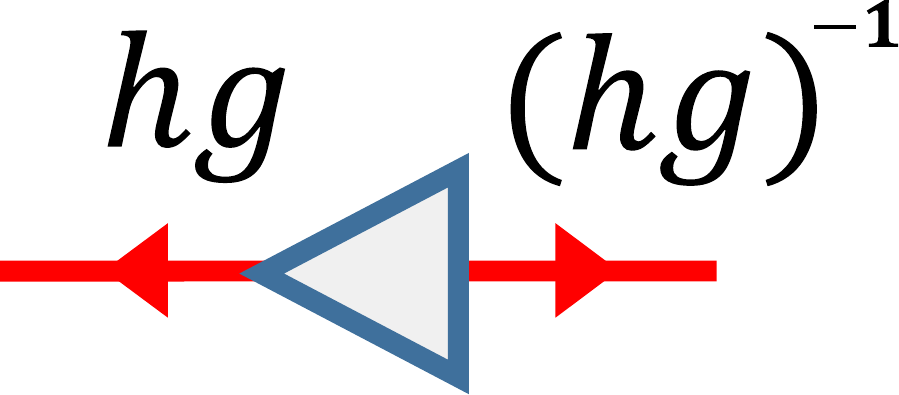}}}
=\raisebox{-.7cm}{\hbox{\includegraphics[height=0.21\linewidth]{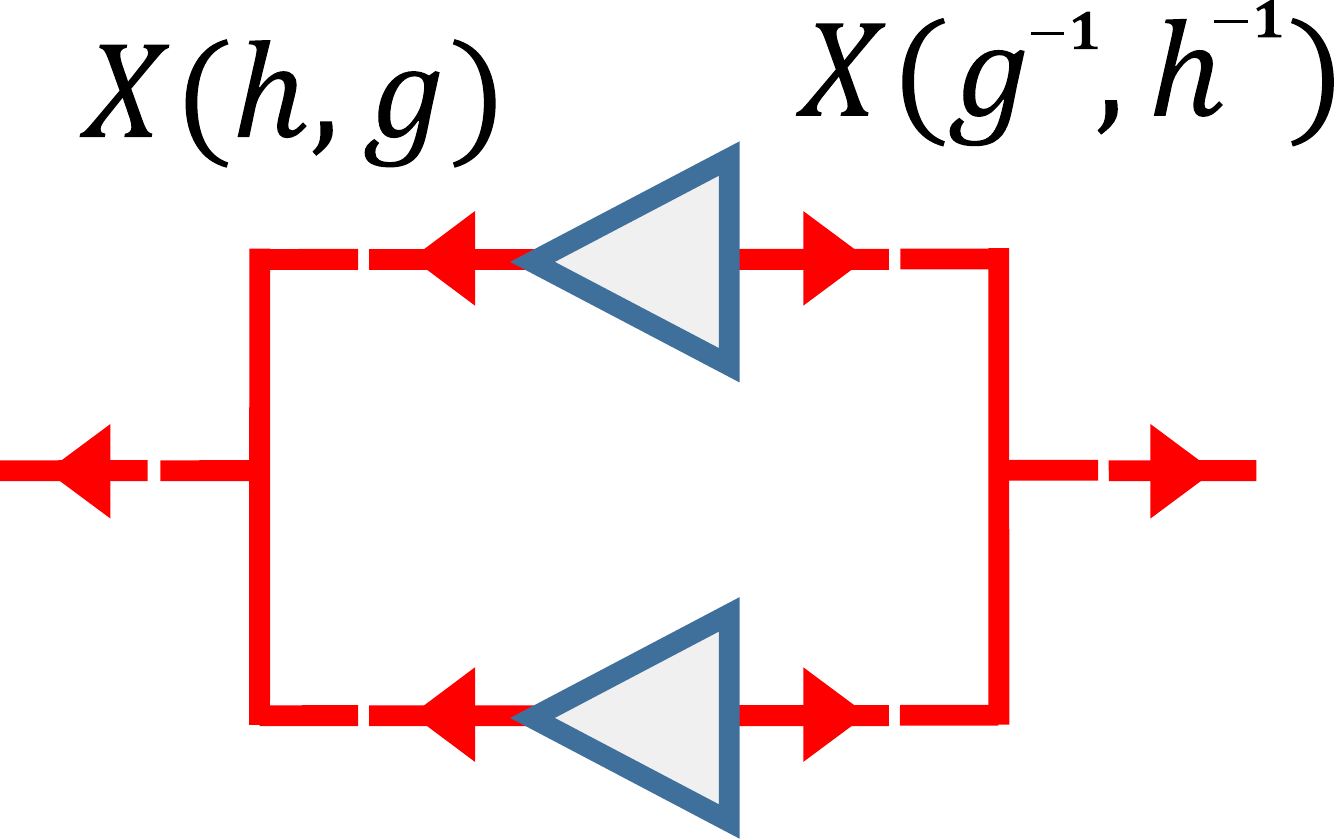}}}
\nonumber \\
= {\chi_g  \chi_h} \ 
\raisebox{-.7cm}{\hbox{\includegraphics[height=0.21\linewidth]{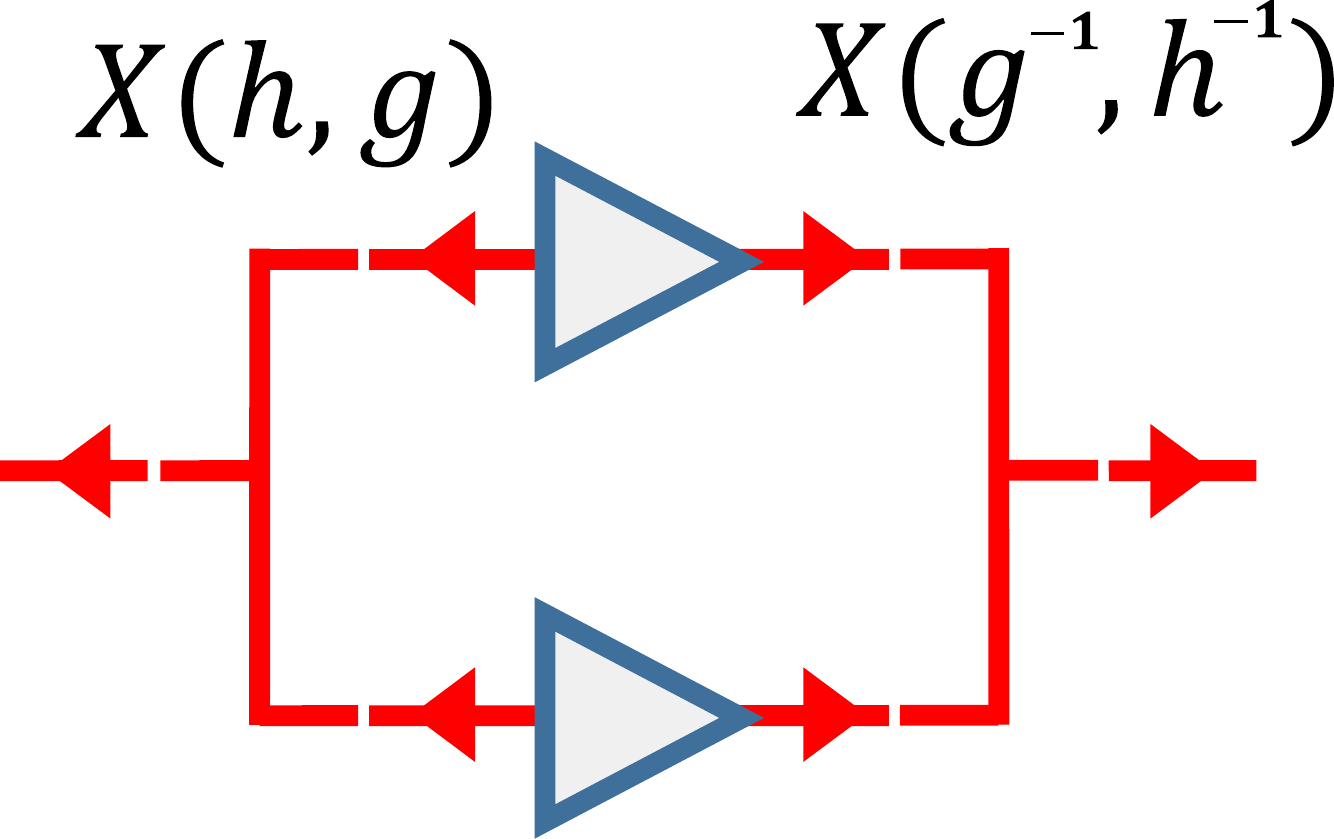}}}
\nonumber \\
= \frac{\chi_g \chi_h \beta(h^{-1},g^{-1})}{\chi_{gh}} \ 
\raisebox{-.16cm}{\hbox{\includegraphics[height=0.1\linewidth]{Figures/newfig45}}}
\label{na17}
\end{align}
 leads to the relation on $\varepsilon$ 
\begin{align}
\varepsilon(k)\varepsilon(h)\varepsilon(hk)=1 \label{na18}
\end{align}
after several applications of the 3-cocycle condition for $\alpha$. 

Using Eq.\eqref{n29} we find
\begin{align}
\raisebox{-.88cm}{\hbox{\includegraphics[height=0.3\linewidth]{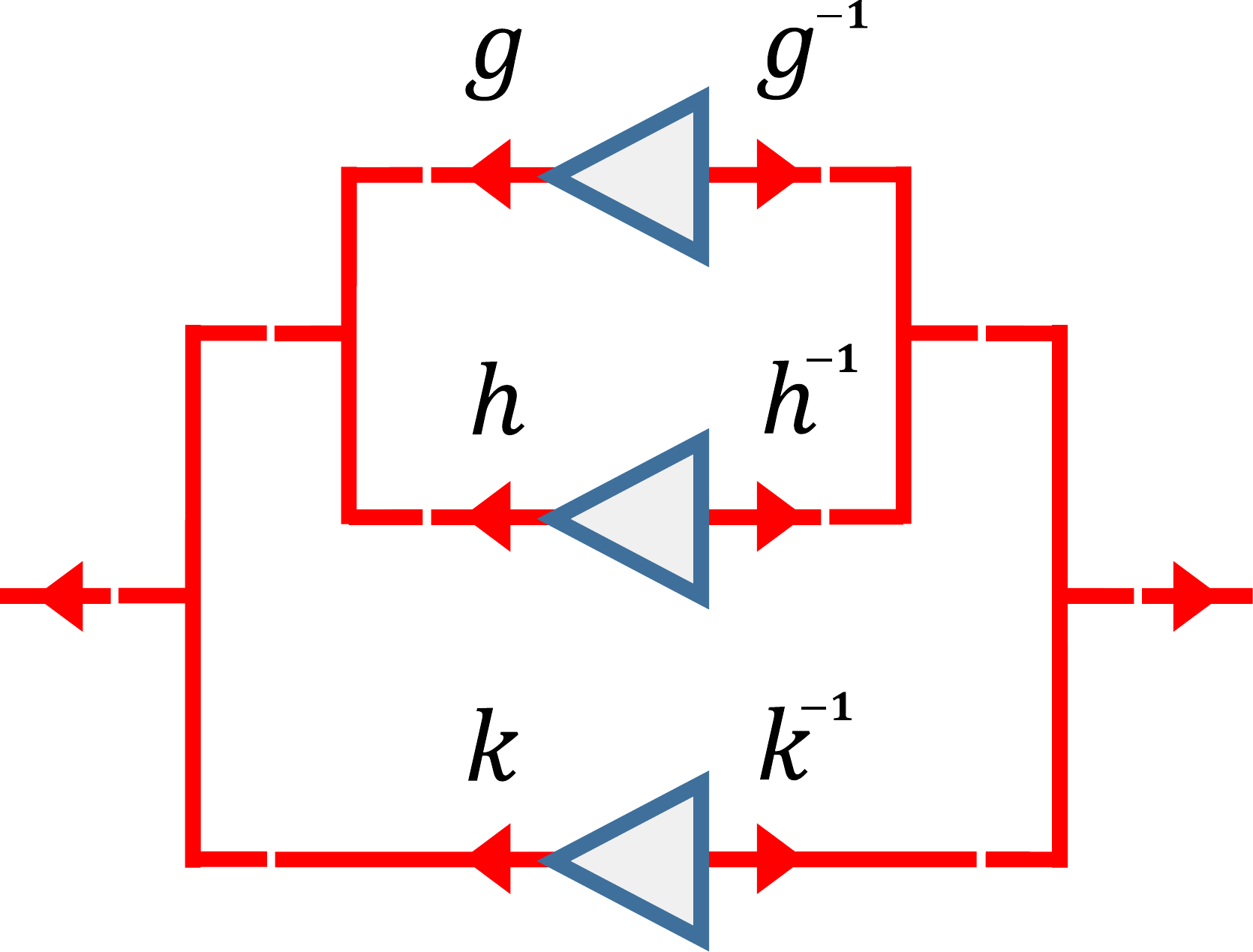}}}
&= \frac{\alpha(g^{-1},h^{-1},h^{-1})}{\alpha(k,h,g)} \nonumber \\
&\times \raisebox{-1.34cm}{\hbox{\includegraphics[height=0.3\linewidth]{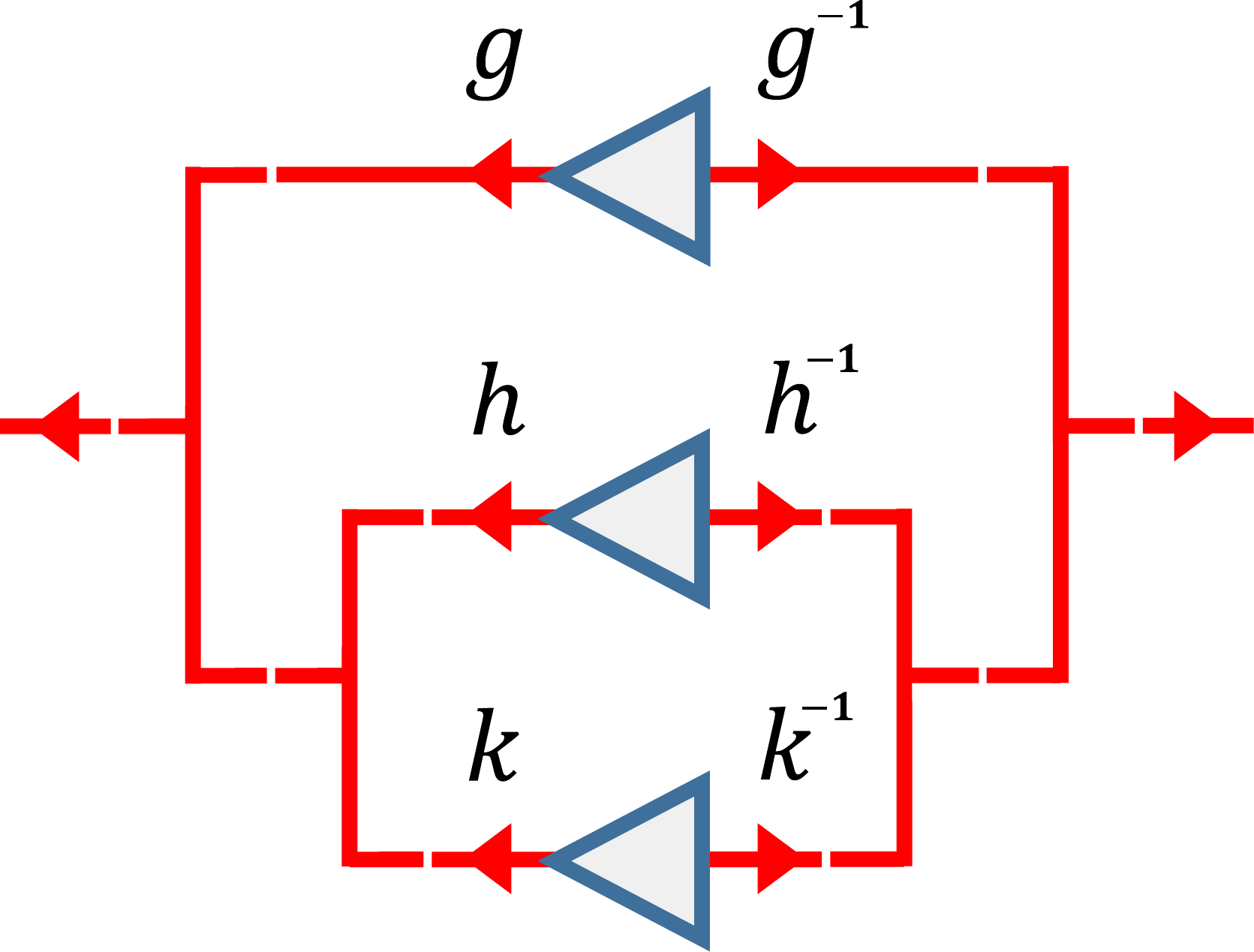}}}
\label{na19}
\end{align}
applying Eq.\eqref{na15} twice to both sides yields the further constraint on $\beta$
\begin{equation} \label{na20}
d\beta (a,b,c):=\frac{\beta(a,b)\beta(ab,c)}{ \beta(b,c) \beta(a,bc)}  =\frac{\alpha(a,b,c)}{\alpha(c^{-1},b^{-1},a^{-1})}
\end{equation}
hence $\alpha$ forms a potential obstruction to $\beta$ being a 2-cocycle.
Note that $\tilde \beta$ also satisfies Eq.\eqref{na20} as a consequence of the 3-cocycle condition for $\alpha$ and hence the function $\theta(a,b):=\varepsilon(a)\varepsilon(b)$ satisfies the 2-cocycle condition $d\theta(a,b,c) = 1$. This 2-cocycle condition, together with Eq.\eqref{na18}, implies that $\varepsilon(a)=\varepsilon(c),\,\forall a,c\in \mathsf{G}$ and since $\varepsilon(1)=1$ consequently $\varepsilon \equiv 1$ is the constant function. This of course implies $\beta\equiv \tilde \beta$ which is the desired result
\begin{align}\label{betafin}
 \beta(g,h)=\frac{\alpha(h,g,g^{-1})}{\alpha(hg,g^{-1},h^{-1})}.
\end{align}

\section{Crossing tensors} \label{newapp2}

In this Appendix we define four crossing tensors and demonstrate that they are related by phases involving only the 3-cocycle of the MPO representation. We proceed to define a composition operation on the crossing tensors and calculate the resulting crossing tensor. Building upon this result we determine the transformation of a crossing tensor under the global symmetry. Finally we calculate the effect of modular transformations on the crossing tensors. 

\subsection{Definitions}

We now introduce several different forms for the crossing tensor (see Eq.\eqref{cxy}) that are related by phases which play an important role in our calculations
\begin{align}
\QW{R}{g}{h}:&=X(g,h) X^{+}(h,g) = \ \vcenter{\hbox{\includegraphics[width=0.24\linewidth]{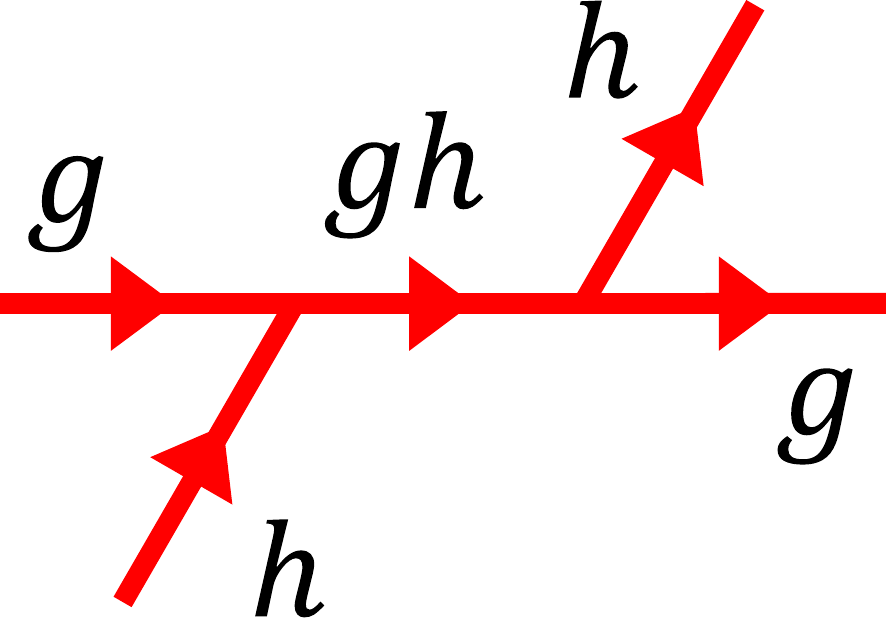}}} \label{na21a} 
\\
\QW{L}{g}{h}:&= X(h,g) X^{+}(g,h) = \ \raisebox{-.62cm}{\hbox{\includegraphics[width=0.24\linewidth]{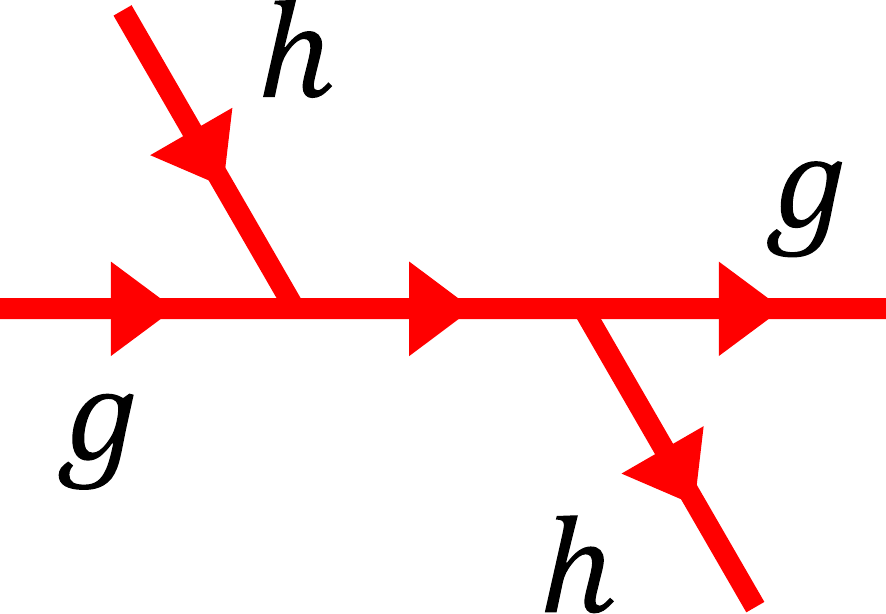}}} \label{na21b} 
\\
\QV{R}{g}{h}:&= X^{+}(gh,h^{-1}) [X^{+}(h,g)\otimes Z_h^{-1}] 
\nonumber \\
&= \ \raisebox{-.94cm}{\hbox{\includegraphics[width=0.24\linewidth]{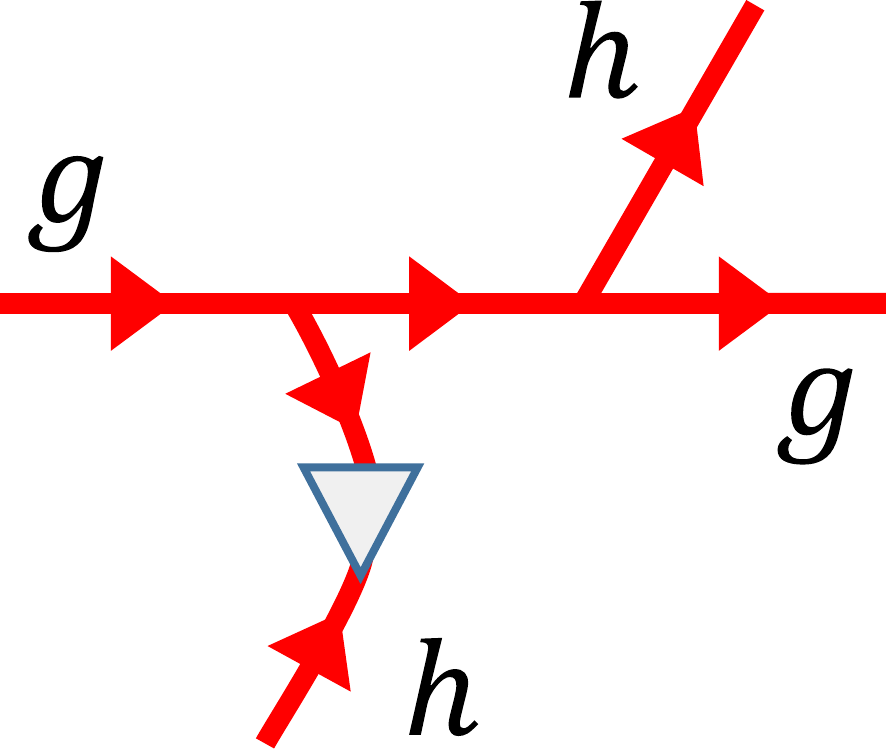}}} \label{na21c}
\\
\QV{L}{g}{h}:&= [X(h,g)\otimes Z_h] X(gh,h^{-1}) 
\nonumber \\
&=\ \raisebox{-.94cm}{\hbox{\includegraphics[width=0.24\linewidth]{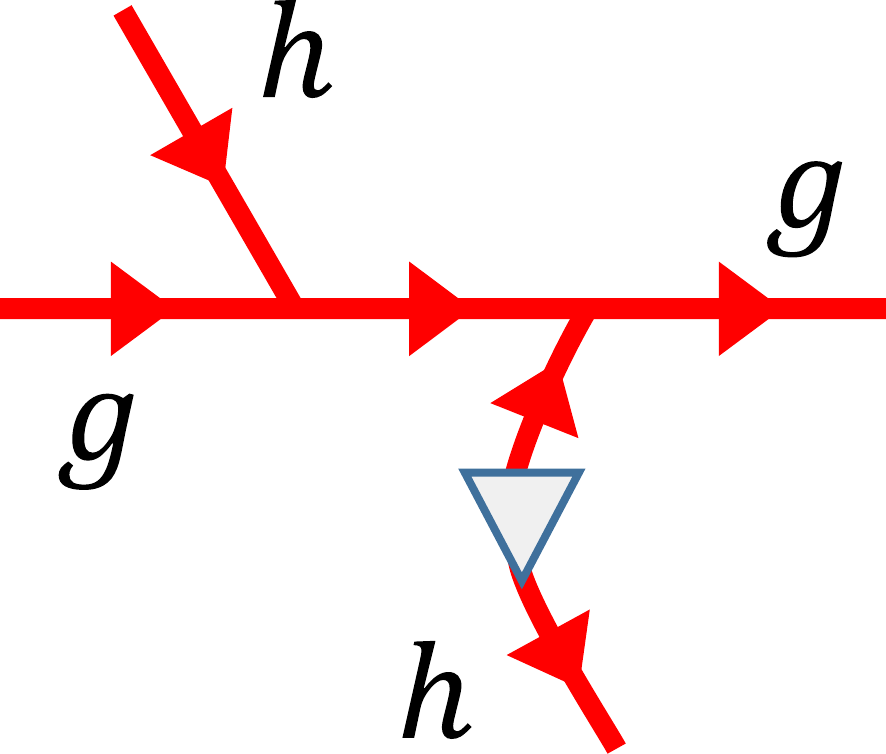}}} 
\, \sim \, \raisebox{-.6cm}{\hbox{\includegraphics[width=0.24\linewidth]{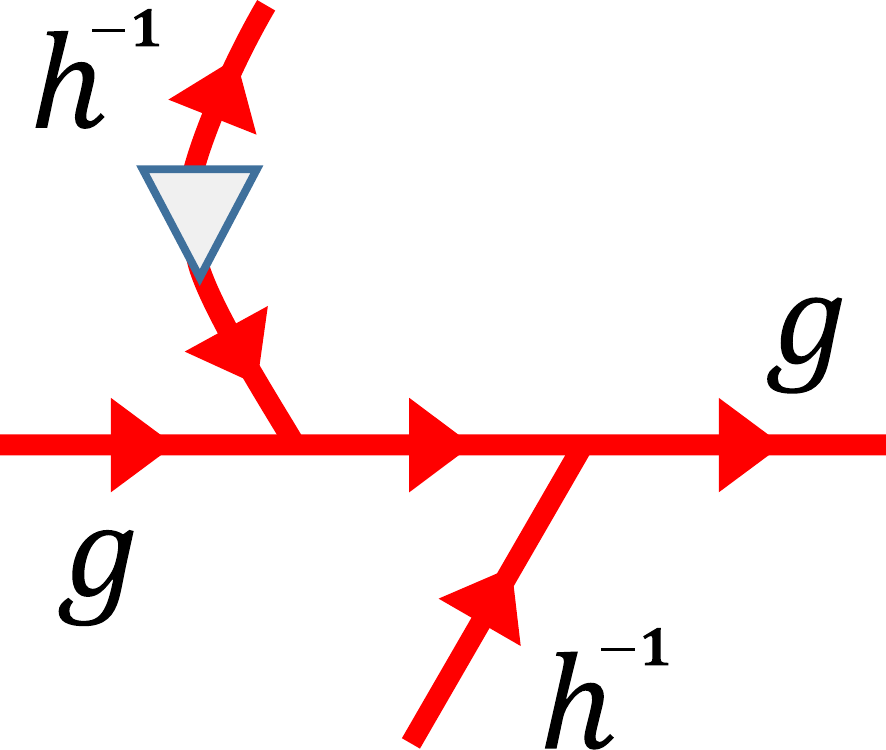}}}
\label{na21d}
\end{align}
note $h\in\mathsf{C}(g)$ and each tensor above is treated as a representative of an equivalence class of all crossing tensors that give rise to equal PEPS. 
Using Eqs.(\ref{n29},\ref{na12}) one finds $\QV{L}{g}{h}=\alpha(g,h,h^{-1})\QW{L}{g}{h}$, $\QW{L}{g}{h}= \slant{\alpha}{1}{g}{h,h^{-1}}^{-1} \QW{R}{g}{h^{-1}}$, and $\QW{R}{g}{h}=\alpha(g,h,h^{-1}) \QV{R}{g}{h}$, i.e.
\begin{align}
\begin{CD}
\QW{L}{g}{h} @> \slant{\alpha}{1}{g}{h,h^{-1}} >> \QW{R}{g}{h^{-1}}\\
@V \alpha(g,h,h^{-1}) VV 			@AA \alpha(g,h^{-1},h) A\\
\QV{L}{g}{h} @>> \omega^g(h,h^{-1}) > \QV{R}{g}{h^{-1}}
\end{CD}
\label{na22}
\end{align}
where 
\begin{align}\label{slant}
\slant{\alpha}{1}{g}{k,h}:=\alpha(g,k,h) \alpha(k,h,g) \alpha^{-1}(k,g,h)
\end{align}
 is the slant product of $\alpha$ (which is a 2-cocycle) and 
\begin{align}\label{omegag}
\omega^g(k,h):=\slant{\alpha}{1}{g}{k,h}\frac{\alpha(g,kh,(kh)^{-1})}{\alpha(g,k,k^{-1})\alpha(g,h,h^{-1})}
\end{align}
 is an equivalent 2-cocycle, i.e. $[\omega^g]=[\slan{\alpha}{1}{g}]$. 
One can easily verify that changing $\alpha$ by a 3-coboundary alters $\slan{\alpha}{1}{g}$ by a 2-coboundary and hence the cohomology class $[\alpha]$ is mapped to $[\slan{\alpha}{1}{g}]$ by the slant product.

\subsection{Composition rule}

There is a natural composition operation on the $\QV{R}{g}{h}$ tensors induced by the action of a global symmetry $U(k)^{\otimes |\man|_v},\, k\in\mathsf{C}(g,h),$ upon a symmetry twisted ground state as follows 
\begin{align}
\QV{R}{g}{k} \times \QV{R}{g}{h} := \, \raisebox{-1.6cm}{\hbox{\includegraphics[width=0.35\linewidth]{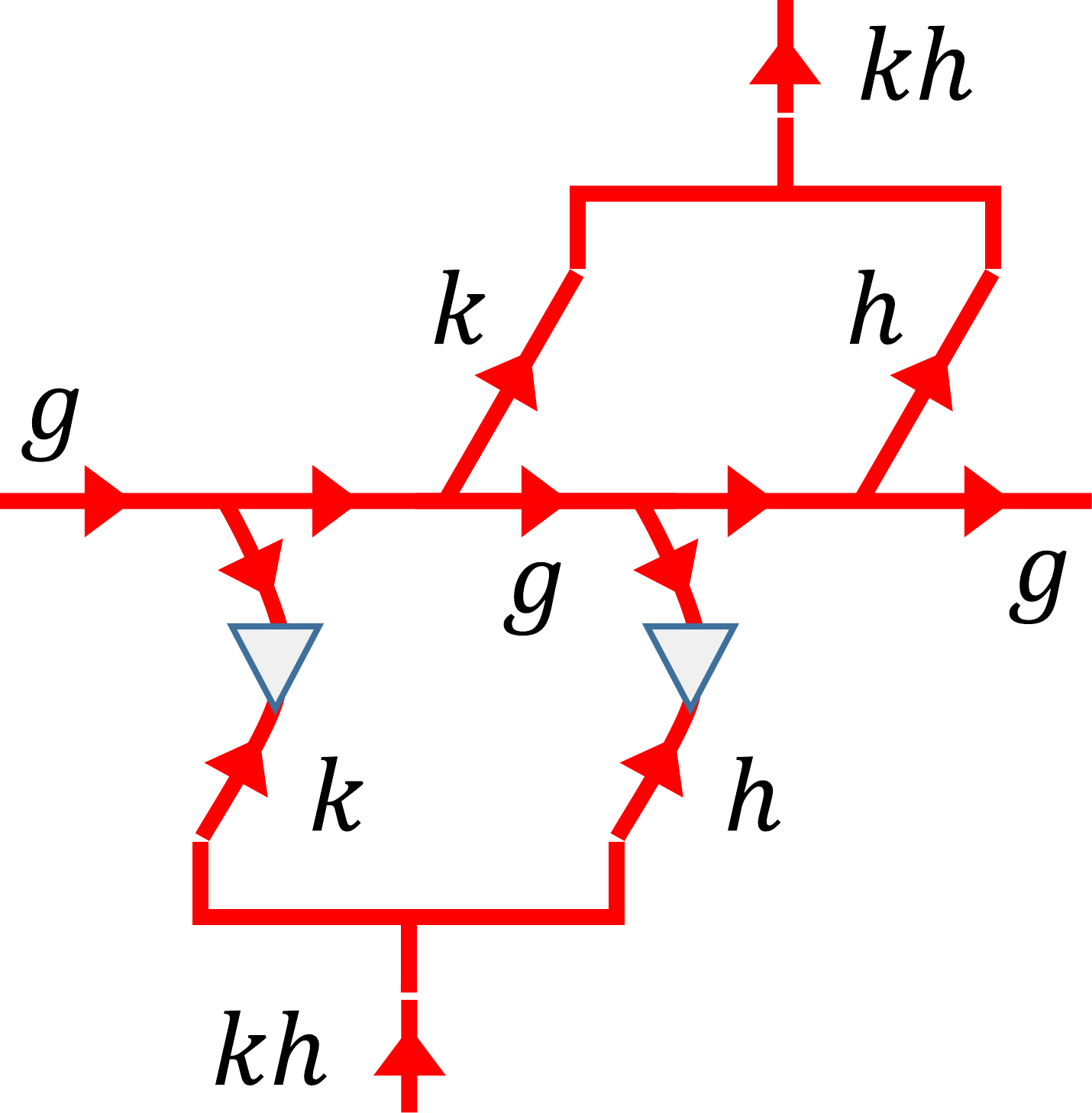}}} \label{na23}
\end{align}
which includes a reduction of the tensors by $X(k,h)$ and $X^+(k,h)$, 
note this product is associative but not commutative. 
The $\QV{R}{g}{h}$ tensors in fact form a projective representation under this composition rule since 
\begin{align}
\begin{CD}
\raisebox{-1.6cm}{\hbox{\includegraphics[width=0.35\linewidth]{Figures/newfig53}}} @>\alpha(k,gh,h^{-1})^{-1}>> \raisebox{-1.6cm}{\hbox{\includegraphics[width=0.35\linewidth]{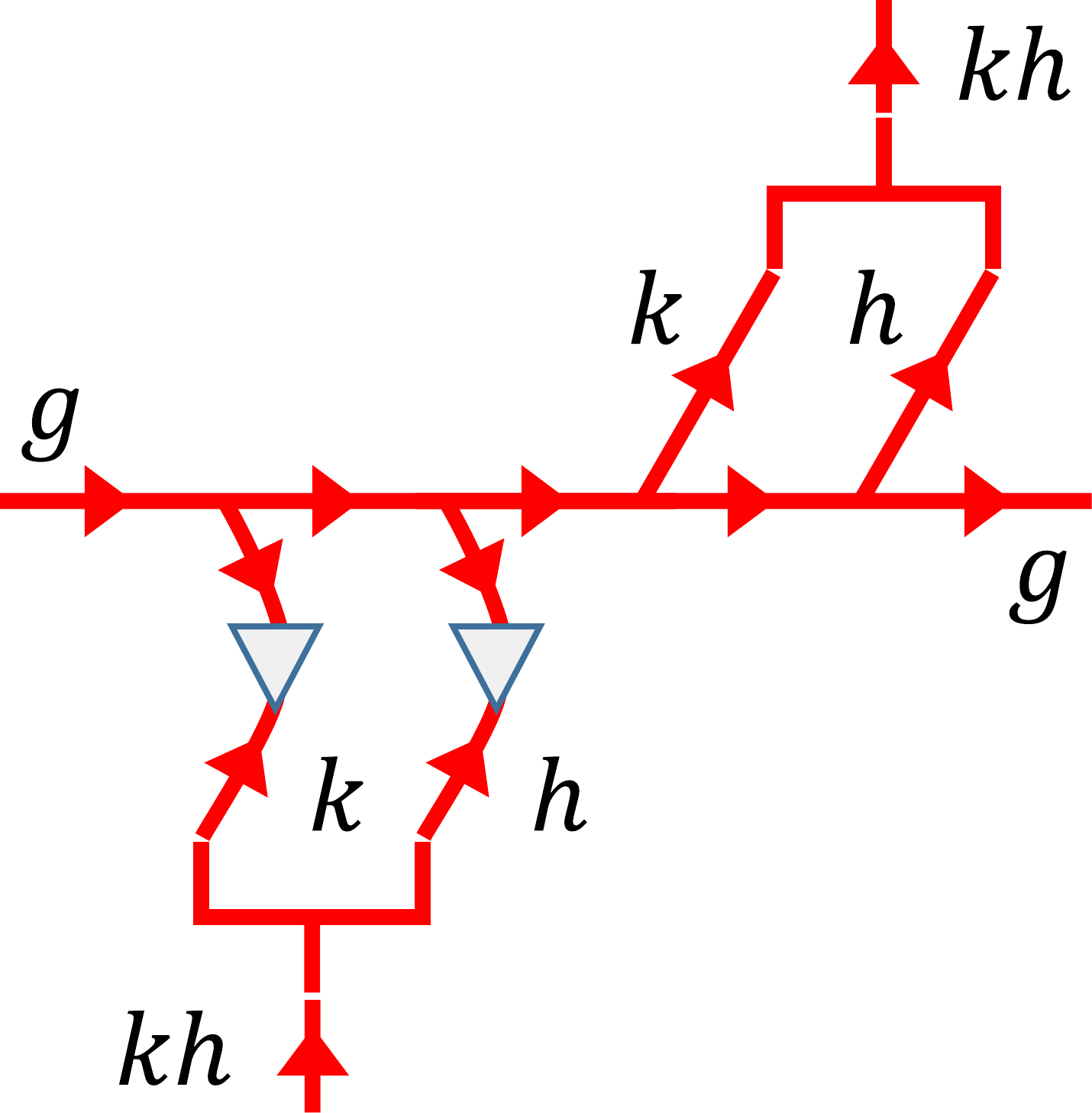}}}\\
@.		@V \alpha(gkh,h^{-1},k^{-1}) VV\\
\raisebox{-2.15cm}{\hbox{\includegraphics[height=0.23\linewidth,angle=90]{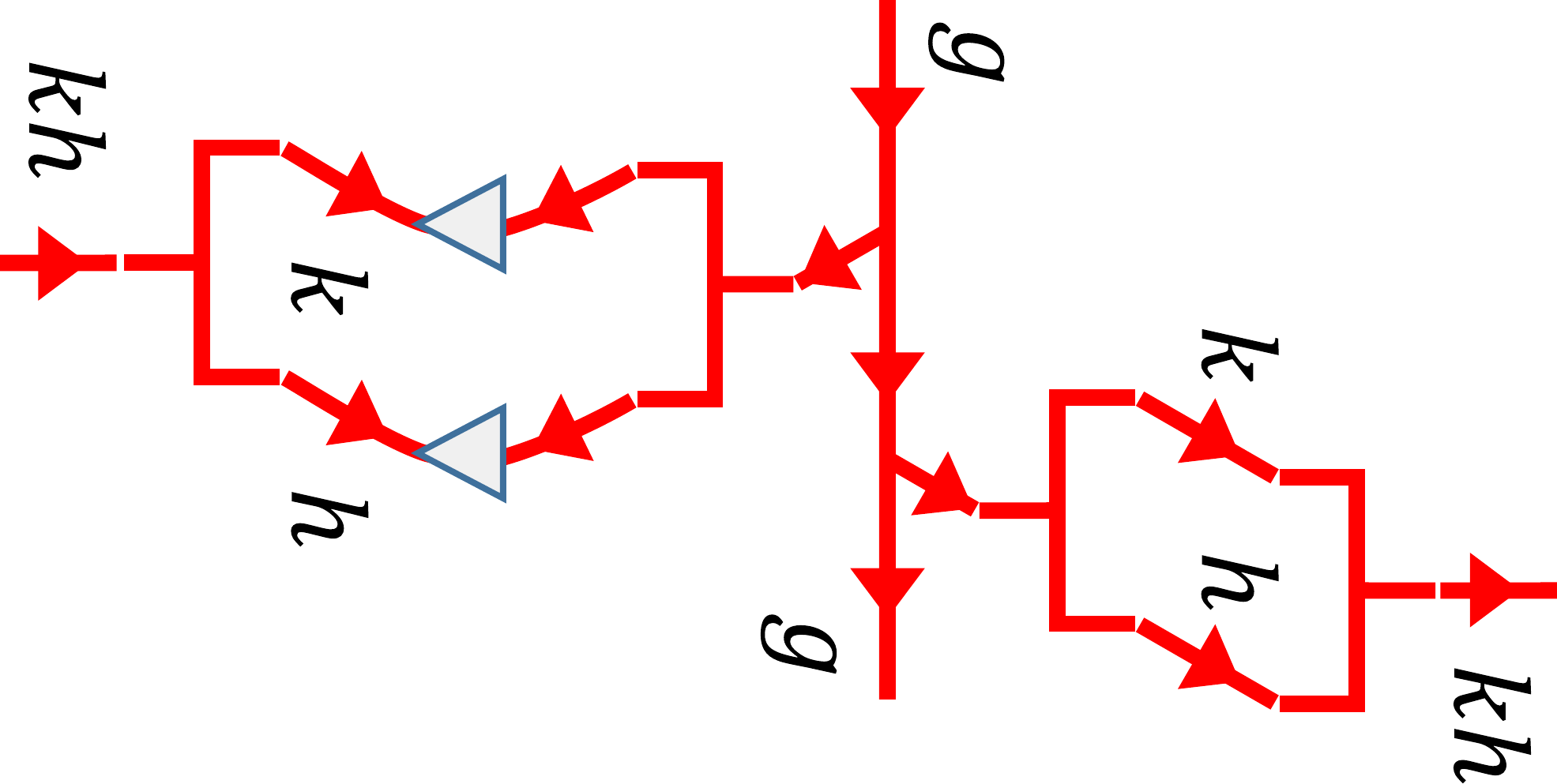}}} @< \alpha(k,h,g)^{-1}<< \raisebox{-2.2cm}{\hbox{\includegraphics[width=0.28\linewidth]{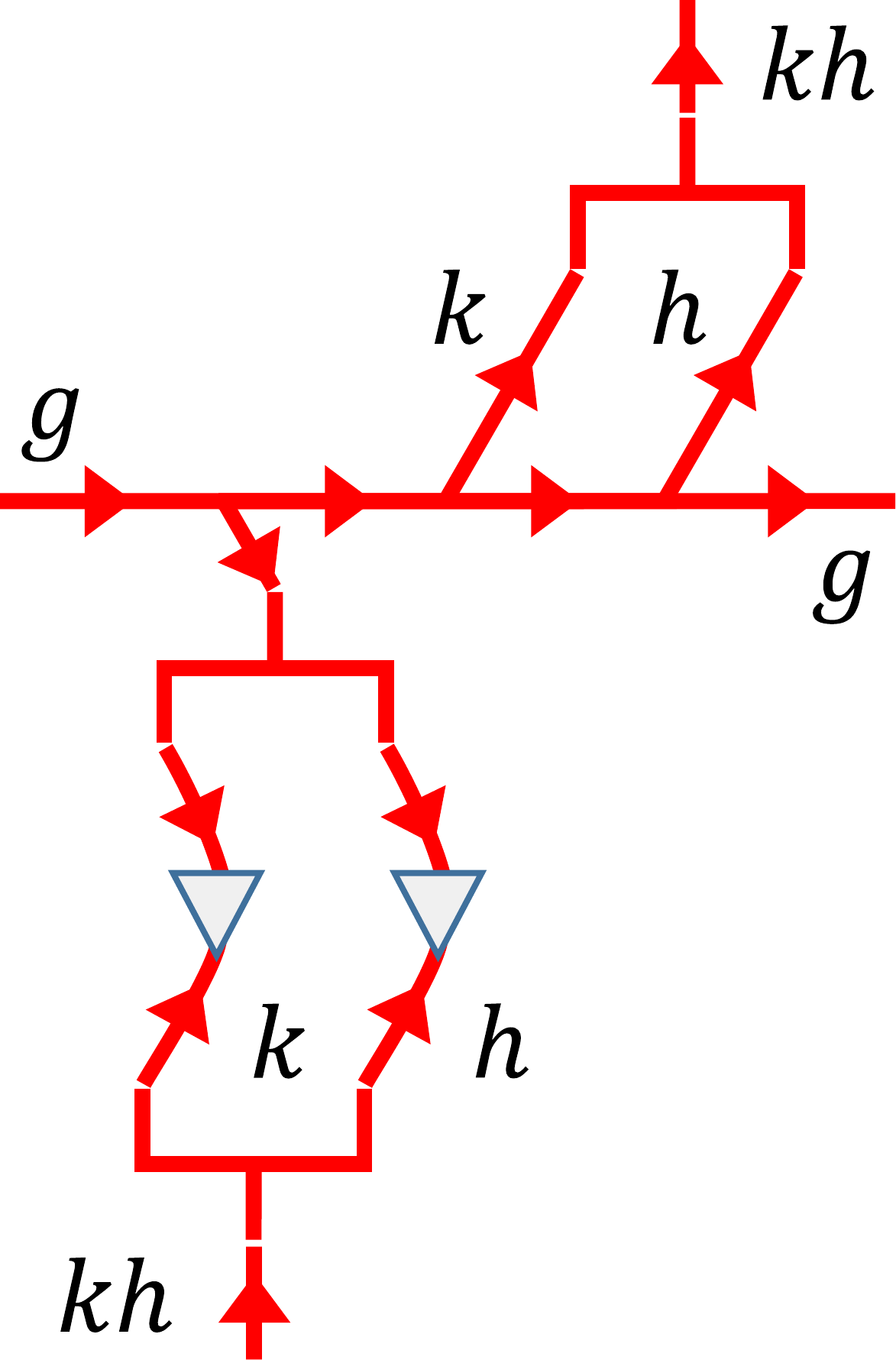}}} \\
@VV \beta(h,k)  V 		@. \\
\raisebox{-.82cm}{\hbox{\includegraphics[width=0.21\linewidth]{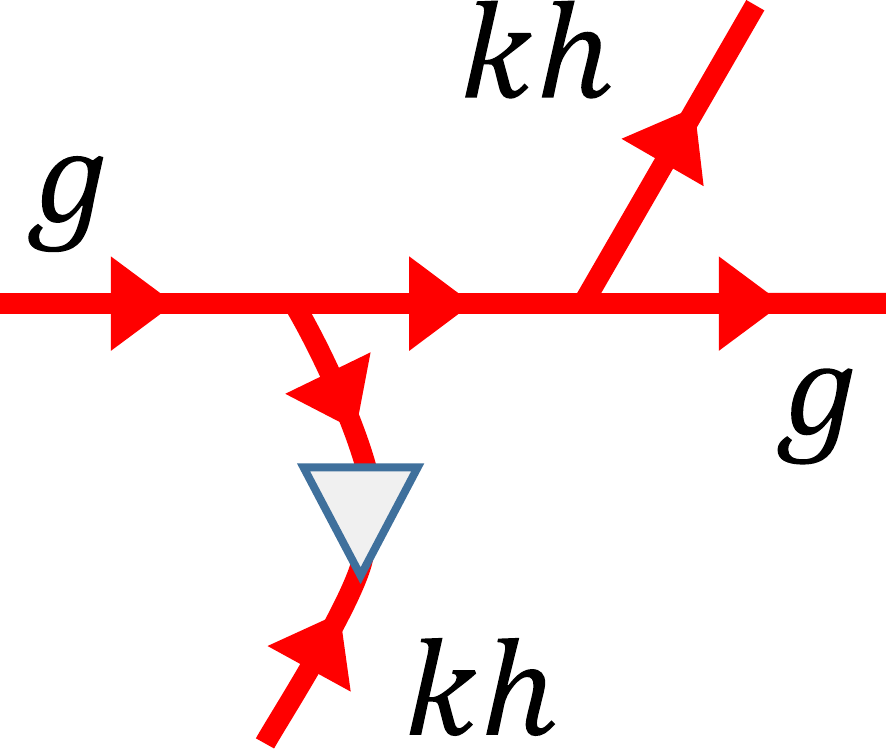}}}
\end{CD}\nonumber
 \end{align}
which yields
\begin{align}\label{na25}
\QV{R}{g}{k} \times \QV{R}{g}{h}=& \frac{\alpha(k,gh,h^{-1})\alpha(k,h,g)}{\alpha(gkh,h^{-1},k^{-1}) \beta(h,k)}
 \QV{R}{g}{kh} 
 \nonumber \\
=& \slant{\alpha}{1}{g}{k,h} \frac{\alpha(g,kh,h^{-1}k^{-1})}{\alpha(g,k,k^{-1}) \alpha(g,h,h^{-1})} 
\nonumber \\
&\frac{\alpha(k,h,h^{-1}k^{-1})}{\alpha(h,h^{-1},k^{-1})\beta(h,k)}
 \QV{R}{g}{kh}
\nonumber  \\
=& \omega^g(k,h) \QV{R}{g}{kh}
\end{align}
after several applications of the 3-cocycle condition for $\alpha$, see Eq.\eqref{cocycle}.

\subsection{Symmetry action}

We are now in a position to calculate the effect of applying a global symmetry $k\in\mathsf{C}(g,h)$ to an $(x,y)$ symmetry twisted SPT PEPS on a torus as follows 
\begin{align} \label{na26}
\vcenter{\hbox{\includegraphics[width=0.38\linewidth]{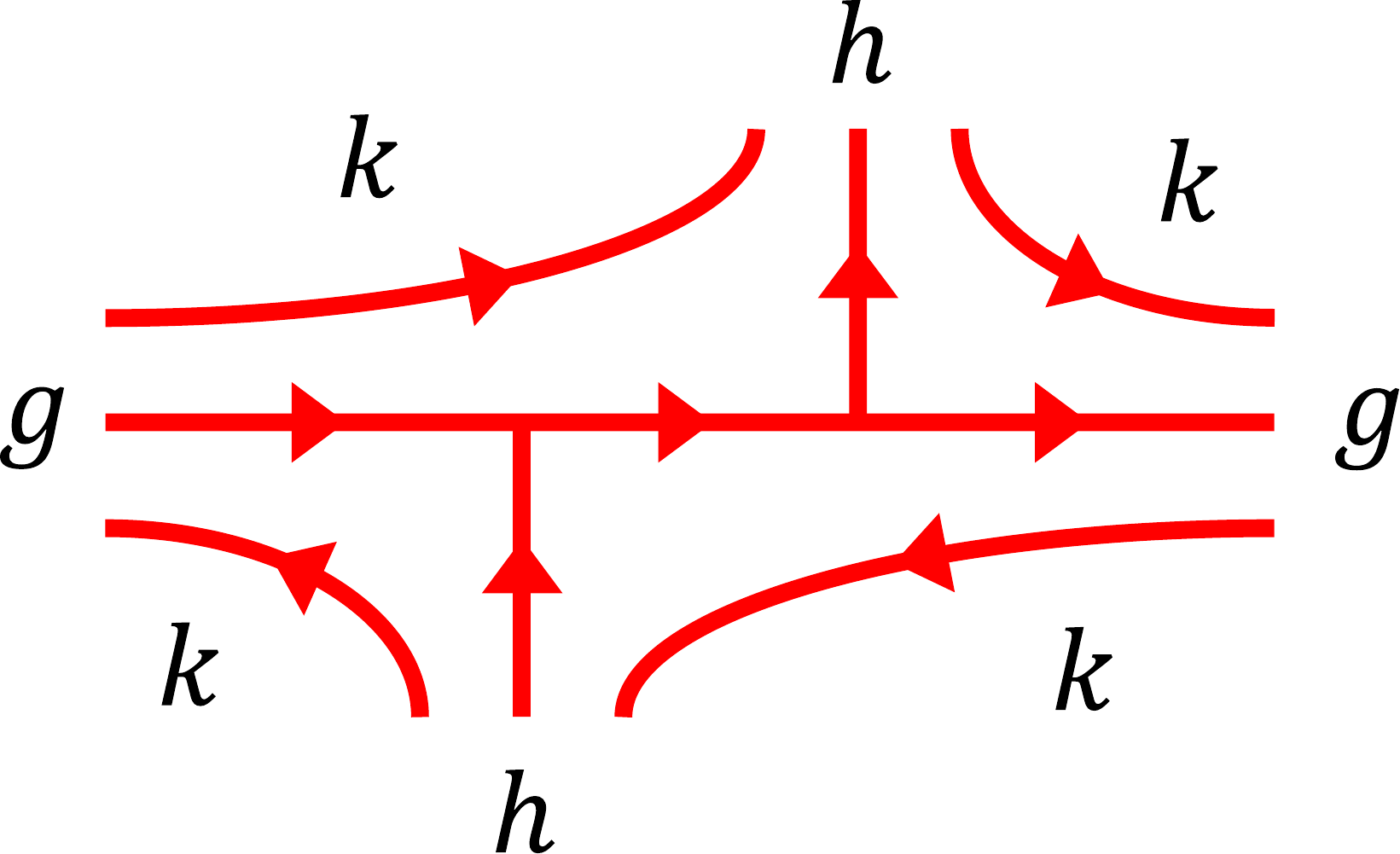}}}
\, \sim \,
\vcenter{\hbox{\includegraphics[width=0.38\linewidth]{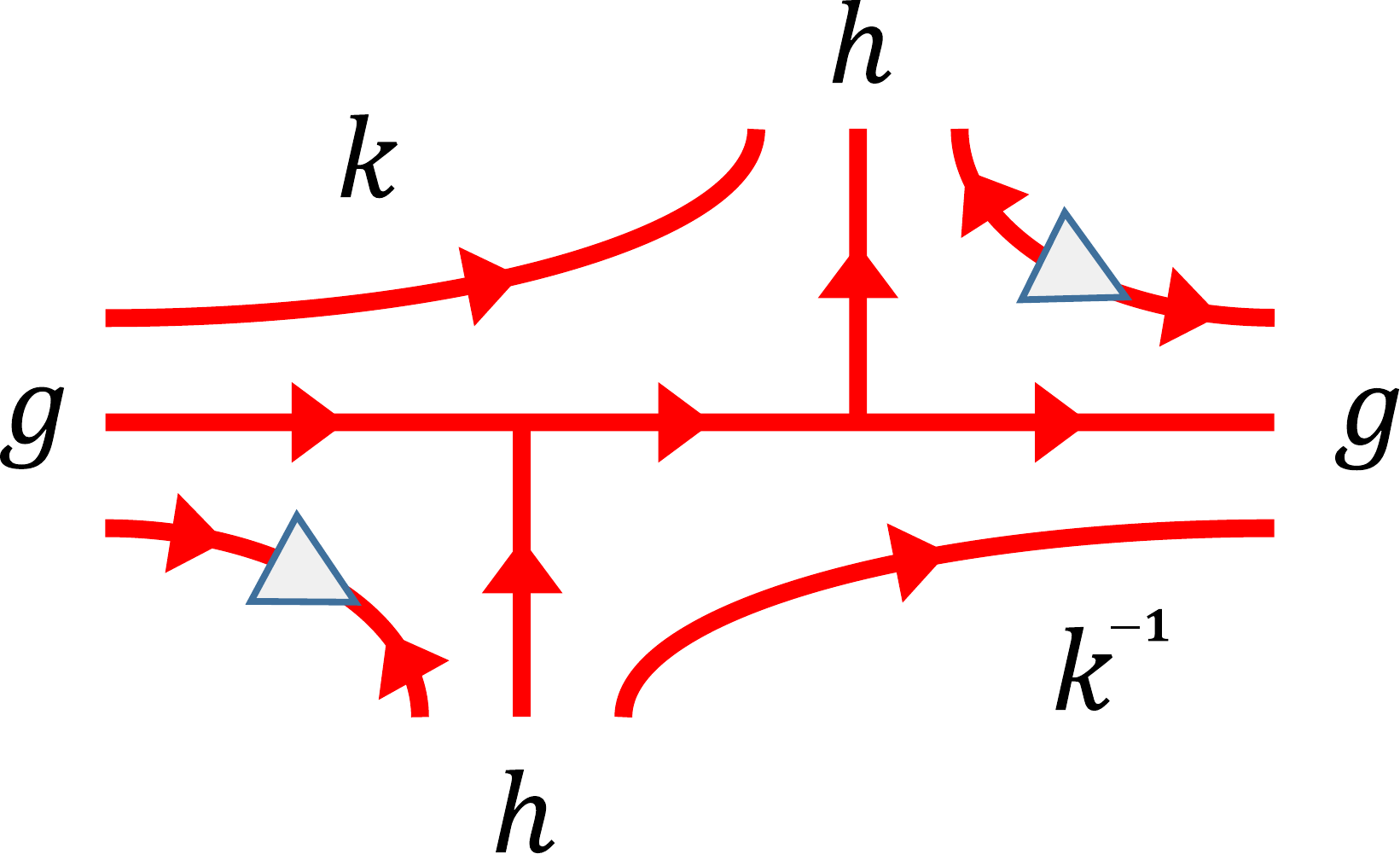}}}
\\
\sim \, \vcenter{\hbox{\includegraphics[width=0.38\linewidth]{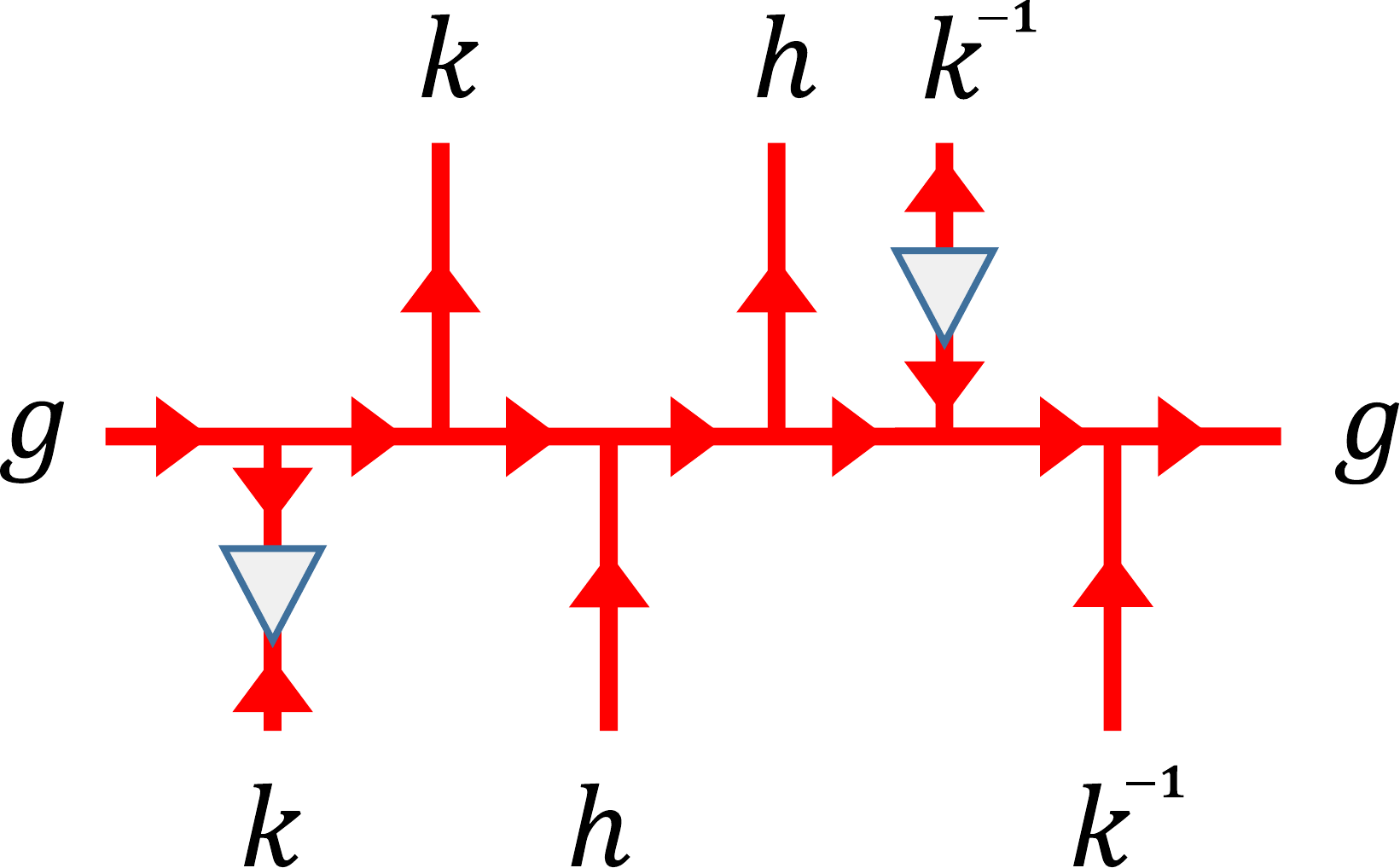}}}
\\
\sim \, \vcenter{\hbox{\includegraphics[width=0.38\linewidth]{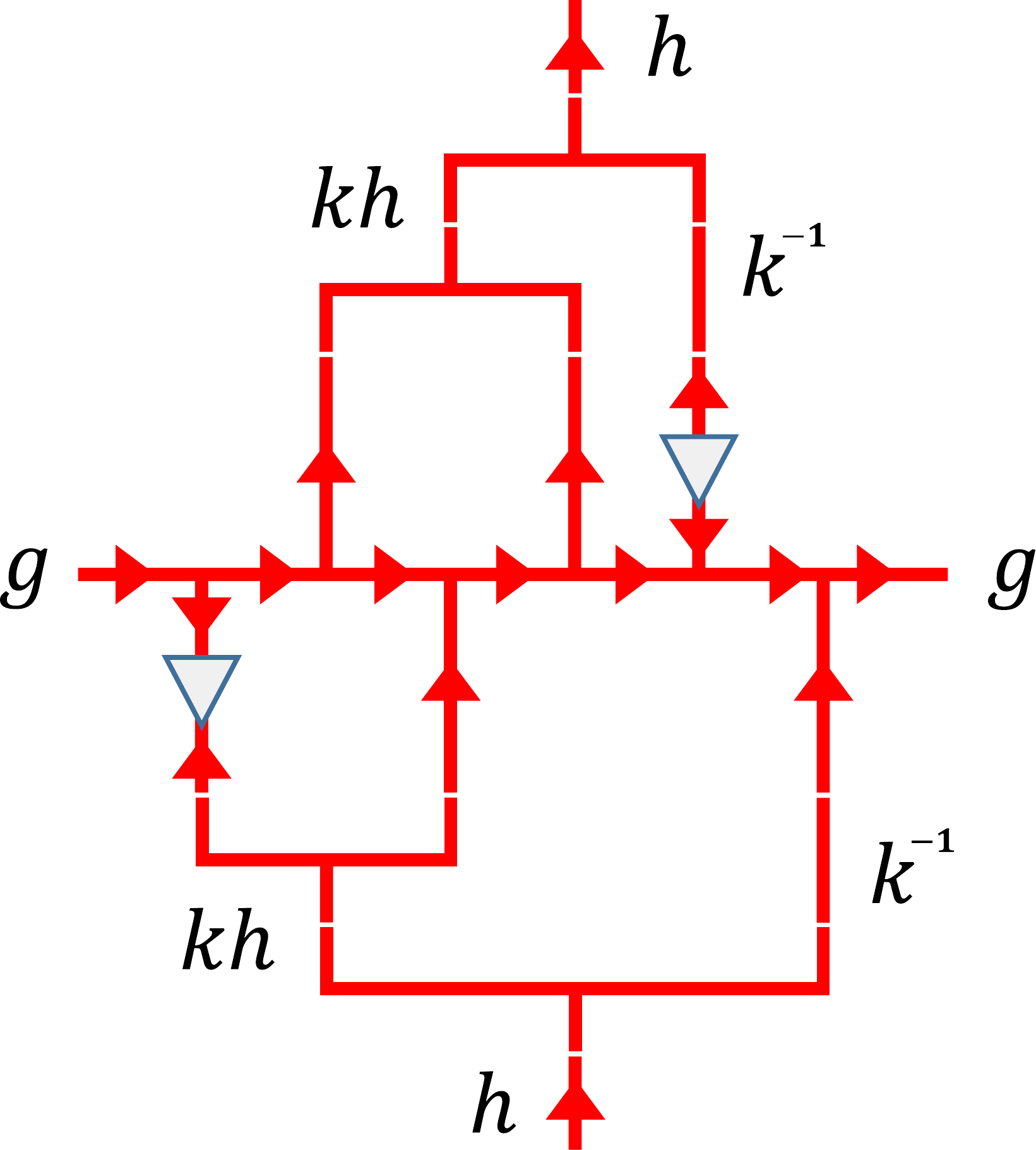}}}
\\
= \QV{R}{g}{k} \times \QW{R}{g}{h} \times  \QV{L}{g}{k}
\\
= \frac{\alpha(g,h,h^{-1}) }{\omega^g(k,k^{-1})} \ \QV{R}{g}{k} \times \QV{R}{g}{h} \times  \QV{R}{g}{k^{-1}}
\\
= \frac{\alpha(g,h,h^{-1}) \omega^g (k,h)}{\omega^g(k,k^{-1})} \ \QV{R}{g}{k h} \times  \QV{R}{g}{k^{-1}}
\\
= \frac{\omega^g (k,h)  \omega^g (kh,k^{-1}) \alpha(g,h,h^{-1})}{\omega^g(k,k^{-1})} \ \QV{R}{g}{h}
\\
= \frac{\omega^g (k,h)}{\omega^g (h,k) } \ \QW{R}{g}{h} 
\end{align}
where we have made use of the 3-cocycle condition on $\alpha$ and the relations from Eq.\eqref{na22}. 
Hence we have found the group action $\pi_k[\cdot]$ induced on the crossing tensor by the physical symmetry to be
\begin{align}
\pi_k [\QW{R}{g}{h} ] &= (\omega^g)^{(h)} (k)^{-1} \, \QW{R}{g}{h}
\nonumber
\\ 
&= \slant{\alpha}{2}{g,h}{k}^{-1} \, \QW{R}{g}{h}
\label{na27}
\end{align} 
where $(\omega^g)^{(h)}$ is the slant product of $\omega^g$ (it is easy to see this equals the coeficient in Eq.\eqref{na26}) and hence a 1D representation of $\mathsf{C}(g,h)$ which 
equals the twice slant product of alpha, i.e. $(\omega^g)^{(h)}=\slan{\alpha}{2}{g,h}$ (since the slant product maps cohomology classes to cohomology classes). 
Now by the orthogonality of characters we have that the projector $\Pi_{g,h}[\cdot]:=\sum\limits_{k\in\mathsf{C}(g,h)} \pi_k[\cdot]$ maps a nonzero $\QW{R}{g}{h}$ to zero iff $\slan{\alpha}{2}{g,h}$ is nontrivial i.e.
\begin{align}\label{na27b}
\Pi_{g,h}[ \QW{R}{g}{h} ]\neq 0 \iff \slan{\alpha}{2}{g,h}\equiv 1 .
\end{align}

\subsection{Modular transformations}
\label{appmodular}

In this section we will calculate the effects of the $S$ and $T$ transformations ($\frac{\pi}{2}$ rotation and Dehn twist respectively) on the crossing tensor $\QW{R}{g}{h}$ which is relevant for both symmetry twisted and topological ground states. We use the following left handed convention
\begin{align}\label{na28}
\begin{CD}
\vcenter{\hbox{\includegraphics[height=.16\linewidth]{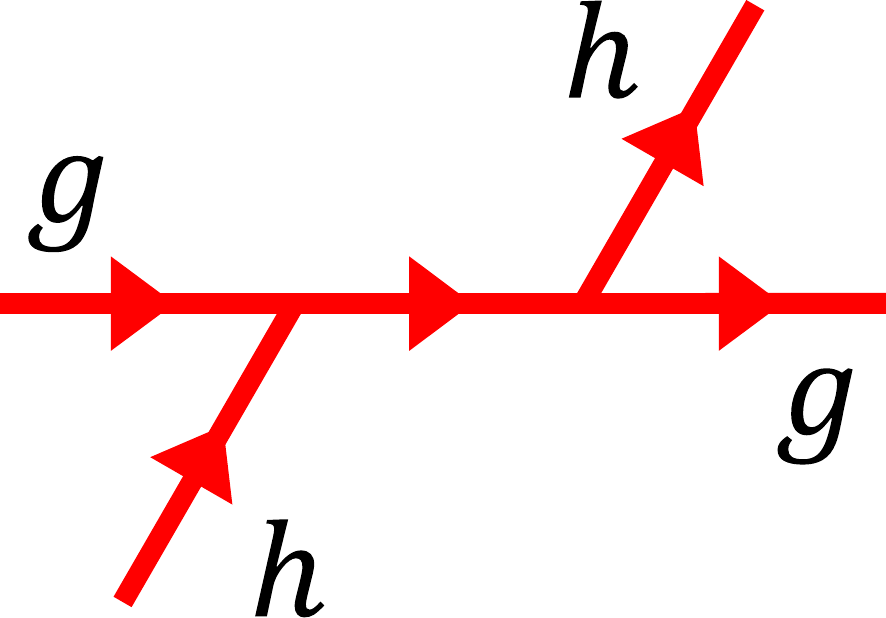}}}  @>S>> \vcenter{\hbox{\includegraphics[width=0.23\linewidth]{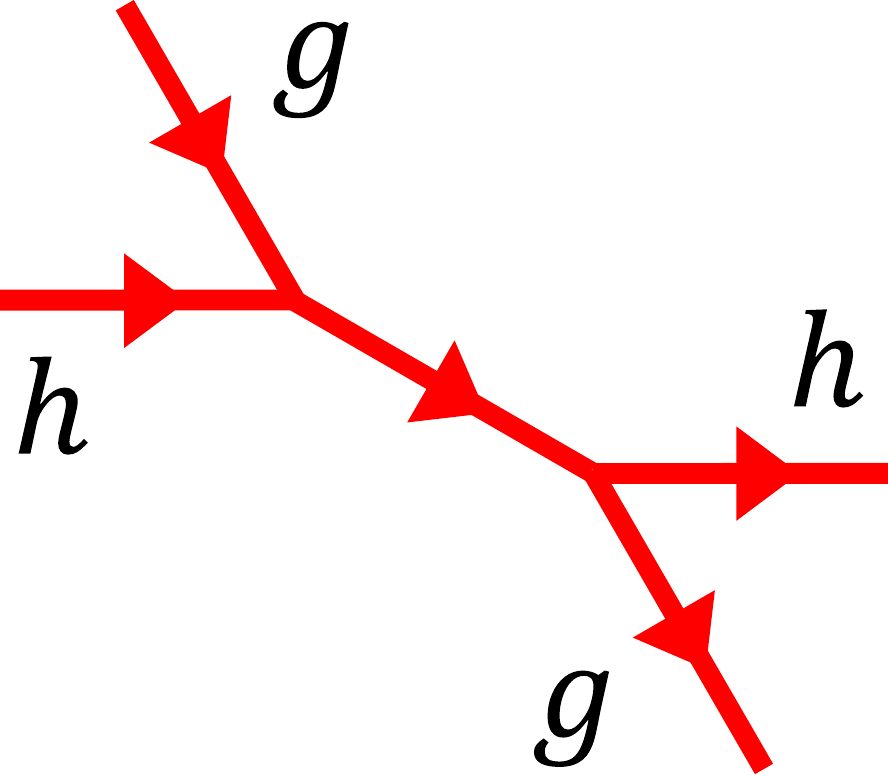}}}
\\
\vcenter{\hbox{\includegraphics[height=.16\linewidth]{Figures/newfig59}}} @> T>> \vcenter{\hbox{\includegraphics[height=0.16\linewidth]{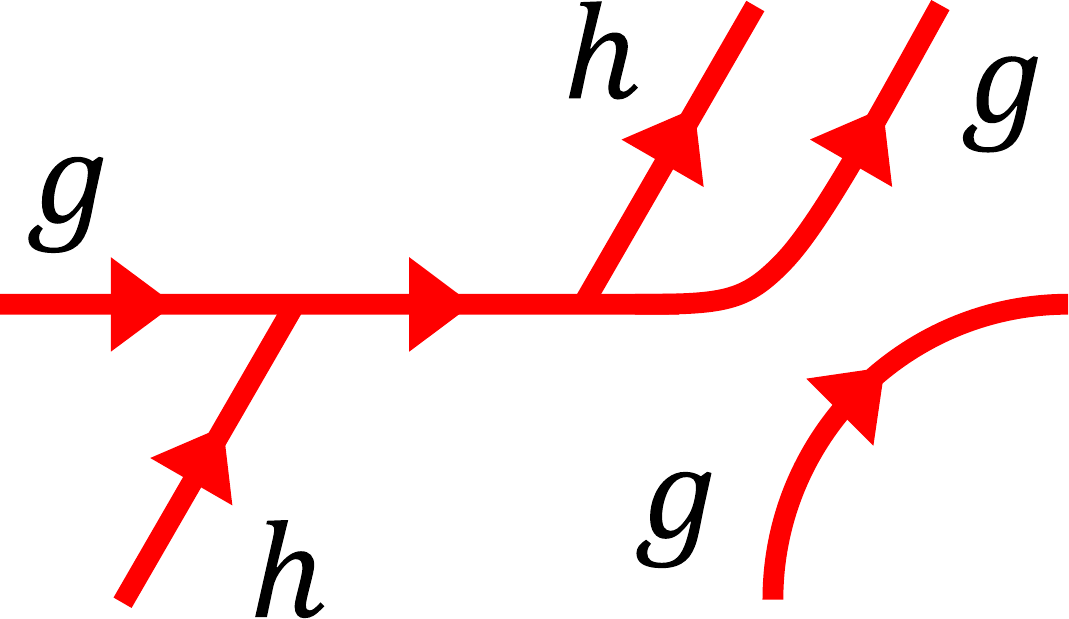}}} 
\end{CD}
\\
\sim  \
\vcenter{\hbox{\includegraphics[width=0.22\linewidth]{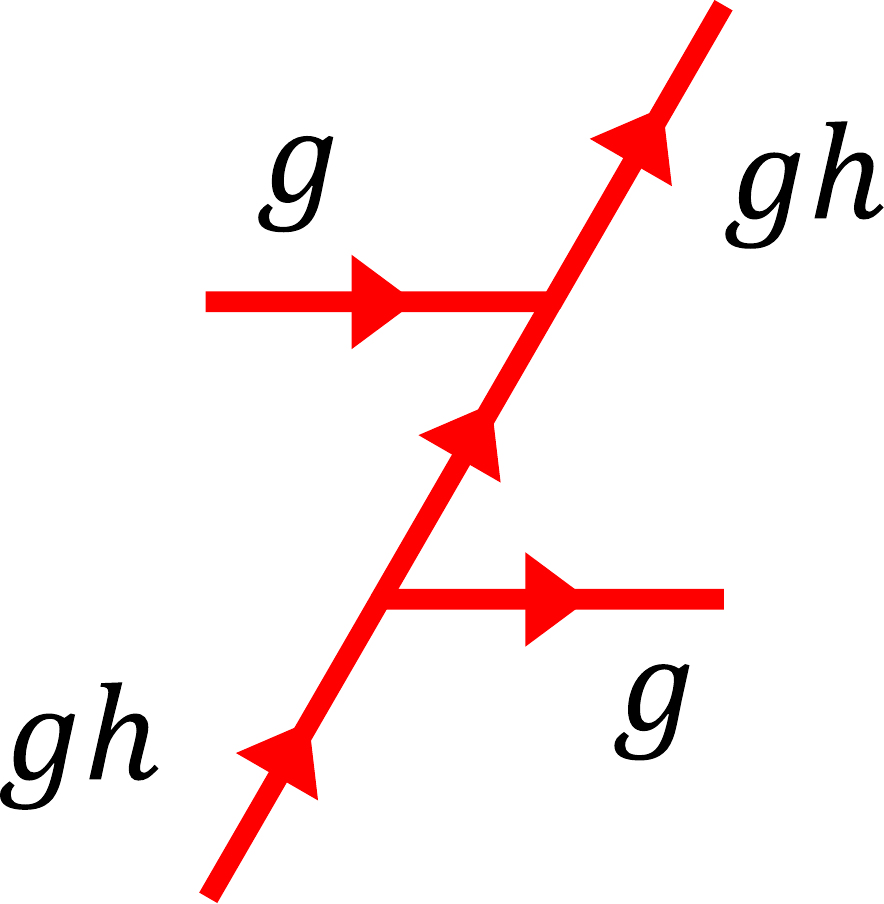}}}. \nonumber
\end{align}
Using Eqs.(\ref{cocycle},\ref{na8},\ref{na11},\ref{na12},\ref{na22}) and the 3-cocycle condition on $\alpha$ we find
\begin{align}\label{na29}
S[ \QW{R}{g}{h} ]&= \slant{\alpha}{1}{h}{g^{-1},g}^{-1} \QW{R}{h}{g^{-1}} 
\\
T[ \QW{R}{g}{h} ]&= \alpha(g,h,g) \QW{R}{g}{gh} 
\end{align}
with these formulas we have explicitly verified that the action of $S$ and $T$ generate a linear representation of the modular group, i.e. they satisfy the relations
$$S^4=\openone,\ (ST)^3=S^2.$$ 
It was sufficient to simply consider the multiplication of these generators since the gauge theories we deal with are doubled topological orders and consequently have zero modular central charge. We do not reproduce the tedious calculation here.

\section{ Gauging SPT PEPS yields topological PEPS}\label{d}
In this appendix we recount the definition of the quantum state gauging procedure of Ref.\cite{Gaugingpaper} and generalize their proof to show that gauging a SPT PEPS results in a MPO-injective PEPS with a projection MPO that has multiple blocks in its canonical form, labeled by the group elements.

\subsection{Quantum state gauging procedure}

Let us first recount the definition of the global projector onto the gauge invariant subspace. This is defined on a directed graph $\fullgraph$ in which the vertices are enumerated and the edges are directed from larger to smaller vertex. To each vertex $v\in\fullgraph$  we associate a Hilbert space $\hilbert_v$ together with a representation $U_v(g)$ of the group $\mathsf{G}$ and to each edge $e\in\fullgraph$ we associate a Hilbert space isomorphic to the group algebra $\hilbert_e\cong \mathbb{C}[\mathsf{G}]$. We define the matter Hilbert space $\hilbert_\mathsf{m}:=\bigotimes_{v\in\fullgraph}\hilbert_v$ and the gauge Hilbert space $\hilbert_\mathsf{g}:=\bigotimes_{e\in\fullgraph}\hilbert_e$ which together form the full Hilbert space $\hilbert_\mathsf{g,m}:=\hilbert_\mathsf{g}\otimes\hilbert_\mathsf{m}$.
The states in $\hilbert_\mathsf{g,m}$ that are relevant for the gauge theory satisfy a local gauge invariance condition at each vertex. Specifically, they lie in the simultaneous $+1$ eigenspace of the following projection operators
\begin{equation}
P_v:=\int \mathrm{d} g_v U_v(g_v) \bigotimes_{e\in E_v^+} R_e(g_v) \bigotimes_{e\in E_v^-} L_e(g_v)\label{constraint}
\end{equation}
where $E_v^+$ ($E_v^-$) is the set of adjacent edges directed away from (towards) vertex $v$. $R(g),L(g)$ are the right and left regular representations, respectively.
The projector onto the gauge invariant subspace is given by $P_\fullgraph:=\prod_v P_v$
and the analogous projector $\mathscr{P}_\subgraph$ for any operator $O$ supported on a subgraph $\subgraph\subset \fullgraph$ (which contains the bounding vertices of all its edges) is defined to be
\begin{align}
\mathscr{P}_\subgraph[O] := &\int  \prod_{v\in\subgraph} \mathrm{d} g_v [\bigotimes_{v\in\subgraph} U_v(g_v) \bigotimes_{e\in\subgraph} L_e(g_{v_e^-})R_e(g_{v_e^+})] 
\nonumber \\
& \, \times O\ [\bigotimes_{v\in\subgraph} U_v(g_v) \bigotimes_{e\in\subgraph} L_e(g_{v_e^-})R_e(g_{v_e^+})]^\dagger \label{operator}
\end{align}
where edge $e$ points from $v_e^+$ to $v_e^-$.

We proceed to describe a gauging procedure for models defined purely on the matter degrees of freedom $\hilbert_\mathsf{m}$. To apply $P_\fullgraph$ and $\mathscr{P}_\subgraph$ we first require a procedure to embed states and operators from $\hilbert_\mathsf{m}$ into $\hilbert_\mathsf{g,m}$. For this we define the gauging map for matter states $\ket{\psi}\in \hilbert_\mathsf{m}$ by 
\begin{equation}
\gauge \ket{\psi} := P[\, \ket{\psi} \bigotimes_{e}\ket{1}_e] \, ,\label{gaugingmap}
\end{equation}
and for matter operators $O \in \mathbb{L}(\hilbert_\mathsf{m})$ acting on a subgraph $\subgraph\subseteq \fullgraph$ (containing all edges between its vertices) by
\begin{equation}
\gaugeoperator_\subgraph[O]:=\mathscr{P}_\subgraph[O\bigotimes_{e\in \subgraph} \ket{1}\bra{1}_e]\,. \label{gaugingoperator}
\end{equation}

\subsection{Gauging SPT PEPS}

In this section we show that a gauged SPT PEPS satisfies the axioms of MPO-injectivity.

Consider a region $\region$ of a SPT PEPS $\ket{\psi}\in\hilbert_\mathsf{m}$ built from local tensor $\peps$. The PEPS map $\peps_\region$ on this region satisfies $\peps_\region^+\peps_\region=P_{\partial\region}$ and hence is injective on the support subspace of a single block projection MPO $P_{\partial\region}=V^{\partial \region}(1)$ given by $\text{supp}(P_{\partial\region})\subseteq(\mathbb{V}_e)^{\otimes L}$ where $\mathbb{V}_e$ denotes the Hilbert space of a virtual index and $L:=|\partial\region|_e$. 

For the gauged PEPS $G\ket{\psi}\in\hilbert_{\mathsf{g,m}}$, the region $\region$ is defined to include only those edges between vertices within $\region$, i.e. excluding the edges $e\in\partial\region$. 
Note our proof is easily adapted to the case where the edge degrees of freedom are `doubled' and absorbed into the neighboring vertex degrees of freedom, as in Section~\ref{exfpspt}. 

The gauged PEPS map on region $\region$, $\peps_\region^\mathsf{g}:(\mathbb{V}_e \otimes \mathbb{C}[\mathsf{G}])^{\otimes L}\rightarrow \mathbb{H}_v^{\otimes |\region|_v}\otimes \mathbb{H}_e^{\otimes |\region|_e}$, naturally decomposes into the original PEPS map and a gauging tensor network operator multiplying the physical degrees of freedom $\peps_\region^\mathsf{g}=G_\region\peps_\region$ where
\begin{align}
G_\region:=\int\prod_{v\in\region}\mathrm{d} g_v \bigotimes_{v\in\region} U_v(g_v) \bigotimes_{e\in\region} \ket{g_{v_e^-}g_{v_e^+}^{-1}}_e \bigotimes_{e\in\partial \region} \bbra{g_{v_e^\pm}}_e
\end{align}
where $v_e^{\pm}\in\region$ denotes the unique vertex in $\region$ adjacent to the edge $e\in\partial \region$. 
\begin{claim}
A generalized inverse of the gauged PEPS is given by $(A_\region^\mathsf{g})^+=\peps_\region^+G_\region^\dagger$ which satisfies $(\peps_\region^\mathsf{g})^+\peps_\region^\mathsf{g}=\frac{1}{|G|} \sum\limits_{g\in\mathsf{G}} V^{\partial \region}(g)\otimes R(g)^{\otimes L}$. Furthermore, the gauged PEPS is MPO-injective with respect to the projection MPO $\frac{1}{|G|}\sum\limits_{g\in\mathsf{G}} V^{\partial \region}(g)\otimes R(g)^{\otimes L}$ which is a sum of single block injective MPOs labeled by $g\in\mathsf{G}$.
\end{claim}
Firstly we have 
\begin{align}
G_\region^\dagger G_\region=&\int\prod_{v\in\region}\mathrm{d} h_v\mathrm{d} g_v \bigotimes_{v\in\region} U_v(h_v^{-1}g_v) \nonumber \\
&\bigotimes_{e\in\region} \braket{h_{v_e^-}h_{v_e^+}^{-1}|g_{v_e^-}g_{v_e^+}^{-1}} \bigotimes_{e\in\partial \region} \bket{h_{v_e^\pm}}\bbra{g_{v_e^\pm}}_e
\nonumber 
\\
=& \int \mathrm{d} g \bigotimes_{v\in\region} U_v(g) \bigotimes_{e\in\partial\region} R_e(g)
\end{align}
since the delta conditions $\braket{h_{v_e^-}h_{v_e^+}^{-1}|g_{v_e^-}g_{v_e^+}^{-1}}$ force $h_{v_e^-}^{-1}g_{v_e^-}=h_{v_e^+}^{-1}g_{v_e^+}$ and hence $h_v^{-1}g_v=:g$ is constant across all $v\in\region$, assuming $\region$ is connected. 
Hence
\begin{align}
\peps_\region^+ G_\region^\dagger G_\region \peps_\region = P_{\partial\region}  \int \mathrm{d} g \ V^{\partial \region}(g) \bigotimes_{e\in\partial\region} R_e(g)
\end{align}
since $U(g)^{\otimes |\region|_v}\peps_\region =\peps_\region V^{\partial \region}(g)$ for a SPT PEPS (see Section~\ref{globalsymmetry}) then the result follows as ${P_{\partial \region} V^{\partial \region} (g) =V^{\partial \region} (g)}$. 

Let us now address the remaining conditions for MPO-injectivity. Most importantly the pulling through condition is easily seen to hold by Eq.\eqref{n23} and since $P_v U^\dagger_v(g) = P_v \bigotimes_{e\in E_v^+} R_e(g) \bigotimes_{e\in E_v^-} L_e(g)$, see Appendix~\ref{e}, Proposition~\ref{prop11} for more detail. The trivial loops condition for the MPO $V^{\partial \region}(g)\otimes R(g)^{\otimes L}$ follows directly from the trivial loops condition for $V^{\partial \region}(g)$ and the convention that $R(g)$ is inverted depending on the orientation of the crossing of the MPO loop with the virtual bond edge of the PEPS graph, see Eqs.\eqref{n23},\eqref{n25}. Finally, as discussed at the end of Appendix~\ref{a} the extended inverse condition is automatically satisfied when the projection MPO has a canonical form with injective blocks~\cite{nick}, which is the case for the MPO $V^{\partial \region}(g)\otimes R(g)^{\otimes L}$.

\section{Generalizing the gauging procedure to arbitrary flat $\mathsf{G}$-connections}\label{e}
In this section we outline a generalization of the gauging procedure defined in Ref.\cite{Gaugingpaper} to arbitrary flat $\mathsf{G}$-connections. For equivalent $\mathsf{G}$-connections the gauging maps are related by local operations while for inequivalent $\mathsf{G}$-connections, which are necessary to construct the full ground space of a gauged model on a nontrivial manifold, the gauging maps are topologically distinct. The gauging maps for nontrivial flat $\mathsf{G}$-connections take inequivalent symmetry twisted states of the initial SPT models to orthogonal ground states of the topologically ordered gauged models.
\subsection{Elementary definitions}
\begin{figure}[ht]
\center
\begin{align*}
a) \vcenter{\hbox{
 \includegraphics[width=0.22\linewidth]{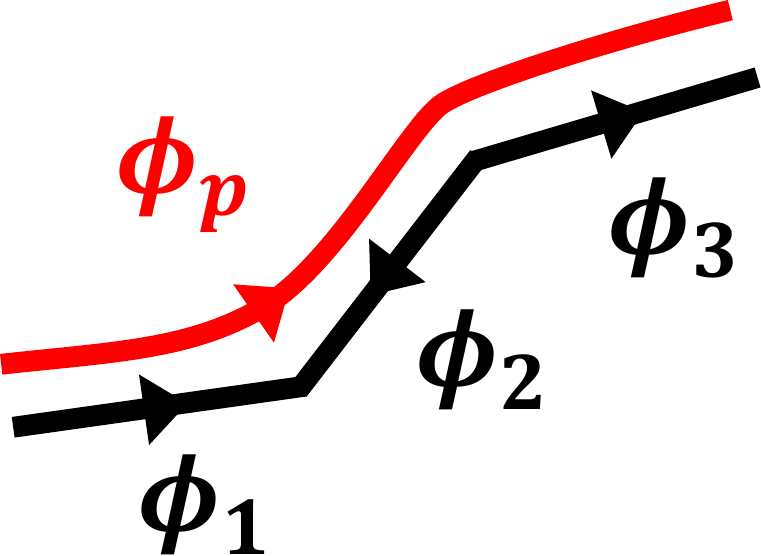}}} \hspace{.5cm} b)\, 
\vcenter{\hbox{\includegraphics[width=0.6\linewidth]{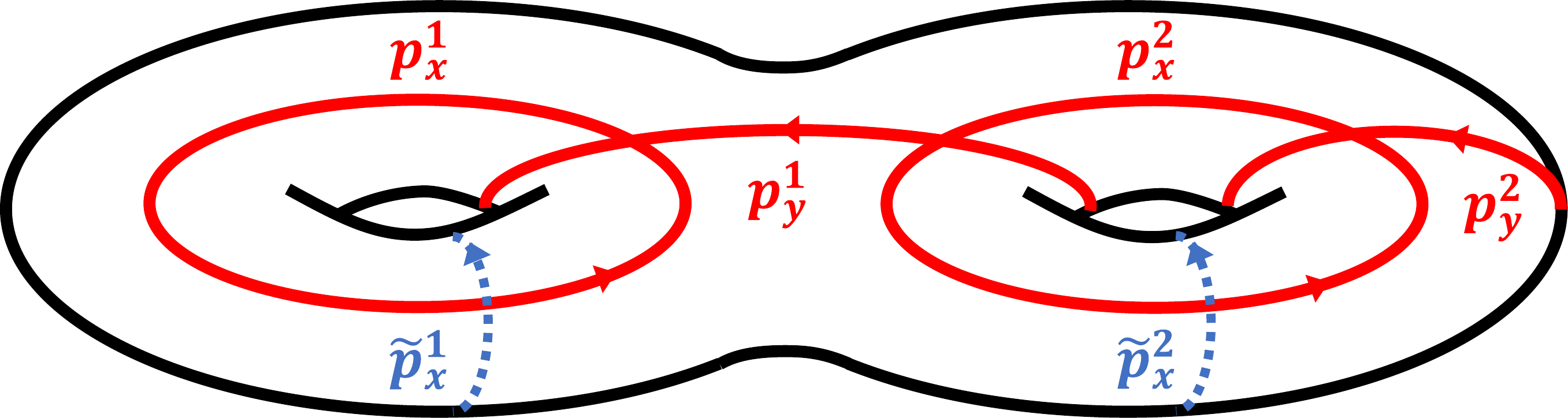}}} 
\end{align*}
\caption{a) A simple example $\con_\epath=\con_3\con_2^{-1}\con_1$. b) Noncontractible cycles of the 2-torus.}\label{e10}
\end{figure}

\begin{definition}
A $\mathsf{G}$-connection $\phi$ on a directed graph $\fullgraph$, embedded in an oriented 2-manifold $\man$, is given by specifying a group element $\con_e\in\mathsf{G}$ for each edge $e\in\fullgraph$.
\begin{align*}
\con:\fullgraph_e &\rightarrow\mathsf{G}
\\
e&\mapsto\con_e 
\end{align*}
\end{definition}
where $\Lambda_e$ is the set of edges in $\Lambda$ and 
$\phi$ can be thought of as a labeling $\{ \con_e \}$ of the edges in $\fullgraph$ by group elements $\con_e\in\mathsf{G}$.
We view these connections as basis states $\ket{\con}:=\bigotimes_e \ket{\con_e}_e \in \mathbb{C}[\mathsf{G}]^{\otimes |\fullgraph_e|}$. 
\\
Each $\mathsf{G}$-connection $\phi$ defines a notion of transport along any oriented path (with origin and end point specified) $\epath\in\fullgraph$ on the edges of the graph, the transport is specified by the group element 
\begin{equation}
\con_\epath:=\prod_{i =|\epath|_e}^{1} \con_{e_i}^{\sigma_i}=\con_{e_{|p|}}^{\sigma_{|p|}} \cdots \con_{e_1}^{\sigma_1}
\end{equation}
where the edges $e_i\in \epath$ are ordered as they occur following $\epath$ along its orientation, and $\sigma_i$ is 1 if the orientation of $e_i$ matches that of $\epath$ and $-1$ if it does not, see Fig.\ref{e10}. Note for paths $\epath^1$, from $v_0$ to $v_1$, and $\epath^2$, from $v_1$ to $v_2$, we have the following relation $\con_{\epath^2}\con_{\epath^1}=\con_{\epath^{12}}$, where $\epath^{12}:=\epath^1\cup\epath^2$ is given by composing paths 1 and 2.

A pair of $\mathsf{G}$-connections $\phi,\varphi$ are considered equivalent if they are related by a sequence of local gauge transformations from the set 
\begin{align}\label{gceq}
&\{\st_v^g:=\bigotimes_{e\in E_v^+} R_e(g)\bigotimes_{e\in E_v^-} L_e(g) \, |\, \forall g\in\mathsf{G},v\in\fullgraph\}
\\
&\text{i.e.}\quad\quad \phi\sim\varphi \iff \ket{\phi}=\prod_i \st_{v_i}^{g_i}\ket{\varphi} . \nonumber
\end{align}
One can easily verify that this constitutes an equivalence relation.
Importantly, this equivalence relation preserves the conjugacy class of the $\mathsf{G}$-holonomy $\con_\epath$ of any closed path $\epath\in\fullgraph$ with a fixed base point.

An important class of connections are the flat $\mathsf{G}$-connections which are defined to have trivial holonomy along any contractible path.
\begin{definition}
A $\mathsf{G}$-connection $\phi$ is flat iff  $\con_\epath=1$ for any closed path $\epath\in\fullgraph$ that is contractible in the underlying manifold $\man$. 
\end{definition}
This definition immediately implies that $\con_\epath=\con_{\epath'}$ for any pair of homotopic oriented paths $\epath,\epath'$ with matching endpoints.
It is easy to see that a $\mathsf{G}$-connection is flat if and only if it satisfies the local condition $\con_{\partial \plaq}=1$ for every plaquette $\plaq$ of the graph $\fullgraph\subset\man$, where $\partial \plaq\subset\fullgraph$ is the boundary of $\plaq$ with the orientation inherited from $\man$. 
Moreover, one can easily verify that flatness is preserved under the equivalence relation \eqref{gceq} and hence the flat $\mathsf{G}$-connections form equivalence classes under this relation. 
Note there can be multiple flat equivalence classes since it is possible for a flat $\mathsf{G}$-connection to have a nontrivial holonomy $\con_\epath\neq1$ along a noncontractible loop $\epath\in\fullgraph\subset \man$.

One can easily show that any contractible region $\subgraph\subseteq \fullgraph\subset\man$ (formed by a set of vertices and the edges between them) of a flat $\mathsf{G}$-connection $\ket{\con}$ can be `cleaned' by a sequence of operations $\prod_i a_{v_i}^{g_i}$, where each $v_i\in\subgraph$, such that the resulting equivalent connection $\ket{\con'}:=\prod_i a_{v_i}^{g_i}\ket{\con}$ satisfies $\con'_e=1,\forall e\in\subgraph$.
\\
Utilizing the cleaning procedure leads one to the following conclusion 
\begin{claim}\label{conl}
The equivalence class $[\con]$ of a flat $\mathsf{G}$-connection $\con$ on an oriented 2-manifold (w.l.o.g. a genus-$n$ torus or $n$-torus) $\man$ is labeled uniquely by the conjugacy class of n pairs of group elements that commute with their neighbors, i.e. $\left\{ [ (x_1,y_1),\dots,(x_n,y_n) ]\, | \,\exists x_i,y_i\in\mathsf{G},x_iy_i=y_ix_i,\right.$ $\left.y_ix_{i+1}=x_{i+1}y_i\right\}$, the set of such labels is henceforth referred to as $\mathcal{I}_\man$.
\end{claim}
The argument proceeds as follows: any $\mathsf{G}$-connection can be `cleaned' onto the set of edges that cross any of the $2n$ closed paths $\{(p^i_x,p^i_y)\}$ in the dual graph $\fullgraph^*$ (where each $(p_x^i,p_y^i)$ and $(p_y^i,p_x^{i+1})$ pair intersect once) that span the inequivalent noncontractible loops of the $n$-torus, see Fig.\ref{e10}. 
Now by the flatness condition the group elements along any loop must be the same (assuming w.l.o.g. the edges on that loop have the same orientation) and the group elements $(x_i,y_i)$ and $(y_i,x_{i+1})$ of each pair of intersecting loops must commute.
 Furthermore, equivalence under the application of $\bigotimes_{v\in\fullgraph} a_v^g,\,\forall g\in\mathsf{G}$ implies that every set of labels in the same conjugacy class are equivalent.

Note there is a uniquely defined set of group elements 
\begin{align}\label{glcon}
\left\{ (x_1,y_1),\dots,(x_n,y_n)\, | \,x_i,y_i\in\mathsf{G},x_iy_i=y_ix_i,\right. \nonumber \\
\left. y_ix_{i+1}=x_{i+1}y_i\right\}
\end{align}
for each flat $\mathsf{G}$-connection $\con$ which are specified by the $\mathsf{G}$-holonomies $x_i:=\con_{\tilde p^i_x},\, y_i:=\con_{\tilde p^i_y}$ of pairs of paths $(\tilde p^i_y,\tilde p^i_x)$ in the graph $\fullgraph$, where $\tilde p^i_x$ is defined to be a path that intersects $p^i_x$ once and all other paths $p^k_y,p^j_x,\, j\neq i$, zero times ($\tilde p^i_y$ is defined similarly). 
Moreover, the conjugacy class $[(x_1,y_1),\dots,(x_n,y_n)]:=\{(x^h_1,y^h_1),\dots,(x^h_n,y^h_n) \, |\, \forall h\in\mathsf{G}\}$ labels the equivalence class $[\con]$ of the $\mathsf{G}$-connection $\con$, where $x^h:=hxh^{-1}$.

For a fixed representative $\gamma=\{(x_i,y_i)\}$ of conjugacy class $[\gamma]\in\mathcal{I}_\man$ and choice of paths $\{(p^i_x,p^i_y)\}$ spanning the inequivalent noncontractible cycles of the $n$-torus, we construct a particularly simple representative flat $\mathsf{G}$-connection as follows
\begin{definition}\label{flaty}
The simple representative flat $\mathsf{G}$-connection $\con^\gamma$ is defined by setting  $\con_e^\gamma:=x_i^{\sigma^i_e}$ if $\epath_x^i$ crosses $e$ and $\con_e^\gamma:=y_i^{\sigma^i_e}$ if $\epath_y^i$ crosses $e$, where $\sigma^i_e$ is +1 if the crossing is right handed and -1 if it is left handed, and otherwise $\con_e^\gamma:=1$ for edges that are not crossed by either $\epath_x^i,\epath_y^i$.
\end{definition}
Note an arbitrary flat connection $\ket{\con}$ is related to some $\ket{\con^\gamma}$ by a sequence of local operations $\ket{\con}=\prod_i a_{v_i}^{g_i} \ket{\con^\gamma}$. 
In particular, the representative connection $\ket{\tilde{\con}^\gamma}$ corresponding to a deformation of the paths $(p^i_x,p^i_y)\mapsto(\tilde p^i_x,\tilde p^i_y)$ that does not introduce additional intersections (a planar isotopy) is related to $\ket{\con^\gamma}$ by a sequence of local operations $\prod_i a_{v_i}^{g_i}$ that implements the deformation.

\begin{figure}[ht]
\center
\includegraphics[width=0.4\linewidth]{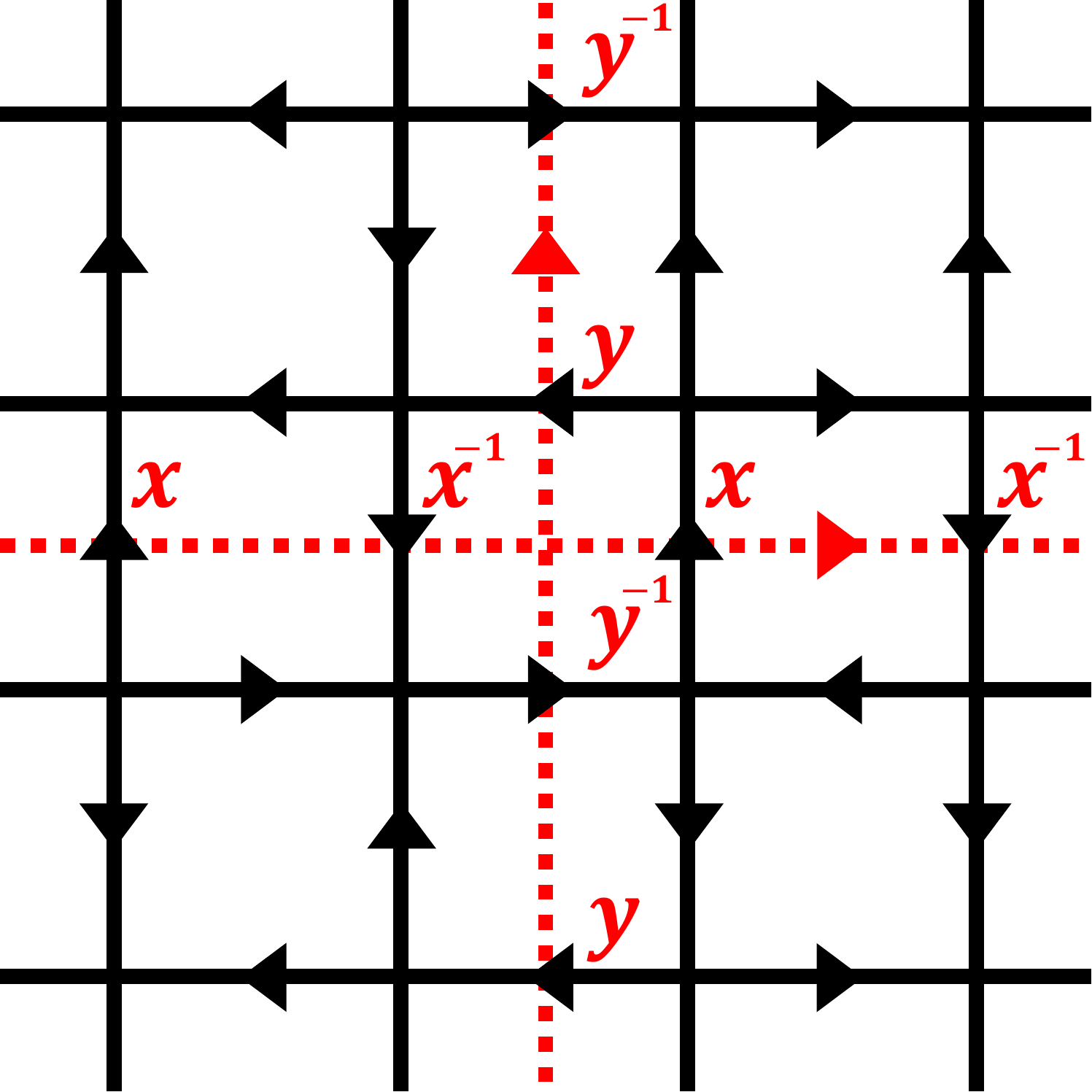}
\caption{ A representative flat $\mathsf{G}$ connection labeled by $(x,y)$.}\label{e11}
\end{figure}

\subsection{Twisting and gauging operators and states}

For any local operator $O$ acting on the matter degrees of freedom in a contractible region $\subgraph\subseteq\fullgraph$ there is a well defined notion of twisting $O$ by a flat $\mathsf{G}$-connection $\con$. Fixing a base vertex $v_0\in\subgraph$ the twisted operator is given by 
\begin{align}
O^\con:=\int\mathrm{d}g\bigotimes_{v\in\subgraph}U_v( \con_{\epath_v} g)\, O \bigotimes_{v\in\subgraph}U_v^\dagger(  \con_{\epath_v} g)
\end{align}
 where $\epath_v$ is any path from $v_0$ to $v$ within $\subgraph$ (the choice does not matter since the connection is flat and $\subgraph$ is contractible). 
The choice of distinguished base vertex $v_0$ is irrelevant since a change $v_0\mapsto v_0'$ can be compensated by shifting $g\mapsto \con_{\epath'}^{-1}g$, where $\epath'$ is a path from $v_0'$ to $v_0$, which has no effect since $g$ is summed over. 
Note this definition of $O^\con$ first projects $O$ onto the space of symmetric operators, hence the sum over $g$ is unnecessary if $O$ is already symmetric.
One can verify that $O^\con$ commutes with the following twisted symmetry $ \bigotimes_{v\in\subgraph}U_v( g^{\con_{\epath_v}}),\,\forall g\in\mathsf{G}$, where $g^h=hgh^{-1}$, independent of the choice of base point $v_0$ and paths $\epath_v\in\subgraph$ from $v_0$ to $v$.

The twisted state gauging map $G_\con$, for a flat $\mathsf{G}$-connection $\con$, is defined by the following action
\begin{align}
&G_\con \ket{\psi}:=P[\, \ket{\psi} \otimes \ket{\con} ]
\nonumber \\
&\phantom{G_\con}
=\int \prod_{v\in\fullgraph} \mathrm{d} g_v [\, \bigotimes_{v\in\fullgraph}U_v(g_v)] \ket{\psi} \bigotimes_{e\in\fullgraph} \ket{g_{v_e^-}\con_eg_{v_e^+}^{-1}}_e
\end{align}
where $\ket{\psi}\in\hilb_\mathsf{m}$ is a state of the matter degrees of freedom.
One can verify that ${G_\con^\dagger G_\con=\int \mathrm{d} g \bigotimes_{v\in\fullgraph}U_v(g^{\con_{\epath_v}}) \prod_i \delta_{gx_i,x_ig}\delta_{gy_i,y_ig}}$ is the projection onto the symmetric subspace of the twisted symmetry, where $(x_i,y_i)$ are the pairs of commuting group elements that label $\con$, see Eq.\eqref{glcon}. The $\delta$ conditions arise since the state overlaps force the conjugation of $g$ by the transport group elements $\con_{\epath_v},\con_{\epath_v'}$ to agree for non homotopic paths $\epath_v,\epath_v'$ from $v_0$ to $v$. These $\delta$ conditions also ensure the choice of fixed base point $v_0$ is irrelevant.
\\
The twisted operator gauging map $\mathscr{G}_\subgraph^\con$ is defined similarly
\begin{align}
\mathscr{G}_\subgraph^\con[O]:=\int\prod_{v\in\subgraph} \mathrm{d} g_v \bigotimes_{v\in\subgraph} U_v(g_v)\, O \bigotimes_{v\in\subgraph} U^\dagger_v(g_v) 
\nonumber \\
\bigotimes_{e\in\subgraph} \ket{g_{v_e^-}\con_e g_{v_e^+}^{-1}}\bra{g_{v_e^-}\con_e g_{v_e^+}^{-1}}
\end{align}
 where $O$ is  an operator that acts on the matter degrees of freedom on sites $v\in\subgraph\subseteq\fullgraph$, and $\subgraph$ is defined to include all the edges between its vertices.
$\mathscr{G}_\subgraph^\con$ is invertible on the space of $\con$-twisted symmetric local operators $O^\con$ in the following sense 
\begin{align}
&\text{Tr}_{e\in\subgraph}[{\mathscr{G}_\subgraph^\con[O^\con]\bigotimes_{e\in\subgraph}\ket{\con_e}\bra{\con_e}_e}]=
\int\prod_{v\in\subgraph} \mathrm{d} g_v \bigotimes_{v\in\subgraph} U_v(g_v) 
\nonumber \\
&\phantom{= \int\prod_{v\in\subgraph}\mathrm{d} g_v \bigotimes_{v\in\subgraph}} 
\times O^\con \bigotimes_{v\in\subgraph} U^\dagger_v(g_v) \prod_{e\in\subgraph} \delta_{g_{v_e^-}\con_e g_{v_e^+}^{-1},\con_e}
\nonumber \\
&\phantom{\int\prod_{v\in\subgraph}}= \int \mathrm{d} g_{v_0} \bigotimes_{v\in\subgraph} U_v(g_{v_0}^{\con_{\epath_v}}) 
\, O^\con \bigotimes_{v\in\subgraph} U^\dagger_v(g_{v_0}^{\con_{\epath_v}})
\nonumber\\
&\phantom{\int\prod_{v\in\subgraph}}=O^\con
\end{align}
where the final equality follows from the twisted symmetry of $O^\con$ and the second equality follows since the $\delta$ conditions force $g_{v_e^-}=g_{v_e^+}^{\con_e}$ which implies, after fixing a base point $v_0\in\fullgraph$, that $g_v=g_{v_0}^{\con_{\epath_v}}$ for any path $\epath_v$ from $v_0$ to $v$ within $\subgraph$ which is assumed to be contractible in the underlying manifold $\man$.

For the twisted gauging procedure we also have a version of Proposition~\ref{p4}, which states the useful equality $\mathscr{G}_\subgraph[O]G=G O$ for symmetric $O$. 
In the twisted case it must be modified in the following way
\begin{claim}\label{prop7}
The identity $\mathscr{G}_\subgraph^\con[O^\con]G_\con=G_\con O^\con$ holds for any symmetric operator $O$.
\end{claim} 
We now proceed to show this
\begin{align}\label{oggo}
&\mathscr{G}_\subgraph^\con[O^\con]G_\con = \int\prod_{v\in\subgraph} \mathrm{d} h_v\bigotimes_{v\in\subgraph} U_v(h_v) O^\con \bigotimes_{v\in\subgraph} U_v^\dagger(h_v) 
\nonumber \\
&\phantom{\mathscr{G}_\con}
\bigotimes_{e\in\subgraph} \ket{h_{v_e^-}\con_eh_{v_e^+}^{-1}}\bra{h_{v_e^-}\con_eh_{v_e^+}^{-1}} \int\prod_{v\in\fullgraph} \mathrm{d} g_v\bigotimes_{v\in\fullgraph} U_v(g_v)
\nonumber \\
&\phantom{\mathscr{G}_\con[O^\con]}
\bigotimes_{e\in\fullgraph} \ket{g_{v_e^-}\con_eg_{v_e^+}^{-1}}
\nonumber \\
&
=\int\prod_{v\in\fullgraph} \mathrm{d} g_v \prod_{v\in\subgraph} \mathrm{d} h_v  \bigotimes_{v\in\fullgraph} U_v(g_v)
\bigotimes_{v\in\subgraph} U_v(g_v^{-1}h_v) \, O^\con
\nonumber \\
& \bigotimes_{v\in\subgraph} U^\dagger_v(g_v^{-1}h_v) 
\prod_{e\in\subgraph} \delta_{(g_{v_e^-}^{-1}h_{v_e^-}),\, (g_{v_e^+}^{-1}h_{v_e^+})^{\con_e}}
\bigotimes_{e\in\fullgraph} \ket{g_{v_e^-}\con_e g_{v_e^+}^{-1}}
\nonumber \\
&
= G_\con O^\con
\end{align}
the last equality follows since the $\delta$ condition forces $g_v^{-1}h_v=(g_{v_0}^{-1}h_{v_0})^{\con_{\epath_v}}$ (for a fixed choice of vertex $v_0$ and path $\epath_v\in\subgraph$ from $v_0$ to $v$ which has no effect on the outcome) implying $\bigotimes_{v\in\subgraph} U_v(g_v^{-1}h_v) = \bigotimes_{v\in\subgraph} U_v(\, (g_{v_0}^{-1}h_{v_0})^{\con_{\epath_v}})$ which is precisely a twisted symmetry that commutes with $O^\con$ to yield the desired result.

For a symmetric local Hamiltonian that has been twisted by a flat $\mathsf{G}$-connection $\con$,  $H_\mathsf{m}^\con=\sum_v h_v^\con$, we define the twisted gauged Hamiltonian $ (H_\mathsf{m}^\con)^{\mathscr{G}^\con} :=\sum_v\mathscr{G}_{\subgraph_v}^\con[h_v^\con]$ in a locality preserving way similar to the untwisted case. With this definition we pose the following proposition
\begin{claim}\label{gceg}
For all flat $\mathsf{G}$-connections $\con$ we have $(H_\mathsf{m}^\con)^{\mathscr{G}^\con}=H_\mathsf{m}^{\mathscr{G}}$.
\end{claim}
 To prove this it suffices to consider a generic local term $h_v^\con$ acting on the subgraph $\subgraph_v$
\begin{align}\label{ghv}
\mathscr{G}_{\subgraph_v}^\con[h_v^\con]
&=\int\prod_{v\in\subgraph} \mathrm{d} g_v \bigotimes_{v\in\subgraph} U_v(g_v) U_v( \con_{\epath_v})\,  h_v \bigotimes_{v\in\subgraph}U_v^{\dagger}( \con_{\epath_v})
\nonumber \\ 
& \phantom{=\int\prod_{v\in\subgraph}}
 U^\dagger_v(g_v) \bigotimes_{e\in\subgraph} \ket{g_{v_e^-}\con_e g_{v_e^+}^{-1}}\bra{g_{v_e^-}\con_e g_{v_e^+}^{-1}}
\nonumber \\ 
&= \int\prod_{v\in\subgraph} \mathrm{d} g_v \bigotimes_{v\in\subgraph} U_v(g_v \con_{\epath_v})\,  h_v \bigotimes_{v\in\subgraph}U_v^\dagger( g_v\con_{\epath_v})
\nonumber \\ 
 &\phantom{=\int\ }
\bigotimes_{e\in\subgraph} \ket{g_{v_e^-}\con_{\epath_{v_e^-}}\con_{\epath_{v_e^+}}^{-1} g_{v_e^+}^{-1}}\bra{g_{v_e^-}\con_{\epath_{v_e^-}}\con_{\epath_{v_e^+}}^{-1} g_{v_e^+}^{-1}}
\nonumber \\ 
&=  \int\prod_{v\in\subgraph} \mathrm{d} \tilde{g}_v \bigotimes_{v\in\subgraph} U_v(\tilde{g}_v )\,  h_v \bigotimes_{v\in\subgraph}U_v^\dagger( \tilde{g}_v)
\nonumber \\ 
 &\phantom{=\int\prod_{v\in\subgraph}}
\bigotimes_{e\in\subgraph} \ket{\tilde{g}_{v_e^-}\tilde{g}_{v_e^+}^{-1}}\bra{\tilde{g}_{v_e^-}\tilde{g}_{v_e^+}^{-1}}
\nonumber \\ 
&= \mathscr{G}_{\subgraph_v}[h_v]
\end{align}
for the first equality we use the symmetry of $h_v$, for the second we use the fact $\con_e=\con_{\epath_{v_e^-}}\con_{\epath_{v_e^+}}^{-1}$, note the choice of base point $v_0$ and paths $\epath_v$ from $v_0$ to $v$ in $\subgraph$ have no effect since $h_v$ is symmetric and $\subgraph$ is contractible, for the third we use the invariance of the Haar measure under the change of group variables $g_v\mapsto\tilde{g}_v:=g_v\con_{\epath_v}$.

\subsection{Gauging preserves the gap and leads to a topological degeneracy}

We are now in a position to prove that gauging a SPT Hamiltonian defined on an arbitrary oriented 2-manifold $\man$ preserves the energy gap, generalizing the proof presented in Section~\ref{gptg}.

The full gauged Hamiltonian is given by ${H_\text{full}:=H_\mathsf{m}^{\mathscr{G}}+\Delta_\mathcal{B} H_\mathcal{B}+\Delta_P H_P}$, see Section~\ref{gptg} for a discussion of each term in the Hamiltonian. Note by Proposition~\ref{gceg} the same full Hamiltonian $H_\text{full}$ is achieved by gauging any $\con$-twist of a given SPT Hamiltonian. 

As argued in Section~\ref{gptg}, for $\Delta_\mathcal{B},\, \Delta_P$ sufficiently large, the low energy subspace of $H_\text{full}$ lies within the common ground space of $H_\mathcal{B}$ and $H_P$. This subspace is spanned by the states $P[\, \ket{\lambda}_\mathsf{m}\otimes \ket{\con}_\mathsf{g}]=G_\con \ket{\lambda}$, where the matter states $\ket{\lambda}$ form a basis of $\mathbb{H}_\mathsf{m}$, and the gauge states $\ket{\con}$ span the flat $\mathsf{G}$-connections.
This leads to a generalization of Proposition~\ref{prop3} to arbitrary 2-manifolds
\begin{claim}\label{prop9}
For an oriented 2-manifold $\man$ the set of states $\{ G_{\con^\gamma}\ket{\lambda} \}$, for $\{\ket{\lambda}\}$ a basis of $\mathbb{H}_\mathsf{m}$ 
and a fixed  choice of representatives $\gamma\in[\gamma]\in\mathcal{I}_\man$, span the common ground space of $H_\mathcal{B}$ and $H_P$.
\end{claim}
Firstly, by Proposition~\ref{conl}, an arbitrary flat connection $\ket{\con}$ is related to $\ket{\con^\gamma},\, \exists[\gamma]\in\mathcal{I}_\man$ by a sequence of local operations $\ket{\con}=\prod_i a_{v_i}^{g_i} \ket{\con^\gamma}$. 
Since $P_v=\int\mathrm{d} g U_v(g) \otimes \st_v^g$ one can easily see $P_v \st_v^g = P_v U^\dagger_v(g)$ and hence we have
\begin{align}
G_{\con}\ket{\psi}_\mathsf{m} &=P[\ket{\psi}_\mathsf{m}\otimes\prod_i a_{v_i}^{g_i}\ket{\con^\gamma}_\mathsf{g}]
\nonumber \\
&=
P[\, [\prod_i U_{v_i}(g_i)]^\dagger\ket{\psi}_\mathsf{m}\otimes\ket{\con^\gamma}_\mathsf{g}]
\nonumber \\
&= G_{\con^\gamma} [\prod_i U_{v_i}(g_i)]^\dagger\ket{\psi}_\mathsf{m}\, .
\end{align}
Therefore the common ground space of $H_\mathcal{B}$ and $H_P$ is spanned by the states $\{G_{\con^\gamma}\ket{\lambda}\}_{(\lambda,\gamma)}$ for a basis $\{\ket{\lambda}\}_\lambda$ of $\mathbb{H}_\mathsf{m}$ and a representative $\gamma$ of each conjugacy class $[\gamma]\in\mathcal{I}_\man$.

We now bring together the definitions and propositions laid out thus far to show the following 
\begin{claim}
{Gauging a gapped SPT Hamiltonian on an arbitrary oriented 2-manifold $\man$ yields a gapped local Hamiltonian with a topology dependent ground space degeneracy}. 
\end{claim}
Let $\ket{\lambda^\gamma}$ denote an eigenstate of the twisted SPT Hamiltonian $H_\mathsf{m}^{\con^\gamma}$ with eigenvalue $\lambda$. From Propositions~\ref{prop7} \&~\ref{gceg} it follows that gauging an eigenstate of a $\con$-twisted SPT Hamiltonian yields an eigenstate of the gauged Hamiltonian, so we have $H_\mathsf{m}^{\mathscr{G}} G_{\con^\gamma}\ket{\lambda^\gamma}=\lambda G_{\con^\gamma}\ket{\lambda^\gamma}$.

If $H_\mathsf{m}$ has a unique ground state $\ket{\lambda_0}$ 
Proposition~\ref{prop9} implies the ground space of the full Hamiltonian $H_\text{full}$ is spanned by the states $\{G_{\con^\gamma}\ket{\lambda^\gamma_0}\}_\gamma$ and its gap satisfies ${\Delta_\text{full}\geq\min(\Delta_m,\Delta_\mathcal{B},\Delta_P)}$

In the above we have assumed that $G_{\con^\gamma}\ket{\lambda^\gamma_0}\neq0$, for some $\gamma$. Note $G\ket{\lambda_0}\neq 0$ always holds for a unique ground state $\ket{\lambda_0}$ of a symmetric Hamiltonian (possibly after rephasing the matrices of the physical group representation which is assumed to have occurred).

We now proceed to show that the ground space degeneracy is equal to 
the number of distinct equivalence classes of symmetry twists which are invariant under the residual physical symmetry.
This relies on the assumption that the distinct symmetry twisted SPT Hamiltonians $H_\mathsf{m}^{\con^\gamma}$ each have a nonzero unique ground state $\ket{\lambda^\gamma_0}$ with the same energy $\lambda_0$. We show this to be the case, when the original frustration free SPT Hamiltonian $H_\mathsf{m}$ has a SPT PEPS ground state, by explicitly constructing tensor network representations of the twisted ground states, see Definition~\ref{stp}.

\begin{claim}\label{overlapprop}
The overlap matrix of the gauged ground states $\overlap{\gamma'}{\gamma}:=\bra{\lambda^{\gamma'}_0}G_{\con^{\gamma'}}^\dagger G_{\con^\gamma}\ket{\lambda^\gamma_0}$  is diagonal, where $\gamma,\gamma'$ are drawn from a fixed set of representatives for the conjugacy classes in $\mathcal{I}_\man$. Furthermore, $\overlap{\gamma'}{\gamma}$ is invariant under a change of representatives and $\overlap{\gamma}{\gamma}=0$ iff $\ket{\lambda^\gamma_0}$ transforms as a nontrivial representation of the physical symmetry action of $\mathsf{C}(\gamma)$.
\end{claim}
The operators $G^\dagger_\varphi G_\con$ that appear in the overlaps of the gauged twisted ground states imply that they are orthogonal. To see this consider the following
\begin{align}
&G^\dagger_\varphi G_\con =\int \prod_{v\in\fullgraph} \mathrm{d} k_v \mathrm{d} g_v \bigotimes_{v\in\fullgraph}U_v(k_v^{-1}g_v)
\nonumber\\ 
\label{GG}
&\phantom{G^\dagger_\varphi G_\con =\int \prod_{v\in\fullgraph}}
\prod_{e\in\fullgraph} \braket{k_{v_e^-}\varphi_e k_{v_e^+}^{-1}|g_{v_e^-}\con_eg_{v_e^+}^{-1}} 
\\
&=\int \mathrm{d} g_{v_0} \bigotimes_{v\in\fullgraph}U_v(\varphi_{p_v} g_{v_0} \con_{p_v}^{-1}) \prod_i \delta_{x_i'g_{v_0},g_{v_0}x_i} \delta_{y_i'g_{v_0},g_{v_0}y_i}
\nonumber
\end{align}
where we have fixed an arbitrary base vertex $v_0$, $p_v$ is any path from $v_0$ to $v$, and $\{(x_i',y_i')\}_i,\{(x_i,y_i)\}_i$ label the connections $\varphi,\con$ respectively. The delta conditions arise since the overlaps in Eq.\eqref{GG} force the transported group element $\varphi_{p_v}g_{v_0} \con_{p_v}^{-1} $ to agree for any choice of path $p_v$ (which may be homotopically distinct).
This implies that $G^\dagger_\varphi G_\con =0$ whenever the labels $\{(x_i',y_i')\}_i,\{(x_i,y_i)\}_i$ fall into distinct equivalence classes of $\mathcal{I}_\man$.
\\
For the particular case of the simple representative $\mathsf{G}$-connections $\con^\gamma$ we have 
\begin{align*}
G_{\con^{\gamma'}}^\dagger G_{\con^{\gamma}}
= \delta_{[\gamma'],[\gamma]} \int \mathrm{d} g \bigotimes_{v\in\fullgraph}U_v(g) \prod_i \delta_{x_i'g,gx_i } \delta_{y_i'g,gy_i } 
\end{align*}
for equivalence classes $[\gamma'],[\gamma]\in\mathcal{I}_\man$.
Furthermore, if $\gamma'\sim\gamma$ then there exists a group element $h\in\mathsf{G}$ such that $(x_i',y_i')=(x_i^g,y_i^g),\, \forall i\, \iff g\in h\, \mathsf{C}(\gamma) $, a left coset of the centralizer of $\gamma=\{x_i,y_i\}_i$.   
In this case
\begin{equation}
G_{\con^{\gamma'}}^\dagger G_{\con^{\gamma}}=\int \mathrm{d} g \bigotimes_{v\in\fullgraph}U_v(g)\, \delta_{g\in h \mathsf{C}(\gamma) }
\end{equation}
and ${H_\mathsf{m}^{\con^{\gamma'}}=U(g)^{\otimes |\fullgraph|_v} H_\mathsf{m}^{\con^\gamma} U^\dagger (g)^{\otimes |\fullgraph|_v} }$ for any ${g\in h\, \mathsf{C}(\gamma) }$, which implies ${\theta_{g}^\gamma \ket{\lambda^{\gamma'}_0}=U(g)^{\otimes |\fullgraph|_v} \ket{\lambda^\gamma_0} }$ for some phase $\theta_{g}^\gamma\in \mathsf{U(1)}$. 
Hence 
$${\bra{\lambda^\gamma_0}G_{\con^\gamma}^\dagger G_{\con^{\gamma}}\ket{\lambda^\gamma_0}=\theta_h^{\gamma} \bra{\lambda^\gamma_0}G_{\con^\gamma}^\dagger G_{\con^{\gamma'}}\ket{\lambda^{\gamma'}_0}\iff [\gamma]=[\gamma']}.$$
Moreover since $\ket{\lambda^\gamma_0} $ is the unique groundstate of a $\mathsf{C}(\gamma)$-symmetric Hamiltonian $\theta_{(\cdot)}^\gamma$ is a 1D representation of $\mathsf{C}(\gamma)$. By the orthogonality of characters we have ${G_{\con^\gamma}^\dagger G_{\con^{\gamma}}\ket{\lambda^\gamma_0}\neq 0\iff\theta_{(\cdot)}^\gamma\equiv 1}$. 
Note $\theta_{(\cdot)}^\gamma\equiv 1$ is in fact a property of a conjugacy class as it does not depend on the choice of representative $\gamma$. 

Consequently the choice of representative symmetry twist $\gamma\in[\gamma]\in\mathcal{I}_\man$ does not matter as all lead to the same gauged state $\ket{\lambda_0,[\gamma]}:=G_{\con^{\gamma}}\ket{\lambda^\gamma_0}$. 
Hence the overlap matrix of the gauged twisted SPT groundstates is given by
\begin{align}
\overlap{\gamma'}{\gamma}&=\braket{\lambda_0,[\gamma']|\lambda_0,[\gamma]}
\nonumber \\
&= \delta_{[\gamma'],[\gamma]} \, \delta_{\theta_{(\cdot)}^\gamma, 1} \frac{|\mathsf{C}(\gamma)|}{|G|} \braket{\lambda^\gamma_0|\lambda^\gamma_0}
\end{align}
and the set of states ${\{\ket{\lambda_0,[\gamma]}|\, {[\gamma]}\in\mathcal{I}_\man,\,\theta^\gamma_{(\cdot)}\equiv 1\}}$ form an orthogonal basis for the ground space of the full gauged Hamiltonian $H_\text{full}$.

\section{Symmetry twists \& monodromy defects }\label{g}
In this appendix we describe a general and unambiguous procedure for applying symmetry twists to SPT PEPS using virtual symmetry MPOs. We furthermore demonstrate that the gauging procedure maps the symmetry MPOs to freely deformable topological MPOs on the virtual level and hence the gauged symmetry twisted PEPS are locally indistinguishable while remaining globally orthogonal, implying that they exhibit topological order. 
We move on to discuss how the same MPOs can be arranged along open paths to describe monodromy defects in SPT PEPS and anyons in the gauged PEPS.
 Moreover, we explicitly calculate the projective transformation of individual monodromy defects under the residual symmetry group using tensor network techniques.

\subsection{Symmetry twisted states}

In this section we discuss the ground states of symmetry twisted Hamiltonians in more detail and show that the PEPS framework naturally accommodates a simple construction of these states.

On a trivial topology a symmetry twist can be applied directly to a state by acting on some region of the lattice with the physical symmetry. For example on an infinite square lattice in the 2D plane a symmetry twist $(x,y)$ along an oriented horizontal and vertical path $p_x,p_y$, in the dual lattice, acts on a state $\ket{\psi}$ via 
\begin{align*}
\ket{\psi}^\phi:&=\int\mathrm{d}g\bigotimes_{v\in\subgraph}U_v( \con_{\epath_v} g) \ket{\psi}
\\
&= \bigotimes_{v\in \mathcal{U}} U_v(x) \bigotimes_{v\in\mathcal{R}} U_v(y) \int\mathrm{d}g\bigotimes_{v\in\subgraph}U_v(g)\ket{\psi}
\end{align*}
where $\con$ is the simple representative connection with label $(x,y)$ on paths $p_x,p_y$, see Definition~\ref{flaty}, and $\region$ is the half plane to the right of $p_y$, $\mathcal{U}$ the half plane above $p_x$, see Fig.\ref{e4}.
Note this definition implicitly projects $\ket{\psi}$ onto the trivial representation and we have $O^\con \ket{\psi}^\con= (O\ket{\psi})^\con$ for symmetric operators $O$. Hence twisting an eigenstate $\ket{\lambda}$ of a SPT Hamiltonian $H_\mathsf{m}$ yields an eigenstate $\ket{\lambda}^\con$ of the twisted Hamiltonian $H_\mathsf{m}^\con$ with the same eigenvalue.
Note $x$ and $y$ must commute for $\con$ to be a flat connection, equivalently if one thinks of first applying the $x$ twist to a symmetric Hamiltonian, then the resulting operator will only be symmetric under the centralizer subgroup of  $x$, $\mathsf{C}(x)\leq\mathsf{G}$, and hence it only makes sense to apply a second twist for an element $y\in \mathsf{C}(x)$. 

The effect of such a symmetry twist on a SPT PEPS $\ket{\psi}$ is particularly simple, it can be achieved by adding the virtual symmetry MPOs $V^{\epath_x}(x)$ and $V^{\epath_y}(y)$ (with inner indices contracted with the four index crossing tensor $Q_{x,y}=\QW{R}{x}{y}$~(\ref{cxy},\ref{na21a}) where $\epath_x,\epath_y$ intersect, see Fig.\ref{e5}) to the virtual level of the PEPS. Let us denote the resulting tensor network state $\ket{\psi^{(x,y)}}$, then by Eq.\eqref{n23} we have $\ket{\psi^{(x,y)}}=\ket{\psi}^\con$. 

\begin{figure}[ht]
\center
\includegraphics[width=0.5\linewidth]{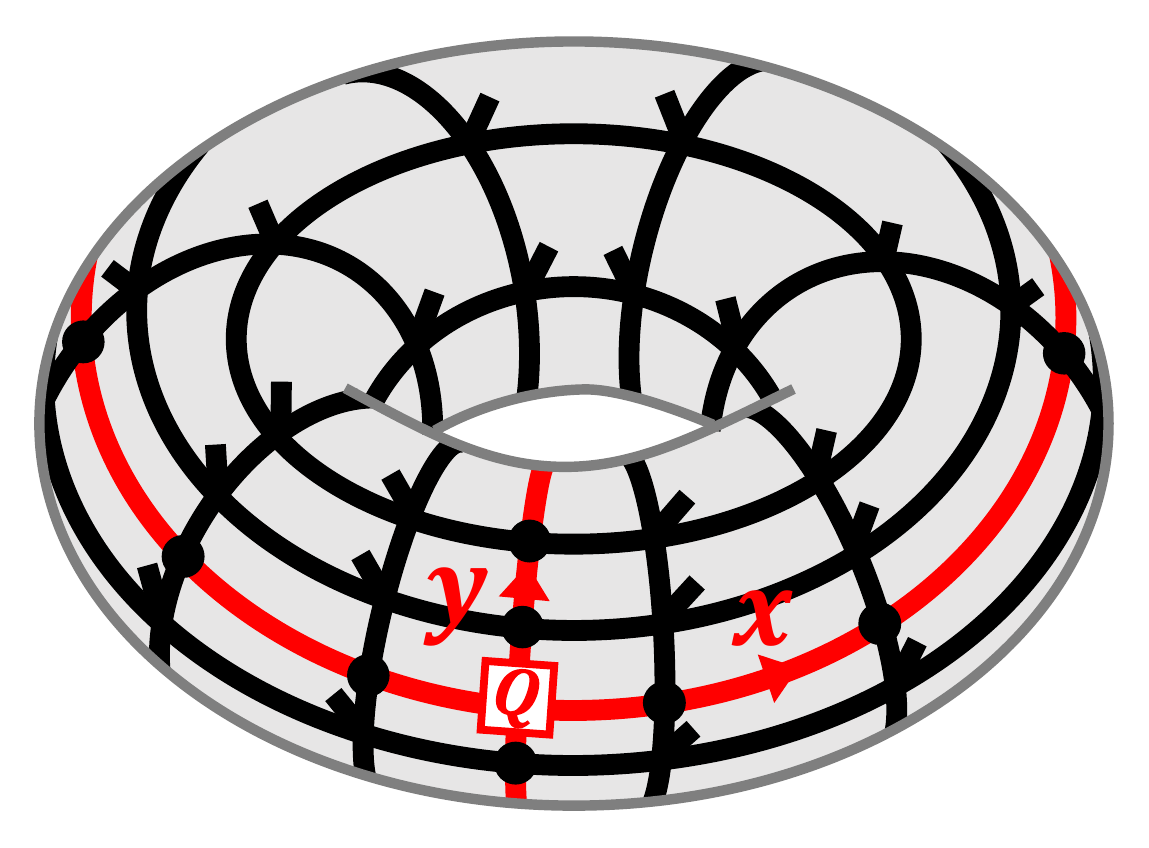}
\caption{ An $(x,y)$ symmetry twisted PEPS on a torus.}\label{e12}
\end{figure}

For nontrivial topologies the symmetry twist on a state $\ket{\psi}^{\phi^\gamma}$ is not well defined in terms of a physical symmetry action since two homotopically inequivalent paths $\epath_v,\epath_v'$ can give rise to distinct transport elements $\phi_{\epath_v}\neq\phi_{\epath_v'}$. Note this problem does not arise when symmetry twisting local operators, such as the terms in a local Hamiltonian, since each operator acts within a contractible region. 
The PEPS formalism yields a simple resolution to this problem since the process of applying a symmetry twist $\con^\gamma$ on the virtual level of a PEPS $\ket{\psi^\gamma}$ remains well defined, see Definition~\ref{stp} and Fig.\ref{e12}.

The general scenario is as follows; we have a local gapped frustration free SPT Hamiltonian $H_\mathsf{m}$ defined on an oriented 2-manifold $\man$ with a SPT PEPS $\ket{\lambda_0}$ as its unique ground state (note SPT PEPS parent Hamiltonians satisfy these conditions) and we want to apply a symmetry twist along paths $\epath_x^i,\epath_y^i$ in the dual graph labeled by $\gamma=\{(x_i,y_i)\}_i$. 
\begin{definition}[Symmetry Twisted SPT PEPS]\label{stp}
For a SPT PEPS $\ket{\psi}$ and a symmetry twist $\gamma$, specified by a set of pairwise intersecting paths in the dual graph $\{\epath_x^i,\epath_y^i\}_i$ and pairwise commuting group elements $\{(x_i,y_i)\}_i$ in $\mathsf{G}$, the symmetry twisted PEPS $\ket{\psi^\gamma}$ is constructed by taking the tensor network for $\ket{\psi}$ with open virtual indices on edges that cross $\{\epath_x^i,\epath_y^i\}_i$ and contracting these virtual indices with the MPOs $\{V^{\epath_x^i}(x_i),V^{\epath_y^i}(y_i)\}_i$. Moreover, at the intersection of the paths $\epath_x^i\cap\epath_y^i$ the internal indices of the MPOs $V^{\epath_x^i}(x_i),V^{\epath_y^i}(y_i)$ are contracted with four index crossing tensors $Q_{x_i,y_i}=\QW{R}{x_i}{y_i}$, defined in Eqs.(\ref{cxy},\ref{na21a}) and similarly with $Q_{y_{i-1},x_i}=\QW{R}{y_{i-1}}{x_i}$ at the intersections $\epath_y^{i-1}\cap\epath_x^{i}$. This is depicted in Fig.~\ref{e12}.
\end{definition}
It follows from Eq.\eqref{n23} and the \emph{zipper} condition~\eqref{a3} for $X(x_i,y_i)$ that the symmetry twisted ground state SPT PEPS $\ket{\lambda_0^{\gamma}}$ is the ground state of the twisted SPT Hamiltonian $H_\mathsf{m}^{\con^\gamma}$. 
More generally for any SPT PEPS $\ket{\psi}$ that is an eigenstate of each local term in $H_\mathsf{m}$,  Eq.\eqref{n23} implies that $\ket{\psi^{\gamma}}$ is an eigenstate of $H_\mathsf{m}^{\con^\gamma}$ with the same eigenvalue (thereby justifying the notation).
Note the twisted SPT PEPS $\ket{\psi^{\gamma}}$ for different choices of representative $\gamma$ from the same conjugacy class $[\gamma]\in\mathcal{I}_\man$ are all related by the action of some global symmetry, which again follows from Eqs.\eqref{n23},\eqref{a3} and Proposition~\ref{conl}.

{\begin{claim}\label{prop10}
A $\gamma$-twisted SPT PEPS $\ket{\psi^\gamma}$ transforms as the following 1D representation 
\begin{equation}
{\theta_{(\cdot)}^\gamma=\slant{\alpha}{2}{x_0,y_0}{\cdot}^{-1}\prod\limits_{i=1} [\slant{\alpha}{2}{y_{i-1},x_i}{\cdot}\slant{\alpha}{2}{x_i,y_{i}}{\cdot}]^{-1}}
\end{equation}
 under the physical action of the residual symmetry group $\mathsf{C}(\gamma)$.
\end{claim}}

The physical action of the symmetry ${U(k)^{\otimes |\man|_v}}$ induces a local action $\pi_k$ on each crossing tensor ${\{\QW{R}{x_i}{y_i},\QW{R}{y_{i-1}}{x_i}\}_i}$ and by Eq.\eqref{na27} we find the combined action to be ${\slant{\alpha}{2}{x_0,y_0}{\cdot}^{-1}\prod\limits_{i=1} [\slant{\alpha}{2}{y_{i-1},x_i}{\cdot}\slant{\alpha}{2}{x_i,y_{i}}{\cdot}]^{-1}}$ as claimed.

\subsection{Topological ground states}

We now show that the twisted gauging procedure maps the virtual symmetry MPO to a freely deformable topological MPO on the virtual level.

{\begin{claim}\label{prop11}
Applying the twisted gauging map $G_{\con^\gamma}$ to a nonzero twisted SPT PEPS $\ket{\psi^\gamma}$ yields the MPO-injective PEPS $G\ket{\psi}$ with a set of freely deformable MPOs joined by crossing tensors, specified by $[\gamma]$, acting on the virtual level. 
The gauged state is zero iff $\ket{\psi^\gamma}$ transforms nontrivially under the residual symmetry group $\mathsf{C}(\gamma)$, this property depends only on $[\gamma]$ and $[\alpha]$.
\end{claim}}

We will first show that the tensor network $G_{\con^{\gamma}}\ket{\psi^{\gamma}}$ is given by contracting the MPOs $[V^{\epath_x^i}(x_i)\bigotimes_{e\in\epath_x^i} R_e(x_i)],[V^{\epath_y^i}(y_i)\bigotimes_{e\in\epath_y^i} R_e(y_i)]$ (contracted with the crossing tensor $Q_{x_i,y_i}=\QW{R}{x_i}{y_i}$ at $p_x^i\cap p_y^i$) with the virtual indices of $G\ket{\psi}$ on edges that cross the paths $\{\epath_x^i,\epath_y^i\}$. 

In general $G_\phi$ is a projected entangled pair operator (PEPO) with vertex tensors $G_\phi^v=\int \mathrm{d} g \, U_v(g) \bigotimes_{e\in E_v} \bbra{g}=G^v$ and edge tensors $G_\phi^e=\int \mathrm{d} g_{v_e^+} \mathrm{d} g_{v_e^-} L_e(g_{v_e^-})R_e(g_{v_e^+})\ket{\phi_e}\otimes\bbra{ g_{v_e^+}}\bbra{ g_{v_e^-}}$~\cite{Gaugingpaper}. 
Furthermore the edge tensors satisfy $G_\phi^e=G_1^e (R(\phi_e)\otimes\openone)=G_1^e (\openone\otimes R^\dagger(\phi_e)\, )$.
Hence the PEPO $G_\phi$ is given by the untwisted gauging map $G$ with the tensor product operators $\{\bigotimes_{e\in\epath_x^i} R_e(x_i),\bigotimes_{e\in\epath_y^i} R_e(y_i)\}$ applied to the virtual indices that cross $\{\epath_x^i,\epath_y^i\}$. 

Eqs.\eqref{n23} and \eqref{a3} together with $P_v \st_v^g = P_v U^\dagger_v(g)$ imply $G_{\con^{\gamma}}\ket{\psi^{\gamma}}=G_{\con^{\tilde\gamma}}\ket{\psi^{\tilde\gamma}}$ for any deformation $\tilde\gamma=\{\tilde\epath_x^i,\tilde\epath_y^i\}$ of the paths $\gamma=\{\epath_x^i,\epath_y^i\}$ that does not introduce additional intersections (a planar isotopy).
This furthermore implies that the MPOs $[V^{\epath}(g)\bigotimes_{e\in\epath} R_e(g)]$ satisfy the pulling through condition of Ref.\cite{MPOpaper} for any path $p$. Consequently, the MPO $\frac{1}{|G|}\sum_g[V^{\epath}(g)\bigotimes_{e\in\epath} R_e(g)]$, that was shown to be the projection onto the injectivity subspace of the gauged PEPS  in Appendix~\ref{d}, also satisfies the pulling through condition.

By Proposition~\ref{overlapprop} the gauged SPT PEPS $G_{\con^\gamma}\ket{\psi^\gamma}$ is zero iff $\theta^\gamma_{(\cdot)}$ is nontrivial, which is a property of the conjugacy class $[\gamma]$. Now by Proposition~\ref{prop10} and the fact that the slant product maps cohomology classes to cohomology classes we have the stated result.

Hence \emph{the nonzero gauged symmetry twisted PEPS ground states $\ket{\lambda_0,[\gamma]}:=G_{\con^{\gamma}}\ket{\lambda^\gamma_0}$ are topologically ordered} 
since the tensors $Q_{x_i,y_i}$ that determine the ground state are locally undetectable, which follows from the pulling through condition satisfied by the topological MPOs and Eq.\eqref{a3}, while for $[\gamma]\neq[\gamma']$ the states are globally orthogonal $\braket{\lambda_0,[\gamma]|\lambda_0,[\gamma']}=0$, as shown above.

{
In fact there is a slight subtlety, as while the reduced density matrices for all $[\gamma],[\gamma']\in\mathcal{I}_\man$ are supported on the same subspace $\rho_\region^{\lambda_0,[\gamma]},\rho_\region^{\lambda_0,[\gamma']}\in \text{Im}(\peps_\region\otimes \peps^\dagger_\region)$ for any contractible region $\region$, they are not necessarily equal~\cite{Ginjectivity} (or even exponentially close in the size of the region). One might also fret over the possibility that the state exhibits spontaneous symmetry breaking. 
\\
However neither of these complications can occur for the gauged symmetry twisted SPT PEPS, since an exact isometric fixed-point SPT PEPS does not exhibit symmetry breaking and is gauged to a topologically ordered fixed-point state which also does not exhibit symmetry breaking (see Section~\ref{exfpspt}). Furthermore the gauging map is gap preserving, hence gauging any SPT PEPS in the same phase as an SPT fixed-point maps it to a topological PEPS in the same phase as the gauged topological fixed-point PEPS.
}

\subsection{Monodromy defects in SPT PEPS}
\begin{figure}[ht]
\center
\begin{align*}
\vcenter{\hbox{
 \includegraphics[width=0.4\linewidth]{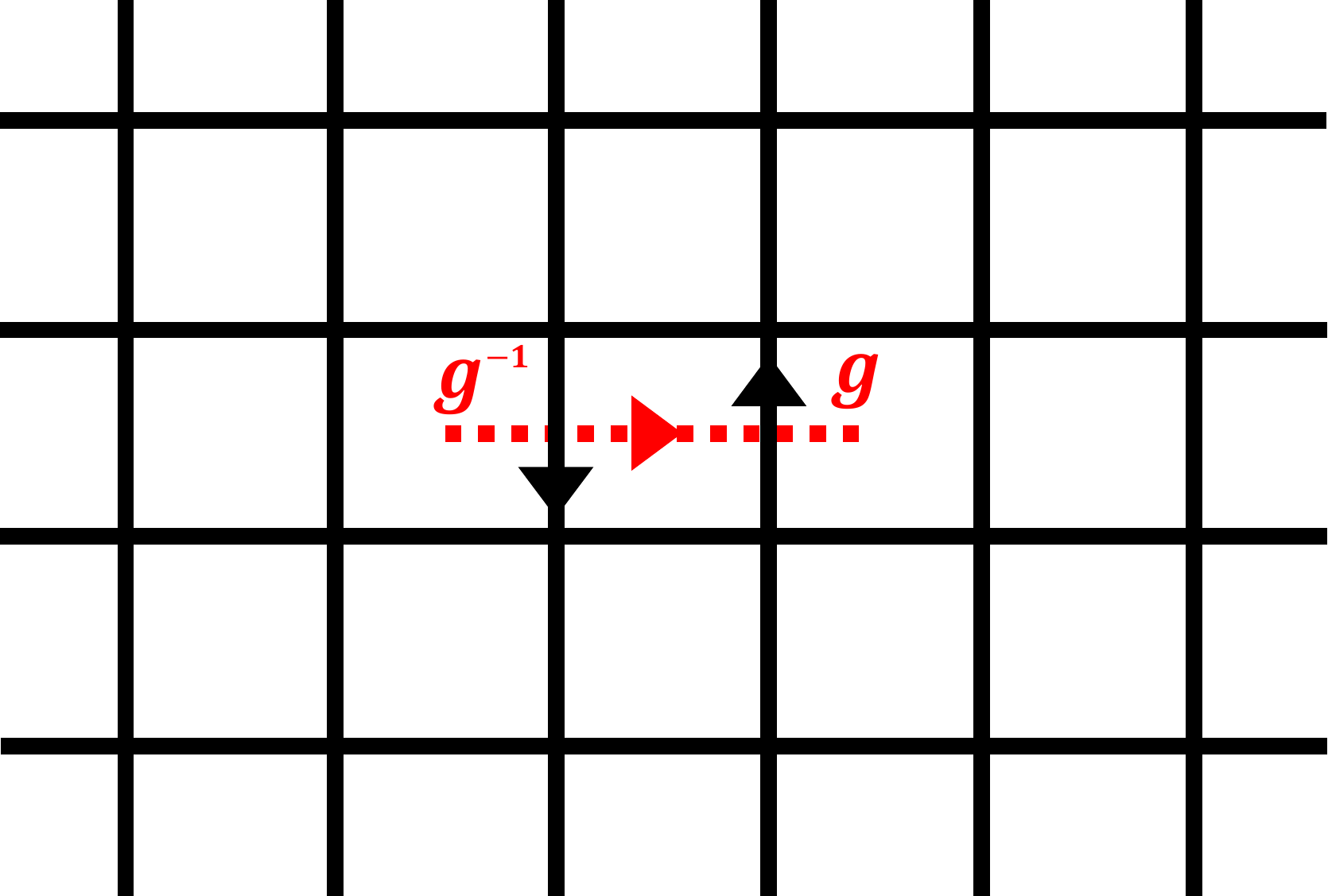}}} \hspace{.7cm}  
\end{align*}
\caption{ A symmetry twist $g$ along an open path.}\label{e13}
\end{figure}

Monodromy defects can be created in a SPT theory by applying a symmetry twist along an open ended path in the dual graph $\epath_g$ from plaquette $q_0$ to $q_1$, specified by a $\mathsf{G}$-connection $\con^{\epath_g}$, where $\con^{\epath_g}_e=1$, for $ e\notin \epath_g$ and $\con^{\epath_g}_e=g^{\sigma_e}$ for $e\in {\epath_g}$ ($\sigma_e$ is +1 if $\epath_{g}$ crosses $e$ in a right handed fashion and -1 for left handed crossings) hence $\con^{\epath_g}$ is flat on every plaquette except $q_0,q_1$, the end points of ${\epath_g}$, see Fig.\ref{e13}. The defect states can be realized as ground states of some twisted Hamiltonians $H_\mathsf{m}^{\con^{\epath_g}}=\sum_{q\in\fullgraph\backslash{\partial {\epath_g}}} h_q^{\con^{\epath_g}}+h'_{q_0}+h'_{q_1}$ where the choice of the end terms $h'_{q_0},h'_{q_1}$ is somewhat arbitrary. 
These monodromy defects can be introduced into a SPT PEPS $\ket{\psi}$ following the framework set up for symmetry twists.
\begin{definition}[Monodromy defected SPT PEPS]
A monodromoy defect specified by $\epath_g$ in a SPT PEPS $\ket{\psi}$ is described by a set of tensor network states parametrized by a pair of tensors $B_0,B_1$ where $B_0:(\mathbb{C}^D)^{\otimes |E_{v_0}|}\otimes\mathbb{C}^\mpod\rightarrow\mathbb{C}^d$ is a local tensor associated to a vertex $v_0\in\partial q_0$ with a set of indices matching those of the tensor $A_{v_0}$, and an extra virtual index of the same bond dimension $\mpod$ as the internal index of the MPO ($B_1$ is defined similarly). 
\\
The monodromy defected tensor network states $\ket{\psi^{\epath_g},B_0,B_1}$ are constructed from the SPT PEPS $\ket{\psi}$ by replacing the PEPS tensors $\peps_{v_0},\peps_{v_1}$ with $B_0,B_1$ and contracting the extra virtual indices thus introduced with the open end indices of the MPO $V^{\epath_g}(g)$ which acts on the virtual indices of the PEPS that cross $\epath_g$. This is depicted in Fig.~\ref{a1}~b).
\end{definition}

This provides an ansatz~\cite{nick} for symmetry twists by choosing appropriate boundary tensors $B_0,B_1$ to close the free internal MPO indices at $q_0,q_1$, the possibility of different boundary conditions corresponds to the ambiguity in the local Hamiltonian terms $h'_{q_0},h'_{q_1}$, see Figs.\ref{e13},\ref{a1}. 
Eq.\eqref{n23} implies that the defect state ansatz $\ket{\psi^{\epath_g},B_0,B_1}$ is in the ground space of the sum of  Hamiltonian terms away from the end points of $\epath_g$, $\sum_{q\in\fullgraph\backslash{\partial {\epath_g}}} h_q^{\con^{\epath_g}}$, for any choice of tensors $B_0,B_1$.

Since the connection $\con^{\epath_g}$ is flat everywhere but $q_0,q_1,$ the gauging map can be applied, in the usual way, to operators that are supported away from these plaquettes.
Hence the twisted gauged defect Hamiltonian is $(H_\mathsf{m}^{\con^{\epath_g}})^{\mathscr{G}^{\con^{\epath_g}}}:=\sum_{q\in\fullgraph\backslash{\partial {\epath_g}}} \mathscr{G}_{\subgraph_q}^{\con^{\epath_g}}[h_q^{\con^{\epath_g}}]+h''_{q_0}+h''_{q_1}$ where again there is an ambiguity in the choice of end terms $h''_{q_0},h''_{q_1}$. 
The SPT PEPS with monodromy defect $\epath_g$ can be gauged via the standard gauging procedure for the $\mathsf{G}$-connection $\con^{\epath_g}$ to yield the tensor network $G_{\con^{\epath_g}}\ket{\psi^{{\epath_g}},B_0,B_1}$.
Similar to the case of symmetry twists on closed paths, the gauged defected SPT PEPS $G_{\con^{\epath_g}}\ket{\psi^{{\epath_g}},B_0,B_1}$ is constructed from the untwisted gauged SPT PEPS $G\ket{\psi}$
by removing the tensors $G^{v_0}A_{v_0},G^{v_1}A_{v_1}$ and replacing them with the pair of tensors $G^{v_0}B_{0},G^{v_1}B_{1}$ connected by a virtual MPO $[V^{{\epath_g}}(g)\bigotimes_{e\in p_g}R_e(g)]$ acting on the virtual indices of the PEPS that cross $\epath_g$. 
Note the dimension of the inner indices of this MPO match the extra indices of $G^{v_0}B_{0},G^{v_1}B_{1}$ since the newly introduced component of the MPO $\bigotimes_{e\in p_g}R_e(g)$ has trivial inner indices. To achieve a more general ansatz one may want to replace $G^{v_0}B_0,G^{v_1}B_1$ by arbitrary tensors $\tilde B_0,\tilde B_1:(\mathbb{C}^D\otimes \mathbb{C}[\mathsf{G}])^{\otimes |E_{v}|}\otimes\mathbb{C}^\mpod\rightarrow\mathbb{C}^d$.

As shown above, the MPO $[V^{{\epath_g}}(g)\bigotimes_{e\in p_g}R_e(g)]$ satisfies the pulling through condition of Ref.\cite{MPOpaper} and hence $G_{\con^{\epath_g}}\ket{\psi^{{\epath_g}},B_0,B_1}=G_{\con^{\epath'_g}}\ket{\psi^{{\epath'_g}},B_0,B_1}$ for $\epath_g'$ an arbitrary, end point preserving, deformation of $\epath_g$.
By Eq.\eqref{ghv} we have $\mathscr{G}_{\subgraph_{q}}^{\con^{\epath_g}}[h_q^{\con^{\epath_g}}]= \mathscr{G}_{\subgraph_{q}}[h_q]$ and hence the gauged defected SPT PEPS $G_{\con^{\epath_g}}\ket{\psi^{{\epath_g}},B_0,B_1}$, for all $B_0,B_1$, is in the ground space of the sum of gauged Hamiltonian terms away from the end points $\sum_{q\in\fullgraph\backslash{\partial {\epath_g}}} \mathscr{G}_{\subgraph_{q}}[h_q]$. 
Consequently $G_{\con^{\epath_g}}\ket{\psi^{{\epath_g}},B_0,B_1}$ must represent a superposition of anyon pairs, localized to the plaquettes $q_0,q_1$, on top of the vacuum (ground state). Furthermore the freedom in choosing $B_0,B_1$ leads to a fully general anyon ansatz within the framework of MPO-injective PEPS~\cite{nick}.

\subsection{Projective symmetry transformation of monodromy defects}
We proceed to show that the internal degrees of freedom of a monodromy defect $\epath_g$ transform under a projective representation of the residual global symmetry group $\mathsf{C}(g)$ via a generalization of the mechanism for virtual symmetry actions in MPS~\cite{SchuchGarciaCirac11, 1Done, 1Dtwo}.

\begin{figure}[ht]
\center
\begin{align*}
a) \vcenter{\hbox{
 \includegraphics[width=0.4\linewidth]{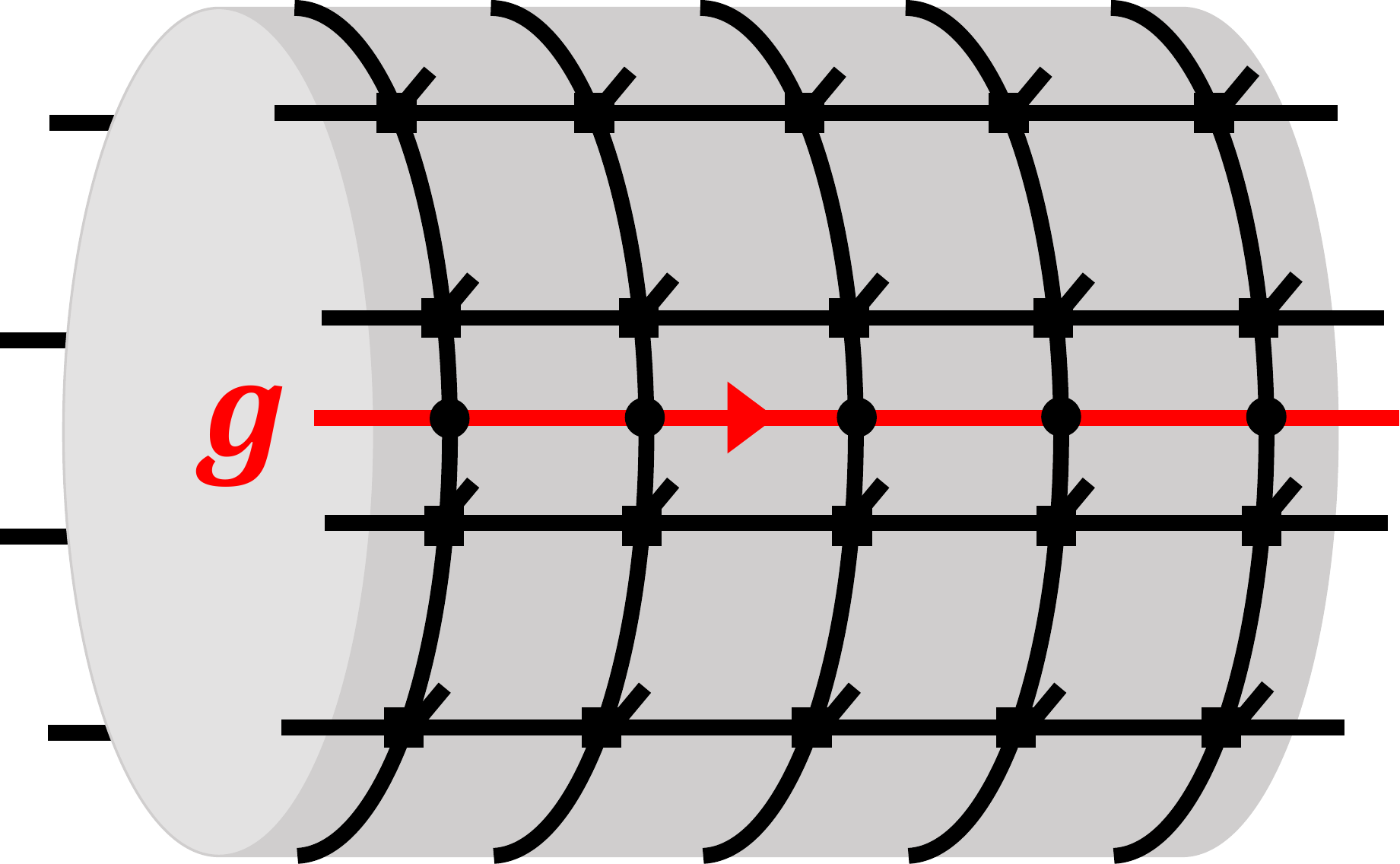}}} \quad b)
\vcenter{\hbox{
\includegraphics[width=0.4\linewidth]{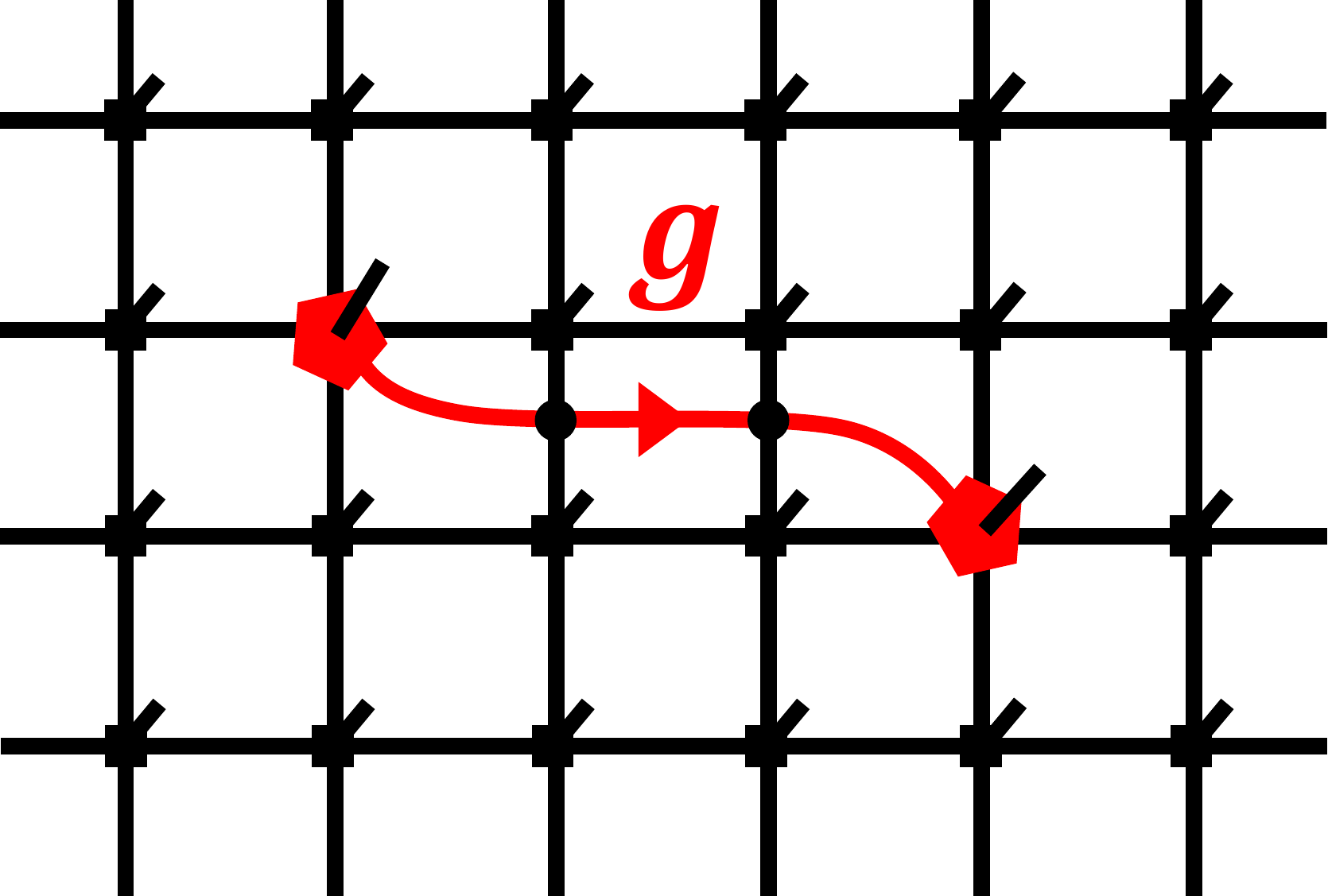}}} 
\end{align*}\caption{ a) A symmetry twisted SPT PEPS on a cylinder. b) A pair of monodromy defects on a sphere.}\label{a1}
\end{figure}

We consider a SPT PEPS on an oriented manifold $\man$ with a twice punctured sphere topology and a symmetry twist $\epath_g$ running from one puncture $\puncture_0$ to the other $\puncture_1$. This captures both the case of a symmetry twisted SPT model defined on a cylinder (when the virtual bonds that enter the punctures are left open), and the case of a pair of monodromy defects on a sphere (when the punctures are formed by removing a pair of PEPS tensors $\peps_{v_0},\peps_{v_1}$ and contracting the virtual indices thus opened with $B_0,B_1$), see Fig.~\ref{a1}. 

The bulk of the symmetry twisted state is invariant under the physical on-site representation $U(h)^{\otimes |\man|_v}$ of $\mathsf{C}(g)\leq\mathsf{G}$, but this may have some action on the virtual indices that enter the punctures. Treating the SPT PEPS on a cylinder of fixed radius as a one dimensional symmetric MPS implies, by well established arguments~\cite{SchuchGarciaCirac11, 1Done}, that the action of the symmetry on the virtual boundaries $\bdry_0^g(h)\otimes \bdry_1^g(h)$ forms a representation, while each individual boundary action $\bdry_0^g(h), \bdry_1^g(h)$ is free to form a projective representation. 

Assuming the symmetry MPOs satisfy the $\emph{zipper}$ condition~\eqref{a3} one can directly calculate the effect that a physical symmetry action $U(h)^{\otimes |\man|_v},\, h\in \mathsf{C}(g)$ has on the virtual boundary, simultaneously demonstrating the symmetry invariance of the bulk.
\begin{align}
\vcenter{\hbox{
\includegraphics[height=0.21\linewidth]{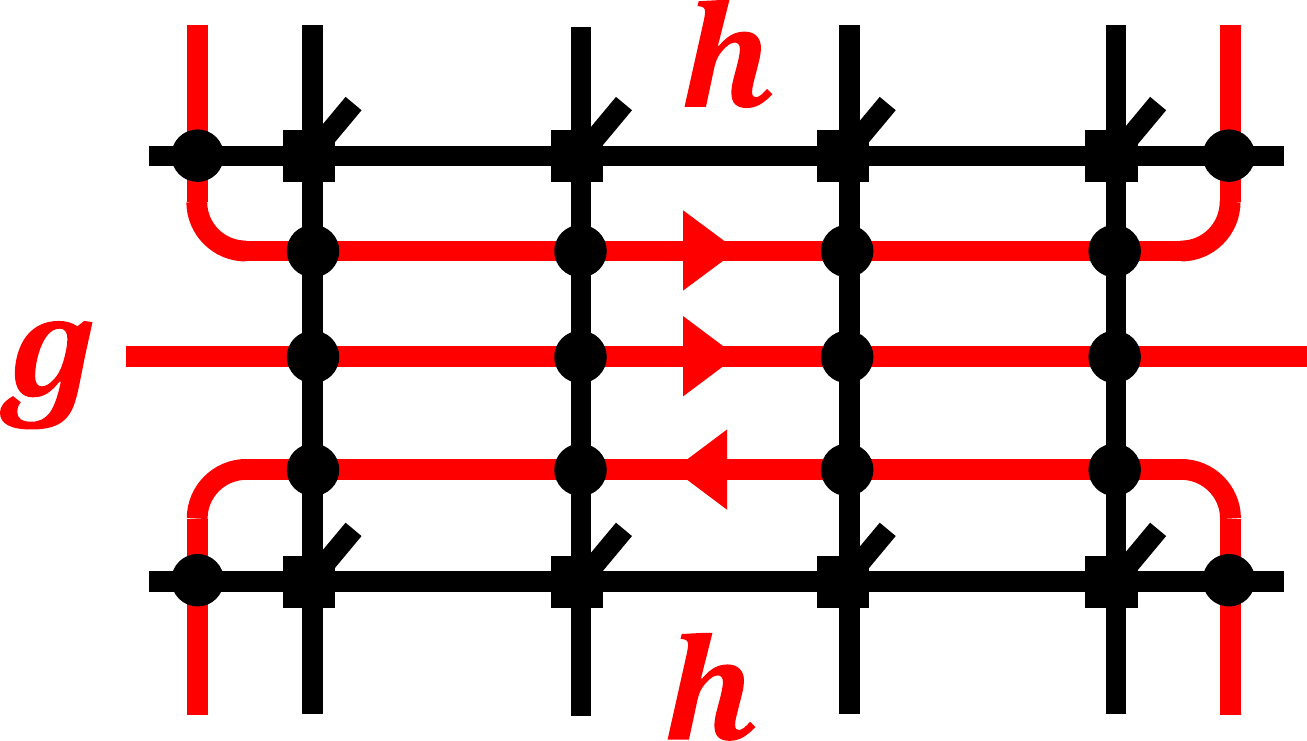}}}
\ = \vcenter{\hbox{
\includegraphics[height=0.21\linewidth]{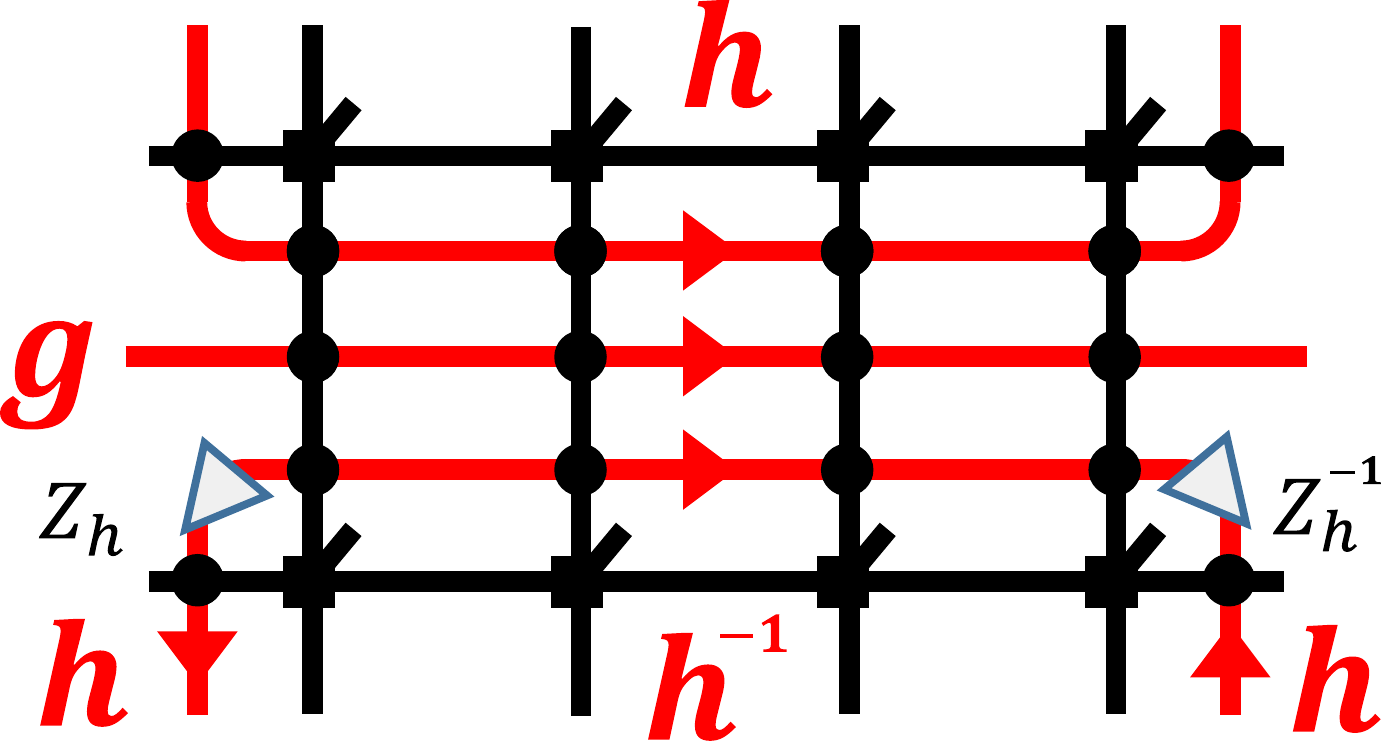}}}
\\ \label{a4}
 = \vcenter{\hbox{
\includegraphics[height=0.205\linewidth]{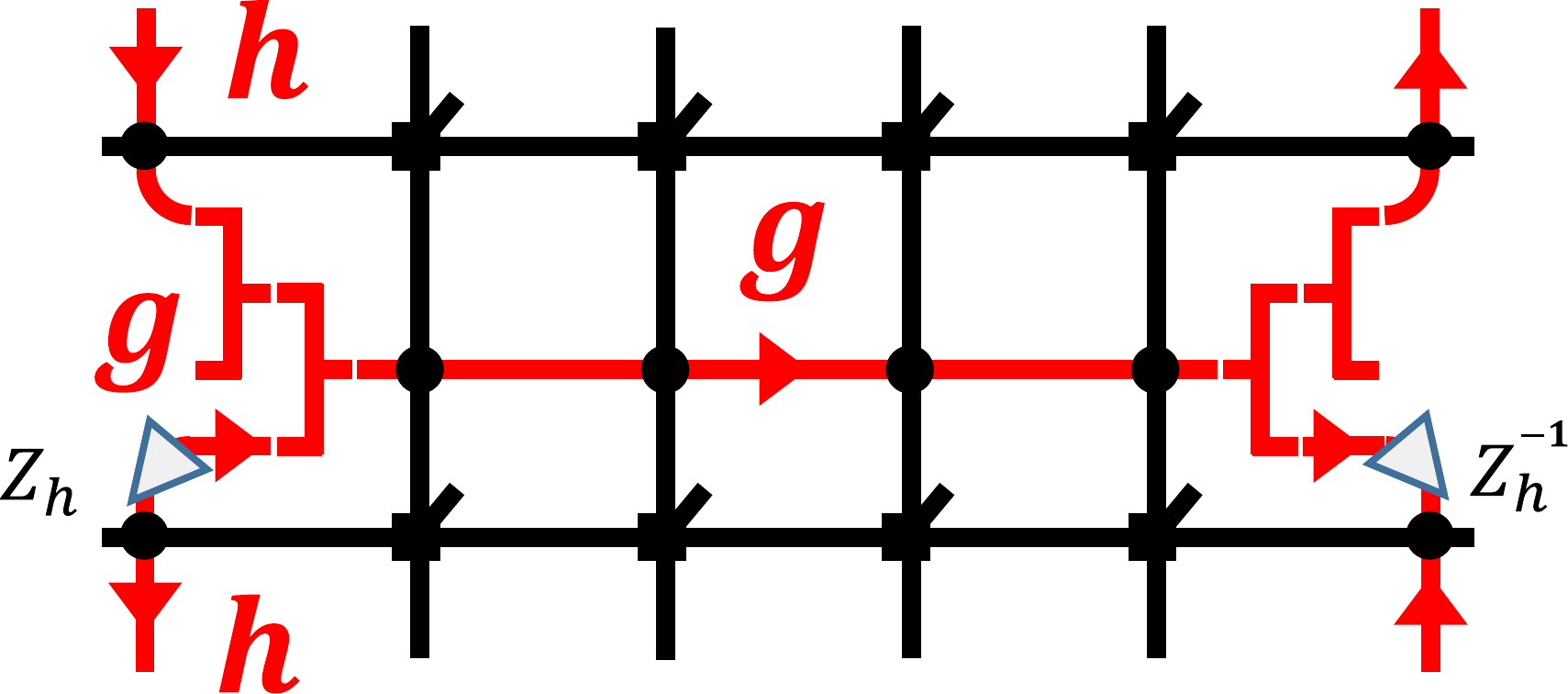}}}
\end{align}
Hence the symmetry action $\bdry^g_1(h)$ (see Eq.\eqref{e8}) on the boundary of a single puncture $\puncture_1$ is given by the MPO $V^{\partial \puncture_1^{-}}(h)$, acting on the virtual indices along $\partial \puncture_1$, contracted with the crossing tensor $\QV{R}{g}{h}$ (see Eq.\eqref{na21c}) acting on the inner MPO index of the symmetry twist $V^{p_g}(g)$ that enters the puncture. Similarly $\bdry^g_0(h)$, acting on the boundary of the other puncture $\puncture_0$, is given by contracting the MPO $V^{\partial \puncture_0^{-}}(h)$ with the crossing tensor $\QV{L}{g}{h}$.

There is a natural composition operation on the crossing tensors $\QV{R}{g}{\cdot}$ that is induced by applying a product of global symmetries $U(k)^{\otimes |\man|_v}U(h)^{\otimes |\man|_v}$ and utilizing the reduction of Eq.\eqref{a4} twice and then zipping the MPOs $V^{\partial \puncture_1^{-}}(k)V^{\partial \puncture_1^{-}}(h)=X(k,h)V^{\partial \puncture_1^{-}}(kh)X^+(k,h)$ by Eq.\eqref{a3}. This is nothing but the product ${\QV{R}{g}{k}\times \QV{R}{g}{h}}$ that was previously defined in Eq.\eqref{na23}.

Since the physical action $U(h)^{\otimes |\man|_v}$ forms a representation of the symmetry group $\mathsf{C}(g)$ the simultaneous virtual action on both boundaries $\puncture_0,\puncture_1$ together $\bdry_0^g(h)\otimes \bdry_1^g(h)$ must also form a representation. However, there is a multiplicative freedom in the multiplication rule of the representation on a single boundary 
$$\bdry^g_1(k)\bdry^g_1(h)=\omega^g_1(k,h)\bdry^g_1(kh)$$ 
(and similarly for $\bdry^g_0(h)$), under the constraint $\omega^g_0(k,h)\omega^g_1(k,h)=1$, allowing the possibility of projective representations. 

Using the result of Eq.\eqref{na25}, ${\QV{R}{g}{k}\times \QV{R}{g}{h}=\omega^g(k,h) \QV{R}{g}{kh}}$, we can pin down the 2-cocycle $\omega^g_1$ explicitly in terms of the 3-cocycle $\alpha$ of the injective MPO representation $V^{\partial \puncture_0^{-}}(\cdot)$ as follows
$\omega^g_1(k,h)=\omega^g(k,h)$ 
(see Eq.\eqref{omegag} for definition of $\omega^g$). 
Hence the cohomology class of the projective representation $\bdry^g_1(\cdot)$ is given by
$$[\omega^g_1(k,h)]=[\slant{\alpha}{1}{g}{k,h}].$$

\begin{figure}[ht]
\center
\begin{align*}
&\vcenter{\hbox{
\includegraphics[width=0.35\linewidth]{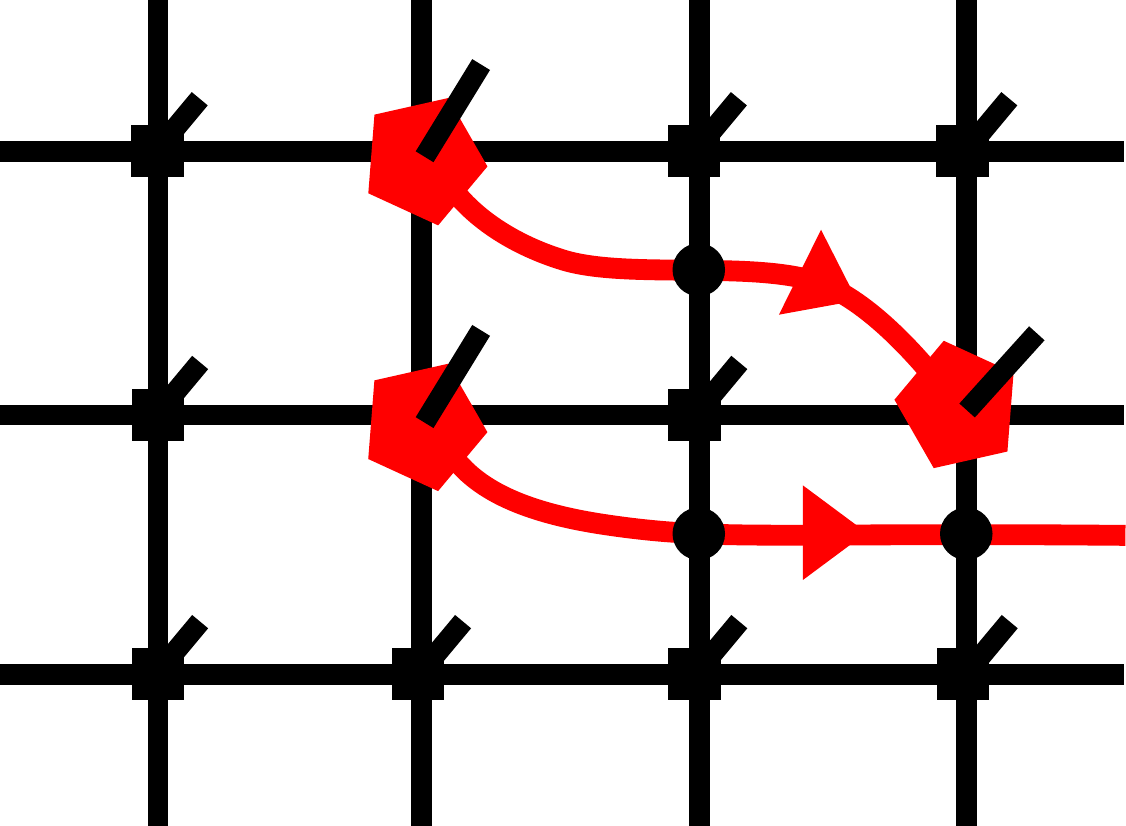}}}   \ \rightarrow
\vcenter{\hbox{
\includegraphics[width=0.35\linewidth]{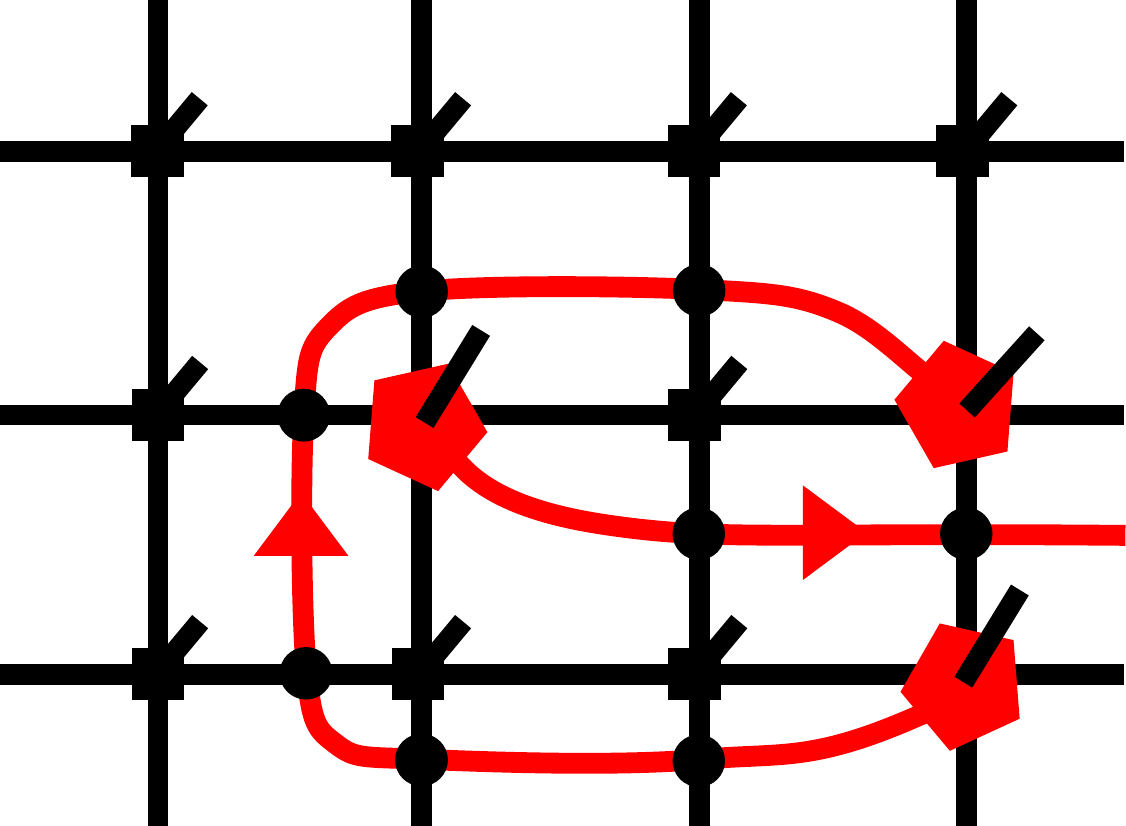}}}  
\\
&\hspace{5.2cm}\downarrow
\\
&\vcenter{\hbox{
\includegraphics[width=0.35\linewidth]{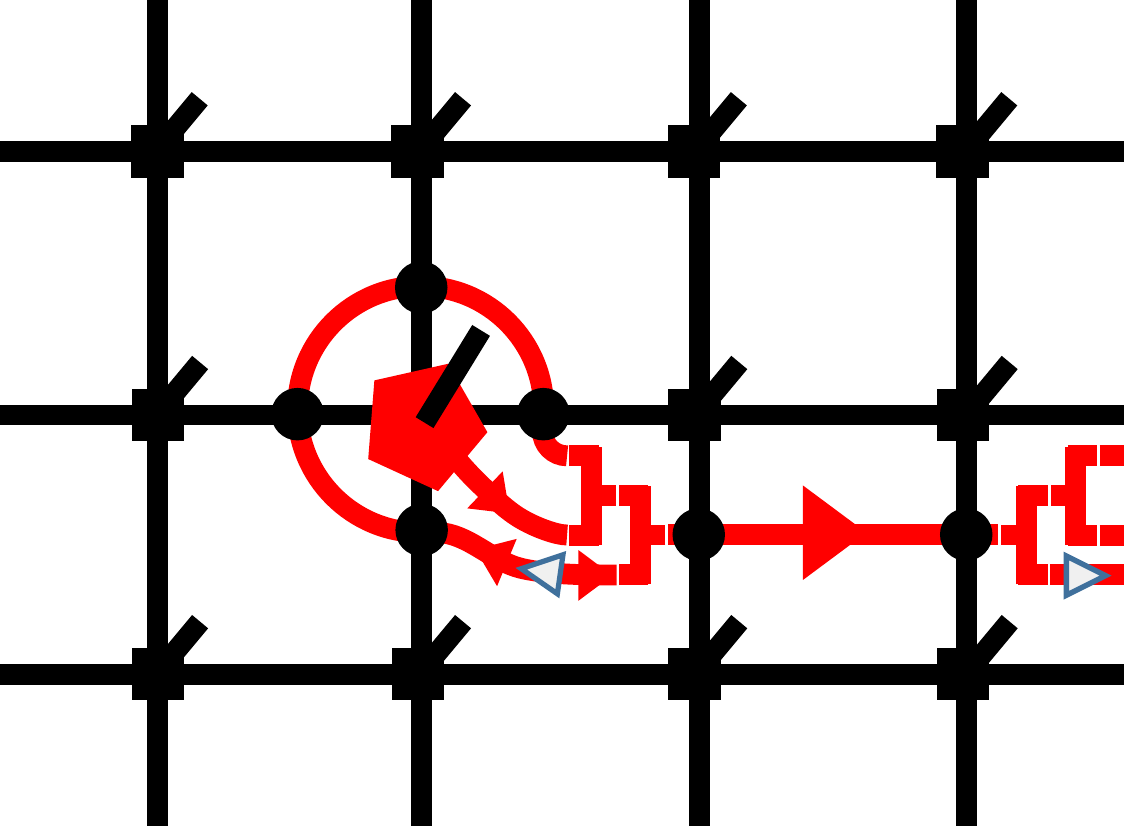}}}  \ =\,
\vcenter{\hbox{
\includegraphics[width=0.35\linewidth]{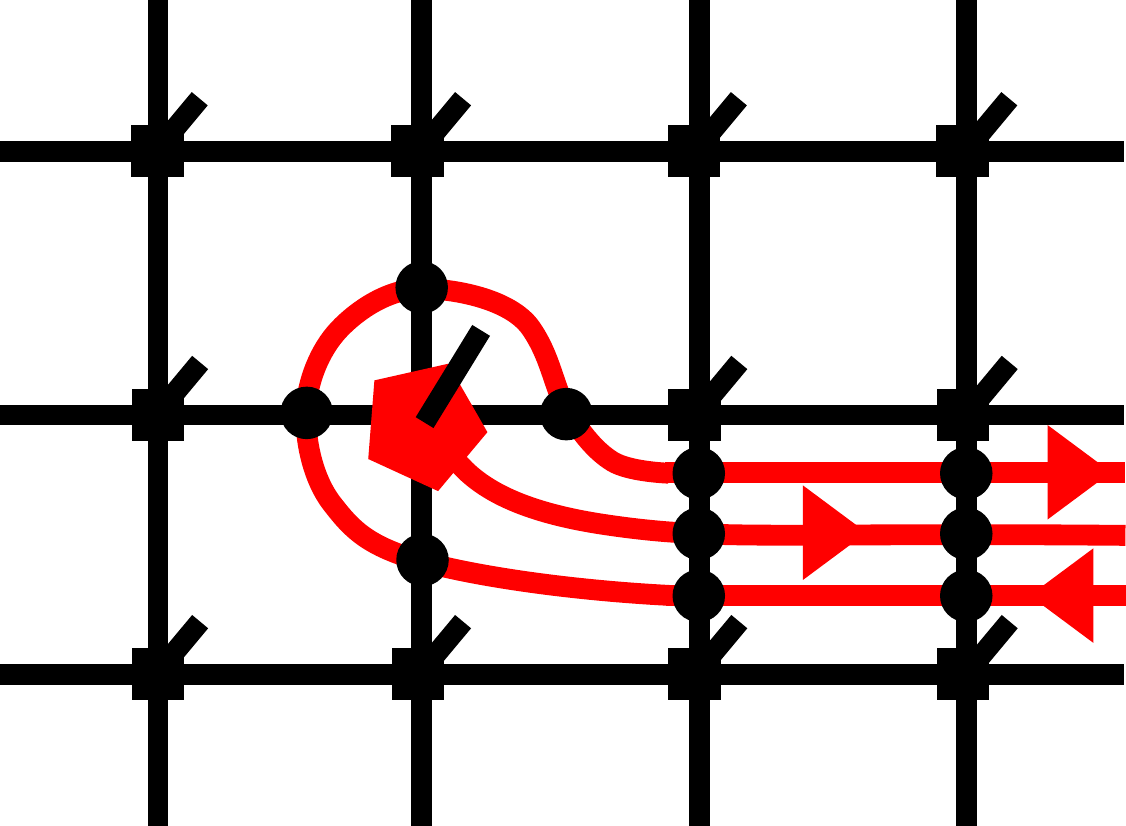}}} 
\end{align*}
\caption{The process used to find the effect of braiding on the internal degrees of freedom of a single anyon.}\label{a14}
\end{figure}

It was shown above that a gauged SPT PEPS with a pair of defects $G_{\con^{\epath_g}}\ket{\psi^{\epath_g},B_0,B_1}$ describes a superposition of anyon pairs in the resulting topological theory. 
The projective transformation of the monodromy defects is intimately related to the braiding of the resulting anyons, which can be inferred from the following process, depicted in Fig.~\ref{a14}. First consider an isolated anyon formed by creating a pair of anyons and then moving the other arbitrarily far away. Next create a second pair and move them to encircle the isolated anyon, at this point one should fuse these anyons, but the full description of such fusion requires a systematic anyon ansatz which is beyond the scope of the current paper (see Ref.\cite{nick}). Instead we drag the pair arbitrarily far away as demonstrated in Fig.~\ref{a14} and use the fact that this can be rewritten as some local action on the internal degrees of freedom of the isolated anyon, plus another locally undetectable action that can be moved arbitrarily far away.

\section{Gauging symmetric Hamiltonians and ground states} \label{gaugingham}
In this appendix we apply the gauging procedure developed in Ref.\cite{Gaugingpaper} to families of trivial and SPT Hamiltonians with symmetric perturbations and find that they are mapped to perturbed quantum double and twisted quantum double models respectively. We then go on to describe gauging the (unperturbed) fixed-point ground states.

\subsection{Gauging the Hamiltonian}

First we apply the gauging procedure to a symmetric Hamiltonian defined on the matter degrees of freedom, each with Hilbert space $\hilbert_v\cong \mathbb{C}[\mathsf{G}]$ and symmetry action $U_v(g)=R_v(g)$, associated to the vertices of a directed graph $\fullgraph$ embedded in a closed oriented 2-manifold $\man$. The Hamiltonian is given by
\begin{equation}
H_\mathsf{m}=\alpha \sum_{v\in\fullgraph} h_v^{0} + \sum_{m\in\mathsf{G}} \beta_m \sum_{e\in\fullgraph} \mathcal{E}_e^m
\end{equation}
the vertex terms are $ h_v^{0}:=\int \mathrm{d} \hat{g}_v \mathrm{d} g_v \ket{\hat{g}_v}\bra{g_v}$ while the edge interaction terms are $\mathcal{E}_e^m:=\int \mathrm{d} g_{v_e^-} \mathrm{d} g_{v_e^+}\,  \delta_{g_{v_e^-}g_{v_e^+}^{-1},m}\, | g_{v_e^-} \rangle \langle g_{v_e^-} | \otimes | g_{v_e^+} \rangle \langle g_{v_e^+}| $.
Each term in this Hamiltonian is symmetric under the group action $\bigotimes_v R_v(g)$. For $\alpha,\beta_m<0$ and $|\alpha|\gg |\beta_m|$ this Hamiltonian describes a symmetric phase with trivial SPT order, while for $|\beta_m| \gg |\alpha|$ the Hamiltonian describes different symmetry broken phases.

We construct the gauge and matter Hamiltonian $H_\mathsf{g,m}$ by first gauging the local terms $h_v^0$, which leaves them invariant $\gaugeoperator_{v}[h_v^0]=h_v^0$. Next we gauge the interaction terms $\mathcal{E}_e^m$ with the gauging map on $\bar{e}$ (the closure of edge $e$) 
\begin{align}
&\gaugeoperator_{\bar{e}}[\mathcal{E}_e^m]=\int \mathrm{d} g_{v_e^-} \mathrm{d} g_{v_e^+} \mathrm{d} h_{v_e^-} \mathrm{d} h_{v_e^+}\  \delta_{g_{v_e^-}g_{v_e^+}^{-1},m} \,  | g_{v_e^-}h^{-1}_{v_e^-} \rangle 
\nonumber \\
&\langle g_{v_e^-}h^{-1}_{v_e^-} | \otimes | h_{v_e^-}h^{-1}_{v_e^+} \rangle \langle h_{v_e^-}h^{-1}_{v_e^+} |_e 
\otimes | g_{v_e^+}h^{-1}_{v_e^+} \rangle \langle g_{v_e^+}h^{-1}_{v_e^+} | .
\nonumber
\end{align}

Finally we consider additional local gauge invariant Hamiltonian terms acting purely on the gauge degrees of freedom: symmetric local fields 
\begin{equation}
\mathcal{F}^{\cc}_e:=\int \mathrm{d} \hat{g}_e \mathrm{d} g_e \, \delta_{\hat{g}_eg_e^{-1}\in \cc} \, \ket{\hat{g}_e}\bra{g_e}
\end{equation}
 where $\cc \in \conj{\mathsf{G}}$ are conjugacy classes of $\mathsf{G}$, and plaquette flux-constraints 
\begin{equation}\label{bp}
\mathcal{B}_p^m:=\int \prod_{e\in \partial p} \mathrm{d} g_e\ \delta_{g^{\sigma^{e_1}}_{e_1}\cdots g^{\sigma^{e_{|\partial p|}}}_{e_{|\partial p|}},m} \, \bigotimes_{e\in \partial p} | g_e \rangle\langle g_e |\, ,
\end{equation}
where each plaquette $p$ has a fixed orientation induced by the 2-manifold $\man$, and the group elements $\{ g_{e_1},\dots,g_{e_{|\partial p|}}\}$ are ordered as the edges are visited starting from the smallest vertex label and moving against the orientation of $\partial p$, then $\sigma^{e_i}=\pm1$ is $+1$ if the edge $e_i$ points in the same direction as the orientation of $p$ and ${-}1$ otherwise. Finally we require that the group elements lie in the center of the group $m\in \mathsf{C}(\mathsf{G})$ which renders the choice of vertex from which we begin our traversal of $\partial p$ irrelevant.

The full gauge and matter Hamiltonian is thus given by 
\begin{align}
H_\mathsf{g,m}=\alpha \sum_{v\in\fullgraph} & h_v^0 + \sum_{m\in\mathsf{G}} \beta_m \sum_{e\in\fullgraph} \gaugeoperator_{\bar{e}}[\mathcal{E}_e^m] + \sum_{\cc\in\conj{\mathsf{G}}} \gamma_\cc \sum_{e\in\fullgraph} \mathcal{F}_e^\cc 
\nonumber \\
&+ \sum_{m\in \mathsf{C}(\mathsf{G})} \varepsilon_m \sum_{p\in\fullgraph} \mathcal{B}_p^m\, .
\end{align}
Note that each term commutes with all local gauge constraints $\{P_v\}$, see Eq.\eqref{constraint}, and the physics takes place within this gauge invariant subspace. 

\subsection{Disentangling the constraints}
To see more clearly that this gauge theory is equivalent to an unconstrained quantum double model we will apply a local disentangling circuit to reveal a clear tensor product structure,  allowing us to `spend' the gauge constraints to remove the matter degrees of freedom.

We define the disentangling circuit to be the product of local unitaries $C_\fullgraph:=\prod_v C_v$, where $C_v:=\int \mathrm{d} g_v \ket{g_v} \bra{g_v}_v \prod_{e\in E_v^+} R_e(g) \prod_{e\in E_v^-} L_e(g)$. Note the order in the product is irrelevant since $[C_v,C_{v'}]=0$. 
This circuit induces the following transformation on the gauge projectors: $C_\fullgraph P_v C_\fullgraph^\dagger=\int \mathrm{d} g R_v(g)$, hence any state $\ket{\psi}$ in the gauge invariant subspace (simultaneous $+1$ eigenspace of all $P_v$) is disentangled into a tensor product of symmetric states on all matter degrees of freedom with an unconstrained state $\ket{\psi'}\in \hilbert_\mathsf{g}$ on the gauge degrees of freedom $C_\fullgraph \ket{\psi} = \ket{\psi'} \bigotimes_v \int \mathrm{d} g_v  \ket{g_v}$.

Now we apply the disentangling circuit to the Hamiltonian $H_\mathsf{g,m}$. First note the pure gauge terms $\mathcal{F}_e^\cc$ and $\mathcal{B}_p^m$ are invariant under conjugation by $C_v$. The vertex terms are mapped to $C_v h_v^0 C_v^\dagger = \int \mathrm{d} g_v R_v(g_v) \bigotimes_{e\in E_v^+} R_e(g_v) \bigotimes_{e\in E_v^-} L_e(g_v)$. Since the disentangled vertex degrees of freedom are invariant under $R_v(g_v)$ we see that this Hamiltonian term acts as $\int \mathrm{d} g_v \bigotimes_{e\in E_v^+} R_e(g_v) \bigotimes_{e\in E_v^-} L_e(g_v)$ on the relevant gauge degrees of freedom. We recognize this as the vertex term of a quantum double model. Finally we examine the transformation of the interaction terms
$C_v \gaugeoperator_{\bar{e}}[\mathcal{E}_e^m] C_v^\dagger = \frac{1}{|\mathsf{G}|}\ket{m}\bra{ m}_e$,
which yield local fields on the gauge degrees of freedom that induce string tension.
Hence we see that the gauge plus matter Hamiltonian after disentangling becomes a local Hamiltonian $H_\mathsf{g}:=C_v H_\mathsf{g,m} C_v^\dagger$ acting purely on the gauge degrees of freedom
\begin{align}
& H_\mathsf{g} = \alpha \sum_{v} \int \mathrm{d} g_v \bigotimes_{e\in E_v^+} R_e(g_v) \bigotimes_{e\in E_v^-} L_e(g_v) + \sum_{m\in\mathsf{G}} \frac{\beta_m }{|\mathsf{G}|} 
\nonumber \\
&\times \sum_{e\in\fullgraph} \ket{m}\bra{ m}_e + \sum_{\cc\in\conj{\mathsf{G}}}\gamma_\cc \sum_{e\in\fullgraph} \mathcal{F}_e^\cc + \sum_{m\in \mathsf{C}(\mathsf{G})} \varepsilon_m \sum_{p\in\fullgraph} \mathcal{B}_p^m
\end{align}
which describes a quantum double model with string tension and flux perturbations. Note that a spontaneous symmetry breaking phase transition in the ungauged model is mapped to a string tension induced anyon condensation transition by the gauging procedure.

\subsection{Gauging nontrivial SPT Hamiltonians}

The gauging procedure extends to nontrivial SPT Hamiltonians which are defined on triangular graphs embedded in closed oriented 2-manifolds $\man$. The only modification required is to replace the trivial vertex terms $h_v^0$ by nontrivial terms $h_v^\alpha$ which are defined by
\begin{equation}\label{hvalpha1}
\int \mathrm{d} \hat{g}_v \mathrm{d} g_v \prod_{v' \in L(v)}\mathrm{d} g_{v'} \prod_{\triangle \in S(v)} \alpha_{\triangle} \, \ket{\hat{g}_v}\bra{g_v} \bigotimes_{v'\in L(v)} \ket{g_{v'}}\bra{g_{v'}} 
\end{equation}
where $S(v)$ is the star of $v$, $L(v)$ is the link of $v$ and $\alpha_\triangle\in\mathsf{U}(1)$ for plaquette $\triangle$, whose vertices are given counterclockwise (relative to the orientation of the 2-manifold) by $v,\ v',\ v'',$ is defined by the 3-cocycle $\alpha_{\triangle}:=\alpha^{\sigma_\pi}(g_1g_2^{-1},g_2g_3^{-1},g_3g_4^{-1})$ where $(g_1,g_2,g_3,g_4):=\pi (\hat{g}_v,g_v,g_{v'},g_{v''})$ for $\pi$ the permutation that sorts the group elements into ascending vertex label order (with the convention that $\hat{g}_v$ immediately precedes $g_v$) and $\sigma_\pi=\pm 1$ is the parity of the permutation.
The terms $h_v^{\alpha}$ are clearly symmetric under global right group multiplication and are seen to be Hermitian since conjugation inverts the phase factor $\alpha_\triangle$ and interchanges the role of $\hat{g}_v$ and $g_v$ which inverts the parity of $\pi$ thereby compensating the conjugation of $\alpha_\triangle$.

We apply the gauging map on the region $\bar{S}(v)$ (the closure of the star of $v$) to $h_v^\alpha$
\begin{align}
\gaugeoperator_{\bar{S}(v)}[h_v^\alpha] 
=& \int  \mathrm{d} \hat{g}_v \mathrm{d} g_v \mathrm{d} h_v \prod_{v' \in L(v)}\mathrm{d} g_{v'} \mathrm{d} h_{v'} \prod_{\triangle \in S(v)} \alpha_{\triangle}
\nonumber \\
\phantom{\gaugeoperator_{\bar{S}(v)}[h_v^\alpha] =} 
&\ket{\hat{g}_v h_v^{-1}} \bra{g_v h_v^{-1}} \bigotimes_{v'\in L(v)} \ket{g_{v'}h_{v'}^{-1}}\bra{g_{v'}h_{v'}^{-1}} 
\nonumber \\
 \phantom{\gaugeoperator_{\bar{S}(v)}[h_v^\alpha] =}
&\bigotimes_{e\in \bar{S}(v)} | {h_{v_e^-}h^{-1}_{v_e^+}} \rangle \langle {h_{v_e^-}h^{-1}_{v_e^+}} |_e
\end{align}
followed by the disentangling circuit $C_\fullgraph \gaugeoperator_{\bar{S}(v)}[h_v^\alpha] C_\fullgraph^\dagger$ which yields
\begin{align}
& \int  \mathrm{d} \hat{g}_v \mathrm{d} g_v \mathrm{d} h_v \prod_{v' \in L(v)}\mathrm{d} g_{v'} \mathrm{d} h_{v'} \prod_{\triangle \in S(v)} \alpha_{\triangle} \, \ket{\hat{g}_v h_v^{-1}}\bra{g_v h_v^{-1}}
\nonumber \\
& \bigotimes_{v'\in L(v)} \ket{g_{v'}h_{v'}^{-1}}\bra{g_{v'}h_{v'}^{-1}} \bigotimes_{e\in E_v^+} | {g_{v_e^-}\hat{g}^{-1}_{v}} \rangle \langle {g_{v_e^-}g^{-1}_{v}} | 
\nonumber \\
&  \ \bigotimes_{e\in E_v^-} | {\hat{g}_{v}g^{-1}_{v_e^+}} \rangle \langle {g_{v}g^{-1}_{v_e^+}} |
\bigotimes_{e\in L(v)} | {g_{v_e^-}g^{-1}_{v_e^+}} \rangle \langle {g_{v_e^-}g^{-1}_{v_e^+}}| .
\end{align}
Note, importantly, the phase functions $\alpha_\triangle$ now depend only on the gauge degrees of freedom. Finally in Eq.\eqref{appehv} we rewrite the pure gauge Hamiltonian terms without reference to the matter degrees of freedom, which become irrelevant as the matter degrees of freedom in any gauge invariant state are fixed to be in the symmetric state $\int \mathrm{d} g_v \ket{g_v}_v$ by the disentangling circuit
\begin{align}
& \int  \mathrm{d} \hat{g}_v \mathrm{d} g_v  \prod_{v' \in L(v)}\mathrm{d} g_{v'} \prod_{\triangle \in S(v)} \alpha_{\triangle}  \bigotimes_{e\in E_v^+} | {g_{v_e^-}\hat{g}^{-1}_{v}} \rangle \langle {g_{v_e^-}g^{-1}_{v}} |
\nonumber \\
& \phantom{=\int} \bigotimes_{e\in E_v^-} | {\hat{g}_{v}g^{-1}_{v_e^+}} \rangle \langle {g_{v}g^{-1}_{v_e^+}} |
\bigotimes_{e\in L(v)} | {g_{v_e^-}g^{-1}_{v_e^+}} \rangle \langle {g_{v_e^-}g^{-1}_{v_e^+}}| 
\label{appehv}
\end{align}
This can be recognized as the vertex term of a 2D twisted quantum double model (the lattice hamiltonian version of a twisted Dijkgraaf Witten theory for the group $\mathsf{G}$ and cocycle $\alpha$).

\subsection{Gauging SPT groundstates}
In this section we apply the gauging procedure directly to the ground states of the nontrivial SPT Hamiltonian that was defined in Eq.\eqref{hvalpha1}. These ground states are constructed using the following local circuit~\cite{else2014classifying}
\begin{equation}\label{dalpha}
D_\alpha:=\int \prod_{v\in\fullgraph}\mathrm{d} g_v \prod_{\triangle\in \fullgraph} \tilde{\alpha}_{\triangle} \bigotimes_{v\in\fullgraph} \ket{g_v}\bra{g_v}
\end{equation}
where $\tilde{\alpha}_\triangle\in\mathsf{U}(1)$ is a function of the degrees of freedom on plaquette $\triangle$, whose vertices are given counterclockwise, relative to the orientation of the 2-manifold $\man$, by $v,\ v',\ v''$ (note the choice of starting vertex is irrelevant) and is defined by a 3-cocycle as follows $\tilde \alpha_{\triangle}:=\alpha^{\sigma_\pi}(g_1g_2^{-1},g_2g_3^{-1},g_3)$ where $(g_1,g_2,g_3):=\pi (g_v,g_{v'},g_{v''})$ with $\pi$ the permutation that sorts the group elements into ascending vertex label order and $\sigma_\pi=\pm 1$ is the parity of the permutation (equivalently the orientation of $\triangle$ embedded within the 2-manifold $\man$).
Note $D_\alpha$ is easily expressed as a product of commuting 3-local gates.

To define SPT fixed-point states we start with the trivial state $\spt{0}:=\bigotimes_v \int \mathrm{d} g_v \ket{g_v}_v$, which is easily seen to be symmetric under global right group multiplication. One can also check that $D_\alpha$ is symmetric under conjugation by global right group multiplication by utilizing the 3-cocycle condition satisfied by each $\tilde{\alpha}_\triangle$. With this we define nontrivial SPT fixed-point states $\spt{\alpha}:=D_\alpha \spt{0}$, which are symmetric by construction. To see that $\spt{\alpha}$ is the ground state of the SPT Hamiltonian $\sum_v h_v^{\alpha}$ we note $h_v^\alpha=D_\alpha h_v^0 D_\alpha^\dagger$ which again is proved using the 3-cocycle condition. 

We will now gauge the SPT fixed-point states by applying the state gauging map to $D_\alpha$, since the input variables of the circuit carry the same information as the virtual indices of the fixed-point SPT PEPS we hope this makes the correspondence between the two pictures more clear
\begin{align}
\gauge D_\alpha = \int \prod_{v\in\fullgraph}\mathrm{d} g_v \mathrm{d} h_v \prod_{\triangle\in \fullgraph} \tilde{\alpha}_{\triangle} \, \bigotimes_{e\in \fullgraph} | h_{v_e^-}h^{-1}_{v_e^+} \rangle  
\nonumber \\
\bigotimes_{v\in\fullgraph} \ket{g_v h_v^{-1}}\bra{g_v} .
\end{align}
Under the local disentangling circuit this transforms to
\begin{align}
C_\fullgraph \gauge D_\alpha =\int \prod_{v\in\fullgraph}\mathrm{d} g_v \prod_{\triangle\in \fullgraph} \tilde{\alpha}_{\triangle} \, \bigotimes_{e\in\fullgraph} & | g_{v_e^-}g^{-1}_{v_e^+} \rangle \bigotimes_{v\in\fullgraph}\bra{g_v} 
\nonumber \\ \label{gaugeda}
&\bigotimes_{v\in\fullgraph} \int  \mathrm{d} k_v \ket{k_v}  ,
\end{align}
where it is clear that the matter degrees of freedom have been disentangled into symmetric states and the cocycles $\tilde{\alpha}_{\triangle}$ depend on both the group variables on the edges and the inputs on the vertices (which correspond to the PEPS virtual degrees of freedom).

The explicit connection to the fixed-point SPT PEPS is made by replacing the basis at each vertex $\ket{g_v}_v$ by an analogous basis of the diagonal subspace of variables at each plaquette surrounding the vertex $\bigotimes_{\triangle \in S(v)} \ket{g_v}_{\triangle,v}$. This  construction lends itself directly to a PEPS description where a tensor is assigned to each plaquette of the original graph (i.e. the PEPS is constructed on the dual graph). This in turn is why we must apply a seemingly modified version of the gauging operator of Ref.\cite{Gaugingpaper} to gauge the PEPS correctly and we note that on the subspace where redundant variables are identified the modified PEPS gauging operator becomes identical to the standard gauging operator.

\end{document}